\documentclass{andp2012}
\usepackage[english]{babel}

\allowdisplaybreaks

\usepackage{preamble}

\newcommand{\newlychanged}[1]{{#1}} 

\begin{abstract}

Conceptually, the Matsubara formalism (MF), using imaginary frequencies, and the Keldysh formalism (KF), formulated in real frequencies, give equivalent results for systems in thermal equilibrium. The MF has less complexity and is thus more convenient than the KF. However, computing dynamical observables in the MF requires the analytic continuation from imaginary to real frequencies. The analytic continuation is well-known for two-point correlation functions (having one frequency argument), but, for multipoint correlators, a straightforward recipe for deducing all Keldysh components from the MF correlator had not been formulated yet. Recently, a representation of MF and KF correlators in terms of formalism-independent partial spectral functions and formalism-specific kernels was introduced by Kugler, Lee, and von Delft
[Phys.~Rev.~X 11, 041006 (2021)]. We use this representation to formally elucidate the connection between both formalisms.
We show how a multipoint MF correlator can be analytically continued to recover all partial spectral functions and yield all Keldysh components of its KF counterpart. The procedure is illustrated for various correlators of the Hubbard atom.
\end{abstract}
\shortabstract

\begin{document}
	
\title{Analytic continuation of multipoint correlation functions}

\author{Anxiang Ge*\textsuperscript{1$\star$}}
\author{Johannes Halbinger \textsuperscript{1$\star$}}
\author{Seung-Sup B.~Lee \textsuperscript{2,3}}
\author{Jan von Delft \textsuperscript{1}}
\author{Fabian B.~Kugler \textsuperscript{4,5}}

\maketitle

\begin{affiliations}
    \textsuperscript{1} Arnold Sommerfeld Center for Theoretical Physics, Center for NanoScience, and Munich Center for Quantum Science and Technology, Ludwig-Maximilians-Universit\"at M\"unchen, 80333 Munich, Germany

    \textsuperscript{2} Department of Physics and Astronomy, Seoul National University, Seoul 08826, Korea

    \textsuperscript{3} Center for Theoretical Physics, Seoul National University, Seoul 08826, Korea
    
    \textsuperscript{4} Center for Computational Quantum Physics, Flatiron Institute, 162 5th Avenue, New York, NY 10010, USA

    \textsuperscript{5} Department of Physics and Astronomy and Center for Materials Theory, Rutgers University, Piscataway, NJ 08854, USA

    \textsuperscript{$\star$} \newlychanged{These authors contributed equally.}
    
\end{affiliations}


\section{Introduction}

Multipoint correlation functions, or correlators for short, are central objects of investigation in many-body physics.
The fermionic one-particle or two-point (2p) correlator describes the propagation of a single particle, containing information on the spectrum of single-particle excitations.
The two-particle or four-point ($4$p) correlator is associated with the effective interaction between two particles.
Interesting observables, like optical and magnetic response functions, can be deduced from it.
Additionally, the closely related 4p vertex, obtained by amputating all four external legs, is an essential ingredient in numerous many-body methods such as the functional renormalization group \cite{Metzner2012}, the parquet formalism \cite{Bickers2004}, and diagrammatic extensions of dynamical mean field theory \cite{Rohringer2018}.

The most common framework for studying systems in thermal equilibrium at temperature $T=1/\beta$  is the imaginary-time Matsubara formalism (\MF/) \cite{Abrikosov1975}.
It exploits the cyclicity of the trace and the fact that the statistical weight of a thermal state for a Hamiltonian $\mH$, $e^{-\beta \mH}$, corresponds to a time-evolution  $e^{-\i\mH t}$ along the imaginary axis of the time argument.
After a so-called Wick rotation, $t\rightarrow -\i\tau$, the correlators are \newlychanged{well-defined on the interval $\tau \in [-\beta,\beta]$ and there satisfy  (anti)periodicity relations with period $\beta$. Correspondingly, they can be expressed through a Fourier series using a discrete set of imaginary frequencies, the so-called Matsubara frequencies, ensuring this (anti)periodicity.}
Due to this periodicity, the Fourier transform of a \MF/ correlator is a function  defined on a discrete set of imaginary frequencies, so-called Matsubara frequencies.
To obtain a correlator of real times or real frequencies, one has to ``unwind'' the Wick rotation by performing a suitable analytic continuation.
Numerically, however, the analytic continuation to real frequencies is a highly challenging problem \cite{Cuniberti2001,Gubernatis1991}.

The Keldysh formalism (\KF/) is another established framework \cite{kamenev2011field}.
Unlike the \MF/, it is not restricted to thermal equilibrium. 
It directly works with real times and frequencies, obviating the need for an analytic continuation.
However, this comes at the cost of an increased complexity: the \KF/ is formulated on a doubled time contour, and an $\ell$-point ($\ell$p) function has $2^\ell$ components \cite{Keldysh1964,Schwinger1961}. 
By contrast, every \MF/ correlator is just a single function.

In thermal equilibrium, both MF and KF must in principle yield identical results for exact computations of any physical observable---the two formalisms only differ in the computational route to arrive at the result. In practice, though, it may be useful to transition from one formalism to the other, in order to exploit advantages from one or the other. The connection between the MF and KF by means of analytic continuation is well known for 2p functions, which effectively depend on a single time or frequency argument, see e.g.\ Refs.\ \cite{Altland2010,Baym1961,Negele1998}. For higher-point functions, progress has been made by various authors: 
Eliashberg discussed the analytic continuation of a specific 4p correlator from the MF to real frequencies
\cite{Eliashberg1962}. 
Evans \cite{Evans1990} and Kobes \cite{Kobes1990,Kobes1991} 
studied the correspondence between both formalisms 
for 3p correlators in Refs.~\cite{Evans1990,Kobes1990,Kobes1991}. Evans then considered $\ell \ge 4$ multipoint correlators and showed that 
fully retarded and fully advanced Keldysh components can be obtained from analytic continuations of MF correlators \cite{Evans1992}. 
Weldon conducted a thorough analysis of real-frequency $\ell$p functions and proved that these KF components are in fact the only ones that can be identified with an analytically continued MF function \cite{Weldon2005a,Weldon2005b}.
Taylor extended Evans' results to arbitrary Keldysh components of the fermionic 4p correlator,
assuming the absence of 
so-called anomalous terms in the MF correlator \cite{Taylor1993}.
(Anomalous terms can arise if the Lehmann representation of a correlator involves vanishing eigenenergy differences and zero bosonic Matsubara frequencies.) Guerin derived analogous results from diagrammatic arguments \cite{Guerin1994a,Guerin1994b}.
 
In this paper, we solve the problem of analytic continuation of multipoint functions from the MF to the KF in full generality:  We develop a strategy for analytically continuing an arbitrary MF $\ell$p correlator $G$ (including anomalous terms) to all $2^\ell$ components of the corresponding KF correlator $G^{\bs{k}}$ as functionals of $G$, i.e.\ $G^{\bs{k}}= G^{\bs{k}}[G]$. We exemplify the procedure for the most relevant cases $\ell \in \{ 2,3,4 \}$.

Our strategy builds upon the spectral representation of general $\ell$p correlators introduced in Ref.~\cite{Kugler2021}. There, the computation of MF and KF correlators is split into two parts: the calculation of formalism-independent but system-dependent partial spectral functions (PSFs), and their subsequent convolution with formalism-dependent but system-independent kernels. 
The main message of the present paper is that individual PSFs can be retrieved from 
the MF correlator, demonstrating the direct link between both formalisms.

\bsubeq \label{eq:nutshell}
In a nutshell, both MF and KF correlators have spectral representations involving sums over permutations of their constituent operators of the form
\begin{alignat}{2}
G(\i \bsomega)  & = \sum_p G_p(\i \bsomega_p), & \quad
G_p (\mi\bsomega_p)  &= (K \ast S_p)(\mi \bsomega_p), 
\\
G^{\bs{k}}(\bsomega) & = \sum_p G^{\bs{k}_p}_p(\bsomega_p) 
, & \quad G^{\bs{k}_p}_p(\bsomega_p)  &= (K^{\bs{k}_p} \ast S_p)(\bsomega_p). \hspace{-1cm} \phantom{.}
\end{alignat} 
Here, the summands $G_p$ and $G_p^{\bs{k}_p}$ are real-frequency convolutions (denoted by $\ast$) of MF or KF kernels, $K$ or $K^{\bs{k}_p}$, with PSFs $S_p$. Importantly, the MF and KF correlators depend on the \textit{same} PSFs, $G = G[S_p]$ and $G^{\bs{k}} = G^{\bs{k}}[S_p]$. 
The key insight of this work is that the so-called regular part of the \textit{partial} MF correlator $G_p$, 
denoted $\tG_p$, can be expressed as an imaginary-frequency convolution (denoted by $\star$) of a kernel and the \textit{full} MF correlator:
\bal
\tG_p(\i\bsomega_p) = (\tK \ast S_p)(\i \bsomega_p) = (K \star G)(\i \bsomega_p)+\mcO(\tfrac{1}{\beta}).
\eal
\esubeq
(The $\mcO(\tfrac{1}{\beta})$ terms can be identified analytically and discarded.) From this, we can extract $S_p$ as a functional of $G$, 
thus inverting the relation $G[S_p] \rightarrow S_p[G]$. That enables
us to express KF through MF correlators, $G^{\bs{k}} = G^{\bs{k}}[G]$.

Our analysis not only provides relations between functions in the \MF/ and the \KF/, but also between different Keldysh components of the \KF/ correlator.
As an application of our general results, we derive a complete set of generalized fluctuation-dissipation relations (gFDRs) for 3p and 4p functions. 
These reproduce the results of Wang and Heinz \cite{Wang2002} for real fields and the generalization to fermionic ones \cite{Jakobs2010a}.
Moreover, we give a comprehensive discussion of the role of anomalous terms during analytic continuation and in gFDRs. Prior discussions of these topics have often neglected anomalous terms; indeed, their presence is acknowledged only in few works, \newlychanged{such as Refs.~\cite{Stevens1965,Kwok1969,Watzenboeck2022}.
As an example of their physical importance, we mention that Ref.~\cite{Watzenboeck2022} analyzed anomalous terms for the Mott--Hubbard metal-insulator transition in the Hubbard model using the dynamical mean-field theory and detected a degeneracy in the insulating regime by means of a finite anomalous term.} 

Conceptually, the Matsubara formalism (MF), using imaginary frequencies, and the Keldysh formalism (KF), formulated in real frequencies, give equivalent results for systems in thermal equilibrium. The MF has less complexity and is thus more convenient than the KF. However, computing dynamical observables in the MF requires the analytic continuation from imaginary to real frequencies. The analytic continuation is well-known for two-point correlation functions (having one frequency argument), but, for multipoint correlators, a straightforward recipe for deducing all Keldysh components from the MF correlator had not been formulated yet. Recently, a representation of MF and KF correlators in terms of formalism-independent partial spectral functions and formalism-specific kernels was introduced by Kugler, Lee, and von Delft.
Regarding the number of independent components in the KF, one observes a general trend, obeyed by the known results for $\ell\in\{2,3,4\}$:
Due to the doubled time contour, there are $2^\ell$ Keldysh components.
In the Keldysh basis, $2^\ell - 1$ of them are nonzero, and $\ell$ are fully retarded components.
Now, there are $2^{\ell-1}$ gFDRs ($2, 4, 8$ for $\ell=2,3,4$).
Thus, the number of independent Keldysh components is $2^{\ell-1} - 1$
($1, 3, 7$ for $\ell=2,3,4$).
It follows that, for $\ell \geq 4$, the fully retarded components do not suffice to encode the entire information of the Keldysh correlator.

The rest of the paper is organized as follows:  In \Sec{sec:spectralrep}, we summarize the most important points of the spectral representation of $\ell$p MF and KF correlators introduced in Ref. \cite{Kugler2021} (\Secs{sec:PSF}, \ref{sec:MF}, and \ref{sec:KF}) and then introduce our general recipe for the analytic continuation of arbitrary $\ell$p correlators (\Secs{sec:Bridge_between_MF_KF} and \ref{eq:Matsubarasums}). 
This recipe is applied to the 2p case in \Sec{sec:Analytic_cont_2p} and, after the investigation of analytic properties of regular $\ell$p MF correlators in \Sec{sec:analytic_regions_and_discontinuities}, also to the 3p and 4p cases in \Secs{sec:Analytic_cont_3p} and \ref{sec:Analytic_cont_4p}.
The results also lead to gFDRs between different Keldysh components of the KF correlator. 
In \Sec{sec:HA}, we perform explicit analytic continuations from MF to KF correlators for the Hubbard atom. The Hubbard atom is a good example for a system with anomalous contributions and, here, serves as a simple, exactly solvable model with just the right degree of complexity for illustrating our approach.
\SEC{sec:oguris_formula} presents another application of our continuation formulas, namely for the computation of vertex corrections to susceptibilities. We conclude in \Sec{sec:Conclusion}.

In App.~\ref{app:MFKernels} and \ref{app:PSF_discussion}, we give details on the MF kernels and PSFs used in calculations throughout the paper.
\APP{sec:App_3p_calculations} is devoted to detailed calculations concerning the analytic continuation of 3p correlators.
In \App{sec:4p_PCF_app}, we extend insights from 2p and 3p results to deduce the relation between 4p PSFs and analytically continued MF correlators.
The spectral representations of various useful combinations of analytically continued MF correlators and anomalous parts are presented in \App{app:add_spectral_rep}.
\APP{app:simplifying_KF_correlators} 
expresses the spectral representation of KF correlators in a form especially suited for deriving their connection to MF functions. 
In \App{sec:App_consistency_checks}, we check the consistency of our results for PSFs by using equilibrium properties. 
Finally, \App{sec:App_HA_simplifications} gives details about simplifications used for the analytic continuation of Hubbard atom correlators and includes full lists of the especially important fermionic 4p KF correlators.

\section{Spectral representations of Matsubara and Keldysh correlators}
\label{sec:spectralrep}

To make our presentation self-contained, we summarize the key elements of the conventions and results of Ref.~\cite{Kugler2021} for common notions (\Sec{sec:PSF}), the MF (\Sec{sec:MF}), and the KF (\Sec{sec:KF}). 
Table~\ref{tab:glossary_for_Gs} provides an overview of our symbols for correlators and their contributions.
Our general strategy for the analytic continuation from MF to KF correlators is described in Secs.~\ref{sec:Bridge_between_MF_KF} and \ref{eq:Matsubarasums}.

\subsection{Formalism-independent expectations values}
\label{sec:PSF}

Consider a tuple of $\ell$ operators $\bs{O}
=(\mO^1, \dots, \mO^\ell)$ at real times $\bs{t} = (t_1, \dots, t_\ell)$, obeying the Heisenberg time evolution $\mO^i(t_i) = e^{\i\mH t_i} \mO^i e^{-\i\mH t_i}$ for a given Hamiltonian $\mH$. 
$\bs{\mO}$ may include an even number of fermionic operators and any number of bosonic operators. 
Time-ordered products of such tuples, defined below, involve permuted tuples $\bs{\mO}_p = ( \mO^{\oli{1}}, \dots, \mO^{\oli{\ell}})$ and $\bs{t}_p = (t_{\oli{1}}, \dots, t_{\oli{\ell}})$, where 
$p = (\bonetol)$ denotes  the permutation of indices that replaces $i$ by $p(i)= \oli{i}$. 
If $\ell = 3$ and $p=(\oli{1}\oli{2}\oli{3})$ is chosen as $(312)$, e.g., then $\bs{t}_p = (t_{\oli{1}}, t_{\oli{2}}, t_{\oli{3}}) = 
(t_3,t_1,t_2)$.
Thermal expectation values of permuted tuples 
are denoted by
\bal
\mc{S}_p[\bs{\mO}_p](\bs{t}_p) 
& =
\zeta_p 
\Big\langle 
\prod_{i=1}^\ell \mO^{\oli{i}}(t_{\oli{i}}) 
\Big\rangle. 
\label{eq:PSF_definition_time}
\eal
For later convenience, the definition includes a sign factor $\zeta_p$ which equals $-1$ if 
the permutation from $\bs{\mO}$ to $\bs{\mO}_p$ involves an odd number of transpositions of fermionic operators; otherwise $\zeta_p=1$. 
We will often suppress the operator arguments $[\bs{\mO}_p]$ for brevity, since the subscript on 
$\mc{S}_p$ specifies their order.
The real-frequency Fourier transform of $\mc{S}_p 
(\bs{t}_p)$ defines the so-called \textit{partial
spectral function} (PSF) 
\begin{subequations}
\bal
\mc{S}_p 
(\bs{\varepsilon}_p)
& = 
\int_{-\infty}^\infty \frac{\mathrm{d}^\ell t_p}{(2\pi)^\ell}\, e^{\i\bs{\varepsilon}_{p}\cdot\bs{t}_{p}} \mc{S}_p 
(\bs{t}_p)  . 
\eal
Here, $\bs{\varepsilon}_p
= (\varepsilon_{\oli{1}}, \dots, \varepsilon_{\oli{\ell}})$ is a permuted version of $\bs{\varepsilon} = (\varepsilon_1, \dots, \varepsilon_\ell)$, a tuple of continuous, real-frequency variables. 
We strictly associate each (integration) variable, such as $t_i$, $\varepsilon_i$, with the operator $\mO^i$ carrying the same index. Time-translational invariance of $\mc{S}_p(\bs{t}_p)$ implies energy conservation for $\mc{S}_p(\bs{\varepsilon}_p)$, which is expressed as
\bal
\mc{S}_p
(\bs{\varepsilon}_p)
& = 
\delta(\varepsilon_{\oli{1} \dots \oli{\ell}}) \, S_p(\bs{\varepsilon}_p) . 
\label{eq:PSF_definition}
\eal
\end{subequations}
Here, $\varepsilon_{\oli{1}\dots\oli{i}} = \varepsilon_{\oli{1}}+\dots+\varepsilon_{\oli{i}}$ is a shorthand for
a frequency sum. We call it bosonic/fermionic if the frequencies $(\varepsilon_{\oli{1}},\dots,\varepsilon_{\oli{i}})$ are associated with an even/odd number of fermionic operators, i.e., if the sign
$\zeta^{\oli{1}\dots \oli{i}} = \zeta^{\oli{1}} \dots \zeta^{\oli{i}}$ equals $\pm 1$ (with $\zeta^j = \pm 1$ for bosonic/fermionic operators $\mO^j$).
The function $\mc{S}_p$ (calligraphic type) on the left of
\Eq{eq:PSF_definition} is non-zero only if 
its arguments satisfy ``energy conservation'',
$\varepsilon_{\oli{1}\dots\oli{\ell}}=0$; for $S_p$ (italic type) on the right,
this condition on $\bs{\varepsilon}_p$ is understood to hold by definition, e.g., by setting $\varepsilon_{\oli{\ell}} = -\varepsilon_{\oli{1}\dots\oli{\ell-1}}$.
This convention  for frequency arguments of functions typeset in calligraphics or italics also holds for the correlators,  $\mc{G}$ vs.\ $G$,  and kernels, $\mc{K}$ vs.\ $K$, defined below.

PSFs whose arguments are cyclically related are proportional to each other.
For two cyclically related permutations,
say $p = (\oli{1} \dots \oli{\lambda-1}\, \oli{\lambda} \dots \oli{\ell})$ and $p_\lambda = (\oli{\lambda} \dots \oli{\ell}\, \oli{1} \dots \oli{\lambda-1})$, the cyclicity of the trace of operator products ensures the equilibrium condition (called cyclicity relation in Ref.~\cite{Kugler2021})
\bal \label{eq:equilibrium:cyclic_PSFs}
S_{p}(\bs{\varepsilon}_{p}) = \zeta_p \zeta_{p_\lambda} e^{\beta \varepsilon_{\oli{1} \dots \oli{\lambda - 1}}} S_{p_\lambda}(\bs{\varepsilon}_{p_\lambda}), \quad \zeta_p \zeta_{p_\lambda} = \zeta^{\oli{1} \dots \oli{\lambda-1}} .
 \eal

Explicit Lehmann-type representations for PSFs in terms of a complete set of eigenenergies and eigenstates of $\mH$ are given in Refs.~\cite{Kugler2021,Lee2021} and exploited for numerical computations; however, they 
are not needed in this work.
Here, it suffices to assume that $S_p(\bs{\varepsilon}_p)$ may contain sums over Dirac delta functions and a part that is (piece-wise) continuous in its arguments.
For future reference, we split it into \textit{regular} and 
\textit{anomalous} parts, 
\bal
\label{eq:splitS_pregular-anomalous}
S_p(\bs{\varepsilon}_p)
= \tS_p(\bs{\varepsilon}_p) + \anomS_p(\bs{\varepsilon}_p) , 
\eal
where the anomalous part, $\anomS_p$, comprises all terms 
containing \textit{bosonic} Dirac $\delta(\varepsilon_\bonetoi)$
factors (i.e.\ ones having bosonic arguments)
setting $\varepsilon_\bonetoi=0$, while $\tS_p$ contains everything else (including fermionic Dirac deltas). We will see later that $\anomS_p$ gives rise to anomalous contributions to MF correlators, whereas $\tS_p$ does not.

\newlychanged{
In the ensuing analysis, we make \textit{no} assumptions on the behavior of the PSFs (apart from cyclicity).
Thus, our analysis is equally applicable to finite systems or infinite systems in the thermodynamic limit, and whether or not an ordered phase is present. Any such information is fully encoded in the PSFs.
}

\subsection{Matsubara formalism}
\label{sec:MF}

A $\ell$p MF correlator $\mc{G}$ is defined as a thermal expectation value of time-ordered operator products of the form
\bal 
\mc{G}(\bs{\tau}) = (-1)^{\ell-1} 
\Big\langle 
\mc{T} \prod_{i=1}^\ell \mO^i(-\mi \tau_i) 
\Big\rangle ,
\eal 
where $\mc{T}$ denotes time-ordering along the imaginary time axis (see \Fig{fig:time_ordering_shifts}(a)). 
This time-ordering ensures that $\mc{G}(\bs{\tau})$ is periodic 
under $\tau_i\rightarrow\tau_i+\beta$
if $\mO^i$ is bosonic, and 
anti-periodic if $\mO^i$ is fermionic. 
Therefore, it suffices to confine all times to the interval $\tau_i \in[0,\beta)$, and the Fourier transform of a MF correlator is defined as 
\bal 
\label{eq:FT_MF_corr}
\mc{G}(\i \bsomega) &= \int_0^\beta \mathrm{d}^\ell \tau\, e^{\i \bsomega \cdot \bs{\tau}} \mc{G}(\bs{\tau}) = \beta \delta_{\i \omega_{1 \dots \ell}} G(\i \bsomega) , 
\eal
where $\bs{\omega} = (\omega_1, \dots, \omega_\ell)$ is a tuple of discrete Matsubara frequencies (as indicated by 
the $\mi$ in the argument of $\mc{G}(\mi \bsomega)$), with $\omega_{i}$ bosonic/fermionic
if $\mO^i$ is bosonic/fermionic. 
On the right, $\delta$ is the Kronecker delta for Matsubara frequencies, $\delta_{\i \omega = 0}=1$ and $\delta_{\mi \omega \neq 0}=0$. 
In \Eq{eq:FT_MF_corr}, it enforces ``energy conservation'',
$\i \omega_{1 \dots \ell} = 0$. This condition originates from time translation invariance of $\mc{G}(\bs{\tau})$; it is understood to hold for the argument of $G(\mi \bs{\omega})$ by definition.

\bsubeq\label{eq:defMFGtau}
As shown in Ref.~\cite{Kugler2021}, it is possible to cleanly separate the analytical properties of correlators from the dynamical properties of the physical system of interest by expressing time-ordered products as sums over $\ell!$ parts, reflecting the
$\ell!$ possible ways of ordering the time arguments: 
\bal
\mc{G}(\bs{\tau}) 
&= 
\sum_p \mc{G}_p(\bs{\tau}_{p}) 
, \label{eq:defMFGtau-a} 
\\ 
\label{eq:defMFGtau-b}
\mc{G}_p(\bs{\tau}_p) &=  \mc{K}(\bs{\tau}_p) \mc{S}_{p}
(-\i\bs{\tau}_p) , 
\\
\mc{K}(\bs{\tau}_p) 
&= 
\prod_{i=1}^{\ell-1} \left[ -\theta(\tau_{\oli{i}} - \tau_{\oli{i+1}}) \right]
. \label{eq:def_MF_kernel_ellp} 
\eal 
\esubeq
Each \textit{partial correlator} $\mc{G}_p(\bs{\tau}_p)$ 
is a product of two factors: 
$\mc{S}_p
(-\i\bs{\tau}_p)$, a thermal expectation value of
imaginary-time operators obtained by Wick rotation of \Eq{eq:PSF_definition_time}; 
and a kernel $\mc{K}(\bs{\tau}_p)$, a product of Heaviside step functions enforcing
time ordering: for given $\bs{\tau}$, only that partial correlator $\mc{G}_p(\bs{\tau}_p) $ in \Eq{eq:defMFGtau-a} is nonzero for which the permuted tuple $\bs{\tau}_p$ is time-ordered.
$\mc{K}$ is independent of the system and operators under consideration; all system-specific dynamical information is encoded in the PSFs $\mc{S}_p$. Note that 
the (anti)periodic properties of $\mc{G}(\bs{\tau})$ 
under $\tau_i \to \tau_i + \beta$ do not hold for the individual partial correlators $\mc{G}_p(\bs{\tau}_p)$;
they emerge only once these are summed over all permutations, \Eq{eq:defMFGtau-a}.

\bsubeq
The product form of \Eq{eq:defMFGtau-b} for 
$\mc{G}_p(\bs{\tau}_p)$ in the time domain implies that, in the Fourier domain, $\mc{G}(\mi \bs{\omega})$ can be expressed as a sum over convolutions:
\label{subeq:MF_kernel_FT} 
\bal
\label{eq:MF_kernel_FT-a} 
\mc{G}(\i\bs{\omega}) 
&= 
\sum_p \mc{G}_p(\i \bsomega_p) , 
\\
\mc{G}_p(\i \bsomega_p) 
 & = 
\int_0^\beta \mathrm{d}^\ell \tau_{p}\, e^{\i \bsomega_{p} \cdot \bs{\tau}_{p}} \mc{G}_p(\bs{\tau}_p) 
\label{eq:FT_MF_define-Matsubara-transform}
\\ 
&=
\bigl[ \mc{K} \ast S_p\bigr]
 (\i \bsomega_p) .  
 \label{eq:FT_MF_corr_as_convolution}
\eal
Here, the convolution $\ast$ is defined as
\bal
\label{eq:define-convolution}
\bigl[ \mc{K} \ast S_p\bigr]
 (\i \bsomega_p) &= 
\int_{-\infty}^{\infty} \! \mathrm{d}^{\ell} \varepsilon_p\,
\delta(\varepsilon_\bonetol)
\mc{K}(\i\bs{\omega}_p \!-\! \bs{\varepsilon}_p) S_p(\bs{\varepsilon}_p) 
,
\eal
\esubeq
\bsubeq\label{eq:Kernel-define}
where $\bs{\varepsilon}_p$ satisfies $\varepsilon_\bonetol=0$ (due to \Eq{eq:PSF_definition}), and the transformed kernel is defined as follows, with $\bs{\Omega}_p = \i\bs{\omega}_p \!-\! \bs{\varepsilon}_p$:
\bal
\label{eq:Kernel-define-a}
\mc{K}(\bs{\Omega}_p)
&=
\int_0^\beta \mathrm{d}^\ell \tau_p\, 
e^{\bs{\Omega}_p \cdot \bs{\tau}_p} 
\mc{K}(\bs{\tau}_p)
\\
&= \beta \delta_{\Omega_{1 \dots \ell}} 
K(\bs{\Omega}_p)
+ \mc{R}(\bs{\Omega}_p) . 
\label{eq:Kernel-define-b}
\eal
\esubeq
\bsubeq
\label{eq:decomposeG_p}
In the second line, $\mc{K}$ has been split into two contributions: $\beta\delta_{\Omega_{1\dots\ell}}$ times a \textit{primary part} $K$, 
with $\Omega_{1\dots\ell} = 0$ understood for its argument, and a \textit{residual part}, $\mc{R}$ not containing $\beta\delta_{\Omega_{1\dots\ell}}$.
Using  $\delta_{\Omega_{1\dots\ell}} =\delta_{\i\omega_{1\dots\ell}}$ (since $\varepsilon_{1\dots \ell}=0$),
each partial correlator $\mc{G}_p(\mi \bs{\omega}_p)$ can correspondingly be split into primary and residual parts,
\bal
\label{eq:decomposeG_p-a}
\mc{G}_p(\i\bs{\omega}_p)
& = \beta \delta_{\omega_{1\dots \ell}} 
{G}_p(\i\bs{\omega}_p) + \mc{G}^{\mc{R}}_p(\i \bsomega_p) , 
\\
\label{eq:decomposeG_p-b}
G_p(\i \bsomega_p) &=  
\bigl[ K \ast S_p \bigr](\i\bs{\omega}_p)
,
\eal
\esubeq
with $\mi \omega_{1\dots \ell}=0$ understood for the argument of $
G(\mi \bs{\omega}_p)$, and $\mc{G}^{\mc{R}}_p = [ \mc{R} \ast S_p ]$.
Since $\mc{K}(\bs{\tau}_p)$ and $\mc{G}_p(\bs{\tau}_p)$ 
lack the (anti)periodicity properties of $\mc{G}(\bs{\tau})$, the residual parts $\mc{R}(\bs{\Omega}_p)$ and $\mc{G}^{\mc{R}}_p(\i \bsomega_p)$ are nonzero \textit{per se}. However, inserting
\Eq{eq:decomposeG_p-a} 
into \Eq{eq:MF_kernel_FT-a} and noting  from \Eq{eq:FT_MF_corr} that $\mc{G}(\mi \bs{\omega})$ 
is proportional to $\beta \delta_{\mi \omega_{1 \dots \ell}}$, one concludes that
\bal
\label{eq:ResidualPartsCancel}
{G}(\i\bs{\omega})  = \sum_p G_p(\i \bsomega_p) 
\eal
and $\sum_p \mc{G}^{\mc{R}}_p(\i \bsomega_p) = 0$.
Thus, the \textit{full} (summed over $p$) MF 
correlator $G$ involves only primary parts 
$G_p$; the residual parts
$\mc{G}^{\mc{R}}_p$ cancel out 
in the sum over all permutations. In the 
discussions below, we will therefore focus only on 
the primary parts $K$ and $G_p$ (as done in Ref.~\cite{Kugler2021}), ignoring 
the residual parts $\mc{R}$ and $\mc{G}^{\mc{R}}_p$ for now.
They will make a brief reappearance in \Sec{sec:Bridge_between_MF_KF}, where we establish the connection between MF and KF correlators. 

\bsubeq
Explicit expressions for the primary kernel $K$ were derived in Refs.~\cite{Kugler2021,Halbinger2023}
and are collected in App.~\ref{app:MFKernels}. 
Here, we just remark that $K$ can be split into a \textit{regular} kernel $\tK$ and an 
\textit{anomalous} kernel $\hK$:
\label{eq:MF_kernel-compact}
\bal
K(\bsOmega_p) & = 
\begin{cases}
\tK(\bsOmega_p) \qquad &\tn{if } \prod_{i=1}^{\ell-1} \Omega_{\oli{1} \dots \oli{i}} \neq 0, \\
\hK(\bsOmega_p) \qquad &\tn{else},
\end{cases}
\label{eq:def_general_kern-compact} 
\\
\tK(\bsOmega_p) &= \prod_{i=1}^{\ell-1} \frac{1}{\Omega_{\oli{1} \dots \oli{i}} }.  \label{eq:def_reg_kern-compact}
\eal
\esubeq
The regular kernel $\tK$ will play a crucial role for the analytic continuation of MF to KF correlators, since the latter can be expressed through kernels having the same structure as $\tK$ (see \Eq{eq:FTretkern} below). 
The anomalous kernel $\hK$ is nonzero only if we have $\Omega_{\oli{1} \dots \oli{i}} =0$ for one or more values of $i < \ell$, requiring both $\mi \omega_{\oli{1} \dots \oli{i}} =0$ \textit{and} $\varepsilon_{\oli{1} \dots \oli{i}} =0$.
The first condition requires $\i\omega_{\oli{1}\dots\oli{i}}$ to be bosonic
(with $\zeta^{\oli{1}\dots \oli{i}} = +1$).
The second condition requires the PSF $S_p(\bsvarepsilon_p)$ to have an
anomalous contribution $\anomS_p(\bsvarepsilon_p)$
containing terms proportional to a bosonic Dirac $\delta(\varepsilon_{\oli{1} \dots \oli{i}})$; then (and only then), the $\varepsilon_p$ integrals in the convolution $K * S_p$ receive a finite contribution from the point $\varepsilon_{\oli{1} \dots \oli{i}} =0$.
(See App.~\ref{app:PSF_decomposition} for a further discussion of this point.)

\bsubeq
\label{eq:MF_G-compact}
The regular/anomalous distinction made
for the kernel
implies, via \Eqs{eq:decomposeG_p-b} and \eqref{eq:ResidualPartsCancel}, a corresponding
decomposition of the full  MF correlator $G$ into regular ($\tG$) and anomalous ($\hG$) parts:
\bal
G(\i \bsomega) &= \tG(\i \bsomega) + \hG(\i \bsomega) ,
\label{eq:MF_G-compact-a-2}
\\
\tG(\i \bsomega) &= \sum_p \tG_p(\i \bsomega_p), 
\label{eq:MF_tG_psum_tGp}
\\ 
\label{eq:MF_G-compact-b-star}
\tG_p(\i \bsomega_p) &= \bigl[ \tK \ast S_p \bigr](\i\bs{\omega}_p) 
\\ 
& = 
\int_{-\infty}^{\infty} \mathrm{d}^{\ell} \varepsilon_p\, \delta(\varepsilon_\bonetol)
\prod_{i=1}^{\ell-1} \frac{{S}_p(\bs{\varepsilon}_p)}
{\i\omega_{\oli{1} \dots \oli{i}} - \varepsilon_{\oli{1} \dots \oli{i}}} 
\label{eq:MF_G-compact-b}.
\eal
\esubeq
The \textit{regular partial correlators} $\tG_p$, constructed via the regular kernel $\tK$, will be the central objects for the analytic continuation from MF to KF correlators, as discussed in  \Sec{sec:Bridge_between_MF_KF} below. Their sum over all permutations defines the \textit{regular full correlator} $\tG$. The \textit{anomalous full correlator} $\hG$ collects all other contributions to $G$; these contain one (or multiple) factors
$\beta\delta_{\i\omega_{\oli{1}\dots \oli{i}}}$ with $i < \ell$, i.e.\ they involve vanishing partial frequency sums (see  App.~\ref{app:kernel_alt} for details). The contribution of $\hG$ to MF-to-KF analytical continuation has been rather poorly understood to date. In this work, we fully clarify how it enters: not directly, but indirectly, in that the central objects $\tG_p(\i\bs{\omega}_p)$ can be expressed explicitly through the full $G = \tG + \hG$ via imaginary-frequency convolutions of
the form $[K \star G](\i\bs{\omega}_p)$ (see \Eq{eq:tildeGpstart} below). There, $\hG$ must not be neglected.

\subsection{Keldysh formalism}
\label{sec:KF}

\begin{figure*}[t!]
    \centering
    \includegraphics[width=1 \textwidth]{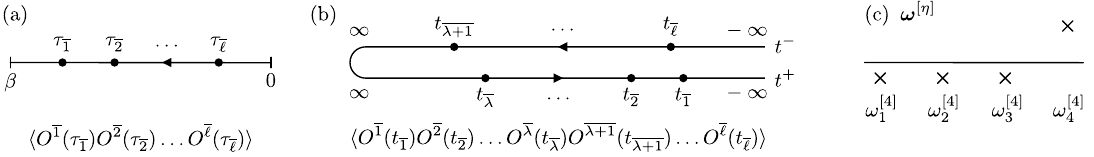}
    \caption{(a) MF imaginary-time ordering: operators are arranged such that they are time-ordered (larger times to the left). 
    (b) KF real-time Keldysh ordering: operators are arranged such that all (forward-branch) times $t^-$
    appear to the right of all (backward-branch) times $t^+$, 
    with $t^-$ times time-ordered (larger ones to the left)
    and $t^+$ times anti-time-ordered (smaller ones to the left).
    (c) Depiction of imaginary shifts of frequencies $\omega^{[\eta]}_i = \omega_i + \i \gamma^{[\eta]}_i$ with $i \in \{ 1, 2, 3, 4 \}$ and $\eta=4$ according to \Eq{eq:imshifts}.}
    \label{fig:time_ordering_shifts}
\end{figure*}

\bsubeq\label{eq:KF_contour_ordering}
A KF $\ell$p correlator in the \textit{contour basis} is defined as
\bal 
\mc{G}^{\bs{c}}(\bs{t}) & = (-\i)^{\ell-1} 
\Big\langle 
{\mc{T}}_c \prod_{i=1}^{\ell} \mO^i(t_i^{c_i}) 
\Big\rangle \, , 
\label{eq:KF_contour_ordering-a}
\\
\label{eq:KF_contour_ordering-b}
& = \sum_p \mc{K}^{\bs{c}_p} (\bs{t}_p) \mc{S} (\bs{t}_p)
\, . 
\eal
\esubeq
Here, $\mc{T}_c$ denotes contour ordering
on the Keldysh contour (see \Fig{fig:time_ordering_shifts}(b)),
and $t^{c_i}_i$ are real times.
They carry a tuple of
contour indices $\bs{c} = (c_1, \dots, c_\ell)$ 
with $c_i = -$ or $+$ if operator $\mO^i$ resides on the forward (upper) or
backward (lower) branch of the Keldysh contour, respectively. 
Equation~\eqref{eq:KF_contour_ordering-b} is a permutation decomposition of the KF correlator $\mc{G}^{\bs{c}}(\bs{t})$, analogous to 
\Eq{eq:defMFGtau-b} for $\mc{G}(\bs{\tau})$ in the MF.
Importantly, it employs the \textit{same} PSFs $\mc{S} (\bs{t}_p)$ as there (which is why the KF and MF formalisms have the same physical information content). The Keldysh
kernel $\mc{K}^{\bs{c}_p} (\bs{t}_p) $ by definition (see Ref.~\cite{Kugler2021} for details) singles out that $p$ for which the operators in $\mc{S}_p(\bs{t})$ are contour ordered.

The Fourier transform of the KF correlator is
\bal \label{eq:KF_FT}
\mc{G}^{\bs{c}}(\bsomega) = \int \mathrm{d}^\ell t\, e^{\i \bsomega \cdot \bs{t}}\, \mc{G}^{\bs{c}}(\bs{t}) = 2\pi \delta(\omega_{1 \dots \ell}) G^{\bs{c}}(\bsomega)
.
\eal
Here, the Dirac $\delta(\omega_{1 \dots \ell})$, 
following from time translation invariance, enforces 
$\omega_{1 \dots \ell} = 0$; this condition is understood for 
the argument of $G^{\bs{c}}(\bsomega)$ by definition.

We now switch to the Keldysh basis. There, correlators $\mc{G}^{\bs{k}}(\bs{\omega})$ carry a tuple of Keldysh indices, $\bs{k} = k_1 \dots k_\ell$, with 
$k_i\in\{1,2\}$. They are obtained by applying a linear transformation $D$ to each contour index, 
\bal
\mc{G}^{\bs{k}}(\bs{\omega})
= \tfrac{1}{2}
\sum_{c_1,\dots,c_\ell}
\prod_{i=1}^\ell [D^{k_ic_i}] \mc{G}^{\bs{c}} (\bs{\omega})
, \quad D^{k_i c_i} = (-1)^{k_i \delta_{c_i,+}}
\vspace{-2mm} 
\label{eq:Keldysh_rotation_prefactor}
\eal
(This convention differs by a prefactor from Ref.~\cite{Kugler2021},
with $\mc{G}^{\bs{k}}_{\mathrm{here}} = 2^{\ell/2-1}\mc{G}^{\bs{k}}_{\mathrm{there}}$, 
to avoid a proliferation of factors of $2^{\ell/2 - 1}$ in later sections.)
One thus obtains
\bsubeq
\label{eq:KF_corr_time_k}
\bal
\label{eq:KF_corr_time_k-1}
G^{\bs{k}}(\bs{\omega}) &
= \sum_p G_p^{\bs{k}_p}(\bs{\omega}_p) , 
\\
G_p^{\bs{k}_p}(\bs{\omega}_p) &
= 
\big(K^{\bs{k}_p} \ast S_p\big)(\bs{\omega}_p) 
\label{eq:KF_corr_time_k-2}
\\ & = 
\int\!\! \tn{d}^{\ell} \varepsilon_p 
\delta(\varepsilon_\bonetol)
K^{\bs{k}_p}(\bsomega_p-\bs{\varepsilon}_p) S_p(\bs{\varepsilon}_p) .
\eal
\esubeq
Remarkably, the same convolution structure emerges as 
for the MF correlator $G(\mi \bsomega)$ (\Eq{eq:decomposeG_p-b}), for the same reason (Fourier transforms of products yield convolutions). But now the frequency arguments are real,
and the kernel $K^{\bs{k}_p}(\bsomega_p)$ carries Keldysh indices, with $\bs{k}_p = k_{\oli{1}} \dots k_{\oli{\ell}}$
a permuted version of the \textit{external} Keldysh index 
$\bs{k}$ on $G^{\bs{k}}$.

An explicit expression for this kernel, derived in Ref.~\cite{Kugler2021}, is given in \Eqs{eq:KF_spec_rep_eta_alpha} below. There, 
an alternative notation for Keldysh indices is employed. 
Each Keldysh index $\bs{k}$, being a list with entries $1$ or $2$, is represented as a list $\bs{k} = [\eta_1 ... \eta_\alpha]$, where $\alpha$ is the total number of $2$'s in $\bs{k}$ and $\eta_i \in \{1, \dots, \ell\}$ denotes the position of the $i$th $2$ in $\bs{k}$ in increasing order; e.g., $\bs{k} = 1212 = [24]$. Similarly, permuted Keldysh indices
are represented as $\bs{k}_p = [\heta_1 ... \heta_\alpha]$,
where $\heta_i$ denotes the position of the $i$th 2 in $\bs{k}_p$. Its values can be deduced from the old $\eta_j$'s as follows: a 2 in slot $\eta_j$ of $\bs{k}$ is moved
by the permutation $p$ to the new slot $\mu_j = p^{-1}(\eta_j)$; denoting the list of new 2-slots by $[\mu_1 ... \mu_\alpha]$ and arranging it in increasing order yields the desired $[\heta_1 ... \heta_\alpha]$. Note also that 
since $\heta_j \in \{ p^{-1}(\eta_1), ..., p^{-1}(\eta_\alpha) \}$, we have $\oli{\heta}_1 \in \{\eta_1, \dots, \eta_\alpha\}$; hence, $\oli{\heta}_j$ is an element of the list specifying the \textit{external} Keldysh index  $\bs{k} = [\eta_1 ... \eta_\alpha]$. This will be crucial below. We illustrate  these conventions
for the permutation $p=(4123)$ and $\bs{k}=1212 = [24]$. Then,  $\bs{k}_p = 2121$, $[\mu_1 \mu_2] = [31]$ and $\bs{k}_p = [\heta_1 \heta_2] = [13]$; moreover, $\oli{\heta}_1 = \oli{1} = 4$ and 
$\oli{\heta}_2 = \oli{3} = 2$ are both elements of 
$\bs{k}=[24]$.

\bsubeq
\label{eq:KF_spec_rep_eta_alpha}
Expressed in this notation, \Eqs{eq:KF_corr_time_k} read
\bal
\label{eq:KF_spec_rep_etas-full} 
G^{[\eta_1\dots\eta_\alpha]} 
(\bsomega) & = \sum_p G_p^{[\heta_1\dots\heta_\alpha]} (\bsomega_p) ,
\\
G_p^{[\heta_1 ... \heta_\alpha]}(\bsomega_p) 
&=
\Big[
    K^{[\heta_1\dots\heta_\alpha]} \ast S_p
\Big](\bsomega_p), 
\label{eq:KF_spec_rep_etas} 
\eal
with the  permuted Keldysh kernel
$K^{[\heta_1\dots\heta_\alpha]}$
given by  \cite{Kugler2021}
\bal
\label{eq:KFkerneletas}
K^{[\heta_1 ... \heta_\alpha]}(\bsomega_p) &= \sum_{j=1}^\alpha (-1)^{j-1} K^{[\heta_j]}(\bsomega_p) \, , 
\\
\label{eq:FTretkern}
K^{[\eta]}(\bs{\omega}_p) 
&=
 \prod_{i=1}^{\ell-1} \frac{1}{\omega^{[\oli{\eta}]}_{\oli{1}\dots\oli{i}} }
\, . 
\eal
\esubeq
\EQs{eq:KF_spec_rep_eta_alpha} compactly express all partial correlators $G^{\bs{k}_p}_p = G_p^{[\heta_1 ... \heta_\alpha]}$, and hence also the full KF correlator $G^{\bs{k}} = G^{[\eta_1 ... \eta_\alpha]}$, through a set of $\ell$ so-called \textit{fully retarded kernels} $K^{[\eta]}$.
These are defined by \Eq{eq:FTretkern} and depend on just a single index $\eta$, which takes the value $\heta_j$ in \Eq{eq:KFkerneletas}. The superscript on the frequencies occurring therein denotes imaginary shifts $\omega_i \rightarrow \omega^{[\eta]}_i = \omega_i + \i \gamma^{[\eta]}_i$, with $\gamma^{[\eta]}_i\in \mathbbm{R}$ chosen such that $\gamma^{[\eta]}_{i\neq\eta}<0$, $\gamma^{[\eta]}_{\eta}>0$, and $\omega_{1\dots\ell}=\omega_{1\dots\ell}^{[\eta]}=0$. 
Shifts of precisely this form are needed to regularize the Fourier integrals expressing $\mc{K}^{\bs{k}_p}(\bsomega_p)$ through $\mc{K}^{\bs{k}_p}(\bs{t}_p)$.
\newlychanged{Indeed, for infinitesimal $\gamma_i^{[\eta]}$ each factor in \Eq{eq:FTretkern} is the Fourier transform of a step function,
\bal
\pm \i \int_{\mathbbm{R}}\mathrm{d}t\, \theta(\pm t) e^{\i\omega t}  
&= 
\frac{1}{\omega \pm \i 0^+}    
= 
\tn{P}\Big(\frac{1}{\omega}\Big) \mp \i\pi \delta(\omega) ,
\label{eq:FT_stepfunctions}
\eal
giving the kernels both principal-value $\tn{P}$ and Dirac-$\delta$ contributions.
}
We choose the same convention as in Ref.~\cite{Kugler2021},
\bal \label{eq:imshifts}
\gamma_{i\neq \eta}^{[\eta]} = - \gamma_0, \quad \gamma_{\eta}^{[\eta]} = (\ell-1)\gamma_0,
\eal
see Fig. \ref{fig:time_ordering_shifts}(c), with $\gamma_0$ taken to be infinitesimal, $\gamma_0 = 0^+$, for analytical considerations.
Below, we also use the shorthand $\omega_{i\dots j}^\pm=\omega_{i\dots j}\pm\i0^+$ to indicate infinitesimal imaginary shifts for sums of frequencies.

Comparing the fully retarded kernel $K^{[\eta]}$ of 
\Eq{eq:FTretkern} with the regular Matsubara kernel $\tK$ of \Eq{eq:def_reg_kern-compact}, we find that the former
is the analytic continuation of the latter:
\bal 
\label{eq:MF-KF-continuation-Kernel}
K^{[\eta]}(\bs{\omega}_p) = \tK\big(\mi \bsomega_p \to \bsomega^{[\oli{\eta}]}_p\big). 
\eal
This remarkable relation between MF and KF kernels constitutes the nucleus from which we will develop our strategy for obtaining KF correlators via analytic continuation of MF correlators.
\newlychanged{
Here, we just note that, by \Eqs{eq:def_reg_kern-compact} and  
\eqref{eq:FT_stepfunctions}, the 
analytical continuation of the \textit{regular} MF kernel
on the right of \Eq{eq:MF-KF-continuation-Kernel}  generally yields 
both principal-value and Dirac-$\delta$ contributions. 
By contrast, we will find below that 
the analytic continuation of \textit{anomalous} MF kernels yields solely Dirac-$\delta$ contributions in KF correlators [cf. \Eqs{eq:overview_AC_3p} and \eqref{eq:overview_AC_4p}].}

Two well-known statements on general $\ell$p correlators follow immediately from \Eqs{eq:KF_spec_rep_eta_alpha}. First, for $\alpha = 0$, they imply $G^{[]}=G^{1\dots1}=0$. Second,  for $\alpha = 1$, 
we have $\oli{\heta}_1 = \eta_1$. Thus, 
$K^{[\heta]}(\bs{\omega}_p) = \tK(\bsomega^{[\eta]}_p)$ 
by \Eq{eq:MF-KF-continuation-Kernel}, and \Eq{eq:KF_spec_rep_etas}  yields
\bal 
\label{subeq:Analytic_cont_fully_ret}
G_p^{[\heta]}(\bsomega) & = [ \tK \ast S_p] (\bsomega^{[\eta]}_p)
= \tG_p(\mi \bsomega_p \to \bsomega^{[\eta]}_p) .
\eal
For the second step, we evoked \Eq{eq:MF_G-compact-b-star}. 
Importantly, the superscript on 
$ \bsomega^{[\eta]}_p$ on the right,
which specifies its imaginary frequency shifts, 
is fully determined by the external Keldysh index $\eta$ and \textit{not} dependent on $p$.  
It thus remains unchanged throughout the 
sum on $p$ in \Eq{eq:KF_corr_time_k-1} for the full
correlator $G^{[\eta]}(\bsomega)$, which hence can be expressed as
\bal
\label{eq:Analytic_cont_fully_ret}
G^{[\eta]}(\bsomega)
 &
 = \tG(\i \bsomega \rightarrow \bsomega^{[\eta]}).
\eal
The fully retarded ($\alpha=1$) components of
KF correlators are therefore fully determined, via analytic continuation, by the \textit{regular} parts of  MF correlators. Conversely, anomalous parts of MF correlators can only influence Keldysh components with $\alpha \ge 2$.

For later use, we also define primed partial correlators
\bsubeq \label{eq:KF_fully_adv}
\bal
G'^{[\eta_1 \dots \eta_\alpha]} (\bsomega) &= \sum_p G_{p}^{\prime[\heta_1 ... \heta_\alpha]}(\bsomega_p), \\
G_{p}^{\prime[\heta_1 ... \heta_\alpha]}(\bsomega_p) 
&= 
\left[ \big(K^{[\heta_1 ... \heta_\alpha]}\big)^* \ast S_p \right](\bs{\omega}_p)
.
\eal
\esubeq
They differ from the unprimed correlators of \Eq{eq:KF_spec_rep_etas} by the complex conjugation of the kernel, replacing $\omega_i + \i \gamma^{[\eta]}_i$ by $\omega_i - \i \gamma^{[\eta]}_i$, with $\gamma^{[\eta]}_i$ still determined by the rule \Eq{eq:imshifts}. For $\alpha = 1$, the corresponding  $G'^{[\eta]}$ will be called \textit{fully advanced} correlators.
For fully retarded or advanced correlators, 
$G^{[\eta]}$ or $G'^{[\eta]}$, \textit{all} frequencies $\omega_{i\neq \eta}$ acquire negative or positive imaginary shifts, respectively.
Note that primed correlators $G^{\prime\bs{k}}$ may differ from complex conjugated correlators $G^{*\bs{k}}$ as the complex conjugation generally affects the PSFs, too.

This concludes our summary of the results of Ref.~\cite{Kugler2021} needed for present purposes. In the next section, we introduce a general strategy for expressing KF correlators through
analytically-continued MF correlators. It is well-known how to do this for all components of 2p correlators, and, as discussed above, for the fully retarded and advanced components of $\ell$p correlators. Our goal is a strategy applicable for all components of $\ell$p correlators.

\begin{table}[t]
    \centering
    \begin{tabular}{c l}
      Symbol            &    Description  \\
      \hline
      $G$   &   full MF correlator, \Eqs{subeq:MF_kernel_FT}  \\
      $G_p$ &   partial MF correlator, \Eq{eq:decomposeG_p-b}   \\
      \multilinebox{
      $\tG$, $\hG$ \\} &   \multilinebox{
                                                regular and anomalous part of the MF\\
                                                correlator, \Eqs{eq:MF_G-compact-a-2} and \eqref{eq:MF_correlators_general_form} }  \\
      \multilinebox{
      $\tG_p$ \\}&  \multilinebox{regular part of the partial MF correlator, \\
                                \Eq{eq:MF_G-compact-b-star} } \\
      \multilinebox{
      $\hG_i$, $\hG_i^\withDelta$, $\hG_i^\noDelta$ \\}  &   \multilinebox{
                                            further decomposition of the anomalous\\ 
                                            MF correlator, \Eqs{eq:Anom_corr_general_form} and \eqref{eq:3p_Gf_Gd_decomp} } \\
      \hline
      \multilinebox{$\tG_{\zcheck}$, $\hG_{i; \zcheck}$\\ \\}   &   \multilinebox{
                        shorthand for analytic continuations\\
                        of the regular/anomalous MF correlator, \\
                        see \Sec{sec:analytic_regions_tG} } \\
      \multilinebox{$\tG^{\omega}_{\zcheck^r}$, $\hG^{\omega}_{i; \zcheck^r}$\\}    & \multilinebox{
                        discontinuities of the regular/anomalous\\
                         MF correlator, \Eq{eq:def_discontinuity_general_ell} 
                        }  \\
      \hline
      $G^{\bs{k}}$, $G^{[\eta_1\dots\eta_\alpha]}$ &   Keldysh correlator, \Eqs{eq:KF_spec_rep_eta_alpha} \\
      $G'^{\bs{k}}$, $G'^{[\eta_1\dots\eta_\alpha]}$ &   primed Keldysh correlator, \Eqs{eq:KF_fully_adv}
    \end{tabular}
    \caption{
    Overview of notation for correlators and their contributions. 
    In the top, we list symbols for the MF correlator and its contributions, then, notation for analytic continuations and discontinuities, and, lastly, notation for Keldysh correlators. 
    }
    \label{tab:glossary_for_Gs}
\end{table}

\subsection{The bridge between the MF and KF formalisms}
\label{sec:Bridge_between_MF_KF}

\bsubeq\label{subeq:KF_corr_through_partial_corr}
\EQ{subeq:Analytic_cont_fully_ret}, expressing KF partial correlators through MF partial correlators for $\alpha=1$, has a counterpart for arbitrary $\alpha$, obtained via \Eqs{eq:KF_spec_rep_eta_alpha}, \eqref{eq:MF-KF-continuation-Kernel}, and \eqref{eq:MF_G-compact-b-star}:
\bal 
\label{eq:KF_corr_through_partial_corr-a}
G_p^{[\heta_1 ... \heta_\alpha]}(\bsomega_p) 
& =
\sum_{j=1}^\alpha (-1)^{j-1}\, 
\big[ \tK \ast S_p \big] \bigl(\bsomega^{[\oli{\heta}_j]}_p\bigr)
\\[-2mm]
\label{eq:KF_corr_through_partial_corr}
&= 
\sum_{j=1}^\alpha (-1)^{j-1}\, \tG_p\big(\mi \bsomega_p \to \bsomega^{[\oli{\heta}_j]}_p\big),
\eal
\esubeq
with $\oli{\heta}_j \in \{ \eta_1, \dots, \eta_\alpha \}$. 
This is already one of our main results: The partial correlators serve as a bridge between the MF and KF.
All components of the partial 
KF correlator $G^{\bs{k}_p}_p = G_p^{[\heta_1 ... \heta_\alpha]}$ can be obtained by taking linear combinations of analytic continuations of partial regular  MF correlators, $\tG_p\big(\mi \bsomega_p \to \bs{\omega}_p^{[\oli{\heta}_j]}\big)$. The external Keldysh indices $\bs{k} = [\eta_1 ... \eta_\alpha]$ and the permutation $p$ together specify the imaginary frequency shifts, encoded in 
$\bsomega^{[\oli{\heta}_j]}_p$, to be used.

\bsubeq \label{eq:KF_corr_through_full_corr}
Equation~\eqref{eq:Analytic_cont_fully_ret}, expressing the full ($p$-summed) KF  correlators through MF ones for $\alpha=1$, does not have a counterpart for 
$\alpha >1$. Then, the full correlators, given by 
\bal 
\label{eq:KF_corr_through_full_corr-a}
G^{[\eta_1 ... \eta_\alpha]}(\bsomega) 
& 
= \sum_p
\big[ K^{[\heta_1 ... \heta_\alpha]} \ast S_p \big] \bigl(\bsomega_p \bigr)
\\[-3mm]
\label{eq:KF_corr_through_full_corr-b}
&= 
\sum_p \sum_{j=1}^\alpha (-1)^{j-1}\, \tG_p\big(\mi \bsomega_p \to \bsomega^{[\oli{\heta}_j]}_p\big),
\eal
\esubeq
involve a sum $\sum_j$. 
The $\oli{\heta}_j$ indices on the right now depend on $p$, so that the imaginary frequency shifts vary from one permutation to the next. As a result, the full $G^{[\eta_1 \dots \eta_\alpha]}$, unlike $G^{[\eta]}$,
does not depend on a single set of frequency shifts and 
cannot be directly expressed through a mere analytic continuation of $\tG(\mi \bsomega)$. 
Instead, \Eq{eq:KF_corr_through_full_corr-b}
requires separate knowledge of each individual  $\tG_p(\mi \bsomega_p)$. 
Most computational methods capable of computing 
the full MF correlator $G(\i\bs{\omega})$ do not have 
access to the separate partial MF correlators $\tG_p(\i\bs{\omega}_p)$. In the following, we therefore develop a strategy for extracting the partial MF correlators $\tG_p(\mi \bsomega_p)$ from a full MF correlator $G(\i\bs{\omega})$ given as input, assuming the latter to be known analytically. By writing the resulting functions 
$\tG_p(\mi \bsomega)$ in the form $[\tK\ast S_p](\mi \bsomega)$, one can deduce
explicit expressions for the PSFs $S_p[G]$ as functionals of the input $G$. 
By inserting these $S_p$ into \Eq{eq:KF_corr_through_full_corr-a},
one obtains $G^{[\eta_1 ... \eta_\alpha]}[G]$ as a functional of $G$,
thereby achieving the desired MF-to-KF analytic continuation.

We start in the MF time domain. There,
a specific partial MF correlator $\mc{G}_p(\bs{\tau}_p)$ can be obtained from the full $\mc{G}(\bs{\tau}) = \sum_p \mc{G}_p(\bs{\tau}_p)$ (\Eqs{eq:defMFGtau}) using the projector property of MF kernels in the time domain, $\mc{K}(\bs{\tau}_p) \mc{K}(\bs{\tau}_{p'}) = (-1)^{\ell-1}\mc{K}(\bs{\tau}_p)$ if $p=p'$ and $0$ otherwise. Hence, we can express the partial correlator as 
\bal
\label{eq:G_pfromG}
\mc{G}_{p}(\bs{\tau}_{p}) = (-1)^{\ell-1} \mc{K}(\bs{\tau}_{p}) \mc{G}(\bs{\tau}).
\eal
\bsubeq\label{subeq:define-discrete-Fourier-transform}
Computing the discrete Fourier transform
of \Eq{eq:G_pfromG} according to \Eq{eq:FT_MF_define-Matsubara-transform}, we obtain 
\bal
\label{subeq:define-discrete-Fourier-transform-1}
\mc{G}_{p}(\i\bsomega_p) & = 
\big[\mc{K} \star G\bigr] (\i\bsomega_p) ,
\eal
with the imaginary-frequency convolution $\star$ defined as 
\bal
\label{subeq:define-discrete-Fourier-transform-2}
\big[\mc{K} \star G\bigr] (\i \bsomega_p) 
& = \tfrac{1}{(-\beta)^{\ell-1}} \sum_{\i\bs{\omega}'_p} 
\delta_{\i\omegap_\bonetol}
\mc{K}(\i\bs{\omega}_p-\i\bs{\omega}'_p) G(\i\bs{\omega}') .
\eal
\esubeq
We will typically sum over the 
$\ell-1$ independent Matsubara frequency variables 
$\i {\omega'}_\bonetoi$,  with $i\in \{1, \dots, \ell-1 \}$. Note that the arguments of $G(\mi \bsomega')$ appear in \textit{unpermuted} order, but are to be viewed as functions
of the summation variables, i.e.,
$\i\bsomega'= \i\bsomega'(\bsomega'_p)$. 
We will often make this
explicit using the notation $G_{\i\bs{\omega}'_p}=G(\i \bsomega'(\bsomega'_p))$, where the subscript is a label indicating the $\ell - 1$ independent frequencies chosen to parametrize $\i \bsomegap$. Consider, e.g., $\ell = 3$ and choose
$ \i\omega_{\oli{1}}$, $\i\omega_{\oli{12}}$
as summation variables. For the 
permutation $p=(132)$, the correlator
is then represented as 
$G_{\i \omega_{\oli{1}}, \i\omega_{\oli{12}}} = G_{\i \omega_{1}, \i\omega_{13}} = G(\i \bsomega(\i\omega_{1}, \i\omega_{13})) = G(\i \omega_1, -\i \omega_{13}, \i \omega_{13} - \i \omega_1)$.

Using \Eq{eq:decomposeG_p-a} for $\mc{G}_p(\mi \bsomega_p)$ and \Eq{eq:Kernel-define-b} for $\mc{K}(\mi \bsomega_p)$ in \Eq{subeq:define-discrete-Fourier-transform-1}, we obtain 
\bal \nonumber 
& \beta \delta_{\i \omega_{1 \dots \ell}} G_p(\i\bs{\omega}_p) + G^{\mc{R}}_p(\i\bs{\omega}_p) 
\\ \label{eq:Gpstartingpoint}
& \;\; =
\beta \delta_{\i \omega_{1 \dots \ell}} \big[ K  \star G \big] (\i\bs{\omega}_p)
+ \big[ \mc{R}  \star G \big] (\i\bs{\omega}_p).
\eal
By construction, neither $G_p^{\mc{R}}$ nor $\mc{R}$ contain an overall factor of $\beta$; in this sense, 
they are $\mc{O}(\beta^0)$. Likewise, $\mc{R} \star G $ 
is $\mc{O}(\beta^0)$, for reasons explained below. 
Moreover, recall that MF-to-KF continuation via \Eq{eq:KF_corr_through_full_corr-b} requires only the regular part $\tG_p(\i \bsomega_p)$. We avoid anomalous contributions to $G_p(\i\bs{\omega}_p)$ in \Eq{eq:Gpstartingpoint} by imposing the condition $\i \omega_{\oli{1}\dots\oli{i}} \neq 0$ on the external frequencies. 
Setting $\i \omega_{1 \dots \ell}=0$, we conclude that
\bal 
\nonumber
 \tG_p(\i\bs{\omega}_p) + \mcO\bigl(\tfrac{1}{\beta}\bigr)
&=
\big[ K \star  G \big] (\i\bs{\omega}_p) , 
\qquad
(\i \omega_{\oli{1}\dots{\oli{i}}} \neq 0,\ \forall i < \ell) 
\\ \label{eq:tildeGpstart}
& \hspace{-0.5cm} = \tfrac{1}{(-\beta)^{\ell-1}} \sum_{\i\bs{\omega}'_p} 
\delta_{\i\omegap_{\oli1\dots\oli\ell}} K(\i\bs{\omega}_p-\i\bs{\omega}'_p) G_{\i\bs{\omega}'_{p}} .
\eal
To find $\tG_p(\i\bs{\omega}_p)$, we should thus 
compute $K \star  G$ with $\i \omega_{\oli{1}\dots{\oli{i}}} \neq 0$ and retain only the $\mc{O}(\beta^{0})$ terms, 
ignoring all $\mcO(1/\beta^{j\ge1})$ contributions. 
Note, however, that the full information on $K$ and $G$, including both regular and anomalous terms, is needed on the right-hand side to obtain $\tG_p$ on the left. 

Equation~\eqref{eq:tildeGpstart} is an important intermediate result. It provides a recipe
for extracting partial regular MF correlators from the full MF correlator by performing Matsubara sums $\sum_{\i\bs{\omega}'_p}$. 
After performing the sums, the final results will
be analytically continued to yield $\tG_p(\i\bs{\omega}_p\to \bsomega_p^{[\eta]})$
through which all Keldysh correlators can be expressed (\Eq{eq:KF_corr_through_full_corr-b}). However, we choose 
to fully evaluate the Matsubara sums \textit{before}
performing this analytic continuation. The reason is
that we will evaluate the sums using contour integration and contour deformation. For the latter step, it is convenient
if the arguments of $\tG_p(\i\bs{\omega}_p)$ all lie safely 
on the imaginary axis, where they do not impede contour deformation.

\subsection{Converting Matsubara sums to contour integrals}
\label{eq:Matsubarasums}

Next, we discuss three technical points relevant for performing Matsubara sums explicitly. To be concrete, we illustrate our general statements for the case $\ell=2$. Other cases are discussed in subsequent sections.

(i) \textit{Singularity-free kernels:}
The argument of  the kernel $K(\bsOmega_p)$ in \Eq{eq:tildeGpstart}  has the form $\bsOmega_p = \i\bsomega_p-\i\bsomega'_p$. This is always bosonic, being the difference of two same-type Matsubara frequencies. The Matsubara sums $\sum_{\mi \bsomega'_p}$ will thus contain 
terms with  $\Omega_{\oli{1} \dots \oli{i}} = 0$. 
To facilitate dealing with these, we assume that the kernel has been expressed in ``singularity-free'' form, where case distinctions ensure that factors of $1/\Omega_{\bonetoi}$ occur only if $\Omega_{\bonetoi} \neq 0$. This is  possible for the presented correlators, as shown in Ref.~\cite{Halbinger2023} and discussed in App.~\ref{app:explicit_MF_kernel_formulas}. 
These case distinctions are expressed via the symbol 
\bal
\label{eq:defineDeltaSymbol}
\Delta_{\Omega_{\bonetoi}} = 
\begin{cases}
\frac{1}{\Omega_\bonetoi} & \text{if} \;\;
\Omega_\bonetoi \neq 0 ,
\\
0 & \text{if} \;\;
\Omega_\bonetoi = 0 . 
\end{cases}
\eal
Thus, $K(\bs \Omega_p)$ is assumed to contain $1/\Omega_\bonetoi$ only via $\Delta_{\Omega_\bonetoi}$. 
A sum over a $\Delta$ symbol becomes a restricted sum, 
lacking the summand for which $\Delta=0$. 
For $\ell = 2$, e.g., we have 
$K(\bsOmega_p)  = 
\Delta_{\Omega_{\oli{1}}} - \tfrac{1}{2} \beta \delta_{\Omega_{\oli{1}}}$ 
(see \Eq{eq:2p_kern_Sbierski}), so that \Eq{eq:tildeGpstart} yields
\bal
\underset{\mi \omega_{\oli{1}} \neq 0}{\tG_p(\i\bs{\omega}_p)} + 
\mcO\bigl(\tfrac{1}{\beta}\bigr) & \stackrel{\ell=2}{=}
\frac{1}{(-\beta)} \sum_{\i{\omega}'_{\oli{1}}}^{\neq \i\omega_{\oli{1}}}
\frac{ G_{\mi \omegap_{\oli{1}}}}{\mi \omega_{\oli{1}} - \mi \omegap_{\oli{1}}}
+ \frac{G_{\mi {\omega}_{\oli{1}}} }{2}  .
\label{eq:sum_g2-example}  
\eal
This involves a restricted sum and an $\mc{O}(\beta^0)$ term resulting from 
$ \beta \delta_{\Omega_{\oli{1}}}$ collapsing the
sum $\tfrac{1}{(-\beta)}\sum_{\mi \omegap_{\oli{1}}}$ in \Eq{eq:tildeGpstart}.

(ii) \textit{$\beta\delta$ expansion of $G$:}
To facilitate the identification 
 of the leading-in-$\beta$ contributions to \Eq{eq:tildeGpstart}, we assume that the anomalous $\hG$ contribution to $G_{\mi \bsomega'_p} = (\tG + \hG)_{\mi \bsomega'_p}$ 
has been expressed as an expansion in powers of $\beta \delta_{\i{\omega}'_{\oli{1} \dots \oli{i}}}$. 
Such a $\beta \delta$ expansion is always possible for the correlators under consideration in this work, as discussed in \App{app:kernel_alt}.
Whenever $\beta \delta_{\i\omegap}$ appears in a Matsubara sum $\tfrac{1}{(-\beta)}\sum_{\i\omegap}$, the sum collapses and their $\beta$ factors cancel. (This cancellation
is why $\mc{R}\star G$ in \Eq{eq:Gpstartingpoint} is $\mc{O}(\beta^0)$, as stated above, even if $G$ contains anomalous terms.)  
For $\ell=2$, e.g.,
we have $G_{\i \omegap_{\oli{1}}} = \tG_{\i \omegap_{\oli{1}}} + \beta \delta_{\i \omegap_{\oli{1}}}\, \hG_{\oli{1}}$, with $\tG_{\i \omegap_{\oli{1}}}$ singularity-free at all Matsubara frequencies $\i \omegap_{\oli{1}}$ and $\hG_{\oli{1}}$ a constant (see \Eq{eq:Corr_Func_gen_2p}). Thus, \Eq{eq:sum_g2-example}  becomes
\bal 
\label{eq:example-Gp-explicit-2}
\underset{\mi \omega_{\oli{1}} \neq 0}{\tG_p(\i\bs{\omega}_p)}
+ \mcO\bigl(\tfrac{1}{\beta}\bigr) & \stackrel{\ell=2}{=}
\frac{1}{(-\beta)} \sum_{\i\omegap_{\oli{1}}}^{\neq \i\omega_{\oli{1}}}
\frac{\tG_{\mi \omegap_{\oli{1}}}}{\i \omega_{\oli{1}} - \i \omegap_{\oli{1}}} 
+ \frac{\tG_{\mi {\omega}_{\oli{1}}} }{2}  
 - \frac{\hG_{\oli{1}}}{\mi \omega_{\oli{1}}}  . 
\eal 
Here, the condition $\mi \omega_{\oli{1}} \neq 0$ on the left
was evoked 
to replace $\tfrac{1}{2}G_{\mi {\omega}_{\oli{1}}}$ by $\tfrac{1}{2}\tG_{\mi {\omega}_{\oli{1}}}$
on the right.

\bsubeq\label{subeq:Matsubarasums}
(iii) \textit{Converting sums to integrals:}
By restricting or collapsing Matsubara sums containing $\Delta$ or $\delta$ factors, one can ensure that the remaining sums are all of the form $\frac{1}{(-\beta)}\sum_{\i \omega'} f(\mi \omega')$ or 
$\frac{1}{(-\beta)}\sum_{\i \omega'}^{\neq \mi \omega} f(\mi \omega')$,
where $f(z)$, viewed as a function of $z\in\mathbbm{C}$,
is \textit{analytic} at each $\mi \omega'$
visited by the sum. 
(More precisely, for each $\mi \omega'$ in the sum, $f(z)$ is analytic in 
an open domain containing that $\mi \omega'$.) 
We express such sums in standard fashion as contour integrals:
\bal 
\label{eq:sumtocont} 
\frac{1}{(-\beta)} \sum_{\i \omega'} f(\i \omega')  & = \ointctrclockwise_z 
n_z f(z)\, , 
\\ 
\label{eq:sumtocont-restricted} 
\frac{1}{(-\beta)} \! \sum_{\i \omega'}^{\neq \mi \omega} f(\i \omega') & = \ointctrclockwise_z 
n_z f(z) - \underset{z=\mi \omega}{\mathrm{Res}} \big( n_z f(z) \bigr).
\eal
\esubeq
Here, $\ointctrclockwise_{z} = \ointctrclockwise \frac{dz}{2\pi \i}$
denotes counterclockwise integration around all 
points $\mi \omega'$ visited by the sum, and 
$n_z$ is a Matsubara weighting function (\MWF{}).
We choose it as
\bal
\label{eq:MWF_def}
& n_z = \frac{\zeta}{e^{-\beta z} - \zeta}
= \frac{1}{(-\beta)} \frac{1}{z-\i \omega'} - \frac{1}{2} + \mc{O}(z-\i \omega') ,
\eal
with $\zeta=\pm$  for bosonic/fermionic $\i \omega'$. 
($n_z$ is related to standard Fermi and Bose distribution functions by 
$-\zeta(1+n_z) = 1/(e^{\beta z} - \zeta)$.)
The Laurent expansion on the right of \Eq{eq:MWF_def} shows that $n_z$ has first-order poles with residues $1/(-\beta)$ at all Matsubara frequencies $\i \omega'$.
Therefore, the integral $\ointctrclockwise_z$ along a contour including all $\mi \omega'$ frequencies recovers the unrestricted Matsubara sum of \Eq{eq:sumtocont} (see left parts of Figs.~\ref{fig:contour}(b) and (c)).
For the restricted sum of \Eq{eq:sumtocont-restricted},
the first term on the right represents
an unrestricted sum, i.e.\ the restricted sum plus a contribution from $\mi \omega' = \mi \omega$, and the residue correction subtracts the latter. For example, consider the case, needed below, that 
$f(\mi \omega') = \tilde f(\mi \omega')/(\mi \omega - \mi \omega')$, 
with $\tilde f(z)$ analytic at $z=\mi \omega$. 
Then, $n_z f(z)$ has a pole of second order at $\mi \omega$, with 
\bal
\nonumber
\underset{z=\mi \omega}{\mathrm{Res}} 
\Big( \frac{n_z  \tilde f(z)}{\mi \omega - z} \Bigr)
& = \Bigl(\partial_z \big[(\mi \omega-z) n_z \tilde f(z) \big]
\Bigr)_{z \to \mi \omega} 
\\ 
\label{eq:residue} & 
= \tfrac{1}{2} \tilde f(\mi \omega)
+ \tfrac{1}{\beta} \Bigl( \partial_z \tilde f(z) \Bigr)_{z \to \mi \omega} \, . 
\eal

\bsubeq\label{sub:illustrate-cancellations}
Note that \Eqs{subeq:Matsubarasums} remain
valid under shifts of the MWF by a constant, $n_z \to n_z + c$. We purposefully exploited this freedom to choose $n_z$ to have $-\frac{1}{2}$ as the second term in the Laurent expansion. The reason is that this leads to a convenient cancellation between terms arising from a $\delta$ in $K$ and residue corrections arising from $\Delta$ restrictions. For example, when evaluating the Matsubara sum in 
\Eq{eq:example-Gp-explicit-2} using 
\Eqs{eq:sumtocont-restricted} with $f(\mi \omega') = \tG_{\mi 
\omega'}/(\mi \omega - \mi \omega')$, 
we obtain:
\bal \nonumber
& \underset{\mi \omega_{\oli{1}} \neq 0}{\tG_p(\i\bs{\omega}_p)}
+ \mcO\bigl(\tfrac{1}{\beta}\bigr) \\ 
\label{eq:tG12_after_deformation-a}
& \stackrel{\ell=2}{=} 
\ointctrclockwise_{z_{\oli{1}}} \frac{ n_{z_{\oli{1}}} \tG_{z_{\oli{1}}} }{\i \omega_{1}-z_{\oli{1}}} 
- \, \underset{z_{\oli{1}} = \i \omega_{\oli{1}}}{\tn{Res}} 
\left(\frac{ n_{z_{\oli{1}}} \tG_{z_{\oli{1}}} }{\i \omega_{\oli{1}}-z_{\oli{1}}} \right)
+ \frac{\tG_{\i \omega_{\oli{1}}}}{2}   - \frac{\hG_{\oli{1}}}{ \i \omega_{\oli{1}} }
\\
& \stackrel{\phantom{\ell=2}}{=} 
\ointctrclockwise_{z_{\oli{1}}} \frac{ n_{z_{\oli{1}}} \tG_{z_{\oli{1}}} }{\i \omega_{1}-z_{\oli{1}}} 
- \frac{1}{\beta} \Bigl( \partial_{z_{\oli{1}}} \tG_{z_{\oli{1}}} \Bigr)_{z_{\oli{1}} \to \mi \omega_{\oli{1}}}
   - \frac{\hG_{\oli{1}}}{ \i \omega_{\oli{1}} } .
\label{eq:tG12_after_deformation-b}
\eal
\esubeq
The $\frac{1}{2}\tG_{\i \omega_{\oli{1}}}$ term in \Eq{eq:tG12_after_deformation-a} conveniently cancels a contribution from the residue correction, evaluated using \Eq{eq:residue}. This cancellation results from our choice of $n_z$
having $-\frac{1}{2}$ in its Laurent expansion. (Similar cancellations occur for 
$\ell>2$; see, e.g.,\ \App{app:3p_Gp_regular_cont}.)
The $- \frac{1}{\beta} \bigl( \partial_z \tG_z \bigr)_{z \to \mi \omega}$
term in \Eq{eq:tG12_after_deformation-b} is an example of an $\orderbeta$ contribution that arises from $K\star G$ but is not part of $\tG_p$. 

Having worked through the example of $\ell=2$, we conclude this section with
some general remarks about \Eq{eq:tildeGpstart} for $\tG_p$.
Once the Matsubara sums from the imaginary-frequency convolution $K \star G$ have
 been expressed through contour integrals, one obtains the general form \footnote{We verified this form explicitly for $\ell \le 7$, using expressions for $K$ 
 derived in Ref.~\cite{Halbinger2023}, but expect it to hold for arbitrary $\ell$.}
\bal \label{eq:tG_p_cont_ints}
&\underset{\mi \omega_\bonetoi \neq 0}{\tG_p(\i \bsomega_p)} + \orderbeta \nn
&= \ointctrclockwise_{z_{\oli{1}}} \dots \ointctrclockwise_{z_{\oli{1} \dots \oli{\ell - 1}}} \tK(\i \bsomega_p - \bs{z}_p)\, n_{z_{\oli{1}}} \dots n_{z_{\oli{1} \dots \oli{\ell-1}}}\, \tG_{z_{\oli{1}}, \dots, z_{\oli{1} \dots \oli{\ell-1}}}
\nn
& \hsp + \tn{contributions from $\hG$},
\eal
Here, the $(\ell-1)$-fold contour integrals involve only 
the \textit{regular} part, $\tG$, of the full MF correlator. 
Its anomalous part, $\hG$, comes with factors $\beta \delta$
that collapse one or multiple sums in \Eq{eq:tildeGpstart}. Therefore, 
contributions from $\hG$ to $\tG_p$ contain at most $\ell - 2$ contour integrals. 

The next step, discussed in detail in \Sec{sec:2p_partial_correlators}, is to 
deform the integration contour in such a way that
it runs infinitesimally above and below the real axis. 
The anomalous contributions from $\hG$ can then be reincorporated
into the real integrals using bosonic Dirac delta functions. As a result,
one recovers precisely the form $\tG_p = \tK \ast S_p$ of the spectral representation \eqref{eq:MF_G-compact-b}: regular kernels $\tK$ convolved with other functions, built from \MWFs{} and analytic continuations of the various components of $\tG$ and $\hG$, the latter multiplied by bosonic Dirac $\delta$ functions. These other functions can thus be identified with the PSFs $S_p = \tS_p + \anomS_p$, now expressed through analytic continutions of $G$.
This clarifies, on a conceptual level, how the information contained in the full MF correlator $G$ needs to be repackaged to obtain PSFs, and the explicit formulas for $\ell=2,3,4$ in \Eqs{eq:Sp_2p_final}, \eqref{eq:Sp_3p_final}, and \eqref{eq:4p:Sp} constitute the main results of this paper. 
These, in turn, can then be used to obtain KF correlators via \Eq{eq:KF_corr_through_full_corr-a}.

To summarize, the MF-to-KF analytic continuation of arbitrary $\ell$p correlation functions
can be achieved via the following three-step strategy:\label{p:3step-recipe}
\begin{itemize}
    \item[Step 1.] \textit{Matsubara summation through contour integration:} 
    Insert the MF kernel $K$ (expressed in singularity-free form) and the MF correlator $G$ (expressed as a $\beta \delta$ expansion), including all regular and anomalous contributions, into \Eq{eq:tildeGpstart}
    for $\tG_p$. Restrict or collapse Matsubara sums containing $\Delta$ or $\beta \delta$  factors and express the remaining sums through contour integrals using \Eqs{subeq:Matsubarasums}, to arrive at \Eq{eq:tG_p_cont_ints}. 
    
    \item[Step 2.] \textit{Extraction of PSFs:} Deform the contours away from the imaginary axis to run along the real axis, while carefully tracking possible singularities of the MF correlators. Reincorporate anomalous contributions
    via bosonic Dirac delta functions. This results in a spectral representation 
    of the form $\tG_p = \tK \ast S_p$. From this, read off
    the PSFs $S_p[G]$, expressed through products of MWFs and MF correlators, analytically continued to real frequencies (see, e.g., \Eq{eq:Sp_2p_final}).

    \item[Step 3.] \textit{Construction of KF correlators:} Construct the full KF correlator $G^{[\eta_1 \dots \eta_\alpha]}$, involving a sum $\sum_p $ over terms of the form $\bigl[K^{[\heta_1 \dots \heta_\alpha]}  \ast S_p\bigr](\mi \bsomega_p)$ (\Eq{eq:KF_corr_through_full_corr-a}). 
    Simplify the kernels $K^{[\heta_1 \dots \heta_\alpha]}$ via a set of kernel identities (see, e.g., \Eqs{subeq:K12-kernels-explicit}) and combine terms with similar structure from the sum $\sum_p$. Insert into the resulting expressions the PSFs from Step 2, and then compute the integrals involved in the $\ast$ convolution. This leads to equations expressing KF correlators through analytically continued MF correlators, $G^{[\eta_1\dots\eta_\alpha]}[G]$.
\end{itemize}
The result of Step~2 already constitutes an analytic continuation since the PSFs $S_p$ suffice to construct the KF correlators via the spectral representation.
Step 3 serves to give direct relations between both formalisms.

In \App{app:consistency_full_recovery_of_Sp}, we follow an independent approach and use the equilibrium condition to explicitly perform the following consistency check: given an arbitrary set of PSFs $S_p$ as input, compute the MF correlator $G = \sum_p K * S_p$ and verify that the formulas $S_p[G]$ correctly recover the input PSFs from $G$, giving $S_p[G]=S_p$.
This consistency check is presented for general 2p and 3p and for fermionic 4p correlators.

The next sections are devoted to 
explicitly working out the details of this strategy.
To demonstrate its basic ideas, we first revisit the well-known 2p case in the following section.
Though that is textbook material, we present it in a manner that readily generalizes to the higher-point correlators
discussed in subsequent sections:  3p correlators in \Sec{sec:Analytic_cont_3p} and 4p correlators in \Sec{sec:Analytic_cont_4p}.

\section{Analytic continuation of 2p functions}
\label{sec:Analytic_cont_2p}

In this section, we carry through the strategy outlined in \Sec{eq:Matsubarasums} to obtain the MF-to-KF analytic continuation in the well-known 2p case. 
While our strategy may seem more cumbersome than traditional textbook discussions (see, e.g., Ref.~\cite{Wen2004}), it has the merit of readily generalizing to $\ell>2$.
We first recapitulate the spectral representation and analytic properties of general 2p MF correlators (\Sec{sec:2p_MF_correlators}). Then, we express the PSFs in terms of analytically continued MF correlators (\Sec{sec:2p_partial_correlators}).
Finally, we use these to recover familiar expressions for the retarded, advanced, and Keldysh components of the KF 2p correlator (\Sec{sec:2p_Keldysh_correlator}).

\subsection{Analytic properties of the 2p MF correlator}
\label{sec:2p_MF_correlators}
We begin by reviewing well-known analytical properties of
the $2$p MF correlator. This also serves to give concrete examples for our notational conventions.

$G(\i \bsomega) = G(\i\omega_1,\i\omega_2)$ explicitly depends on one Matsubara frequency, $\i\omega_1$ or $\i\omega_2$,  while the other frequency is fixed by energy conservation, $\i \omega_{12} = 0$.
Since we want to compute \Eq{eq:tildeGpstart} for arbitrary permutations $p=(\oli{12})$, it proves useful to develop an unbiased notation for the frequency dependence.
The chosen explicit frequency dependence is indicated by a subscript in $G_{\i\omega_{\oli{1}}}$, such that 
$G_{\i\omega_1}=G(\i\bs\omega(\omega_1)) = G(\i\omega_1,-\i\omega_1)$ and 
$G_{\i\omega_2}=G(\i\bs\omega(\omega_2)) = G(-\i\omega_2, \i\omega_2)$. The most general form of $G_{\i\omega_{\oli{1}}}$, covering both fermionic and bosonic cases, reads
\bal \label{eq:Corr_Func_gen_2p}
    G(\i \bsomega(\omega_{\oli{1}})) = G_{\i \omega_{\oli{1}}} = \tG_{\i \omega_{\oli{1}}} + \beta \delta_{\i \omega_{\oli{1}}}\, \hG_{\oli{1}},
\eal
in agreement with the general form \Eq{eq:MF_correlators_general_form}.
The regular part, $\tG_{\i \omega_{\oli{1}}}$, is 
singularity-free for all $\mi \omega_{\oli{1}}$, including 
0. 
$\hG_{\oli{1}}$ denotes the anomalous part, a constant, contributing only for $\i \omega_{\oli{1}} = 0$. The relation $G_{\i \omega_1} = G_{\i \omega_2}$ enforces $\hG_1 = \hG_2$.

\begin{figure*}[t!]
    \centering
    \includegraphics[width=1 \textwidth]{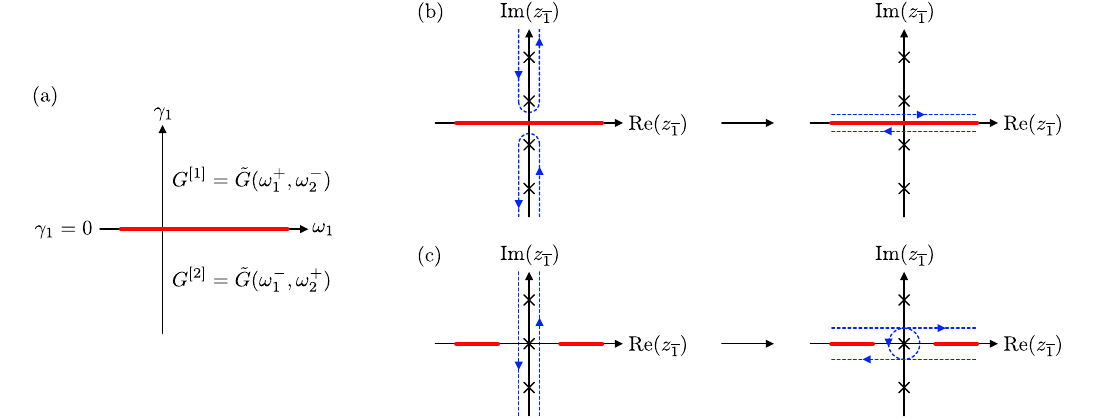}\vspace{-5mm}
    \caption{(a) Analytic regions of a regular 2p MF correlator as a function of a complex frequency $\omega_1 + \i \gamma_1$ with $\omega_1, \gamma_1 \in \mathbbm{R}$. The thick, red line on the real axis depicts a possible branch cut of the correlator. (b,c) Contours to evaluate the Matsubara summation in the (b) fermionic and (c) bosonic case, see \Eqs{eq:Contour_deform_fermion} and \eqref{eq:Contour_deform_boson}, respectively. Crosses indicate the poles of the \MWF{} $n_{z_{\oli{1}}}$ at the Matsubara frequencies on the imaginary axis. The dashed blue contours, initially enclosing all Matsubara frequencies, are deformed away from the imaginary axis to run infinitesimally above and below the real axis. In the bosonic 2p case (c), the branch cut does not extend to $z_{\oli{1}} = 0$ as the correlator, by definition, is free of any singularities at vanishing Matsubara frequencies.
    }
    \label{fig:contour}
\end{figure*}

One of the next steps involves the deformation of the integration contour 
$\ointctrclockwise_{z_{\oli{1}}}$ from the imaginary axis toward the real axis. This requires knowledge of the analytic structure of the MF correlator. It can be made explicit via the spectral representation of $G_{z_1}$ (\Eqs{eq:MF_G-compact}), 
with the PSFs $S_p$ viewed as input. For the regular part, we obtain
\bal
\label{eq:tG_2p_specRep}
\tG_{ z_1 } 
&= 
\!\!\int\!\!\! \newlychanged{\md}^2\varepsilon\, \delta(\varepsilon_{12})
\Bigg[\frac{S_{(12)}(\varepsilon_1)}{z_1 - \varepsilon_1} +  \frac{S_{(21)}(\varepsilon_2)}{-z_1 - \varepsilon_2} \Bigg]
=
\!\!\int\!\!\! \newlychanged{\md}\varepsilon_1\, \frac{S_\mathrm{std}(\varepsilon_1)}{z_1 - \varepsilon_1} .
\eal
Here, we introduced the 
``standard'' spectral function $S_\mathrm{std}$, given by a  commutator of PSFs resulting from the sum over the two permutations
$p=(12)$ and $(21)$:
\bsubeq
\label{eq:Sstd_PSF_anticom_2p}
\bal
\label{eq:Sstd}
S_\mathrm{std}(\varepsilon_1) &= S_{[1,2]_-}(\varepsilon_1,-\varepsilon_1) = S_{(12)}(\varepsilon_1) - S_{(21)}(-\varepsilon_1), \\
\label{eq:PSF_anticom_2p}
S_{[1,2]_\pm}(\bs{\varepsilon}) &= S_{(12)}(\varepsilon_1) \pm S_{(21)}(\varepsilon_2) .
\eal
\esubeq
Here, $\varepsilon_{12}= 0$ is understood for the argument of $S_{[1,2]_\pm}(\bs{\varepsilon})$. 
For PSF (anti)commutators, we always display the unpermuted $\bs{\varepsilon}$ and insert the permuted $\bs{\varepsilon}_p$ only for individual PSFs, as done on the right of \Eq{eq:PSF_anticom_2p}.
Evidently, $\tG_{ z_1 }$ has poles (or branch cuts for continuous spectra) whenever the denominator $z_1 - \varepsilon_1$ vanishes. This can happen only if $\tn{Im}( z_1 ) = 0$ (or, more generally, $\tn{Im}( z_{\oli{1}} ) = 0$), indicated in Fig. \ref{fig:contour} by thick, red lines on the real axis. Hence, the upper and the lower complex half plane are analytic regions of $\tG_{z_1}$, separated by a branch cut at $\Im(z_1)=0$.

\subsection{Extraction of PSFs from partial MF correlators}
\label{sec:2p_partial_correlators}
    
In \Sec{eq:Matsubarasums}, we expressed the regular
partial MF correlators $\tG_p(\mi \bsomega_p)$ for $\ell=2$ in terms of a contour integral $\ointctrclockwise_{z_{\oli{1}}}$ involving the regular MF correlator $\tG_{z_{\oli{1}}}$, see \Eq{eq:tG12_after_deformation-b}. That amounted to 
Step~1 of the 3-step strategy. 
Turning to Step~2,
we write $\tG_p (\mi \bsomega_p)$ 
in the form of a convolution $[\tK \ast S_p] (\mi \bsomega_p)$,
from which we then read out expressions for the PSFs $S_p[G]$.

To this end, we exploit the analyticity of $\tG_{z_{\oli{1}}}$ in the upper and lower half-plane to deform the contours in $\ointctrclockwise_{z_{\oli{1}}}$ 
from enclosing the imaginary axis to running infinitesimally above and below the 
branch cut. We denote the corresponding integration variables along the branch cut by ${\varepsilon_{\oli{1}}^{\pm} = \varepsilon_{\oli{1}}\pm \i 0^+}$, with $\varepsilon_{\oli{1}} = \tn{Re}(z_{\oli{1}})$ now being a real variable and $\pm \mi 0^+$ infinitesimal shifts. 

We discuss the cases of fermionic or bosonic frequencies separately. 
For fermions, the contour deformation of 
$\ointctrclockwise_{z_{\oli{1}}}$ in \Eq{eq:tG12_after_deformation-b} 
is straightforward and yields (see Fig. \ref{fig:contour}(b))
    \bal \label{eq:Contour_deform_fermion}
    \ointctrclockwise \frac{ \newlychanged{\md} z_{\oli{1}} }{2\pi \i} \frac{ n_{z_{\oli{1}}} \tG_{z_{\oli{1}}} }{\i \omega_{\oli{1}}-z_{\oli{1}}} 
    = 
    \int_{-\infty}^\infty \frac{\newlychanged{\md}\varepsilon_{\oli{1}}}{2\pi \i} \frac{ n_{\varepsilon_{\oli{1}}} \tG^{\varepsilon_{\oli{1}}} }{\i \omega_{\oli{1}}-\varepsilon_{\oli{1}}}. 
    \eal
Here, we defined $\tG^{\varepsilon_{\oli{1}}} = \tG_{\varepsilon_{\oli{1}}^+} - \tG_{\varepsilon_{\oli{1}}^-}$ as the discontinuity of $\tG_{z_{\oli{1}}}$ across the branch cut at $\Im(z_{\oli{1}})=0$. 
Moreover, we extended the subscript notation introduced after \Eq{subeq:define-discrete-Fourier-transform-2} to real frequencies with infinitesimal imaginary shifts. (This notation is further discussed after \Eq{eq:Sp_2p_final}.) 

In the bosonic case, the pole at $z_{\oli{1}} = 0$ has to be treated separately
(see Fig.~\ref{fig:contour}(c)):
\bal \label{eq:Contour_deform_boson}
\ointctrclockwise \frac{ \tn{d}z_{\oli{1}} }{2\pi \i} \frac{ n_{z_{\oli{1}}} \tG_{z_{\oli{1}}} }{\i \omega_{\oli{1}}-z_{\oli{1}}} 
&=
\pint{-\infty}{\infty} \frac{\tn{d}\varepsilon_{\oli{1}}}{2\pi \i} \frac{ n_{\varepsilon_{\oli{1}}} \tG^{\varepsilon_{\oli{1}}} }{\i \omega_{\oli{1}}-\varepsilon_{\oli{1}}} 
+ 
\, 
\underset{z_{\oli{1}} = 0}{\mathrm{Res}} \Bigg( \frac{ n_{z_{\oli{1}}} \tG_{z_{\oli{1}}} }{\i \omega_{\oli{1}}-z_{\oli{1}}} \Bigg)
\nn
&= \pint{-\infty}{\infty} \frac{\tn{d}\varepsilon_{\oli{1}}}{2\pi \i} \frac{ n_{\varepsilon_{\oli{1}}} \tG^{\varepsilon_{\oli{1}}} }{\i \omega_{\oli{1}}-\varepsilon_{\oli{1}}} + \orderbeta.
\eal
Here, $\pinttext$ indicates a principal-value integral. The residue evaluates to a contribution of order $\orderbeta$ as the bosonic \MWF{} $n_{z_{\oli{1}}}$ is the only factor having a pole at $z_{\oli{1}} = 0$, with residue $1/(-\beta)$ there (remember that $\i \omega_{\oli{1}} \neq 0$). Combining \Eqs{eq:Contour_deform_fermion}, \eqref{eq:Contour_deform_boson},
 \eqref{eq:tG12_after_deformation-b} and omitting $\orderbeta$ terms, we finally find
    \bal \label{eq:tG_2p_p12}
    \tG_{p}(\i \bsomega_p) = \int_{\varepsilon_{\oli{1}}} \frac{ n_{\varepsilon_{\oli{1}}} \tG^{\varepsilon_{\oli{1}}} }{\i \omega_{\oli{1}}-\varepsilon_{\oli{1}}} - \frac{\hG_{\oli{1}}}{ \i \omega_{\oli{1}} } = \int_{\varepsilon_{\oli{1}}} \frac{ n_{\varepsilon_{\oli{1}}} \tG^{\varepsilon_{\oli{1}}} + \hdelta(\varepsilon_{\oli{1}}) \hG_{\oli{1}}}{\i \omega_{\oli{1}}-\varepsilon_{\oli{1}}} .
    \eal
On the right, we absorbed the anomalous $\hG$ contribution
into the integral, defining
$\hdelta(\varepsilon_{\oli{1}}) = -2\pi \i\, \delta(\varepsilon_{\oli{1}})$. 
Moreover, we introduced the symbol $\int_{\varepsilon_i}$ as
\bal 
\label{eq:def_int}
\int_{\varepsilon_i} ... =
\begin{cases}
{\displaystyle \int_{-\infty}^{\infty} \frac{\tn{d}\varepsilon_i}{2\pi \i} ...} \quad &
\begin{tabularx}{\linewidth}{XX}
for fermionic $\varepsilon_i$ \newline
or anomalous frequency,
\end{tabularx} \\[15pt]
{\displaystyle{\pint{-\infty}{\infty}} \frac{\tn{d}\varepsilon_i}{2\pi \i} ... }\quad &
\begin{tabularx}{\linewidth}{XX}
for bosonic $\varepsilon_i$ \newline
and regular frequency.
\end{tabularx}
\end{cases}
\eal
We call a frequency $\varepsilon_i$ \textit{anomalous} if it 
is directly set to zero by a Dirac $\hdelta(\varepsilon_i)$ in the integrand,
and \textit{regular} otherwise. Since the anomalous contribution arose from
a Kronecker $\delta_{\mi \omega_{\oli{1}}}$, we arrive at a rule of thumb:
when performing Matsubara sums via contour integration and contour deformation
to the real axis, Kronecker deltas with Matsubara arguments lead to Dirac deltas with real arguments.

Importantly, \Eq{eq:tG_2p_p12} has precisely the same form as \Eq{eq:MF_G-compact-b} for $\ell = 2$, with the correspondence
    \bal \label{eq:Sp_2p_final}
    (2\pi \i) S_p(\varepsilon_{\oli{1}}) = n_{\varepsilon_{\oli{1}}} \tG^{\varepsilon_{\oli{1}}} + \hdelta(\varepsilon_{\oli{1}}) \hG_{\oli{1}}.
    \eal
This remarkable formula is the first central result of this section: it shows that a suitable analytic continuation of the MF correlator $G(\i\bsomega)$, combined with a MWF, fully determines the PSF and thus, via the spectral representation \Eqs{eq:KF_corr_through_full_corr-a}, the KF correlator $G^{\bs{k}}$. It also clarifies the role of anomalous contributions. In subsequent sections, we will find analogous results for $\ell = 3,4$.

To conclude this section, we elaborate on the meaning of
the super- and supscript notation used above.
The discontinuity in \Eq{eq:Sp_2p_final}, $\tG^{\varepsilon_{\oli{1}}} = \tG_{\varepsilon^+_{\oli{1}}} - \tG_{\varepsilon^-_{\oli{1}}}$, consists of analytically continued MF correlators, $\tG(\i \bsomega) \rightarrow \tG(
\bs{z})$. Here, the entries of $
\bs{z} = (\varepsilon^\pm_1, \varepsilon^{\mp}_2)$ are infinitesimally shifted by $+ \i 0^+$ or $-\i0^+$, but constrained by energy conservation, $\varepsilon_{12} = 0$. The subscript on $\tG_{\varepsilon^\pm_{\oli{1}}}$ 
has the same meaning as for imaginary frequencies (see paragraph after \Eq{subeq:define-discrete-Fourier-transform-2}): it indicates the chosen explicit (real-)frequency dependence of $\tG(
\bs{z})$, i.e., $\tG_{\varepsilon^\pm_{\oli{1}}} = \tG\big(
\bs{z}(\varepsilon^\pm_{\oli{1}}) \big)$, uniquely determining the imaginary shifts in each entry of $
\bs{z}$.
To be explicit, we have  
\bsubeq
\label{eq:explicit_discontinuity_tGi}
\bal
\label{eq:explicit_discontinuity_tG1}
\tilde{G}^{\varepsilon_1}
&=
\tilde{G}(\varepsilon_1^+,-\varepsilon_1^+)-\tilde{G}(\varepsilon_1^-,-\varepsilon_1^-),
\\
\label{eq:explicit_discontinuity_tG2}
\tilde{G}^{\varepsilon_2}
&=
\tilde{G}(-\varepsilon_2^+,\varepsilon_2^+)-\tilde{G}(-\varepsilon_2^-,\varepsilon_2^-).
\eal
\esubeq
Since $\varepsilon_2=-\varepsilon_1$ (energy conservation) and  
hence $\varepsilon_2^+=-\varepsilon_1^-$, 
we have $\tG^{\varepsilon_1} = -\tG^{\varepsilon_2} = \tG^{-\varepsilon_2}$.
(Check for negative superscripts: $\tG^{-\varepsilon_2} = \tG_{(-\varepsilon_2)^+} - \tG_{(-\varepsilon_2)^-}= \tG_{-\varepsilon_2^-} - \tG_{-\varepsilon_2^+} = -\tG^{\varepsilon_2}$.)

For illustration, we give explicit formulas for $S_p$ for the permutations $p=(12)$ and $p=(21)$,
\bal
\nonumber
(2 \pi \mi) S_{(12)}(\varepsilon_1)
& =
n_{\varepsilon_1} [\tilde{G}(\varepsilon_1^+,-\varepsilon_1^+)-\tilde{G}(\varepsilon_1^-,-\varepsilon_1^-)]
+\hdelta(\varepsilon_1) \hG_1.
\\
\label{eq:S21_2p_final}
(2 \pi \mi) S_{(21)}(\varepsilon_2)
& =
n_{\varepsilon_2} [\tilde{G}(-\varepsilon_2^+,\varepsilon_2^+)-\tilde{G}(-\varepsilon_2^-,\varepsilon_2^-)]
+\hdelta(\varepsilon_2) \hG_2,
\eal
where we inserted \Eq{eq:explicit_discontinuity_tGi} for the discontinuities. 
The anomalous contributions satisfy $\hG_1=\hG_2$ (as explained after \Eq{eq:Corr_Func_gen_2p}) and exist only for bosonic correlators ($\zeta=1$).
Energy conservation $\varepsilon_2=-\varepsilon_1$ then gives
\bal
\label{eq:cyc_rel_2p}
(2\pi\i)S_{(21)}(-\varepsilon_1)
& =
n_{-\varepsilon_1} [\tilde{G}(\varepsilon_1^-,-\varepsilon_1^-)-\tilde{G}(\varepsilon_1^+,-\varepsilon_1^+)] + \hdelta(\varepsilon_1) \hG_2\nn
&= \zeta e^{-\beta \varepsilon_1} (2\pi \i)S_{(12)}(\varepsilon_1).
\eal 
For the last step, we used the identity $- n_{-\varepsilon_1} = \zeta e^{-\beta \varepsilon_1} n_{\varepsilon_1}$.  
As a useful consistency check, we note that \EQ{eq:cyc_rel_2p} 
corresponds to the equilibrium condition  \Eq{eq:equilibrium:cyclic_PSFs} for PSFs (with $p = (21)$, 
$p_{\lambda} = (12)$ there, implying $\zeta_p = \zeta$, 
$\zeta_{p_{\lambda}} = +1$ and $\varepsilon_{\oli{1}} = \varepsilon_2 = -\varepsilon_1$, $\varepsilon_{p_\lambda(1)} = \varepsilon_1$). 

Expressing the standard spectral function $S_\mathrm{std}(\varepsilon_1)$
from \Eq{eq:Sstd} in terms of \Eq{eq:Sp_2p_final},
we 
find
\bal
\label{eq:Sstd_as_discontinuity}
(2\pi\i) S_\mathrm{std}(\varepsilon_1) 
&=
n_{\varepsilon_1} \tG^{\varepsilon_1} + \hdelta(\varepsilon_1) \hG_{1}
-
n_{-\varepsilon_1} \tG^{-\varepsilon_1} - \hdelta(-\varepsilon_1) \hG_{2}
\nn
& =
n_{\varepsilon_1} \tG^{\varepsilon_1}
-
n_{-\varepsilon_1} \tG^{-\varepsilon_1} 
=
(n_{\varepsilon_1}+n_{-\varepsilon_1}) \tG^{\varepsilon_1}
\nn
& =
-\tG^{\varepsilon_1}
,
 \eal
where we used $\tG^{-\varepsilon_1} = - \tG^{\varepsilon_1}$.
Thus, the discontinuity $\tG^{\varepsilon_{\oli{1}}}$ in the PSFs~\eqref{eq:Sp_2p_final} encodes $S_\mathrm{std}(\varepsilon_1)$.
Conversely, however, $S_\mathrm{std}(\varepsilon_1)$ retains only the discontinuity $\tG^{\varepsilon_{\oli{1}}}$ in the PSFs~\eqref{eq:Sp_2p_final}, while 
the information on the MWF and the anomalous part, both contained in the $S_p$ \eqref{eq:S21_2p_final}, is lost.
In \App{app:consistency_full_recovery_of_Sp}, we use \Eq{eq:Sstd_as_discontinuity} and the equilibrium condition to explicitly perform the following consistency check:
given an arbitrary set of PSFs as input, compute the MF correlator 
$G = \sum_pK \ast S_p$ and verify that \Eq{eq:Sp_2p_final} for $S_p$ correctly recovers the input PSFs.

\subsection{Keldysh correlator}
\label{sec:2p_Keldysh_correlator}

Next, we turn to Step~3 of our 3-step strategy: we use the 
PSFs obtained above to explicitly construct the Keldysh
components $G^{[1]}$, $G^{[2]}$, and $G^{[12]}$, expressed through analytically continued MF correlators.
As the structure of KF correlators becomes more intricate with an increasing number of 2's in the Keldysh component, denoted by $\alpha$ in \Eqs{eq:KF_spec_rep_eta_alpha}, we discuss the different values of $\alpha$ separately in the following and throughout the rest of the paper.

\subsubsection{Keldysh components $G^{[\eta]}$}
\label{sec:2p_Gretarded}

 For $\alpha=1$, the fully retarded or fully advanced Keldysh components 
 $G^{[\eta]}$ can be deduced from the regular part of MF correlators alone
 (\Eq{eq:Analytic_cont_fully_ret}). Here, 
 we follow the alternative and equivalent strategy
 of Step 3: we insert the PSFs from \Eq{eq:Sp_2p_final}
 into the spectral representation  \eqref{eq:KF_corr_through_full_corr-a}:
\bsubeq\label{subeq:derive-G-eta-by-analytic-continuation}
\bal 
G^{[\eta]}(\bsomega) &= 
\sum_p \bigl[K^{[\heta]} \ast S_p\bigr] (\bsomega_p) 
= \sum_p \bigl[\tK \ast S_p\bigr] \bigl(\bsomega^{[\eta]}_p\bigr)
\\ & = 
\int \newlychanged{\md}^2\varepsilon\, \delta(\varepsilon_{12}) \left( \frac{S_{(12)}(\varepsilon_1)}{\omega^{[\eta]}_{1} - \varepsilon_1} + \frac{S_{(21)}(\varepsilon_2)}{\omega^{[\eta]}_{2} - \varepsilon_2} \right) \\
\label{eq:Geta_2p_firststep}
&=
\int \newlychanged{\md}\varepsilon_1 \frac{S_{[1,2]_-}(\varepsilon_1, -\varepsilon_1)}{\omega^{[\eta]}_{1} - \varepsilon_1} .
\eal
\esubeq
Here,  we used
$\omega^{[\eta]}_{2}=-\omega^{[\eta]}_{1}$ (\Eq{eq:imshifts})
and that the sum over both permutations, 
$p=(12)$ and $(21)$, leads to
the appearance of the PSF commutator $S_{[1,2]_-}$ (equalling $S_\mathrm{std}$, cf.\ \Eq{eq:Sstd_PSF_anticom_2p}).

\begin{figure*}[t!]
		\centering
		\includegraphics[width=1 \textwidth]{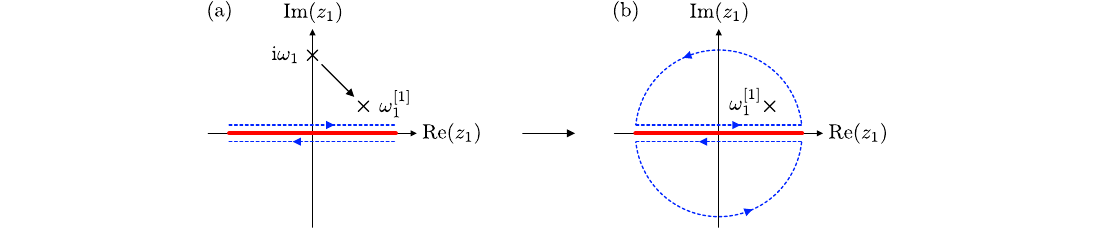}
		\caption{(a) Analytic continuation of the Matsubara frequency $\i \omega_1 \rightarrow \omega^{[1]}_1
        = \omega_1 + \mi \gamma_0$ in \Eqs{subeq:derive-G-eta-by-analytic-continuation} 
        for fermionic frequencies. The imaginary part of the external frequency $\omega^{[1]}_1$ has to be larger than the imaginary parts of $\varepsilon^\pm_1$ used to integrate infinitesimally above and below the real axis. The transition from (a) to (b) illustrates the closing of the contour in the upper/lower half-planes to evaluate the integral in \Eq{eq:Geta_2p_before_contdef}. As the integrand is independent of the fermionic \MWF{} $n_{\varepsilon_1}$, the only contribution to the integral originates from the simple pole at $z_1 = \omega^{[1]}_1$. 
        } 
		\label{fig:Analytic_cont_and_def}
	\end{figure*}

Before proceeding, a general remark is in order: When the external variables $\bsomega^{[\eta]}_p$ appear in $\ast$ convolution integrals such as $\int_{\varepsilon_1}$ in \Eqs{subeq:derive-G-eta-by-analytic-continuation}, it is essential to maintain the hierarchy $\gamma_0 \gg 0^+$ for the infinitesimal imaginary shifts $\pm \mi \gamma_0$ and $\pm \mi 0^\pm$ contained in the external frequencies $\bsomega^{[\eta]}_p$ and the integration variables $\varepsilon^\pm_1$, respectively. The reason is that the contour deformation from $\ointctrclockwise_{z_1}$ to $\int_{\varepsilon_1}$ \vspace{-1mm} has been performed \textit{before} the analytic continuation $\mi \bsomega_p \to \bsomega_p^{[\eta]}$ underlying \Eqs{eq:KF_corr_through_full_corr} and leading to  
\Eq{subeq:derive-G-eta-by-analytic-continuation} (see \Fig{fig:Analytic_cont_and_def}(a)). This hierarchy is particularly relevant for principle-value integrals 
$\pinttext$ (needed below); these exclude an interval $[-0^+, 0^+]$ around the origin, and $\gamma_0$ must lie outside this interval.

Inserting $S_{[1,2]_-}(\varepsilon_1, -\varepsilon_1) = S_\mathrm{std}(\varepsilon_1) = \tG^{\varepsilon_1}/(-2\pi\i)$ (from \Eqs{eq:Sstd} and \eqref{eq:Sstd_as_discontinuity}), we find
\bal \label{eq:Geta_2p_before_contdef}
G^{[\eta]}(\bsomega) 
= - \int_{\varepsilon_1} \frac{ \tG^{\varepsilon_{1}} }{\omega^{[\eta]}_{1}-\varepsilon_{1}} 
= - \int_{\varepsilon_1} \frac{ \tG_{\varepsilon_{1}^+} - \tG_{\varepsilon_{1}^-} }{\omega^{[\eta]}_{1}-\varepsilon_{1}} 
= \tG_{\omega^{[\eta]}_1} . 
\eal
Importantly, no MWFs $n_{\varepsilon_1}$ occur in \Eq{eq:Geta_2p_before_contdef}. For the last step, we were thus able to close the forward (backward) integration contour 
involving $\tG_{\varepsilon_{1}^+}$ ($\tG_{\varepsilon_{1}^-}$)
in the upper (lower) half-plane. 
We then used Cauchy's integral formula for the simple pole at $\omega^{[\eta]}_1$ (see Fig. \ref{fig:Analytic_cont_and_def}(b)).
Equation~\eqref{eq:Geta_2p_before_contdef} expresses the fully retarded
Keldysh correlators through analytic continuations of MF correlators, $G^{[\eta]}[G]$, as desired. To make contact with standard notation, we recall that the retarded and advanced 2p components are given by $G^R = G^{21} = G^{[1]}$ and $G^A = G^{12} = G^{[2]}$.
Reinstating frequency dependencies, with $\omega^{[1]}_1  
= \omega_1 + \mi \gamma_0 \equiv \omega_1^+$ and $\omega^{[2]}_1 
= \omega_1 - \mi \gamma_0 \equiv \omega_1^-$, we get
\bal \label{eq:2p_ret_adv}
G^{R}(\bsomega) = \tG(\omega^+_1, \omega^-_2), \quad
G^{A}(\bsomega) = \tG(\omega^-_1, \omega^+_2).
\eal
This implies the well-known relation
\bal
\label{eq:GRprimed_equals_GA}
G'^{ R}(\bsomega) = G^A(\bsomega), 
\quad
G'^{ A}(\bsomega) = G^R(\bsomega).
\eal

    \subsubsection{ Keldysh component $G^{[12]}$}
\label{sec:2p_G22}

For $\alpha=2$, both Keldysh indices equal 2, 
$G^{22} = G^{[12]} $.
Then, the spectral representation in \Eq{eq:KF_corr_through_full_corr-a}
requires the kernel (\Eq{eq:KFkerneletas})
\bal
\label{eq:K12=K1-K2}
K^{[\heta_1 \heta_2]}(\bsomega_p) = \bigl(K^{[\heta_1]} \!-\! K^{[\heta_2]}\bigr)(\bsomega_p)
= \tK \bigl(\bsomega_p^{[\bar{\heta}_1]}\bigr)
\!-\! \tK\bigl(\bsomega_p^{[\bar{\heta}_2]}\bigr),  
\eal
for the case $[\eta_1 \eta_1] = [12] = [\heta_1 \heta_2] $.
Evaluating this for $p=(12)$ and $(21)$, we find
\bsubeq\label{subeq:K12-kernels-explicit}
\bal \nonumber
K^{[12]}(\bs{\omega}_{(12)}) &= \tK(\bs{\omega}^{[1]}_{(12)}) - \tK(\bs{\omega}^{[2]}_{(12)}) \\
&= \frac{1}{\omega^{[1]}_1} - \frac{1}{\omega^{[2]}_1} = \frac{-2i\gamma_0}{\omega^2_1 + \gamma^2_0} = \hdelta_{\gamma_0}(\omega_1),
\label{eq:2p_kern_22_p}
\\
\nonumber 
K^{[12]}(\bs{\omega}_{(21)}) &= \tK(\bs{\omega}^{[2]}_{(21)}) - \tK(\bs{\omega}^{[1]}_{(21)}) \\
&= \frac{1}{\omega^{[2]}_2} - \frac{1}{\omega^{[1]}_2} = \frac{-2i\gamma_0}{\omega^2_2 + \gamma^2_0} = \hdelta_{\gamma_0}(\omega_1).
\eal
\esubeq
On the right, we introduced a Lorentzian representation of a 
broadened Dirac delta function:
    \bal     \label{eq:hdelta_identity}
    \hdelta_{\gamma_0}(x)
    &= \frac{-2\i \gamma_0}{x^2+\gamma_0^2}, 
    \quad
    \lim_{\gamma_0 \rightarrow 0^+} \hdelta_{\gamma_0}(x) 
    = 
    -2\pi \i \delta(x) = \hdelta(x). 
    \eal

Finally, we obtain $G^{[12]}$ by convolving the kernels \eqref{subeq:K12-kernels-explicit} with the PSFs \eqref{eq:Sp_2p_final} according to \Eq{eq:KF_corr_through_full_corr-a}:
\bal
\nonumber
G^{[12]} & = 
\sum_p \bigl[K^{[\hat 1 \hat 2]} \ast S_p\bigr] (\bsomega_p)
\\
&
= \int_{\varepsilon_1} 
(2\pi \i) S_{[1,2]_+}(\varepsilon_1,-\varepsilon_1) \,
\hdelta_{\gamma_0}(\omega_1 - \varepsilon_1) 
\nn
&
= \int_{\varepsilon_1} \left[ (1+2n_{\varepsilon_1}) \tG^{\varepsilon_1} + 2 \hdelta(\varepsilon_{1}) \hG_{1} \right] \hdelta_{\gamma_0}(\omega_1 - \varepsilon_1)
\nn
&=
N_{\omega_1} \tG^{\omega_1} + 4 \pi \i\, \delta(\omega_1) \hG_1.
\label{eq:2p_G22}
\eal
For the last step we defined 
    \bal
    \label{eq:def_statistical_N}
    N_{\omega_i} = 
    - 1 - 2 n_{\omega_i} = \coth[\beta \omega_i /2 ]^{\zeta^i}.
    \eal
For bosonic correlators, $N_{\omega_1}$ is singular at $\omega_1 = 0$, 
so that a principle-value integral is implied in  \Eq{eq:2p_G22}.
Then, the product $N_{\omega_1} \tG^{\omega_1}$ should be evaluated via the limit $(N_{\omega_1} \tG^{\omega_1})_{\omega_1 \rightarrow 0}$. More precisely, three limits are involved: $0^+$, $\gamma_0$, and 
$\omega_1$ should all be sent to zero, while respecting $0^+ \ll \gamma_0 \ll |\omega_1|$ (see discussion after \Eq{subeq:derive-G-eta-by-analytic-continuation}).
In the following, we suppress the subscript $\gamma_0$ in \Eq{eq:hdelta_identity} and always take $\gamma_0\rightarrow 0^+$ after evaluating a principal-value integral (if present).

Summarizing, all Keldysh components can be expressed 
through analytically continued MF functions.
Comparing \Eqs{eq:2p_G22} and \eqref{eq:Corr_Func_gen_2p}, we find that the anomalous part, $\hG_1$, enters $G^{[12]}$ with a prefactor of $4\pi\i\delta(\omega_1)$.
    Using our previous results from \Eq{eq:2p_ret_adv}, yielding $\tG^{\omega_1} = G^{R}(\omega_1) - G^{A}(\omega_1)$, and defining $G^{[12]} = G^{K}$, the above relation \eqref{eq:2p_G22} can be identified as the FDR
    \bal \label{eq:2p_FDR}
    G^{K}(\omega_1) = N_{\omega_1} \left[ G^{R}(\omega_1) - G^{A}(\omega_1) \right] + 4 \pi \i\, \delta(\omega_1)\, \hG_1.
    \eal
Hence, the way in which anomalous MF terms appear in KF correlators is via Keldysh correlator $G^K$.
The anomalous term contributes only if $\omega_1$ is bosonic and vanishes.

We will refer to general relations between components of KF correlators in thermal equilibrium as \textit{generalized fluctuation-dissipation relations} (\mbox{gFDRs}).
Equations~\eqref{eq:GRprimed_equals_GA} and \eqref{eq:2p_FDR} constitute the two \mbox{gFDRs} available for $\ell=2$. 
In the absence of anomalous contributions, they reduce the three nonzero KF components to a single independent one (typically chosen as $G^R$).

\section{Analytic regions and discontinuities of the MF correlator}
\label{sec:analytic_regions_and_discontinuities}

Step 2 of our three-step strategy, the extraction of PSFs, requires knowledge of possible singularities of the MF correlators. In the 2p case, for $\tG_{z_1}$, a branch cut divides the complex $z_1$ plain into two analytic regions  (see \Fig{fig:contour}(a)), 
and the discontinuity across the branch cut is given by the difference of the analytic continuations $\tG_{\omega_1^\pm}$. In this section, we generalize the concepts of and notations for branch cuts, analytic regions, and discontinuities to general $\ell$, enabling a concise discussion of the analytic continuation of 3p and 4p MF correlators in \Secs{sec:Analytic_cont_3p} and \ref{sec:Analytic_cont_4p}, respectively. We focus on the regular parts $\tG$ of the MF correlators; the anomalous parts will be discussed separately in the sections for $\ell = 3$ and $4$.

\subsection{Analytic regions of $\tG(\bs{z})$}
\label{sec:analytic_regions_tG}

Possible singularities of the regular part can be inferred from the spectral representation in \Eq{eq:MF_G-compact-b} 
\bal 
\label{eq:regular_G_spectalRep_lp}
\tG(\bs{z}) 
&= 
\int \newlychanged{\md}^\ell \varepsilon_p\, \delta(\varepsilon_{ \oli{1} \dots \oli{\ell} })
\sum_p  
\frac{ S_p(\bs{\varepsilon}_p)}{\prod_{i=1}^{\ell-1}(z_{\oli{1}\dots\oli{i}} - \varepsilon_{\oli{1}\dots\oli{i}}) }
,
\eal
with $z_i = \omega_i + \i \gamma_i$ and $z_{1\dots\ell}=0$. Singularities can be located at the points where the imaginary part of the denominator vanishes, defining branch cuts by the condition
\bal
\label{eq:condition_branchcut}
\Im(z_{I}) = \gamma_I = 0, 
\eal
where $z_I = \sum_{i\in I} z_i$ denotes a frequency sum over the elements of a non-empty subset $I\subseteq \{1,\dots,\ell\}$.
In total, condition \eqref{eq:condition_branchcut} defines $2^{\ell-1}-1$ distinct branch cuts since frequency conservation implies $\Im(z_{I}) = - \Im(z_{I^c})$ where $I^c=\{1,\dots,\ell\}\backslash I$ is the complement of $I$, so that $\Im(z_{I})=0$ and $ \Im(z_{I^c})=0$ describe the same branch cut.
The branch cuts divide $\mathbbm{C}^\ell$ into regions of analyticity (regions without singularities), each corresponding to one particular analytic continuation of $\tG$.

We henceforth focus on the case, needed for 
\Eq{eq:KF_corr_through_full_corr-b}, that all arguments of $\tG(\bs{z})$ are real, up to infinitesimal shifts.
To be specific, we take the imaginary shifts of the frequency sums $z_I$ to be infinitesimal, $|\gamma_I|=0^+$ (with signs determined via 
conventions described below). Then,
$\tG(\bs{z})$ is a function of $\ell-1$ independent real frequencies $\omega_i$, and the analytic region is indicated by including the $2^{\ell-1}-1$ shift directions $\gamma_I=\pm0^+$ in the argument of $\tG(\bs{z})$. Thus, for 2p, 3p, and 4p correlators, we need 1, 3, and 7 imaginary parts, respectively (see examples below for $\ell=3,4$ in \Eqs{eq:3p_examples_subscripts} and \eqref{eq:4p_examples_subscripts}).

For a compact presentation of our results, it is convenient to introduce notation that specifies all imaginary shifts via 
a $(\ell-1)$-tuple $\zcheck$ whose components $\check{z}_i=\check{\omega}_i+ \i\check{\gamma}_i$ are frequency sums of the form $\check{z}_i = z_I$.
Then, the argument of $\tG(\bs{z})$ is expressed
as $\bs{z}(\zcheck)$, and the imaginary shifts of 
$\bs{z}$ are determined by those chosen for $\zcheck$.
We will specify the $\ell-1$ independent frequencies $\zcheck$ chosen to parametrize $\bs{z}(\zcheck)$ using subscripts,
$\tG_{\zcheck} = \tG(\bs{z}(\zcheck))$,
extending the subscript notation developed in \Sec{sec:2p_MF_correlators} for $\ell = 2$ 
to the regular parts of $\ell$p correlators. 
To uniquely determine the imaginary shifts in 
$z_I(\zcheck)$, and hence the analytic region
for $\tG_{\zcheck}$, we implicitly assign imaginary shifts to all $\check{z}_i$ via the rule
\bal
\label{eq:standard_imaginary_shifts}
2 |\check{\gamma}_{i-1}|\leq |\check{\gamma}_{i}|
,
\quad \quad \quad
\tn{for } 1 < i < \ell
.
\eal
It ensures that the imaginary part of any $\Im{z_I}$ is always nonzero, and that its sign is specified uniquely through the sign choices made for the shifts  $\pm |\check{\gamma}_i|$. We specify these sign choices 
using superscripts on the corresponding real frequencies $\check{\omega}_i$, writing $\check{z}_i = \check{\omega}_i^\pm = \check{\omega}_i
\pm \mi |\check{\gamma}_i|$.

\paragraph{\indent Examples for $\bs{\ell = 3}$:} For $\ell=3$, the branch cuts are given by $\gamma_1=0$, $\gamma_2=0$, and $\gamma_3=0$, see \Fig{fig:analyticregions}. Therefore, three imaginary parts are required to uniquely identify one analytic region for a regular MF correlator $\tG(\bs{z})$, with $\bs{z} = (z_1, z_2, z_3)$ and $ z_i = \omega^\pm_i$.
Consider, e.g., the set of independent frequencies $\zcheck=(\omega^+_1,\omega^-_2)$  with infinitesimal imaginary shifts fulfilling \Eq{eq:standard_imaginary_shifts}. It yields the analytic continuation
(see \Fig{fig:analyticregions} for the labels of analytic regions):
\bsubeq
\label{eq:3p_examples_subscripts}
\bal
\tG_{\omega^+_1,\omega^-_2} 
&= 
\tG(\omega^+_1, \omega^-_2,-\omega^-_{12}) = \tG(\omega^+_1, \omega^-_2,\omega^+_{3})=
G^{\prime[2]}(\bsomega). 
\eal
The third argument, $z_{3} = -z_{12} = -\check{z}_1 - \check{z}_2
=-\omega_1^+ - \omega_2^-$,  
has a positive imaginary shift since $\Im(z_{3}) = -\Im(|\check{\gamma}_1| - | \check{\gamma}_2|) >0$, by \Eq{eq:standard_imaginary_shifts}.
By contrast, for $\zcheck=(\omega^-_2,\omega^+_1)$, we obtain
\bal
\label{eq:example-2-ell=3}
\tG_{\omega^-_2,\omega^+_1}
&= 
\tG(\omega^+_1, \omega^-_2,-\omega^+_{12}) = \tG(\omega^+_1, \omega^-_2,\omega^-_{3})=
G^{[1]}(\bsomega)
.
\eal
\esubeq
Evidently, $\tG_{\omega^-_2,\omega^+_1} \neq \tG_{\omega^+_1,\omega^-_2}$, because switching $\omega^+_1 \leftrightarrow \omega^-_2$ in the argument list of $\zcheck$ also switches the relative magnitudes of their imaginary parts, due to  \Eq{eq:standard_imaginary_shifts}.

Note that the representation via subscripts is not unique. For instance, $G^{[1]}(\bsomega)$ can also be written as $\tG_{\omega^+_{12}, \omega^+_{1}}$, since 
the subscript $\zcheck = (\omega^+_{12}, \omega^+_{1})$ yields $
\bs{z}(\zcheck) = (\omega^+_{1}, \omega^+_{12} - \omega^+_{1}, -\omega^+_{12})
= (\omega^+_1, \omega^-_2,\omega^-_{3})$, matching the arguments 
found in \Eq{eq:example-2-ell=3}. For the last step, the sign of the imaginary
shift of the second argument follows from  $\Im(z_2) = 
\Im(\omega^+_{12} - \omega^+_{1}) = |\check{\gamma}_1| - 
|\check{\gamma}_2| < 0$.

\paragraph{\indent Example for $\bs{\ell = 4}$:} For $\ell=4$, the branch cuts are located at vanishing $\gamma_1$, $\gamma_2$, $\gamma_3$, $\gamma_4$, $\gamma_{12}$, $\gamma_{13}$, and $\gamma_{14}$, see \Fig{analyticregions4p}. 
Thus, seven imaginary parts are needed to uniquely identify one analytic region for a regular MF correlator $\tG(\bs{z})$.  We therefore write its argument as $\bs{z} = (z_1, z_2, z_3, z_4; z_{12}, z_{13}, z_{14})$, with $z_I = \omega_I^\pm$, also listing the arguments after the semicolon since the signs of their imaginary parts are needed to fully specify the analytic region. 
Consider, e.g., $\zcheck = (\omega^+_{13}, \omega^-_2, \omega^+_3)$. Then, $z_4 = - z_{123} = - \check{z}_1 - \check{z}_2
= - \omega^+_{13} - \omega^-_2 = - \omega^-_{123} = \omega^+_4$, $z_{12} =
\check{z}_1 + \check{z}_2 - \check{z}_3 = \omega_{13}^+ + \omega_2^- - \omega_3^+ =
\omega_{12}^-$, and $z_{14} = - z_{23} = -z_2 - z_3 = - \omega_2^- - \omega_3^+ =
- \omega_{23}^+$; the signs of the imaginary shifts on the right sides follow via \Eq{eq:standard_imaginary_shifts}.
We thus obtain 
\bal
\tG_{\omega^+_{13}, \omega^-_2, \omega^+_3} 
=&
\tG(\omega^-_1, \omega^-_2, \omega^+_3, -\omega^-_{123}; \omega^-_{12}, \omega^+_{13}, -\omega^+_{23}) 
\nn
=& \tG(\omega^-_1, \omega^-_2, \omega^+_3, \omega^+_{4}; \omega^-_{12}, \omega^+_{13}, \omega^-_{14}) 
= C^{(34)}_{\tn{IV}} . 
\label{eq:4p_examples_subscripts}
\eal
In the last line, the frequency arguments were expressed through those used to label the analytic regions in \Fig{analyticregions4p}.
\subsection{Discontinuities of $\tG(\bs{z})$}

The discontinuity of $\tG({\bs{z}})$ across a given branch cut, defined by $\Im z_I= \gamma_I =0$, quantifies the difference between two neighboring analytic regions, $R_+$ and $R_-$, separated by $\gamma_I =0$. 
We denote this discontinuity by $\tG({\bs{z}^{R_+}})-\tG({\bs{z}^{R_-}})$.
Explicitly, we have opposite imaginary shifts $\gamma_I$ in the analytic regions, $\gamma^{R_+}_I = 0^+ = - \gamma^{R_-}_I$, and equivalent shifts for all other $\gamma_J^{R_+}=\gamma_J^{R_-}$ with $J\subsetneq\{1,\dots,\ell\}$ and $J\neq I$.
To describe this discontinuity using $\zcheck$ notation, we  write
$\zcheck^{R_\pm} = (\check{z}^{R_\pm}_1, \zcheck^{\tn{r}})$, where the first variable is chosen as the one whose imaginary part changes sign
across the branch cut, $\check{z}_1^{R_\pm}=\omega_I^\pm$, and $\zcheck^{\tn{r}}$ denotes a tuple of $\ell - 2$ other, independent frequencies,
with imaginary shifts given by the prescription \eqref{eq:standard_imaginary_shifts}.
Then, extending the superscript notation from \Sec{sec:2p_MF_correlators}, 
we can express the discontinuity of $\tG({\bs{z}})$ across $\Im z_I = 0$ as
\bal
    \label{eq:def_discontinuity_general_ell}
    \tG^{\omega_I}_{\zcheck^{\tn{r}}} 
    &
    = 
    \tG_{\omega_I^+,\zcheck^{\tn{r}}} 
    -
    \tG_{\omega_I^-,\zcheck^{\tn{r}}}
    = 
    \tG_{\zcheck^{R_+}}
    -
    \tG_{\zcheck^{R_-}}.
\eal

Similarly, we define consecutive discontinuities across two branch cuts, $\gamma_I = 0$ and $\gamma_J=0$, to be evaluated as
\bal
    \label{eq:def_discontinuity_consecutive_general_ell}
    \tG^{\omega_{I},\omega_{J}}_{\check{z}_3,\dots,\check{z}_{\ell-1}} 
    &
    = 
    \tG^{\omega_{I}}_{\omega_{J}^+,\check{z}_3,\dots,\check{z}_{\ell-1}} 
    -
    \tG^{\omega_{I}}_{\omega_{J}^-,\check{z}_3,\dots,\check{z}_{\ell-1}} 
    ,
\eal
where we have $\check{z}_1=\omega_I^\pm$ and $\check{z}_2=\omega_J^\pm$.

\paragraph{\indent Examples for $\bs{\ell = 3}$:} For a discontinuity across $\gamma_2 = 0$ and $\zcheck^{\tn{r}} = \omega^+_1$, we find
\bal \label{eq:ACof3point:branchcut_single}
\tG^{\omega_2}_{\omega^+_1} &= \tG_{\omega^+_2,\omega^+_1} - \tG_{\omega^-_2,\omega^+_1} \nn
&= \tG(\omega^+_1, \omega^+_2, -\omega^+_{12}) - \tG(\omega^+_1, \omega^-_2, -\omega^+_{12}) \nn
&= \tG(\omega^+_1, \omega^+_2, \omega^-_{3}) - \tG(\omega^+_1, \omega^-_2, \omega^-_{3}) \nn
&= 
G'^{[3]}(\bsomega) - G^{[1]}(\bsomega) 
.
\eal
Two consecutive discontinuities across, e.g., $\gamma_1 = 0$ and $\gamma_2 = 0$ yield
\bal
\tG^{\omega_1,\omega_2} &= \tG^{\omega_1}_{\omega^+_2} - \tG^{\omega_1}_{\omega^-_2} = \tG_{\omega^+_1,\omega^+_2} - \tG_{\omega^-_1,\omega^+_2} - \tG_{\omega^+_1,\omega^-_2} + \tG_{\omega^-_1,\omega^-_2} 
\nn
&= 
\tG(\omega_1^+,\omega_2^+,-\omega_{12}^+) - \tG(\omega_1^-,\omega_2^+,-\omega_{12}^+)
\nn
&\hsp
- \tG(\omega_1^+,\omega_2^-,-\omega_{12}^-) + \tG(\omega_1^-,\omega_2^-,-\omega_{12}^-)
\nn
&= 
G'^{[3]} - G^{[2]} - G'^{[2]} + G^{[3]} 
.
\eal

\paragraph{\indent Example for $\bs{\ell = 4}$:} The discontinuity for, e.g., $\gamma_{123} = 0$ and $\zcheck^{\tn{r}} = (\omega^+_3, \omega^-_1)$ evaluates to
\bal
\tG^{\omega_{123}}_{\omega^+_3, \omega^-_1} &= \tG_{\omega^+_{123},\omega^+_3, \omega^-_1} - \tG_{\omega^-_{123},\omega^+_3, \omega^-_1} \nn
&= \tG(\omega^-_1, \omega^+_2, \omega^+_3, - \omega^+_{123}; \omega^-_{12}, \omega^-_{13}, - \omega^-_{23}) \nn
&\hsp -\tG(\omega^-_1, \omega^+_2, \omega^+_3, - \omega^-_{123}; \omega^-_{12}, \omega^-_{13}, - \omega^-_{23}) \nn
&= C^{(23)}_{\tn{I}} - C^{(234)}.
\eal

\section{Analytic continuation of 3p correlators}
\label{sec:Analytic_cont_3p}

The notation introduced in the previous section enables a concise discussion of the analytic continuation of 3p MF correlators in the following. \SEC{sec:3p_analytic_regions} is devoted to the general structure of these correlators and the connection of their analytical continuations to 3p PSFs. In contrast to the $2$p case, the derivation of these PSFs, constituting Steps 1 and 2 of our three-step strategy, is discussed in \App{sec:DerivationPCF3p}; in the main text, we merely state the final result. In \Sec{sec:3p_Keldysh_correlators}, we show that the PSFs yield all components of the KF correlator as linear combinations of analytically continued MF correlators.

\subsection{Extraction of PSFs}
\label{sec:3p_analytic_regions}

\begin{figure}[t!]
    \centering
    \includegraphics{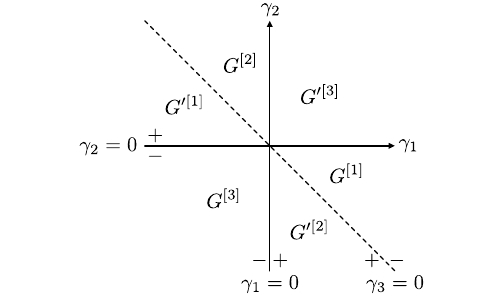}
    \caption{Regions of analyticity  of regular 3p MF correlators. Lines with $\gamma_i=0$ denote possible branch cuts of the correlators. (Figure adapted from Ref.~\cite{Evans1992}.) We label each region by that specific Keldysh correlator, $G^{[\eta]}$ or $G'^{[\eta]}$, whose imaginary shifts $\gamma_i$ lie within that region: For $G^{[1]}$, only $\omega_1$ has a positive imaginary shift, i.e., $\gamma_1 > 0$, $\gamma_2<0$, and $\gamma_3<0$, implying $G^{[1]}(\bsomega) = \tG(\omega^+_1,\omega^-_2,\omega^-_3)$. Primed correlators (\Eqs{eq:KF_fully_adv}) have inverted imaginary shifts, such that $ G'^{[1]}(\bsomega) = \tG(\omega^-_1, \omega^+_2,\omega^+_3)$.}
    \label{fig:analyticregions}
\end{figure}

A general 3p correlator can be decomposed into a regular and various anomalous parts (see \Eq{eq:MF_correlators_general_form} and \App{sec:appstruct3pcorr}): 
\bal \label{eq:3p_corr_MF_general}
G(\i \bsomega(\omega_{\oli{1}},\omega_{\oli{2}})) 
&=
G_{\i \omega_{\oli{1}}, \i \omega_{\oli{2}}} 
\nn 
&= \tG_{\i \omega_{\oli{1}}, \i \omega_{\oli{2}}} + \beta \delta_{\i\omega_{\oli{1}}}\, \hG_{{\oli{1}};\i \omega_{\oli{2}}} + \beta \delta_{\i\omega_{\oli{2}}}\, \hG_{{\oli{2}};\i\omega_{\oli{1}}} 
\nn
&
\hsp+ \beta \delta_{\i\omega_{\oli{12}}}\, \hG_{{\oli{12}};\i \omega_{{\oli{1}}}} + \beta^2\, \delta_{\i\omega_{\oli{1}}}\, \delta_{\i\omega_{\oli{2}}}\, \hG_{{\oli{1}},{\oli{2}}}.
\eal
Here, $\tG$ denotes the regular part, whereas $\hG_i$ represents the anomalous part w.r.t.\ frequency $\i \omega_i$, i.e., $\hG_i$ comes with a factor of $\beta\delta_{\i\omega_i}$ and is independent of $\i\omega_i$. $\hG_{1,2}$ is anomalous w.r.t.\ all frequencies and is a frequency-independent constant. 
(Note that, e.g., $ \beta \delta_{\i \omega_{3}} \hG_3$ can be written as $\beta \delta_{\i \omega_{12}} \hG_{12} $ in the $\beta \delta$ expansion in \Eq{eq:3p_corr_MF_general}, implying relations like $\hG_{12} = \hG_3$. This unbiased notation allows us to write formulas that hold for any permutation $p$.)

The full correlator $G$ as well as the components $\tG$ and $\hG_i$ are, by definition, 
singularity-free for all 
Matsubara frequencies.
For the anomalous contributions, we further have the decomposition
\bal \label{eq:3p_Gf_Gd_decomp}
\hG_{3;\i\omega_1}
& = 
\hG^{\noDelta}_{3;\i\omega_1} + \Delta_{\i \omega_1} \hG^{\withDelta}_{3;1} 
, 
\eal
where $\Delta_{\i\omega_i}$ is defined in \Eq{eq:defineDeltaSymbol} for a purely imaginary $\Omega_i=\i\omega_i$.
Here, $\hG^{\withDelta}_{3;1}$ comprises all terms proportional to a $\Delta_{\i\omega_1}$ symbol, and  $\hG^{\noDelta}_{3;\i\omega_1}$ contains the rest.
Analogous definitions hold for all anomalous terms $\hG_{i}$, see \App{sec:appstruct3pcorr} for a detailed discussion. The distinction between $\hG^{\noDelta}_i$ and $\hG^{\withDelta}_i$ is only needed if all three  operators are bosonic, in which case all anomalous terms in \Eq{eq:3p_corr_MF_general} can occur. For two fermionic and one bosonic operator, all following results equally hold by replacing $\hG^{\noDelta}_i \rightarrow \hG_i$ and $\hG^{\withDelta}_i \rightarrow 0$.

In \App{sec:DerivationPCF3p}, we show that the PSFs can be expressed via analytic continuations of the general constituents of the 3p correlator [\Eq{eq:3p_corr_MF_general}]:
\bal \label{eq:Sp_3p_final}
&(2\pi \i)^2 S_p(\varepsilon_{\oli{1}}, \varepsilon_{\oli{2}}) \nn
&= 
n_{\varepsilon_{\oli{1}}} n_{\varepsilon_{\oli{2}}} \tG^{\varepsilon_{\oli{2}},\varepsilon_{\oli{1}}} + n_{\varepsilon_{\oli{1}}} n_{\varepsilon_{\oli{12}}} \tG^{\varepsilon_{\oli{12}},\varepsilon_{\oli{1}}} + \hdelta(\varepsilon_{\oli{1}}) n_{\varepsilon_{\oli{2}}} \hG^{\noDelta;\varepsilon_{\oli{2}}}_{\oli{1}} \nn
&\hsp + \hdelta(\varepsilon_{\oli{2}}) n_{\varepsilon_{\oli{1}}} \hG^{\noDelta;\varepsilon_{\oli{1}}}_{\oli{2}} + \hdelta(\varepsilon_{\oli{3}}) n_{\varepsilon_{\oli{1}}} \hG^{\noDelta;\varepsilon_{\oli{1}}}_{\oli{3}} \nn
&\hsp + \hdelta(\varepsilon_{\oli{1}}) \hdelta(\varepsilon_{\oli{2}}) \left( \hG_{\oli{1},\oli{2}} - \tfrac{1}{2} \hG^{\withDelta}_{\oli{3};\oli{1}} \right).
\eal
This is our main result for $\ell=3$.
Explicit expressions of the PSFs for individual permutations are obtained by inserting the permuted indices into the above equation. In \Eqs{eq:tG_discont_analytic_regions}, we provide an overview of all possibly occurring discontinuities expressed through the analytic regions in \Fig{fig:analyticregions}.
As for 2p PSFs, we provide a consistency check of \Eq{eq:Sp_3p_final} in \App{sec:App_consistency_checks}.

\subsection{3p Keldysh correlators}
\label{sec:3p_Keldysh_correlators}

In the following two sections, we demonstrate how to construct KF correlators as linear combinations of 
analytically continued MF correlators using the PSFs in \Eq{eq:Sp_3p_final}, corresponding to Step 3 of our strategy.
For $\alpha=1$, \Eq{eq:Analytic_cont_fully_ret} gives the analytic continuation of $G$ to fully retarded components $G^{[\eta]}$ for general $\ell$. Therefore, we directly consider the more challenging cases of $\alpha = 2, 3$ in \Secs{sec:3p_KF_twoetas} and \ref{sec:3p_KF_threeetas}, respectively.
Lastly, in \Sec{sec:3p_overview_and_FDRs} we provide an overview of all Keldysh components and present gFDRs.

\subsubsection{Keldysh components $G^{[\eta_1 \eta_2]}$}
\label{sec:3p_KF_twoetas}

To recapitulate, in \Sec{sec:2p_G22} we performed manipulations on the level of the Keldysh kernels for $\ell = 2$ and $\alpha = 2$ by using the identity \eqref{eq:hdelta_identity}, which directly allowed us to evaluate the convolution with the PSFs. Even though the kernels for $\ell = 3$ are more complicated due to an additional factor in the denominator (see \Eq{eq:FTretkern}), similar manipulations 
are presented in \App{app:rewritingkerneltwo2s_3p} for the Keldysh component $G^{212} = G^{[13]}$. There, it is shown that simplifications of the 3p KF kernel $K^{[\heta_1 \heta_2]}$ (\Eq{eq:KFkerneletas}) yield
\bal \label{eq:G13_int_form}
&G^{[13]}(\bsomega) 
=
\int_{\varepsilon_1, \varepsilon_2} \hdelta(\omega_1 - \varepsilon_1) \frac{(2\pi\i)^2}{\omega^-_2 - \varepsilon_2} S_{[1,[2,3]_- ]_+}(\varepsilon_1,\varepsilon_2,-\varepsilon_{12})  
\nn
&
- \int_{\varepsilon_1, \varepsilon_{2}} \hdelta(\omega_{12} - \varepsilon_{12}) \frac{(2\pi\i)^2}{\omega^-_2 - \varepsilon_2} S_{[[1,2]_-,3]_+}(\varepsilon_1,\varepsilon_2, -\varepsilon_{12}).
\eal
Similarly to the 2p case, we always display the unpermuted $\bs{\varepsilon}$ for PSF (anti)commutators and insert permuted $\bs{\varepsilon}_p$ only for individual PSFs, implying, e.g., $S_{2[3,1]_\pm}(\bs{\varepsilon}) = S_{(231)}(\varepsilon_{2}, \varepsilon_{3})  \pm  S_{(213)}(\varepsilon_{2}, \varepsilon_{1})$.
For the integrations in \Eq{eq:G13_int_form}, we fixed the two independent frequencies $\varepsilon_1$ and $\varepsilon_2$ as integration variables. 
We thus obtain, e.g.,
\bal \label{eq:PDF_anticomm_G212}
&S_{[1,[2,3]_-]_+}(\bs{\varepsilon})
=S_{1[2,3]_-}(\bs{\varepsilon}) + S_{[2,3]_-1}(\bs{\varepsilon}) \nn
&= S_{(123)}(\varepsilon_1, \varepsilon_2) - S_{(132)}(\varepsilon_1, \varepsilon_{3}) + S_{(231)}(\varepsilon_2,\varepsilon_{3})
 \nn
 & \phantom{=}
 - S_{(321)}(\varepsilon_{3},\varepsilon_2).
\eal
with $\varepsilon_3 = -\varepsilon_{12}$ being understood.

To relate the KF to the MF correlator, we insert \Eq{eq:Sp_3p_final} into the PSF (anti)commutators of \Eq{eq:PDF_anticomm_G212} and simplify the results
using relations for the discontinuities such as $\tG^{\varepsilon_2, \varepsilon_3} = - \tG^{\varepsilon_2, \varepsilon_1}$. Such identities follow by explicitly expressing  the discontinuities in terms of $G^{[\eta]}$ and $G^{\prime[\eta]}$ correlators (see \Eqs{eq:tG_discont_analytic_regions}).
Then, the PSF (anti)commutator in \Eq{eq:PDF_anticomm_G212}, e.g., reads
\bal \label{eq:PSF_comm_123}
(2\pi\i)^2S_{[1,[2,3]_-]_+}(\varepsilon_1,\varepsilon_2,-\varepsilon_{12}) &= N_{\varepsilon_1} \tG^{\varepsilon_1,\varepsilon_2} - 2 \hdelta(\varepsilon_1) \hG^{\noDelta;\varepsilon_2}_1 \nn
    &\hsp - 2 \hdelta(\varepsilon_1) \hdelta(\varepsilon_2) \hG^{\withDelta}_{1;2}.
\eal
Inserting \Eq{eq:PSF_comm_123} (and a similar expression for $S_{[[1,2]_-,3]_+}$, see \Eq{eq:PDF_anticomm_G212_app-b}) into \Eq{eq:G13_int_form} and evaluating one of the integrals via the $\delta$-function, we find
\bal \label{eq:3p_G212_before_cont_deform}
G^{[13]}&(\bsomega) = - N_{\omega_1} \int_{\varepsilon_2} \frac{\tG^{\omega_1,\varepsilon_2}}{\omega^-_2-\varepsilon_2} 
+ 
2 \hdelta(\omega_1) 
\left( 
\int_{\varepsilon_2} \frac{ \hG^{\noDelta;\varepsilon_2}_1}{\omega^-_2-\varepsilon_2} 
- 
\frac{\hG^{\withDelta}_{1;2}}{\omega^-_2} 
\right)
\nn
& \hsp
- N_{\omega_{12}} \int_{\varepsilon_2} \frac{\tG^{\omega_{12},\varepsilon_2}}{\omega^-_2-\varepsilon_2} + 2 \hdelta(\omega_{12}) \left( \int_{\varepsilon_2} \frac{\hG^{\noDelta;\varepsilon_2}_{3}}{\omega^-_2-\varepsilon_2} - \frac{\hG^{\withDelta}_{3;2}}{\omega^-_2} \right).
\eal
Here, it becomes apparent why collecting PSFs in terms of (anti)commutators is beneficial. The integrands in \Eq{eq:3p_G212_before_cont_deform} do not contain any \MWFs{} depending on the integration variable $\varepsilon_2$, so that the only pole away from $\Im(z_2)=0$ comes from the denominators. Consequently, the integrals over $\varepsilon_2$ can be evaluated by closing the forward/backward integration contours in the upper/lower half-planes. 
Then, only the pole at $z_2 = \omega^-_2$ contributes (as illustrated in \Fig{fig:Analytic_cont_and_def} for the integral in \Eq{eq:Geta_2p_before_contdef}), and the final result for the Keldysh correlator $G^{[13]}$ 
reads
\bal
\label{eq:3pt:G_13_from_Giw}
G^{[13]}
=& 
N_{\omega_1} \tG^{\omega_1}_{\omega^-_2}  
+ N_{\omega_{12}} \tG^{\omega_{12}}_{\omega^-_2}
\nn
&
+ 4\pi \i \, \delta(\omega_1) \hG_{1;\omega^-_2}
+ 4\pi \i\, \delta(\omega_{12}) \hG_{3;\omega^-_2}
\nn
=& 
N_{\omega_1} \left( G'^{[2]} - G^{[3]} \right) 
+ N_{\omega_3} \left( G'^{[2]} - G^{[1]} \right) 
\nn
&
+ 4\pi \i \, \delta(\omega_1) \hG^{[3]}_1
+ 4\pi \i \, \delta(\omega_3) \hG^{[1]}_3
.
\eal
Here, we used $N_{\omega_{12}} = - N_{\omega_3}$, expressed $\tG^{\omega_1}_{\omega^-_2}$ and $\tG^{\omega_{12}}_{\omega^-_2}$ in terms of the analytic regions in \Fig{fig:analyticregions}, and defined the shorthand
\bal \label{eq:hG_real_freq}
\hG_{i;\omega^\pm_j} = \hG^{\noDelta}_{i;\omega^\pm_j} + \frac{\hG^{\withDelta}_{i;j}}{\omega^\pm_j}.
\eal
We emphasize that \Eq{eq:hG_real_freq} should not be interpreted as a direct analytic continuation of \Eq{eq:3p_Gf_Gd_decomp}. 
Rather, it can be obtained from \Eq{eq:3p_Gf_Gd_decomp} by replacing $\Delta_{\i\omega_j}\rightarrow 1/(\i\omega_j)$ and only afterwards analytically continuing the resulting expression $\i\omega_j\rightarrow\omega_j^\pm$.
Additionally, we defined the shorthand 
$\hG_i^{[\eta]} = \hG_i(\bsomega^{[\eta]})
$, such that, e.g.,
$\hG_{1;\omega^-_2} = \hG_{1;\omega^+_3} = \hG^{[3]}_1$.
The other two Keldysh components with $\alpha = 2$, $G^{[12]}$ and $G^{[23]}$, can be derived similarly, and their results are shown in \Eqs{eq:FDR_3p_G12} and \eqref{eq:FDR_3p_G23}, respectively.

\subsubsection{Keldysh component $G^{[123]}$}
\label{sec:3p_KF_threeetas}

In this section, 
we relate the Keldysh component $G^{[123]}$ to the analytic continued MF correlator.
In the derivation of \Eq{eq:3p_G212_before_cont_deform}, using the identity \eqref{eq:hdelta_identity} for the $\alpha = 2$ kernel $K^{[\heta_1 \heta_2]}$ was essential.
However, the Keldysh kernel for $G^{[123]}$, $K^{[\heta_1 \heta_2 \heta_3]}$, involves three retarded kernels according to \Eq{eq:KFkerneletas}, impeding the direct application of \Eq{eq:hdelta_identity}. 

In \App{app:rewritingkernelthree3s_3p}, we show that this problem can be circumvented by subtracting a fully retarded component, say, $G^{[3]}$. An analysis of the spectral representation of $G^{[123]} - G^{[3]}$ then leads to 
\bal \label{eq:G222_int_PSF_anticom}
&\frac{1}{(2\pi \i)^2}(G^{[123]} - G^{[3]} )(\bsomega)
\nn
&
= 
\int_{\varepsilon_1,\varepsilon_2} \hdelta(\omega_1 - \varepsilon_1)\, \hdelta(\omega_2 - \varepsilon_2)\, S_{[[1,2]_+,3]_+}(\varepsilon_1,\varepsilon_2,-\varepsilon_{12})
\nn
&
+ \int_{\varepsilon_1,\varepsilon_2} \hdelta(\omega_1 - \varepsilon_1)\,  \frac{1}{\omega^-_2 - \varepsilon_2}\, S_{[1,[2,3]_-]_-}(\varepsilon_1,\varepsilon_2,-\varepsilon_{12})
\nn
&+ \int_{\varepsilon_1,\varepsilon_2} \hdelta(\omega_{2} - \varepsilon_2) \frac{1}{\omega^-_1 - \varepsilon_1}\, S_{[2,[1,3]_-]_-}(\varepsilon_1,\varepsilon_2,-\varepsilon_{12}).
\eal
Similiar to \Eqs{eq:PDF_anticomm_G212} and \eqref{eq:3p_G212_before_cont_deform}, we evaluate the PSF (anti)\-commutators by inserting \Eq{eq:Sp_3p_final} (see \Eq{eq:3p_G222_PSF_anticom_app}), and subsequently evaluate the integrals either via the $\delta$-functions or via Cauchy's integral formula, yielding
\bal
&
(G^{[123]} - G^{[3]} )(\bsomega)
\nn
&=
(1 + N_{\omega_1} N_{\omega_2}) \tG^{\omega_2,\omega_1} 
+ N_{\omega_{12}} N_{\omega_1} \tG^{\omega_{12},\omega_1} +\tG^{\omega_1}_{\omega^-_2} + \tG^{\omega_2}_{\omega^-_1} 
\nn
& \hsp
+ 4\pi \i\, \delta(\omega_1) N_{\omega_2} \hG^{\noDelta;\omega_2}_1 
+ 4\pi \i\, \delta(\omega_2) N_{\omega_1} \hG^{\noDelta;\omega_1}_2 
\nn
& \hsp
+4\pi \i\, \hdelta(\omega_{12}) N_{\omega_1} \hG^{\noDelta;\omega_1}_3 + (4\pi \i)^2 \delta(\omega_1) \delta(\omega_2) \hG_{1,2}.
\eal
A more symmetric form of this result (see \Eq{eq:FDR_3p_G123}) can be obtained by expressing all discontinuities in terms of the analytic regions in \Fig{fig:analyticregions} and applying the identity
\bal \label{eq:3p_dis_func_identity}
1 + N_{\omega_1} N_{\omega_2} + N_{\omega_1} N_{\omega_3} + N_{\omega_2} N_{\omega_3} = 0
,
\eal
which holds for $\ell=3$ due to frequency conservation.

\subsubsection{3p generalized fluctuation-dissipation relations}
\label{sec:3p_overview_and_FDRs}

Expressing all Keldysh components with $\alpha \geq 2$ through analytic continuations of MF correlators is equivalent to relating 
them to fully retarded and advanced components. 
Indeed, as in the 2p case, knowledge of the fully retarded and advanced components \textit{and} the anomalous terms suffices to obtain all Keldysh components,
as brought to bear by the 3p gFDRs (where $N_i = N_{\omega_i}$)
\bsubeq
\label{eq:overview_AC_3p}
\bal
&G^{[12]} = N_1 \left( \tilde{G}'^{[3]} - \tilde{G}^{[2]} \right) + N_2 \left( \tilde{G}'^{[3]} - \tilde{G}^{[1]} \right) 
\nn
&
\hspace{30pt} + 4\pi \i\, \delta(\omega_1) \hG^{[2]}_{1} + 4\pi \i\, \delta(\omega_2) \hG^{[1]}_{2}, \label{eq:FDR_3p_G12} \\
&G^{[13]} = N_1 \left( \tilde{G}'^{[2]} - \tilde{G}^{[3]} \right) + N_3 \left( \tilde{G}'^{[2]} - \tilde{G}^{[1]} \right) 
\nn
&
\hspace{30pt} + 4\pi \i\, \delta(\omega_1) \hG^{[3]}_{1} + 4\pi \i\, \delta(\omega_3) \hG^{[1]}_{3}, \label{eq:FDR_3p_G13} \\
&G^{[23]} = N_2 \left( \tilde{G}'^{[1]} - \tilde{G}^{[3]} \right) + N_3 \left( \tilde{G}'^{[1]} - \tilde{G}^{[2]} \right) 
\nn
&
\hspace{30pt} + 4\pi \i\, \delta(\omega_2) \hG^{[3]}_{2} + 4\pi \i\, \delta(\omega_3) \hG^{[2]}_{3}, \label{eq:FDR_3p_G23} 
\\
\label{eq:FDR_3p_G123}
&G^{[123]} 
=
N_2 N_3 G^{[1]} + N_1 N_3 G^{[2]} +N_1 N_2 G^{[3]} 
\nn
&
+ (1+N_2 N_3) G'^{[1]} + (1+N_1 N_3) G'^{[2]} + (1+N_1 N_2) G'^{[3]} 
\nn
&+ 4\pi \i\, \Bigg[\delta(\omega_1) N_2 \left( \hG^{\noDelta;[2]}_1 -\hG^{\noDelta;[3]}_1 \right) 
+ \, \delta(\omega_2) N_3 \left( \hG^{\noDelta;[3]}_2 -\hG^{\noDelta;[1]}_2 \right) 
\nn
&
+ \delta(\omega_3) N_1\left( \hG^{\noDelta;[1]}_3 -\hG^{\noDelta;[2]}_3 \right) 
\Bigg]
+ (4\pi \i)^2 \delta(\omega_1) \delta(\omega_2) \hG_{1,2}. 
\eal
\esubeq
These gFDRs agree with the results in Ref.~\cite{Wang2002}, and generalize those by also including anomalous contributions.
Applications of these formulas to the Hubbard atom are presented in \Sec{sec:HA}.

\section{Analytic continuation of 4p correlators}
\label{sec:Analytic_cont_4p}

In this section, we demonstrate the MF-to-KF analytic continuation of fermionic 4p correlators. In \Sec{sec:4p_analytic_regions}, we first discuss our convention for labelling analytic regions and provide the expression of PSFs in terms of analytically continued MF correlators. In \Sec{sec:4p_Keldysh_correlators}, we then generalize the key concept for the construction of 3p KF correlators, namely rewriting the KF spectral representation using kernel identities and PSF (anti)commutators, to arbitrary $\ell$, and apply it to the relevant case $\ell = 4$.

\subsection{Analytic regions and extraction of PSFs}
\label{sec:4p_analytic_regions}

\begin{figure*}[t!]
    \centering

    \includegraphics[width=1 \textwidth]{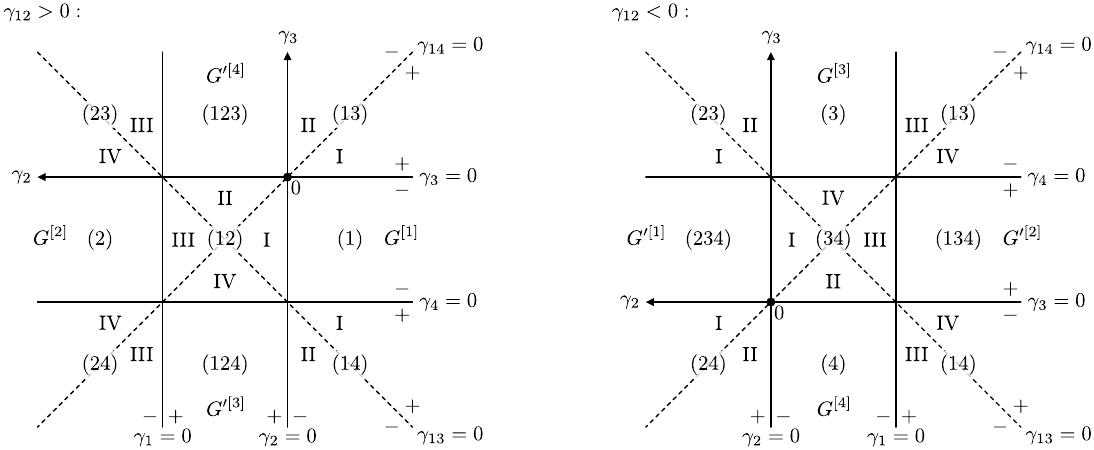}

    \caption{Regions of analyticity of regular \newlychanged{4p MF} correlators (analogous to Ref.~\cite{Eliashberg1962}). Lines with $\tn{Im}\, z_i = \gamma_i=0$ and $\tn{Im}\, z_{ij} = \gamma_{ij}=0$ denote possible branch cuts.
    The rectangular regions are labeled by arabic numbers indicating which $\gamma_i$ are positive; e.g., for region (124), we have $\gamma_1, \gamma_2, \gamma_4 > 0$ but $\gamma_3 < 0$. Consequently, regions composed of one or three arabic numbers correspond to fully retarded or advanced Keldysh components. 
    Regions with two of the $\gamma_i$ positive, like region (12), are further divided into four subregions by the branch cuts in $\gamma_{ij}$ and are distinguished by roman numbers $\tn{I} - \tn{IV}$. 
    }
    \label{analyticregions4p}
  \end{figure*}

As discussed in \Sec{sec:analytic_regions_tG}, the possible singularities of a regular 4p MF correlator are located at seven branch cuts, splitting the complex plane  into a total of 32 regions (see \Fig{analyticregions4p}).
Importantly, for $\ell \ge 4$, only few of these regions correspond to fully retarded or advanced Keldysh components, in contrast to $\ell = 2, 3$.
We label analytic continuations of MF correlators by $C$, e.g.,
\bal
\tG(\omega^+_1, \omega^-_2, \omega^+_3, \omega^-_{4}; \omega^-_{12}, \omega^+_{13}, \omega^-_{14}) = C^{(13)}_{\tn{III}}.
\eal
The superscript of $C^{(13)}_{\tn{III}}$ indicates which $\omega_i$ (with $1 \le i \le 4$) have a positive imaginary shift. 
Analytic regions with two $\omega_i$'s having positive shifts are further divided into four subregions, denoted by roman numbers $\tn{I} - \tn{IV}$ in the subscripts of $C$. This is necessary because for $C^{(13)}_{\tn{III}}$, e.g., the superscripts do not uniquely determine the imaginary parts of $\omega^+_1 + \omega^-_2 = \omega^\pm_{12}$ and $\omega^+_1 + \omega^-_4 = \omega^\pm_{14}$.
Fully retarded or advanced Keldysh components, on the other hand, are directly related to analytic regions, 
$G^{[\eta]}=C^{(i)}$ with $i=\eta$ and 
$G'^{[\eta]}=C^{(ijk)}$ with $i,j,k \neq \eta$, as depicted in \Fig{analyticregions4p}. 

Priming correlators, i.e., complex conjugation of the imaginary parts of frequencies (\Eq{eq:KF_fully_adv}), is directly applicable to the analytic regions. Consider, e.g., $C^{(1)}$, where only $\omega_1$ has a positive imaginary part; then, priming $C^{(1)}$ yields $
\left( C^{(1)} \right)' = \left(  G^{[1]} \right)' = 
G'^{[1]} = C^{(234)}$, where only $\omega_1$ has a negative imaginary part. The roman subscripts are chosen such that they are unaffected by complex conjugation of imaginary parts, so that, e.g., $\left( C^{(14)}_{\tn{II}} \right)' = C^{(23)}_{\tn{II}}$.

Finally, we note that double bosonic discontinuities, e.g., $\tG^{\omega_{13},\omega_{14}}_{\omega_{1}^+}$, vanish since the fermionic 4p kernel contains only one bosonic frequency, see \App{app:general_ell_disc}. 
This implies that not all analytic regions displayed in \Fig{analyticregions4p} are independent, since the following relations hold:
\bal
\label{eq:C_12_I_to_IV}
C^{(ij)}_{\tn{I}} - C^{(ij)}_{\tn{II}} + C^{(ij)}_{\tn{III}} - C^{(ij)}_{\tn{IV}} = 0, \quad\quad \tn{with } 1\leq i<j\leq4.
\eal
The identity for $(ij) = (12)$, e.g., follows from $\tG^{\omega_{13},\omega_{14}}_{\omega_{1}^+} = 0$.

After establishing our convention for labelling analytic regions, we now apply our strategy for the analytic continuation to fermionic 4p MF correlators. Anomalous terms, requiring bosonic Matsubara frequencies, only occur for sums of two fermionic Matsubara frequencies, implying the general form (\Eq{eq:MF_correlators_general_form})
\bal \label{eq:4p_MF_corr_general_form}
G(\i\bsomega(\omega_{\oli{1}}, \omega_{\oli{2}}, &\omega_{\oli{3}}) ) = G_{\i \omega_{\oli{1}}, \i \omega_{\oli{2}}, \i \omega_{\oli{3}}} \nn
&=\tG_{\i \omega_{\oli{1}}, \i \omega_{\oli{2}}, \i \omega_{\oli{3}}} + \beta\delta_{\i \omega_{\oli{12}}}\, \hG_{\oli{12}; \i \omega_{\oli{1}}, \i \omega_{\oli{3}}}
\nn 
&
\hsp + \beta\delta_{\i \omega_{\oli{13}}}\, \hG_{\oli{13}; \i \omega_{\oli{1}}, \i \omega_{\oli{2}}} + \beta\delta_{\i \omega_{\oli{14}}}\, \hG_{\oli{14}; \i \omega_{\oli{1}}, \i \omega_{\oli{2}}}.
\eal
The anomalous terms need not be further distinguished by factors of $\Delta_{\i\omega}$ as in \Eq{eq:3p_Gf_Gd_decomp}, since the remaining frequency arguments are fermionic  ($\i \omega_i \neq 0$).

Using \Eq{eq:4p_MF_corr_general_form}, Steps 1 and 2 of our three-step strategy are discussed in \App{sec:4p_PCF_app}; they yield the PSFs 
\bal \label{eq:4p:Sp}
&(2\pi \i)^3 S_p(\varepsilon_{\oli{1}}, \varepsilon_{\oli{2}}, \varepsilon_{\oli{3}}) 
\nn
&=
n_{\varepsilon_{\oli{1}}}\, n_{\varepsilon_{\oli{2}}}\, n_{\varepsilon_{\oli{3}}}\, \tG^{\varepsilon_{\oli{3}},\varepsilon_{\oli{2}},\varepsilon_{\oli{1}}} 
+ n_{\varepsilon_{\oli{1}}}\, n_{\varepsilon_{\oli{2}}}\, n_{\varepsilon_{\oli{123}}}\, \tG^{\varepsilon_{\oli{123}},\varepsilon_{\oli{2}},\varepsilon_{\oli{1}}} 
\nn
&\phantom{=}
+ n_{\varepsilon_{\oli{1}}}\, n_{\varepsilon_{\oli{2}}}\, n_{\varepsilon_{\oli{13}}}\, \tG^{\varepsilon_{\oli{13}},\varepsilon_{\oli{2}},\varepsilon_{\oli{1}}} 
+ n_{\varepsilon_{\oli{1}}}\, n_{\varepsilon_{\oli{2}}}\, n_{\varepsilon_{\oli{23}}}\, \tG^{\varepsilon_{\oli{23}},\varepsilon_{\oli{2}},\varepsilon_{\oli{1}}} 
\nn
&\phantom{=}
+ n_{\varepsilon_{\oli{1}}}\, n_{\varepsilon_{\oli{12}}}\, n_{\varepsilon_{\oli{3}}}\, \tG^{\varepsilon_{\oli{3}},\varepsilon_{\oli{12}},\varepsilon_{\oli{1}}} 
+ n_{\varepsilon_{\oli{1}}}\, n_{\varepsilon_{\oli{12}}}\, n_{\varepsilon_{\oli{123}}}\, \tG^{\varepsilon_{\oli{123}},\varepsilon_{\oli{12}},\varepsilon_{\oli{1}}} 
\nn
&\phantom{=}
+ n_{\varepsilon_{\oli{1}}}\, n_{\varepsilon_{\oli{3}}}\, \hdelta(\varepsilon_{\oli{12}})\, \hG^{\varepsilon_{\oli{3}},\varepsilon_{\oli{1}}}_{\oli{12}} 
+ n_{\varepsilon_{\oli{1}}}\, n_{\varepsilon_{\oli{2}}}\, \hdelta(\varepsilon_{\oli{13}})\, \hG^{\varepsilon_{\oli{2}},\varepsilon_{\oli{1}}}_{\oli{13}} 
\nn
&\phantom{=}
+ n_{\varepsilon_{\oli{1}}}\, n_{\varepsilon_{\oli{2}}}\, \hdelta(\varepsilon_{\oli{14}})\, \hG^{\varepsilon_{\oli{2}},\varepsilon_{\oli{1}}}_{\oli{14}}.
\eal
This is our main result for $\ell=4$. \EQs{eq:4p:discontinuities} give an overview over all possibly occurring discontinuities expressed through the analytic regions in \Fig{analyticregions4p}.
As for the 2p and 3p cases, we provide a consistency check of \Eq{eq:4p:Sp} in \App{sec:App_consistency_checks}.

To conclude this section, we further comment on properties of the anomalous parts.
As discussed in \App{app:4p_anom_cont}, the anomalous contribution $\hG_{13;\i\omega_1, \i \omega_2}$, e.g., can only depend on the frequencies $\i \omega_1$ and $\i \omega_2$ separately, but not on $\i \omega_{12}$. For anomalos parts, the complex frequency plane is thus divided into only four analytic regions corresponding to the imaginary parts of $\varepsilon^\pm_1$ and $\varepsilon^\pm_3$, in contrast to the six analytic regions for 3p correlators. This directly implies symmetries for discontinuities, such as $\hG^{\varepsilon_{\oli{2}},\varepsilon_{\oli{1}}}_{\oli{13}} = \hG^{\varepsilon_{\oli{1}},\varepsilon_{\oli{2}}}_{\oli{13}}$.
Similarly as for the regular parts, we label analytic continuations of anomalous parts with $\hC$, e.g.,
\bal
\hG_{12;\omega^+_1, \omega^-_3} = \hC^{(14)}_{12},
\eal
with the difference that subscripts indicate the anomalous contributions. Since $\hG_{12;\omega^+_1, \omega^-_3}$ is always multiplied by $\delta(\omega_{12})$, the remaining frequencies must have imaginary parts $\omega^-_2$ and $\omega^+_4$. Accordingly, the superscript of $\hC^{(14)}_{12}$ indicates the positive imaginary shifts of $\omega_1$ and $\omega_4$.

\subsection{4p Keldysh correlators}
\label{sec:4p_Keldysh_correlators}

In this section, we discuss the construction of KF correlators as linear combinations of analytically continued MF correlators. In \Eqs{eq:2p_G22}, \eqref{eq:G13_int_form}, and \eqref{eq:G222_int_PSF_anticom}, we expressed various Keldysh components via a convolution of PSF (anti)commutators with modified KF kernels, which originated from kernel identities presented in \Eqs{subeq:K12-kernels-explicit} and \App{app:simplifying_KF_correlators_3p}. To generalize these insights to arbitrary $\ell$p correlators and to present our results in a concise way, we now introduce further notation.
The goal of this notation is to collect terms which are related to discontinuities, each expressible via a sum over restricted permutations, such as the $\sum_{\oli{I}^1|\oli{I}^2}$ terms in \Eq{KFcorrfunceta123final}.

The set of all indices $L=\{1,\dots,\ell\}$ can be partitioned into $\alpha$ subsets $I^j$ of length $\vert I^j \vert$, such that $L=\bigcup_{j=1}^\alpha I^j$ with $I^j\cap I^{j'} =\emptyset$ for $j \neq j'$ and $\ell = \sum_{j=1}^\alpha \vert I^j \vert$. For a general Keldysh component $[\eta_1 \dots \eta_\alpha]$, we define the subsets $I^j$ to contain at least the element $\eta_j \in I^j$ for all $j \in \{ 1, \dots, \alpha \}$, implying $\vert I^j \vert \ge 1$. For example, a possible choice of the subsets for $\ell = 4$ and $[\eta_1 \eta_2] = [12]$ is given by $I^1 = \{ 1, 3 \} $ and $I^2 = \{ 2, 4 \} $.
With $\sum_{\oli{I}^1|\oli{I}^2}$, we denote sums over \textit{restricted} permutations $p=\oli{I}^1|\oli{I}^2$ for which all indices in subset $I^1$ appear to the left of those in subset $I^2$. Then, in the previous example, $\sum_{\oli{I}^1 \vert \oli{I}^2}$ sums over $\oli{I}^1 \vert \oli{I}^2 \in \{ (1324), (3124), (1342), (3142) \}$. Consequently, we always find $\vert \oli{I}^j \vert = \vert I^j \vert$ and $\eta_j \in \oli{I}^j$ for all $j \in \{ 1, \dots, \alpha \}$. In the following, we denote the elements of $\oli{I}^j$ by $\oli{I}_i^j$ with $i \in \{ 1, \dots, \vert I^j \vert \}$.

We further define the \textit{retarded product kernel}
\bsubeq
\bal
\label{eq:def_productKernel}
\tK_{\oli{I}^1 \vert \dots \vert \oli{I}^\alpha} \Big( \bsomega^{[\eta_1] \dots [\eta_\alpha]}_{\oli{I}^1 \vert \dots \vert \oli{I}^\alpha} \Big) 
&=
\prod_{j=1}^{\alpha-1} \Big[ \hdelta(\omega_{\oli{I}^j}) \Big] \prod_{j=1}^\alpha \Big[ \tK \Big( \bsomega^{[\eta_j]}_{\oli{I}^j} \Big) \Big], 
\\
\label{eq:regular_subkernel}
\tK \big( \bsomega_{\oli{I}^j} \big) &= \prod_{i=1}^{\vert \oli{I}^j \vert - 1} \frac{1}{\omega_{\oli{I}_1^j \dots \oli{I}_i^j}}.
\eal
\esubeq
The regular kernel in the last line is defined according to \Eq{eq:FTretkern} but restricted to the subtuple of frequencies $\bsomega_{\oli{I}^j} = (\omega_{\oli{I}_1^j}, \dots, \omega_{\oli{I}^j_{\scalebox{0.5}{$\vert I^j \vert$}}})$. Additionally, we defined the shorthand $\hdelta(\omega_{\oli{I}^j}) = - 2\pi \i \, \delta(\omega_{\oli{I}^j})$ and $\omega_{\oli{I}^j} = \omega_{I^j} = \sum_{i\in I^j}\omega_{i}$. The superscript on $\bsomega^{[\eta_1] \dots [\eta_\alpha]}_{\oli{I}^1 \vert \dots \vert \oli{I}^\alpha}$ indicates that the frequencies carry imaginary parts $\omega_{i} + \i \gamma^{[\eta_j]}_{i}$ for $i \in \oli{I}^j$ and $j \in \{ 1, \dots, \alpha \}$, such that $\gamma^{[\eta_j]}_{\eta_j}>0$ and $\gamma^{[\eta_j]}_{i\neq \eta_j}<0$. The Dirac delta function also ensures conservation of imaginary parts, $\gamma_{I^j}=0$.

As an example, consider again $\ell = 4$ and $[\eta_1 \eta_2] = [12]$ with $\oli{I}^1 = \{ 3, 1 \}$ and $\oli{I}^2 = \{ 2, 4 \}$. Then, we find
\bal
\tK_{\oli{I}^1 \vert \oli{I}^2} \Big( \bsomega^{[\eta_1] [\eta_2]}_{\oli{I}^1 \vert \oli{I}^2} \Big) &= \hdelta \big( \omega_{\oli{I}^1} \big) \tK \Big( \bsomega^{[\eta_1]}_{\oli{I}^1} \Big) \tK \Big( \bsomega^{[\eta_2]}_{\oli{I}^2} \Big) \nn
&= \hdelta(\omega_{13}) \frac{1}{\omega^{[1]}_3} \frac{1}{\omega^{[2]}_2}.
\eal
The retarded product kernels, together with PSF (anti)\-commutators, constitute the central objects for expressing \Eqs{eq:KF_spec_rep_eta_alpha} in a form particularly suitable for relating KF components to analytically continued MF correlators.

\subsubsection{Keldysh components $G^{[\eta_1 \eta_2]}$}
\label{sec:KF_components_alpha_2}

In \Eqs{eq:PSF_anticom_2p} and \eqref{eq:PDF_anticomm_G212}, we introduced PSF (anti)\-commutators for $\ell = 2$ and $\ell = 3$, respectively. We generalize this notation to arbitrary subsets by defining 
\bal
\label{eq:def_AntiCommutator}
S_{[\oli{I}^1, \oli{I}^2]_\pm}(\bs{\varepsilon}) 
= 
S_{ \oli{I}^1 \vert \oli{I}^2 } \big( \bs{\varepsilon}_{\oli{I}^1 \vert \oli{I}^2} \big) \pm S_{ \oli{I}^2 \vert \oli{I}^1 } \big( \bs{\varepsilon}_{\oli{I}^2 \vert \oli{I}^1} \big),
\eal
where the PSF (anti)commutator takes unpermuted variables $\bs{\varepsilon}$ as its argument.
In \App{app:simplifications_for_G_alpha2}, we then show that Keldysh components with $\alpha = 2$ can be rewritten as
\bal
G^{[\eta_1 \eta_2]}(\bsomega) 
=& 
\sum_{(I^1,I^2) \in \mc{I}^{12}} 
\sum_{ {\oli{I}^1 \vert \oli{I}^2} } 
\Big( 
    \tK_{ \oli{I}^1 | \oli{I}^2 } 
    \!\!\ACast 
    S_{[ \oli{I}^1, \oli{I}^2]_+ } 
\Big)
\Big( \bsomega^{[\eta_1][\eta_2]}_{ \oli{I}^1 | \oli{I}^2 } \Big)
.
\label{KFcorrfunceta123final}
\eal
 Here, $ \mathcal{I}^{12}=\{(I^1,I^2)|\, \eta_1\in I^1,\eta_2\in I^2, I^1\cup I^{2}=L, I^1\cap I^2=\emptyset\}$ is the set of all possibilities to partition $L=\{1,...,\ell\}$ into two non-empty subsets, $I^1$ and $I^2$, such that $\eta_1\in I^1$ and $\eta_2\in I^2$.
The convolution of a kernel with a PSF (anti)commutator is defined as
\bal
&\Big( 
\tK_{ \oli{I}^1 | \oli{I}^2 } 
\!\!\ACast 
S_{[ \oli{I}^1, \oli{I}^2]_\pm } 
\Big) \Big( \bsomega^{[\eta_1][\eta_2]}_{ \oli{I}^1 | \oli{I}^2 } \Big)
\\
&=
\int\tn{d}^{\ell}\varepsilon\,
\delta(\varepsilon_{1\dots\ell})
\tK_{ \oli{I}^1 | \oli{I}^2 } \Big( \bsomega^{[\eta_1][\eta_2]}_{ \oli{I}^1 | \oli{I}^2 } - \bs{\varepsilon}_{ \oli{I}^1 | \oli{I}^2 } \Big) S_{[ \oli{I}^1, \oli{I}^2]_\pm }(\bs{\varepsilon})
\nonumber
.
\eal
Further, as shown in \Eq{eq:G_eta1eta2_any_ell_via_AC}, \Eq{KFcorrfunceta123final} can be expressed in terms of analytically continued Matsubara correlators,
\bal \label{eq:alpha_2_general}
G^{[\eta_1\eta_2]}(\bsomega) 
=&
\sum_{I^1 \in \mc{I}^1} 
\bigg[ N_{\omega_{I^1}} \tG^{\omega_{I^1}}_{\bsomega^*}
+ 4\pi\i\, \delta(\omega_{I^1}) \hG_{I^1; \bsomega^*} \bigg]
,
\eal
with $\mc{I}^1 = \{ I^1 \subsetneq L \vert \eta_1 \in I^1, \eta_2 \notin I^1 \}$ the set of all subtuples of $L$ containing $\eta_1$ but not $\eta_2$.
The $\ell-2$ frequencies in $\bsomega^*=\{\omega_i^-|\, i\neq\eta_1, i\neq\eta_2\}$ all carry negative imaginary shifts, in accordance with the definition of $\bsomega^{[\eta_1\eta_2]}$.
The anomalous part $\hG_{I^1;\bsomega^*} = \hG_{I^1}(\bs{z}(\bsomega^*))$ for complex $\bs{z}$, which is independent of the anomalous frequency $\omega_I$ and parametrized via $\bsomega^*$, is defined as
\bal
\label{eq:definition_analytically_continued_hG_general_ell}
\hG_{I^1;\bsomega^*} =&
\Big[\hG_{I^1}(\i\bsomega) \Big]_{\Delta_{\i\omega}\rightarrow \tfrac{1}{\i\omega}, \i\bsomega\rightarrow\bs{z}(\bsomega^*)}.
\eal
We first replaced the symbol $\Delta_{\i\omega}$ by $1/(\i\omega)$ to obtain a functional form that we  can analytically continue, and then continue it as $\i\bsomega\rightarrow\bs{z}(\bsomega^*)$.
Remarkably, \Eq{eq:alpha_2_general} holds for arbitrary $\ell$, $\eta_1$, and $\eta_2$, and elucidates how anomalous terms enter the Keldysh components with $\alpha=2$. Examples are found in \Eq{eq:2p_G22} for $\ell = 2$, where $[\eta_1 \eta_2] = [12]$, $\mc{I}^1 = \{ 1 \}$, and $\bsomega^*$ is an empty set, or in \Eq{eq:3pt:G_13_from_Giw} for $\ell = 3$, where $[\eta_1 \eta_2] = [13]$, $\mc{I}^1 = \{ 1, 12 \}$, and $\bsomega^* = \omega^-_2$

For $\ell=4$, consider $[\eta_1 \eta_2] = [12]$, implying the set $\mc{I}^1 = \{ 1, 13, 14, 134 \}$ and $\bsomega^* = \omega^-_3, \omega^-_4$. Then, \Eq{eq:alpha_2_general} directly yields
\bal
G^{[12]}(\bsomega) =&
N_1 \tG^{\omega_1}_{\omega_3^-,\omega_4^-}
+
N_{13} \tG^{\omega_{13}}_{\omega_3^-,\omega_4^-}
+
N_{14} \tG^{\omega_{14}}_{\omega_3^-,\omega_4^-}
\nn
&
+
N_{134} \tG^{\omega_{134}}_{\omega_3^-,\omega_4^-}
+
4\pi\i\, \delta(\omega_{13}) \hG_{13; \omega_3^-,\omega_4^-}
\nn
&
+
4\pi\i\, \delta(\omega_{14}) \hG_{14; \omega_3^-,\omega_4^-}.
\label{eq:4p_G12_analytic_reg}
\eal
An expression for $G^{[12]}$ expressed in terms of analytic regions is given in \Eq{eq:G12_anal_regions}. Additionally, a full list of all $G^{[\eta_1 \eta_2]}$ is provided in \Eqs{eq:4p_overview_G12_analytic_reg}--\eqref{eq:4p_overview_G23_analytic_reg} (with relations such as $N_{134} \tG^{\omega_{134}}_{\omega_3^-,\omega_4^-} = -N_{2} \tG^{-\omega_{2}}_{\omega_3^-,\omega_4^-} = N_2 \tG^{\omega_{2}}_{\omega_3^-,\omega_4^-}$ used).

\subsubsection{Other Keldysh components}
\label{sec:4p_correlator_alpha_le_3}

The derivation of $G^{[123]} - G^{[3]}$ in \Sec{sec:3p_KF_threeetas} can be extended to arbitray $\ell$ and $[\eta_1 \eta_2 \eta_3]$ by keeping track of permutations that are cyclically related,
generalizing \Eq{eq:G222_int_PSF_anticom} to (see \App{sec:appeta123} for details)
\bal
\label{eq:GFAlphaThree_general}
&
(G^{[\eta_1 \eta_2 \eta_3]} - G^{[\eta_3]})(\bsomega) 
=
\\
&
\underset{(I^1,I^{23}) \in \mathcal{I}^{1\vert 23}}{\overset{ }{\sum}}  \hspace{10pt}
\underset{\oli{I}^1| \oli{I}^{23}}{\overset{ }{\sum}} 
\left[ \tK_{\ovb{I}^1| \ovb{I}^{23}}
\ACast 
\S{[\ovb{{I}}^1,\ovb{  {I}}^{23}]_-}{} \right]
\Big( \bsomega^{[\eta_1][\eta_3]}_{\ovb{I}^1 | \ovb{I}^{23}} \Big)
\nn
&
+
\underset{(I^2,I^{13}) \in \mathcal{I}^{2\vert 13}}{\overset{ }{\sum}} \hspace{10pt}
\underset{\oli{I}^2| \oli{I}^{13}}{\overset{ }{\sum}} 
\left[ \tK_{\ovb{I}^2 | \ovb{I}^{13}}
\ACast
\S{[\ovb{{I}}^2,\ovb{  {I}}^{13}]_-}{} \right]
\Big( \bsomega^{[\eta_2][\eta_3]}_{\ovb{I}^2 | \ovb{I}^{13}} \Big)
\nn
&+\,
\underset{(I^1,I^2,I^3)\in \mathcal{I}^{123}}{\overset{ }{\sum}} \hspace{3pt} 
\underset{\oli{I}^1|\oli{I}^2|\oli{I}^3}{\overset{ }{\sum}} 
\!\!
\left[
\tK_{\ovb{I}^1|\ovb{I}^2|\ovb{I}^3} 
\ACast
\S{[[\ovb{{I}}^1,\ovb{{I}}^2]_+,\ovb{{I}}^3]_+}{} 
\!\!
\right] 
\!\!
\Big( \bsomega^{[\eta_1][\eta_2][\eta_3]}_{\ovb{I}^1|\ovb{I}^2|\ovb{I}^3} \Big) .
\nonumber
\eal
Here, $ \mathcal{I}^{123}=\{(I^1,I^2,I^3)|\, \eta_1\in I^1,\eta_2\in I^2, \eta^3 \in I^3, I^j\cap I^{j'}=\emptyset \text{ for } j\neq j'\}$ is the set of all possibilities to partition $L=\{1,...,\ell\}$ into three subsets, each of which contains one of the indices $\eta_j\in I^j$. 
The remaining sets are defined as
\bsubeq
\bal
\label{eq:I1vert23}
\mathcal{I}^{1\vert 23} =& \{(I^1,I^{23})|\, {\eta_1\in I^1,\,}\,\, {\eta_2, \eta_3 \in I^{23},}\,\, I^1\cap I^{23} =\emptyset \},
\\
\mathcal{I}^{2\vert 13} =& \{(I^2,I^{13})|\, {\eta_2\in I^2,\,}\,\, {\eta_1, \eta_3 \in I^{13},}\,\, I^2\cap I^{13}=\emptyset  \}.
\eal 
\esubeq
Then, \Eq{eq:G222_int_PSF_anticom} provides an example for $\ell = 3$ and $[\eta_1 \eta_2 \eta_3] = [123]$, where $\mathcal{I}^{1\vert 23} = \{ (1,23) \}$, $\mathcal{I}^{2\vert 13} = \{ (2,13) \}$ and $\mathcal{I}^{123} = \{ (1,2,3) \}$.

For $\ell = 4$, consider $[\eta_1 \eta_2 \eta_3] = [123]$. Compared to the 3p case, the additional index allows for larger sets $\mathcal{I}^{1\vert 23} = \{ (1,234), (14,23) \}$, $\mathcal{I}^{2\vert 13} = \{ (2,134), (24,13) \}$, and $\mathcal{I}^{123} = \{ (1,2,34), (1,24,3), (14,2,3) \}$, resulting in (suppressing the frequency arguments of PSF (anti)commutators)
\bal
\label{eq:4p_G123_explicit}
&
(G^{[123]} - G^{[3]})(\bsomega) =
\tG^{\omega_1}_{\omega_2^-\omega_4^-}
+
\tG^{\omega_{14}}_{\omega_2^-\omega_4^-}
+
\tG^{\omega_2}_{\omega_1^-\omega_4^-}
+
\tG^{\omega_{24}}_{\omega_1^-\omega_4^-}
\nn
&
+
\int_{\varepsilon_1 \varepsilon_2 \varepsilon_3}\,
\bigg[ 
\hdelta(\omega_1-\varepsilon_1)\hdelta(\omega_2-\varepsilon_2) \frac{(2\pi\i)^3}{\omega_3^+-\varepsilon_3}
S_{[[1,2]_+,[3,4]_-]_+} 
\nn
&\hspace{50pt} +
\hdelta(\omega_1-\varepsilon_1)\hdelta(\omega_3-\varepsilon_3) \frac{(2\pi\i)^3}{\omega_2^+-\varepsilon_2}
S_{[[1,[2,4]_-]_+,3]_+} 
\nn
&\hspace{50pt} +
 \hdelta(\omega_2-\varepsilon_2)\hdelta(\omega_3-\varepsilon_3) \frac{(2\pi\i)^3}{\omega_1^+-\varepsilon_1}
 S_{[[[1,4]_-,2]_+,3]_+} \bigg]
.
\eal
Here, we identified the terms in the first line of \Eq{eq:GFAlphaThree_general} with discontinuities (see \App{app:branchcuts_as_PSFs}).
After inserting the PSFs (see \Eqs{eq:4p_123_PSF_anticomm_list})
and performing the remaining integrations using Cauchy's integral formula, we obtain \Eq{eq:G123_eq_ACs}.

For $\alpha \ge 4$, expressing the spectral representation of $G^{[\eta_1 \dots \eta_\alpha]}$ in terms of retarded product kernels and PSF (anti)commutators becomes increasingly challenging. Nevertheless, we provide a formula for $G^{[1234]}$ and $\ell = 4$ in \Eq{eq:4p_alpha_4_spec_rep}, with a list of all relevant PSF (anti)commutators given in \Eq{eq:4p_1234_PSF_anticomm_list}. \EQ{eq:G2222_as_BCs} then displays the result after evaluating all convolution integrals.

\subsubsection{Overview of Keldysh components}
\label{sec:4p_overview_KF_components}

To summarize the results of the previous sections, we give an overview of all Keldysh components with $\alpha>1$:
\begin{widetext}
\bsubeq
\label{eq:overview_AC_4p}
\bal \label{eq:4p_overview_G12_analytic_reg}
G^{[12]}(\bsomega) =&
N_1 \tG^{\omega_1}_{\omega_3^-,\omega_4^-}
+
N_2 \tG^{\omega_2}_{\omega_3^-,\omega_4^-}
+
N_{13} \tG^{\omega_{13}}_{\omega_3^-,\omega_4^-}
+
N_{14} \tG^{\omega_{14}}_{\omega_3^-,\omega_4^-}
+
4\pi\i\delta(\omega_{13}) \hG_{13; \omega_3^-,\omega_4^-}
+
4\pi\i\delta(\omega_{14}) \hG_{14; \omega_3^-,\omega_4^-},
\\
G^{[34]}(\bsomega) =&
N_3 \tG^{\omega_3}_{\omega_1^-,\omega_2^-}
+
N_{13} \tG^{\omega_{13}}_{\omega_1^-,\omega_2^-}
+
N_{14} \tG^{\omega_{14}}_{\omega_1^-,\omega_2^-}
+
N_4 \tG^{\omega_4}_{\omega_1^-,\omega_2^-}
+
4\pi\i\delta(\omega_{13}) \hG_{13; \omega_1^-,\omega_2^-}
+
4\pi\i\delta(\omega_{14}) \hG_{14; \omega_1^-,\omega_2^-},
\\
G^{[13]}(\bsomega) =&
N_1 \tG^{\omega_1}_{\omega_2^-,\omega_4^-}
+
N_{12} \tG^{\omega_{12}}_{\omega_2^-,\omega_4^-}
+
N_{14} \tG^{\omega_{14}}_{\omega_2^-,\omega_4^-}
+
N_3 \tG^{\omega_3}_{\omega_2^-,\omega_4^-}
+
4\pi\i\delta(\omega_{12}) \hG_{12; \omega_2^-,\omega_4^-}
+
4\pi\i\delta(\omega_{14}) \hG_{14; \omega_2^-,\omega_4^-},
\\
G^{[24]}(\bsomega) =&
N_2 \tG^{\omega_2}_{\omega_1^-,\omega_3^-}
+
N_{12} \tG^{\omega_{12}}_{\omega_1^-,\omega_3^-}
+
N_{14} \tG^{\omega_{14}}_{\omega_1^-,\omega_3^-}
+
N_4 \tG^{\omega_4}_{\omega_1^-,\omega_3^-}
+
4\pi\i\delta(\omega_{12}) \hG_{12; \omega_1^-,\omega_3^-}
+
4\pi\i\delta(\omega_{14}) \hG_{14; \omega_1^-,\omega_3^-},
\\
G^{[14]}(\bsomega) =&
N_1 \tG^{\omega_1}_{\omega_2^-,\omega_3^-}
+
N_{12} \tG^{\omega_{12}}_{\omega_2^-,\omega_3^-}
+
N_{13} \tG^{\omega_{13}}_{\omega_2^-,\omega_3^-}
+
N_4 \tG^{\omega_4}_{\omega_2^-,\omega_3^-}
+
4\pi\i\delta(\omega_{12}) \hG_{12; \omega_2^-,\omega_3^-}
+
4\pi\i\delta(\omega_{13}) \hG_{13; \omega_2^-,\omega_3^-},
\\
\label{eq:4p_overview_G23_analytic_reg}
G^{[23]}(\bsomega) =&
N_2 \tG^{\omega_2}_{\omega_1^-,\omega_4^-}
+
N_{12} \tG^{\omega_{12}}_{\omega_1^-,\omega_4^-}
+
N_{13} \tG^{\omega_{13}}_{\omega_1^-,\omega_4^-}
+
N_3 \tG^{\omega_3}_{\omega_1^-,\omega_4^-}
+
4\pi\i\delta(\omega_{12}) \hG_{12; \omega_1^-,\omega_4^-}
+
4\pi\i\delta(\omega_{13}) \hG_{13; \omega_1^-,\omega_4^-},
\\
\label{eq:G123_eq_ACs}
(G^{[123]} - G^{[3]})(\bsomega) 
=&
\left( N_1 N_2 + 1 \right) \tG^{\omega_2,\omega_1}_{\omega^+_3} 
+ N_1 N_{12} \tG^{\omega_{12}, \omega_1}_{\omega^+_3} 
+ N_1 N_3 \tG^{\omega_{3}, \omega_1}_{\omega^+_2} 
+ (N_1 N_{13} - 1) \tG^{\omega_{13}, \omega_1}_{\omega^+_2} 
\nn
&
+ N_2 N_3 \tG^{\omega_{3}, \omega_2}_{\omega^+_1} 
+ (N_2 N_{23} - 1) \tG^{\omega_{23}, \omega_2}_{\omega^+_1} 
+ \tG^{\omega_1}_{\omega^-_2, \omega^+_3} 
- \tG^{\omega_{23}}_{\omega^+_1, \omega^-_2} 
+ \tG^{\omega_2}_{\omega^-_1, \omega^+_3} 
- \tG^{\omega_{13}}_{\omega^+_2, \omega^-_1} 
\nn
&+ 4\pi \i \, \delta(\omega_{12})\, N_1\, \hG^{\omega_1}_{12;\omega^+_3} 
+ 4\pi \i \, \delta(\omega_{13})\, N_1\, \hG^{\omega_1}_{13;\omega^+_2} 
+ 4\pi \i \, \delta(\omega_{14})\, N_2\, \hG^{\omega_2}_{14;\omega^+_1} ,
\\
(G^{[124]} - G^{[4]})(\bsomega) 
= &
\left( N_1 N_2 + 1 \right) \tG^{\omega_2,\omega_1}_{\omega^+_4} + N_1 N_{12} \tG^{\omega_{12}, \omega_1}_{\omega^+_4} + N_1 N_4 \tG^{\omega_{4}, \omega_1}_{\omega^+_2} + (N_1 N_{14} - 1) \tG^{\omega_{14}, \omega_1}_{\omega^+_2} 
\nn
&+ N_2 N_4 \tG^{\omega_{4}, \omega_2}_{\omega^+_1} + (N_2 N_{24} - 1) \tG^{\omega_{24}, \omega_2}_{\omega^+_1} + \tG^{\omega_1}_{\omega^-_2, \omega^+_4} - \tG^{\omega_{24}}_{\omega^+_1, \omega^-_2} + \tG^{\omega_2}_{\omega^-_1, \omega^+_4} - \tG^{\omega_{14}}_{\omega^+_2, \omega^-_1} 
\nn
&+ 4\pi \i \, \delta(\omega_{12})\, N_1\, \hG^{\omega_1}_{12;\omega^+_4} + 4\pi \i \, \delta(\omega_{13})\, N_2\, \hG^{\omega_2}_{13;\omega^+_1} + 4\pi \i \, \delta(\omega_{14})\, N_1\, \hG^{\omega_1}_{14;\omega^+_2} ,
\\
(G^{[134]} - G^{[4]})(\bsomega) 
=& 
\left( N_1 N_3 + 1 \right) \tG^{\omega_3,\omega_1}_{\omega^+_4} + N_1 N_{13} \tG^{\omega_{13}, \omega_1}_{\omega^+_4} + N_1 N_4 \tG^{\omega_{4}, \omega_1}_{\omega^+_3} + (N_1 N_{14} - 1) \tG^{\omega_{14}, \omega_1}_{\omega^+_3} 
\nn
&+ N_3 N_4 \tG^{\omega_{4}, \omega_3}_{\omega^+_1} + (N_3 N_{34} - 1) \tG^{\omega_{34}, \omega_3}_{\omega^+_1} + \tG^{\omega_1}_{\omega^-_3, \omega^+_4} - \tG^{\omega_{34}}_{\omega^+_1, \omega^-_3} + \tG^{\omega_3}_{\omega^-_1, \omega^+_4} - \tG^{\omega_{14}}_{\omega^+_3, \omega^-_1} 
\nn
&+ 4\pi \i \, \delta(\omega_{12})\, N_3\, \hG^{\omega_3}_{12;\omega^+_1} + 4\pi \i \, \delta(\omega_{13})\, N_1\, \hG^{\omega_1}_{13;\omega^+_4} + 4\pi \i \, \delta(\omega_{14})\, N_1\, \hG^{\omega_1}_{14;\omega^+_3} ,
\\
\label{eq:G234_eq_ACs}
(G^{[234]} - G^{[4]})(\bsomega)
=& 
\left( N_2 N_3 + 1 \right) \tG^{\omega_3,\omega_2}_{\omega^+_4} + N_2 N_{23} \tG^{\omega_{23}, \omega_2}_{\omega^+_4} + N_2 N_4 \tG^{\omega_{4}, \omega_2}_{\omega^+_3} + (N_2 N_{24} - 1) \tG^{\omega_{24}, \omega_2}_{\omega^+_3} 
\nn
&+ N_3 N_4 \tG^{\omega_{4}, \omega_3}_{\omega^+_2} + (N_3 N_{34} - 1) \tG^{\omega_{34}, \omega_3}_{\omega^+_2} + \tG^{\omega_2}_{\omega^-_3, \omega^+_4} - \tG^{\omega_{34}}_{\omega^+_2, \omega^-_3} + \tG^{\omega_3}_{\omega^-_2, \omega^+_4} - \tG^{\omega_{24}}_{\omega^+_3, \omega^-_2} 
\nn
&+ 4\pi \i \, \delta(\omega_{12})\, N_3\, \hG^{\omega_3}_{12;\omega^+_2} + 4\pi \i \, \delta(\omega_{13})\, N_2\, \hG^{\omega_2}_{13;\omega^+_3} + 4\pi \i \, \delta(\omega_{14})\, N_2\, \hG^{\omega_2}_{14;\omega^+_4},
\\
\label{eq:G2222_as_BCs}
G^{[1234]}(\bsomega) 
=&
 N_1 \tG^{\omega_1}_{\omega_2^-,\omega_3^-}
 +
 N_2 \tG^{\omega_2}_{\omega_3^-,\omega_{4}^-
}
 +
 N_3 \tG^{\omega_3}_{\omega_1^-,\omega_4^-}
 +
 N_4 \tG^{\omega_4}_{\omega_1^-,\omega_2^-}
 +
 N_3 \tG^{\omega_3,\omega_4}_{\omega_2^+}
 +
 N_2 \tG^{\omega_2,\omega_3}_{\omega_1^+}
 \nn
 &
 +
 N_4 \tG^{\omega_4,\omega_1}_{\omega_3^+}
 +
 N_1 \tG^{\omega_1,\omega_2}_{\omega_4^+}
+ N_2\tG^{\omega_2,\omega_4}_{\omega_3^+}
+ N_4\tG^{\omega_4,\omega_2}_{\omega_3^+}
+ N_1\tG^{\omega_1,\omega_3}_{\omega_4^+}
+ N_3\tG^{\omega_3,\omega_1}_{\omega_4^+}
 \nn
 &
 +(N_1 N_2 N_3 + N_1 + N_3)
    \tG^{\omega_3,\omega_2,\omega_1}
+(N_1 N_2 N_4 + N_4 + N_2)
    \tG^{\omega_{4},\omega_2,\omega_1}
    \nn
    &
+(N_1 N_2 N_{13} + N_1 - N_2)
    \tG^{\omega_{13},\omega_2,\omega_1}
+(N_1 N_2 N_{23})
    \tG^{\omega_{23},\omega_2,\omega_1}
 \nn
 &
+N_1 (1+ N_{12} N_3)
    \tG^{\omega_3,\omega_{12},\omega_1}
+N_1 N_{12} N_4
    \tG^{\omega_{4},\omega_2,\omega_1}
 \nn
 &
 + 4\pi\i N_1N_3 \delta(\omega_{12})\hG_{12}^{\omega_1,\omega_3}
 + 4\pi \i N_1N_2 \left[ \delta(\omega_{13}) \hG_{13}^{\omega_1,\omega_2} + \delta(\omega_{13})\hG_{14}^{\omega_1,\omega_2} \right].
\eal
\esubeq
\end{widetext}
These equations constitute the main results of the MF-to-KF analytic continuation: They relate all components of a fermionic KF 4p correlator to linear combinations of analytically continued regular and anomalous parts of the corresponding MF correlator, expressed in terms of discontinuities and statistical factors $N_i$.

\subsubsection{4p gFDRs}

For 4p correlators, there are several regions of analyticity that cannot be identified with a KF correlator. Therefore, in contrast to $\ell \le 3$, fully retarded and advanced Keldysh components do not suffice to determine all other Keldysh components. Nevertheless, different Keldysh components can be related to each other. We now present the strategy for deriving these gFDRs for the Keldsyh component $G^{[12]}$.

Since every Keldysh component can be represented as a linear combination of analytically continued MF correlators, the analytic regions can serve as a basis to find relations among different Keldysh components. Expressing the discontinuities in \Eq{eq:4p_overview_G12_analytic_reg} via analytic regions, the KF correlator $G^{[12]}$ reads
\bal \label{eq:G12_anal_regions}
G^{[12]} =& N_{1} \big( C^{(12)}_{\tn{III}} - 
G^{[2]} \big) + N_{13} \big( C^{(12)}_{\tn{II}} - C^{(12)}_{\tn{III}} \big)
\nn
& + N_{14} \big( C^{(12)}_{\tn{IV}} - C^{(12)}_{\tn{III}} \big) +
N_{2} \big( C^{(12)}_{\tn{I}} - 
G^{[1]} \big)
\nn
&
+ 4\pi \i \, \delta(\omega_{13})\, \hC^{(12)}_{13} + 4\pi \i \, \delta(\omega_{14})\, \hC^{(12)}_{14},
\eal
where we inserted $G^{[1]} = C^{(1)}$ and $G^{[2]}= C^{(2)}$. Evidently, $G^{[12]}$ cannot be expressed in terms of fully retarded and advanced components only (modulo anomalous terms) due to the occurrence of $C^{(12)}_{\tn{I}/\tn{III}/\tn{IV}}$. However, these analytic regions and the same anomalous contributions appear in the primed KF correlator $G'^{[34]}$ as well:
\bal
\label{eq:G1122primed_as_AC}
G'^{[34]} &= N_{3} \big( C^{(12)}_{\tn{II}} - 
G'^{[4]} \big) + N_{13} \big(  C^{(12)}_{\tn{III}} - C^{(12)}_{\tn{II}} \big)
\nn & + N_{14} \big( C^{(12)}_{\tn{III}} - C^{(12)}_{\tn{IV}}\big) +
N_{4} \big( C^{(12)}_{\tn{IV}} - 
G'^{[3]} \big) 
\nn
&
-
4\pi\i\, \delta(\omega_{13}) \hC^{(12)}_{13}
-
4\pi\i\, \delta(\omega_{14}) \hC^{(12)}_{14}.
\eal
Note that priming the $ \i \delta(\dots)$ factors amounts to complex conjugation, as these arise from the identity \eqref{eq:hdelta_identity}, i.e., $ [\i \delta(\dots)]' = - \i \delta(\dots)$.
Therefore, we make the ansatz of expressing $G^{[12]}$ as a linear combination of $G'^{[34]}$, $G^{[1]}$ , $G^{[2]}$, $G'^{[3]}$, and $G'^{[4]}$, where the coefficients are determined by comparing terms proportional to the same analytic regions. Even though the resulting set of equations is overdetermined (including anomalous contributions, we have ten equations for five coefficients), we find the gFDR
\bsubeq
\label{eq:overview_FDRs_4p}
\bal 
\label{eq:4p_FDR_G12_final}
G^{[12]} =& -N_1 G^{[2]}-N_2 G^{[1]} \nn
&+ 
\frac{N_1+N_2}{N_3+N_4}
\Big[
    G'^{[34]}  +N_3 G'^{[4]}+N_4 G'^{[3]}
\Big].
\eal
The anomalous terms enter the right-hand side only implicitly via $G'^{[34]}$. However, using $\frac{N_1+N_2}{N_3+N_4}\delta(\omega_{13}) = -\delta(\omega_{13})$ and $\frac{N_1+N_2}{N_3+N_4}\delta(\omega_{14}) = -\delta(\omega_{14})$, it is straightforward to show that the $\hC^{(12)}_{13}$ and $\hC^{(12)}_{14}$ contributions in the last line of \Eq{eq:G12_anal_regions} are recovered by the corresponding terms in \Eq{eq:G1122primed_as_AC} via \Eq{eq:4p_FDR_G12_final}.
Conversely, the gFDR for $G^{[34]}$ can be derived from \Eq{eq:4p_FDR_G12_final} by solving for $G'^{[34]}$ and priming all correlators. \newline
\indent The gFDRs for all other Keldysh components with $\alpha \ge 2$ follow from the same strategy: Express Keldysh components in terms of linearly independent analytic regions and find relations between different components by solving a set of equations to determine coefficients. In addition to \Eq{eq:4p_FDR_G12_final}, we then obtain for $\alpha = 2$
 \bal
    G^{[13]} =& -N_1 G^{[3]}-N_3 G^{[1]} 
    \nn&
    + 
    \frac{N_1+N_3}{N_2+N_4}
    \Big[
        G'^{[24]}  +N_2 G'^{[4]}+N_4 G'^{[2]}
    \Big],
    \\
    G^{[14]} =& -N_1 G^{[4]}-N_4 G^{[1]} 
    \nn&
    + 
    \frac{N_1+N_4}{N_2+N_3}
    \Big[
        G'^{[23]}  +N_2 G'^{[3]}+N_3 G'^{[2]}
    \Big],
\intertext{for $\alpha = 3$}
G^{[2 3 4]}
=&
(1+  N_2 N_4 + N_2 N_3 + N_3 N_4)\Gconj^{[1]}
\nn&
-  N_3 N_4 G^{[2]}
 - N_2 N_4 G^{[3]} - N_2 N_3 G^{[4]} 
 \nn&
- N_4 G^{[23]} - N_3 G^{[24]} - N_2 G^{[34]},
\\
G^{[1 3 4]}
=&
(1+  N_1 N_4 + N_1 N_3 + N_3 N_4)\Gconj^{[2]}
\nn&
-  N_3 N_4 G^{[1]}
 - N_1 N_4 G^{[3]} - N_1 N_3 G^{[4]} 
 \nn
&
- N_4 G^{[13]} - N_3 G^{[14]} - N_1 G^{[34]},
\\
G^{[1 2 4]}
=&
(1+  N_1 N_2 + N_1 N_2 + N_2 N_4)\Gconj^{[3]}
\nn&
-  N_2 N_4 G^{[1]}
 - N_1 N_4 G^{[2]} - N_1 N_2 G^{[4]} 
\nn
&
- N_4 G^{[12]} - N_2 G^{[14]} - N_1 G^{[24]},
\\
G^{[1 2 3]}
=&
(1+  N_1 N_2 + N_1 N_3 + N_2 N_3)\Gconj^{[4]}
\nn&
-  N_2 N_3 G^{[1]}
 - N_1 N_3 G^{[2]} - N_1 N_2 G^{[3]} 
 \nn
&
- N_1 G^{[23]} - N_2 G^{[13]} - N_3 G^{[12]},
\intertext{and for $\alpha = 4$}
G^{[1234]} 
=&
2 N_2 N_3 N_4 G^{[1]} 
+
(N_2 N_3N_4 + N_2 + N_3 + N_4) \Gconj^{[1]}
\nn
& 
+
2 N_1 N_3 N_4 G^{[2]}
+
(N_1 N_3N_4 + N_1 + N_3 + N_4) \Gconj^{[2]}
\nn
& 
+ 
2 N_1 N_2 N_4 G^{[3]}
+
(N_1 N_2N_4 + N_1 + N_2 + N_4) \Gconj^{[3]}
\nn
& 
+
2 N_2 N_3 N_4 G^{[4]}
+
(N_1 N_2N_3 + N_1 + N_2 + N_3) \Gconj^{[4]}
\nn
&
+N_3 N_4 G^{[12]}+N_2 N_4 G^{[13]}
+N_2 N_3 G^{[14]}
\nn
&
+N_1 N_4 G^{[23]}
+N_1 N_3 G^{[24]}+N_1 N_2 G^{[34]}.
\eal
\esubeq
These results agree with the FDRs found in Ref.~\cite{Wang2002}, and therefore provide a consistency check for our approach. 
Moreover, we checked that the anomalous parts fulfill the same gFDRs. They enter \Eqs{eq:overview_FDRs_4p} only implicitly through $G^{[\eta_1 \eta_2]}$ and $G'^{[\eta_1 \eta_2]}$ on the right-hand sides, which contain anomalous parts via \Eqs{eq:4p_overview_G12_analytic_reg}--\eqref{eq:4p_overview_G23_analytic_reg}. This is in contrast to the 2p and 3p cases in \Eqs{eq:2p_FDR} and \eqref{eq:overview_AC_3p}, respectively. There, only fully retarded and advanced Keldysh correlators, which solely depend on the regular part $\tG$ of the corresponding MF correlator (see \Eq{eq:Analytic_cont_fully_ret}), occur on the right-hand side, and thus the anomalous parts have to enter the gFDRs explicitly.

\section{Hubbard atom}
\label{sec:HA}

\begin{figure*}[t!]
    \centering
    \includegraphics[width=1 \textwidth]{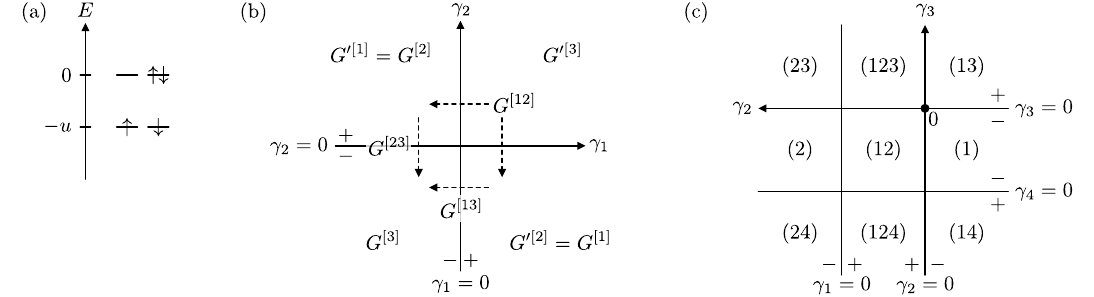}
    \caption{(a) Degenerate energy levels of the half-filled Hubbard atom for $u>0$. 
    (b) Relevant analytic regions of the regular part of the 3p electron-density correlator in \Eq{eq:eeb_corr_MF_explicit}. As the correlator is independent of $\i \omega_3 = -\i \omega_{12}$, there are are no poles on the line $\gamma_3 = 0$ in \Fig{fig:analyticregions}, resulting in $G'^{[1]} = G^{[2]}$ and $G'^{[2]} = G^{[1]}$. The dashed arrows indicate the relevant discontinuities for the different Keldysh components with $\alpha = 2$, see \Eq{eq:overview_AC_3p}.
    (c) Reduced analytic regions of the regular part of the fermionic 4p correlator in \Eq{eq:4p_ud_corr_MF}. The regions labelled by $(3)$, $(4)$, $(34)$, $(134)$, and $(234)$ in \Fig{analyticregions4p} are missing.}
    \label{fig:HA_Energies_3p_analytic_reg}
\end{figure*}

To illustrate the use of our analytic continuation formulas, we consider the Hubbard atom (HA) with the Hamiltonian
\bal 
\mH = U n^{}_\uparrow n^{}_\downarrow - \mu (n^{}_\uparrow + n^{}_\downarrow)
.
\eal 
It describes an interacting system of spin--$\tfrac{1}{2}$ electrons on a single site, created by $d^\dagger_\sigma$, with $n^{}_\sigma = d^\dagger_\sigma d^{}_\sigma$ the number operator for spin $\sigma \in \{ \uparrow, \downarrow \}$. The chemical potential $\mu$ is set to the half-filling value $\mu = u = U/2$ for compact results, where $U$ is the interaction parameter. The Hilbert space of the HA is only four-dimensional, with the site being either unoccupied, $\vert 0 \rangle$, singly occupied, $\vert \! \uparrow \rangle$ or $\vert \! \downarrow \rangle$, or doubly occupied, $\vert \! \uparrow \downarrow \rangle$. The eigenenergies are (see \Fig{fig:HA_Energies_3p_analytic_reg}(a))
\bal
E_0 = E_{\uparrow\downarrow} = 0, \quad E_\uparrow = E_\downarrow = -u
.
\eal
The partition sum evaluates to $Z= \tn{tr}(e^{-\beta \mH}) = 2+2e^{\beta u}$.

This very simple model is interesting as it is accessible via analytically exact computations. It describes the Hubbard model and the single-impurity Anderson model in the atomic limit (where the interaction $U$ dominates over all other energy scales) and can thus serve as a benchmark for numerical methods \cite{Kugler2021,Wallerberger2021,krienParquetlikeEquations2019,krienTilingTriangles2021}. 
Several correlators of the Hubbard atom were computed in the MF and studied extensively, like fermionic 2p (one-particle) and 4p (two-particle) correlators \cite{Hafermann2009,Pairault2000,Rohringer2012,Rohringer2013,Wentzell2016}. 
\newlychanged{Also its 3p MF functions have been computed and applied in previous works \cite{Ayral2015,Ayral2016,vanLoon2018}.} 
The vertex of the Hubbard atom, obtained from the fermionic 4p correlator by dividing out external legs, was used as a starting point for an expansion around strong coupling \cite{Metzner1991,Pairault2000,Rohringer2012,Kinza2013}. Additionally, it was found that (despite the simplicity of the model) the two-particle irreducible (2PI) vertices display a complicated frequency dependence, and their divergencies are subject to ongoing research \cite{Schafer2016,Thunstrom2018,chalupaDivergencesIrreducible2018,pelzHighlyNonperturbative2023}. Such divergencies have been related to the breakdown of the perturbative expansion due to the multivaluedness of the Luttinger--Ward functional \cite{Schafer2016,Gunnarsson2017,Kozik2015,Vuvcivcevic2018} and to the local moment formation in generalized susceptibilities \cite{Chalupa2021,adlerNonperturbativeIntertwining2022}. 

2p and 3p bosonic correlators have gained interest in recent years as well. They describe not only the asymptotic behaviour of the 4p vertex for large frequencies \cite{Wentzell2016} or the interaction of electrons via the exchange of effective bosons \cite{krienSinglebosonExchange2019,gieversMultiloopFlow2022}, but they are also the central objects of linear and non-linear response theory \cite{Kubo1957,kapplNonlinearResponses2022}.

KF correlators for the HA (beyond $\ell=2$) were of smaller interest due to the lack of numerical real-frequency studies. However, substantial progress has been made in this direction \cite{Kugler2021,Lee2021,Ge2023,Taheridehkordi2019,Lihm2023}. 
Hence, we exemplify the analytic continuation from MF to KF correlators on the example of the HA for various correlators of interest. 

One further comment is in order: The following MF correlators are derived by first computing the PSFs, followed by a convolution with the MF kernels. 
From our experience, a direct insertion of these PSFs into the spectral representation of KF correlators yields cluttered expressions, cumbersome to simplify due to the infinitesimal imaginary shifts $\gamma_0$. With the analytic continuation formulas, on the other hand, terms are conveniently preorganized, collecting those contributions with the same imaginary shifts. Additionally, the discontinuities conveniently yield Dirac delta contributions, as we will show below.
In order to derive, e.g., our first results for the 4p correlator, \Eqs{eq:G4p_updown}, it is much more convenient to start from the analytic continuation formulas, \Eqs{eq:overview_AC_4p}, than from the original KF \Eq{eq:KF_spec_rep_eta_alpha}.

For a compact presentation of our results, we distinguish different correlators with operators in subscripts, e.g., ${G[\mO^1,\mO^2](\i \bsomega)} = {G_{\mO^1\mO^2}(\i \bsomega)}$. 
Furthermore, we will make use of the identities (proven in \App{sec:Limit_identities})
\bsubeq
\label{eq:identity_diff} 
\bal
\frac{\omega^+}{(\omega^+)^2 - u^2} - \frac{\omega^-}{(\omega^-)^2 - u^2} &= \frac{\pi}{\i} [ \delta(\omega+u) + \delta(\omega-u) ], \label{eq:identity_diff_1} 
\\
\frac{1}{(\omega^+)^2 - u^2} - \frac{1}{(\omega^-)^2 - u^2} &= \frac{\pi \i}{u} [ \delta(\omega+u) - \delta(\omega-u) ].
\label{eq:identity_diff_2}
\eal
\esubeq
All following correlators refer to the connected part.

\subsection{Examples for $\ell=2$}

\subsubsection{Fermionic 2p correlator}

To begin with, we consider the fermionic 2p correlator (propagator), with $\bs{\mO} = (d^{}_\uparrow, d^\dagger_\uparrow)$. By SU(2) spin symmetry, reversing all spins leaves the correlator invariant. 
As the nonzero matrix elements are $\langle\uparrow\!|d_\uparrow^\dagger|0\rangle=\langle\uparrow\downarrow\!|d_\uparrow^\dagger|\downarrow\rangle=1$ and $\langle0\!|d_\uparrow|\uparrow\rangle=\langle\downarrow\!|d_\uparrow|\uparrow\downarrow\rangle=1$, we can readily compute the PSFs, $S_p$, via Eq.~(22b) in Ref.~\cite{Kugler2021}.
Evaluating the spectral representation yields
\bal
G_{d^{}_\uparrow d^\dagger_\uparrow}(\i \omega) = \frac{\i \omega}{(\i \omega)^2 - u^2} = \tG(\i \omega).
\eal
By construction, there is no anomalous part $\hG_1 = 0$. The retarded and advanced component are directly obtained from \Eq{eq:2p_ret_adv}:
\bal
G^{[1/2]}_{d^{}_\uparrow d^\dagger_\uparrow}(\omega) = \frac{\omega^\pm}{(\omega^\pm)^2 - u^2}.
\eal
The Keldysh component involves the difference of the retarded and advanced component. Via \Eq{eq:identity_diff_1}, one gets
\bal
G^{[12]}_{d^{}_\uparrow d^\dagger_\uparrow}(\omega) = \pi \i\, \th \left[ \delta(\omega+u) - \delta(\omega-u) \right],
\eal
where we used $N_{-\omega} = - N_{\omega}$ and defined $\th = \tanh(\beta u/2)$.

\subsubsection{Density-density correlator}

Our second example is the density-density correlator $\bs{\mO}=(n^{}_\uparrow, n^{}_\downarrow)$. 
The spectral representation in the MF yields a purely anomalous result
\bal
G_{n^{}_\uparrow n^{}_\downarrow}(\i \omega) = \beta \delta_{\i \omega} \tfrac{1}{4} \th = \beta \delta_{\i\omega} \hG_1.
\label{eq:HA_G_nup_ndn_MF}
\eal
\newlychanged{The correlator $G_{n_\uparrow n_\downarrow}$ discussed above describes the linear response of the spin-up occupation to a shift of the spin-down energy level, which lifts the degeneracy of the singly-occupied energy levels in Fig.~\ref{fig:HA_Energies_3p_analytic_reg}(a). For decreasing temperatures, the system becomes increasingly susceptible to such perturbations. This is reflected by the $\beta = 1/T$ divergence  for $T \rightarrow 0$ in the MF correlator of Eq.~\eqref{eq:HA_G_nup_ndn_MF}, and the $\delta(\omega)$ behavior in Eq.~\eqref{eq:HA_G_nup_ndn_KF} for its Keldysh counterpart.}

Using \Eqs{eq:2p_ret_adv} and \eqref{eq:2p_FDR}, the Keldysh components read
\bal 
G^{[1]}_{n^{}_\uparrow n^{}_\downarrow}(\omega) &= G^{[2]}_{n_\uparrow n_\downarrow}(\omega) = 0, \nn
G^{[12]}_{n^{}_\uparrow n^{}_\downarrow}(\omega) &= 4\pi \i\, \delta(\omega) \tfrac{1}{4} \th .
\label{eq:HA_G_nup_ndn_KF}
\eal
We again emphasize the importance of the anomalous term in the \mbox{gFDR}. If it were discarded, the Keldysh component $G^{[12]}_{n^{}_\uparrow n^{}_\downarrow}$ would falsely vanish entirely.

\subsection{Examples for $\ell=3$}

\subsubsection{3p electron-density correlator}
\label{sec:HA_eeb_corr}

Our first example for $\ell=3$ involves the operators $\bs{\mO} = (d^{}_\uparrow, d^\dagger_\uparrow, n^{}_\uparrow)$. 
As only the third operator is bosonic, there is at most one anomalous term if $\i \omega_3 = - \i \omega_{12} = 0$. Indeed, the spectral representation evaluates to
\bal \label{eq:eeb_corr_MF_explicit}
G_{d^{}_\uparrow d^\dagger_\uparrow n^{}_\uparrow}(\i \bsomega) 
& = 
\frac{u^2 - \i \omega_1 \, \i\omega_2}{\left[ (\i \omega_1)^2 - u^2 \right] \left[ (\i \omega_2)^2 - u^2 \right]} 
\nn
& \
+ \beta \delta_{\i \omega_{12},0}\, \frac{u\, \th}{2} \frac{1}{(\i \omega_1)^2 - u^2} 
\nn
& =
\tG(\i \bsomega) + \beta \delta_{\i \omega_{12}} \hG_3(\i \omega_1).
\eal
Since the fully retarded and fully advanced components of the correlator trivially follow from the regular part, we focus on the $\alpha \ge 2$ components in the following. We begin with the Keldysh component $G^{[13]}$ in \Eq{eq:FDR_3p_G13}: The regular part is independent of $\i \omega_3 = -\i \omega_{12}$, such that the discontinuity across $\gamma_{3} = -\gamma_{12} = 0$ vanishes, implying $G'^{[2]} - G^{[1]} = 0$ (see \Fig{fig:HA_Energies_3p_analytic_reg}(b)).
The discontinuity $G'^{[2]} - G^{[3]}$, on the other hand, is nonzero and can be easily evaluated using \Eqs{eq:identity_diff}, leading to (see \App{sec:simp_eeb_corr})
\bal \label{eq:HA_G13_eeb}
G^{[13]}_{d^{}_\uparrow d^\dagger_\uparrow n^{}_\uparrow}(\bsomega) 
& = 
N_1 \left( \tG(\omega^+_1, \omega^-_2) - \tG(\omega^-_1, \omega^-_2) \right) 
\nn
& \
+ 4\pi \i\, \delta(\omega_{12}) \hG_3(\omega^+_1) 
\nn
& = 
\pi \i\, \th \left[ \frac{\delta(\omega_1 - u)}{\omega^-_2 + u} - \frac{\delta(\omega_1 + u)}{\omega^-_2 - u} \right]
\nn
& \
+ 4\pi \i\, \delta(\omega_{12}) \frac{u\, \th}{2} \frac{1}{(\omega^+_1)^2 - u^2}.
\eal
Similarly, the remaining components with $\alpha=2$, as well as the Keldysh component with $\alpha=3$, read
\bal \label{eq:corr_eeb_all}
G^{[23]}_{d^{}_\uparrow d^\dagger_\uparrow n^{}_\uparrow}(\bsomega) 
& = 
\pi\i \, \th \left[ \frac{\delta(\omega_2 - u)}{\omega^-_1 + u} - \frac{\delta(\omega_2 + u)}{\omega^-_1 - u} \right] 
\nn
& \
+ 4\pi \i\, \delta(\omega_{12}) \frac{u\, \th}{2} \frac{1}{(\omega^-_1)^2 - u^2}
,
\nn
G^{[12]}_{d^{}_\uparrow d^\dagger_\uparrow n^{}_\uparrow}(\bsomega) 
& = 
\pi\i \, \th \left[ \frac{\delta(\omega_1 - u)}{\omega^+_2 + u} - \frac{\delta(\omega_1 + u)}{\omega^+_2 - u} \right] 
\nn
& \
+ \pi\i \, \th \left[ \frac{\delta(\omega_2 - u)}{\omega^+_1 + u} - \frac{\delta(\omega_2 + u)}{\omega^+_1 - u} \right]
,
\nn
G^{[123]}_{d^{}_\uparrow d^\dagger_\uparrow n^{}_\uparrow}(\bsomega) &= \frac{u^2 - \omega^+_1 \, \omega^+_2}{\left[ (\omega^+_1)^2 - u^2 \right] \left[ (\omega^+_2)^2 - u^2 \right]}
.
\eal
Here, $G^{[12]}_{d^{}_\uparrow d^\dagger_\uparrow n^{}_\uparrow}$ includes two discontinuities across $\gamma_1 = 0$ and $\gamma_2 = 0$, but no contribution from $\hG_3$, leading to the different structure compared to the other two Keldysh components with $\alpha = 2$.
Surprisingly, $G^{[123]}_{d^{}_\uparrow d^\dagger_\uparrow n^{}_\uparrow}$ is directly determined by $G'^{[3]}$. All other contributions from regular and anomalous parts mutually cancel, see \App{sec:simp_eeb_corr}.

\subsubsection{Three-spin correlator}
\label{sec:three_spin}

3p bosonic correlators are the central objects in non-linear response theory. Here, we consider the correlator for the spin operators $\bs{\mO}=(S_x, S_y, S_z)$, describing second-order changes in the magnetization by applying an external magnetic field. The spin operators are given by
\bal
S_x =& \tfrac{1}{2} \left( d^\dagger_\uparrow d^{}_\downarrow + d^\dagger_\downarrow d^{}_\uparrow \right), \quad S_y = -\tfrac{\i}{2} \left( d^\dagger_\uparrow d^{}_\downarrow - d^\dagger_\downarrow d^{}_\uparrow \right), 
\nn
S_z =& \tfrac{1}{2} \left( n^{}_\uparrow - n^{}_\downarrow \right).
\eal
The spectral representation, using the MF kernel in \Eq{eq:3p_kern_Kugler}, then yields
\bal \label{eq:HA_3spin_corr}
G_{S_x S_y S_z}(\i \bsomega) 
& =
-
\beta \delta_{\i \omega_1} \tilde{Z}\Delta_{\i \omega_2} + \beta \delta_{\i \omega_2} \tilde{Z}\Delta_{\i \omega_1} - \beta \delta_{\i \omega_{12}}\tilde{Z}\Delta_{\i \omega_1} 
\nn
& =
\beta \delta_{\i \omega_1}\, \hG^{\withDelta}_{1}(\i \omega_2) + \beta \delta_{\i \omega_2}\, \hG^{\withDelta}_{2}(\i \omega_1) 
\nn
&
+ \beta \delta_{\i \omega_3}\, \hG^{\withDelta}_{3}(\i \omega_1),
\eal
where $\tilde{Z} = \i e^{\beta u}/(2Z)$. 

From \Eqs{eq:FDR_3p_G12} -- \eqref{eq:FDR_3p_G123}, we deduce the only nonzero Keldysh components as
\bal
G^{[12]}_{S_x S_y S_z}(\bsomega) &= -4\pi \i\, \delta(\omega_1)\, \frac{\tilde{Z}}{\omega^+_2} + 4\pi \i\, \delta(\omega_2) \frac{\tilde{Z}}{\i \omega^+_1}, \nn
G^{[13]}_{S_x S_y S_z}(\bsomega) &= -4\pi \i\, \delta(\omega_1)\, \frac{\tilde{Z}}{\omega^-_2} - 4\pi \i\, \delta(\omega_{12}) \frac{\tilde{Z}}{\i \omega^+_1}, \nn
G^{[23]}_{S_x S_y S_z}(\bsomega) &= 4\pi \i\, \delta(\omega_1)\, \frac{\tilde{Z}}{\omega^-_1} - 4\pi \i\, \delta(\omega_{12}) \frac{\tilde{Z}}{\i \omega^-_1}.
\eal
Even though anomalous parts contribute to $G^{[123]}$ as well, 
they solely originate from the $\hG^{\noDelta}_i$ terms, such that $G^{[123]}$ vanishes in this case.

\subsection{Example for $\ell=4$: Fermionic 4p correlator}
\label{sec:4p_four_electron_HA}

Finally, we consider the 4p correlator $G_{\sigma \sigma'}$ involving the operators $\bs{\mO}=(d^{}_\sigma, d^\dagger_\sigma, d^{}_{\sigma'}, d^\dagger_{\sigma'})$.
Let us showcase the analytic continuation for $G_{\uparrow \downarrow}$, which evaluates in the MF to
\bal \label{eq:4p_ud_corr_MF}
G_{\uparrow \downarrow}(\i \bsomega) 
& = 
\frac{2u \prod_{i=1}^4 (\i \omega_i) + u^3 \sum_{i=1}^4 (\i \omega_i)^2 - 6 u^5}{\prod_{i=1}^4 \left[ (\i \omega_i)^2 - u^2 \right]} 
\nn
& \ +
\frac{u^2 \left[ \beta \delta_{\i \omega_{12}} \th + \beta \delta_{\i \omega_{13}} (\th-1) + \beta \delta_{\i \omega_{14}} (\th+1) \right]}{\prod_{i=1}^4 (\i \omega_i + u)} \nn
& = \tG(\i \bsomega) + \beta \delta_{\i \omega_{12}} \hG_{12}(\i \bsomega) + \beta \delta_{\i \omega_{13}} \hG_{13}(\i \bsomega) 
\nn
& \ + 
\beta \delta_{\i \omega_{14}} \hG_{14}(\i \bsomega).
\eal
We study the analytic continuation to the Keldysh component $G^{[12]}$, expressed in terms of the analytic regions from \Eq{eq:G12_anal_regions}. Since the regular part only depends on the frequencies $\i \omega_i$ individually, the discontinuities across $\gamma_{12}=0$, $\gamma_{13}=0$, and $\gamma_{14}=0$ vanish (\Fig{fig:HA_Energies_3p_analytic_reg}(c)), resulting in
\bal \label{eq:4p_HA_G12_red}
G^{[12]}_{\uparrow \downarrow}(\bsomega) 
=& 
N_{1} \big( C^{(12)}_{\tn{III}} - C^{(2)} \big) +
N_{2} \big( C^{(12)}_{\tn{I}} - C^{(1)} \big) 
\nn
&+ 4\pi \i \, \delta(\omega_{13})\, \hC^{(12)}_{13} + 4\pi \i \, \delta(\omega_{14})\, \hC^{(12)}_{14}.
\eal
The remaining discontinuities can be computed without further complications. From \Eq{eq:4p_ud_corr_MF}, we can already infer some of their structures. Since the regular part has poles at $\i \omega_1 \rightarrow z_1 = \pm u$ (or $\i \omega_2 \rightarrow z_2 = \pm u$), we expect the discontinuity across $\gamma_1 = 0$ (or $\gamma_2 = 0$) to select these poles. Indeed, we find (see \App{sec:4p_Gud_simp})
\bal
G^{[12]}_{\uparrow \downarrow}(\bsomega) 
=& 
2\pi \i\, u\, \th
\frac{\delta(\omega_1 - u) - \delta(\omega_1 + u)}{(\omega^+_2)^2 - u^2}  \left( \frac{1}{\omega^-_{13}} + \frac{1}{\omega^-_{14}} \right) 
\nn
&
+ (1 \leftrightarrow 2) 
+ 4\pi \i\, u^2\, \frac{\delta(\omega_{13})(\th-1) + \delta(\omega_{14})(\th+1)}{\left[ (\omega^+_1)^2 - u^2 \right] \left[ (\omega^+_2)^2 - u^2 \right]}.
\eal
where $1 \leftrightarrow 2$ indicates that indices $1$ and $2$ are exchanged compared to the first term. This expression can be simplified even further by collecting terms proportional to $\th$ and rewriting the $\delta$-functions in the resulting prefactor using \Eqs{eq:hdelta_identity} and \eqref{eq:identity_diff_2}. We eventually obtain
\bal
& G^{[12]}_{\uparrow \downarrow}(\bsomega) 
= 
4\pi \i\, u^2
\frac{ \delta(\omega_{14}) - \delta(\omega_{13}) }{\left[ (\omega^+_1)^2 - u^2 \right] \left[ (\omega^+_2)^2 - u^2 \right]} 
\nn
& \
+ 2u^2 \th \left[ 
\frac{1}{(\omega^+_1)^2 - u^2} 
\frac{1}{(\omega^-_2)^2 - u^2} 
\left( \frac{1}{\omega^-_{23}} + \frac{1}{\omega^-_{24}} \right) - \tn{c.c.} \right],
\eal
where $\tn{c.c.}$ is the complex conjugate. The other Keldysh components follow by similar calculations, see \App{sec:4p_G_results}.

This concludes the section on HA examples for the analytic continuation of multipoint correlators. We again stress the simplicity of the analytic continuation procedure using our results for the Keldysh components expressed through analytic regions.

\section{Vertex corrections to conductance}
\label{sec:oguris_formula}
\begin{figure}[t!]
\centering
\includegraphics{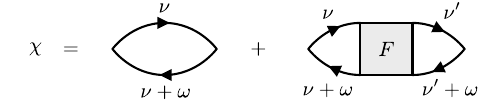}
\caption{Diagrammatic representation of the susceptibility $\chi$ consisting of a bubble and a vertex contribution. 
Lines represent propagators $\propag$, and the square is a vertex $\vertex$.
}
\label{fig:susceptibility_bubble}
\end{figure}

In this section, we consider a specific application of the analytic continuation of 4p functions regarding vertex corrections to the conductivity.
One can deduce vertex corrections to real-frequency susceptibilities either by working directly in the KF or by using the MF and the analytic continuation method.
The latter strategy was pursued by Eliashberg \cite{Eliashberg1962}, converting Matsubara sums into contour integrals and thereby obtaining various vertex contributions which consist of linear combinations of the MF vertex analytically continued to specific regions. 
For the special case of the linear conductance through an interacting region coupled to two noninteracting leads, Oguri \cite{Oguri2001} subsequently found that only one of these many vertex corrections contributes to the final result.
A very similar formula for the linear conductance was later derived by Heyder et al.\ \cite{Heyder2017} with a different line of argument, working entirely in the KF. 
With our insights on 4p analytic continuation and gFDRs, we can demonstrate the equivalence between the results by Oguri and Heyder et al.\ and connect the MF and KF derivations.

A general susceptibility $\chi$ can be expressed as in Fig.~\ref{fig:susceptibility_bubble}. The first  (``bubble'') term merely comprises two 2p correlators. We thus focus on the second term, the vertex correction, which in the MF reads
\bal
\label{eq:vertex_correction_susceptibility_MF}
\chi_{\vertex}(\i\omega) =&
\tfrac{1}{\beta^2} \!\!
\displaystyle{ \sum_{\i \nu, \i \nu'} }
\!
\propag(\i\nu) \propag(\i\nu\!+\!\i\omega) 
F(\i\nu,\i\nu',\i\omega)
\propag(\i\nu') \propag(\i\nu'\!\!+\!\i\omega)
.
\eal
Definitions of the propagator $\propag$ and vertex $F$ can be found in Sec.~III~A of Ref.~\cite{Kugler2021}.
The summand in \Eq{eq:vertex_correction_susceptibility_MF} is the connected 4p correlator. Due to their close relation, the vertex $F$ inherits its analytic properties from the correlator. 
In fact, by a transformation of Keldysh correlators to the $R/A$ basis \cite{Aurenche1992,Eijck1992}, it can be easily shown that our formulas in \Eq{eq:overview_AC_4p} identically hold for $F$, and we thus use the same symbols $C$ to denote analytic continuations of $F$ (see, e.g., \Eq{eq:oguris_formula_vertex}).
Note that the Keldysh indices $1$ and $2$ exchange their meaning for $F$, such that, e.g., a fully retarded component reads $F^{[1]}=F^{1222}$ (while $G^{[1]}=G^{2111}$).

In Ref.~\cite{Eliashberg1962}, Eliashberg converted the Matsubara sums in \Eq{eq:vertex_correction_susceptibility_MF} to contour integrals, thereby analytically continuing the MF functions and picking up contributions from all regions of analyticity (see Fig.~\ref{analyticregions4p}).
In Ref.~\cite{Oguri2001}, Oguri showed that the $\omega$-linear part, needed for the linear conductance (lc), stems from only one function, $\vertex^{(\mathrm{O})}$, see Eq.~(2.34) in Ref.~\cite{Oguri2001}. 
The corresponding vertex correction to the retarded susceptibility reads
\begin{subequations}
\bal
\label{eq:oguris_formula}
\chi_{\vertex,\mathrm{lc}}^R(\omega) 
& =
-\int\!\!\!\int\!\!\frac{\tn{d}\nu\tn{d}\nu'}{(4\pi\i)^2}
\propag^R(\nu+\omega)\propag^A(\nu) \propag^A(\nu') \propag^R(\nu'+\omega)
\nn
& \ \times
\big[\tanh\big(\tfrac{\nu+\omega}{2T}\big) - \tanh\big(\tfrac{\nu}{2T}\big)\big]
F_{\mathrm{O}}(\nu,\nu',\omega)
,
\\
\label{eq:oguris_formula_vertex}
\vertex_{\mathrm{O}}
& =
-N_{\omega_3}C^{(12)}_{\text{II}}
-N_{\omega_4}C^{(12)}_{\text{IV}}
+ N_{\omega_{13}} [C^{(12)}_{\text{II}}-C^{(12)}_{\text{III}}]
\nn
& \
+ N_{\omega_{14}} [C^{(12)}_{\text{IV}}-C^{(12)}_{\text{III}}]
,
\eal
\end{subequations}
where we used 
\bal
\label{eq:freqparametrization_OgurisFormula}
(\omega_1,\omega_2,\omega_3,\omega_4) = (\nu+\omega,-\nu,\nu',-\nu'-\omega)
\eal
as frequency parametrization.
Note that the results by Oguri and Eliashberg differ in their choice of the MWF; \Eq{eq:oguris_formula_vertex} corresponds
to $\mc{T}_{22}$ in Eq.~(12) of Ref.~\cite{Eliashberg1962}.

An analogous result with an independent KF derivation was obtained in Eqs.~(11) and (17) of Ref.~\cite{Heyder2017} by Heyder et al.
There, the vertex correction to the linear conductance corresponds to
\begin{subequations}
\bal
\label{eq:heyders_formula}
\chi_{\vertex,\mathrm{lc}}^R (\omega) 
& =
\int\!\!\!\int\!\!\frac{\tn{d}\nu\tn{d}\nu'}{(4\pi\i)^2}
\propag^R(\nu+\omega)\propag^A(\nu)  
\propag^A(\nu') \propag^R(\nu'+\omega)
\nn
& \ \times
\big[\tanh\big(\tfrac{\nu'+\omega}{2T}\big) - \tanh\big(\tfrac{\nu'}{2T}\big)\big]
\vertex_{\mathrm{H}}(\nu,\nu',\omega) 
,
\\
\label{eq:heyders_formula_vertex}
\vertex_{\mathrm{H}} =& -
\big( \vertex^{[12]} + N_{\omega_1} \vertex^{[2]} + N_{\omega_2} \vertex^{[1]} \big)
.
\eal
\end{subequations}
For an easier comparison with \Eq{eq:oguris_formula}, we here used the $\tanh$ function instead of the Fermi distribution function.
We also absorbed a factor of 2 due to our choice of convention for the Keldysh rotation of multipoint functions (cf.\ \Eq{eq:Keldysh_rotation_prefactor}).

To show that Eqs.~\eqref{eq:oguris_formula} and \eqref{eq:heyders_formula}
are equivalent, we translate the analytic continuations of the MF vertex in \Eq{eq:oguris_formula_vertex} to Keldysh components.
First, we note that the linear combination of terms comprising $\vertex_{\tn{O}}$ in \Eq{eq:oguris_formula_vertex} can also be expressed as follows, using \eqref{eq:G1122primed_as_AC}:
\bal
& \vertex^{\prime[34]} + N_{\omega_3} \vertex^{\prime[4]} + N_{\omega_4} \vertex^{\prime[3]}
\nn
& \ \ =
N_{\omega_3}C^{(12)}_{\text{II}}
+ N_{\omega_4}C^{(12)}_{\text{IV}}
+ N_{\omega_{13}} [ C^{(12)}_{\text{III}} - C^{(12)}_{\text{II}} ]
\nn
& \quad \
+ N_{\omega_{14}} [ C^{(12)}_{\text{III}} - C^{(12)}_{\text{IV}} ]
= - \vertex_{\mathrm{O}}
,
\eal
where we assumed vanishing anomalous parts.
Next, we use the gFDR in \Eq{eq:4p_FDR_G12_final} for vertices, \bal
(N_{\omega_3} & + N_{\omega_4}) 
\big( \vertex^{[12]} + N_{\omega_1} \vertex^{[2]} + N_{\omega_2} \vertex^{[1]} \big)
\nn
& = 
(N_{\omega_1}+N_{\omega_2})
\big( \vertex^{\prime[34]} + N_{\omega_3} \vertex^{\prime[4]} + N_{\omega_4} \vertex^{\prime[3]} \big)
.
\eal
Together with \Eq{eq:freqparametrization_OgurisFormula}, this implies the equivalence of 
\Eqs{eq:oguris_formula} and \eqref{eq:heyders_formula} as
\bal
(N_{\omega_3}+N_{\omega_4})\vertex_{\mathrm{H}} 
& = 
(N_{\omega_1}+N_{\omega_2})\vertex_{\mathrm{O}}.
\eal
With the analytic continuation formulas and the gFDRs, we have thereby shown that both results agree and provided a direct transcription between two independent MF and KF derivations.

\section{Conclusion}
\label{sec:Conclusion}

We showed how to perform the analytic continuation of multipoint correlators in thermal equilibrium from the imaginary-frequency MF to the real-frequency KF. To this end, we used the spectral representation derived in Ref.~\cite{Kugler2021}, separating the correlator into formalism-independent partial spectral functions (PSFs) and formal\-ism-specific kernels.
From this analytical starting point, we showed that it is possible to fully recover all $2^\ell$ components of the $\ell$p KF correlator from the one $\ell$p MF correlator. 
Our main result is that each of the $(\ell!)$ PSFs can be obtained by linear combinations of analytic continuations of the MF correlator multiplied with combinations of Matsubara weighting functions (\MWFs{}). Explicit formulas are given in \Eqs{eq:Sp_2p_final} and \eqref{eq:Sp_3p_final} for arbitrary 2p and 3p correlators, respectively, and \Eq{eq:4p:Sp} for fermionic 4p correlators. For these cases, we additionally derived direct MF-to-KF continuation formulas 
in \Eq{eq:2p_FDR} ($\ell = 2$), 
\Eqs{eq:overview_AC_3p} ($\ell = 3$), and 
\Eqs{eq:overview_AC_4p} ($\ell = 4$),
complementing the general \Eq{eq:Analytic_cont_fully_ret} for any $\ell$.

We approached the problem of analytic continuation by comparing the spectral representations of general $\ell$p MF ($G$) and KF ($G^{[\eta_1 \dots \eta_\alpha]}$) correlators and by identifying the regular partial MF correlators, $\tG_p$, as the central link between them. 
A key insight was that the partial MF correlators can be obtained by an imaginary-frequency convolution of MF kernels with the full MF correlator, $\tG_p(\i \bsomega_p) + \orderbeta = (K \star G)(\i \bsomega_p)$. Building on this formula, we developed a three-step strategy for the MF-to-KF analytic continuation, applicable to arbitrary $\ell$p correlators and explicitly presented in the aforementioned cases $\ell \le 4$. In the first step, we used the kernel representation of Ref.~\cite{Halbinger2023} to express the Matsubara sums, inherent in the imaginary-frequency convolution, through contour integrals enclosing the imaginary axis. In the second step, we deformed the contours toward the real axis, carefully tracking possible singularities of the MF correlator. This resulted in a spectral representation $\tG_p(\i \bsomega_p) = (\tK \ast S_p)(\i \bsomega_p)$, which allowed us to extract the PSFs, $S_p[G]$, as functionals of the regular and the various anomalous parts of $G$ multiplied with \MWFs{}. In the third and final step, we simplified the spectral representation for the KF components $G^{[\eta_1 \dots \eta_\alpha]}$, inserted the PSFs from the second step, and evaluated all real-frequency integrals to express the KF correlators as linear combinations of analytically continued MF correlators.

In our analysis, we explicitly considered so-called anomalous parts of the MF correlator which can occur, e.g., for conserved quantities or in finite systems with degenerate energy eigenstates. 
\newlychanged{The analytical continuations of these terms do not contribute to fully retarded correlators, but they do contribute to other components of the KF correlator. In the KF, the notion of ``anomalous terms'' is not needed; instead, the corresponding contributions are included via $\delta$-terms in the kernels, see \Eq{eq:FT_stepfunctions} and \Eqs{subeq:K12-kernels-explicit}-\eqref{eq:2p_G22} for $\ell=2$.}

Exploiting the relations between KF correlators and analytically continued MF functions, we derived generalized fluctuation-dissipation relations (gFDRs) for 3p and 4p correlators, Eqs.~\eqref{eq:overview_AC_3p} and \eqref{eq:overview_FDRs_4p}, 
establishing relations between the different KF components. 
We thereby reproduced the results of Refs.~\cite{Wang2002,Jakobs2010a},
while additionally including the anomalous terms.

\newlychanged{
We expect that similar results can be obtained for multipoint ($\ell>2$) out-of-time-ordering correlators (OTOCs) \cite{Larkin1969}
which generalize the KF by additional copies of the Keldysh contour.
Multipoint OTOCs, too, can be written as a sum over permutations of PSFs and kernels which encode the ordering on the desired number of branches. Importantly, the PSFs arising in this manner are precisely the same as those used in this work. Hence, the steps presented in Secs.~\ref{sec:2p_Keldysh_correlator}, \ref{sec:3p_Keldysh_correlators} and \ref{sec:4p_Keldysh_correlators} should be generalizable to multipoint OTOCs. 
Expressing the PSFs in terms of analytically continued MF correlators, analogous calculations would then reveal direct MF-to-OTOC continuation formulas.
We leave this to future work.
}

As an application of our results, we considered various correlators of the Hubbard atom. Starting from their MF expressions, we calculated all components of the corresponding KF correlators using analytic continuation. For the fermionic 4p correlator, a full list of all Keldysh components for the two relevant spin configurations is given in \Eqs{eq:G4p_updown} and \eqref{eq:G4p_upup}.

We further used our formulas to find KF expressions of the MF results derived in Refs.~\cite{Eliashberg1962,Oguri2001} for the linear conductance through an interacting system. There, the authors showed that only few analytic continuations of the vertex function are required for the vertex corrections to the linear conductance. 
Similar results were derived in Ref.~\cite{Heyder2017} working entirely in the KF.
We reproduced their real-frequency results by analytic continuation and could thus provide a direct transcription between two independent derivations in the MF and the KF.

For future investigations, it would be interesting to apply our formulas in conjunction with the algorithmic Matsubara integration technique \cite{Taheridehkordi2019}. There, the evaluation of Feynman diagrams yields an exact symbolic expression for $G(\i\bsomega)$ that can be readily continued to full Keldysh correlators or to PSFs. 
If, by contrast, the Matsubara results are only available as numerical data, the numerical analytic continuation is an ill-conditioned problem.
Nevertheless, recent advances suggest that it can possibly be tamed to some \newlychanged{extent} by exploiting further information on mathematical properties of the function \cite{Fei2021a,Fei2021b,Huang2023}.

Numerically representing multipoint MF correlators is another fruitful direction to explore.
References~\cite{Shinaoka2017,Kaye2022} showed that 2p MF functions can be represented compactly by a suitable basis expansion. 
Yet, for multipoint functions, Ref.~\cite{Wallerberger2021} found that the overcompleteness of the basis hinders an extraction of the basis coefficients by projection.
Here, a numerical counterpart of our method for recovering individual PSFs $S_p$ (or partial correlators $G_p$) from a full correlator $G(\i\bsomega)$ might be helpful. 
Finally, our formulas might also be useful for evaluating diagrammatic relations typically formulated for correlators while using the PSFs as the main information carriers.
\newlychanged{For recent developments regarding the numerical computation of MF or KF multipoint correlators  using symmetric improved estimators, see Ref.~\cite{Lihm2023}.}

\medskip
\textbf{Acknowledgements} \par 
AG, JJH, and JvD were supported by the Deutsche Forschungsgemeinschaft under Germany’s Excellence Strategy EXC-2111 (Project No.~390814868), and the Munich Quantum Valley, supported by the Bavarian state government with funds from the Hightech Agenda Bayern Plus. 
JJH acknowledges support by the International Max Planck Research School for Quantum Science and Technology (IMPRSQST). SSBL is supported by a New Faculty Startup Fund from Seoul National University, and also by a National Research Foundation of Korea (NRF) grant funded by the Korean government (MSIT) (No. RS-2023-00214464).
FBK acknowledges support by the Alexander von Humboldt Foundation through the Feodor Lynen Fellowship. The Flatiron Institute is a division of the Simons Foundation.

\medskip
\textbf{Conﬂict of Interest} \par
The authors declare no conﬂict of interest.

\medskip
\textbf{Data Availability Statement} \par
Research data are not shared.

\medskip

\bibliographystyle{MSP}
\bibliography{bibliography}

\appendix

\section{MF kernels}
\label{app:MFKernels}

This appendix is devoted to a discussion of 
the full primary MF kernel $K$, including both
regular and anomalous terms.  
It is defined via \Eq{eq:Kernel-define-a} for 
the MF kernel $\mc{K}(\bsOmega_p)$. 
In Ref.~\cite{Kugler2021}, it was shown that it can be computed via
\bal
\label{eq:kernel_integral}
\mc{K}(\bsOmega_p) &= \int_0^\beta \newlychanged{\md}\tau'_{\oli{\ell}}\, e^{\Omega_{\oli{1} \dots \oli{\ell}} \tau'_{\oli{\ell}}} \prod^{1}_{i=\ell-1} \Bigg[- \int_0^{\beta - \tau'_{\oli{i+1}\dots \oli{\ell}} } \md\tau'_{\oli{i}}\, e^{\Omega_{\oli{1}\dots\oli{i}}\tau'_{\oli{i}}} \Bigg] \nn
&= \beta \delta_{\Omega_{\oli{1} \dots \oli{\ell}}} K(\bsOmega_p) + \mc{R}(\bsOmega_p).
\eal
The residual part $\mc{R}$ is not of interest, for reasons explained after \Eq{eq:ResidualPartsCancel}.
The primary part $K(\bsOmega_p)$ is obtained\footnote{In Ref.~\cite{Kugler2021}, after Eq.~(41), it was stated that the lower boundary terms yield $K$; that is incorrect---they yield only $\tK$.} by collecting all contributions multiplying $\beta \delta_{\Omega_{\oli{1} \dots \oli{\ell}}}$,
and its argument satisfies $\Omega_{\oli{1} \dots \oli{\ell}} =0$
by definition. Before presenting explicit 
expressions for $K$, let us briefly
recall where it is needed in the main text. 

The analytical continuation of MF to KF correlators, based on
$\tG_p(\mi \bsomega_p \to \bsomega^{[\eta_j]})$ (\Eq{eq:KF_corr_through_partial_corr}), utilizes regular partial MF correlators, $\tG_p (\mi \bsomega_p)= [\tK \ast S_p](\mi \bsomega_p)$ (\Eq{eq:MF_G-compact-b-star}). These are expressed through regular MF kernels $\tK(\bsOmega_p)$ having a simple product form, $\prod_{i=1}^{\ell-1} \Omega^{-1}_{\oli{1}\dots \oli{i}}$,
with $\Omega_{\oli{1}\dots \oli{\ell}} = 0$ understood.
The more complicated primary kernel $K(\bsOmega_p)$ is defined implicitly via \Eq{eq:Kernel-define-a}. It includes both regular and anomalous parts, the latter involving vanishing partial frequency sums, $\Omega_{\oli{1}\dots \oli{i}} = 0$ with $i < \ell$. The primary kernel arises in two distinct contexts, involving either (i) imaginary-frequency convolutions $\star$ or (ii) real-frequency convolutions $\ast$, with different requirements for the bookkeeping of anomalous contributions.  We discuss them in turn. 

(i) For a specified permutation $p$, the regular partial $\tG_p(\mi \bsomega_p)$ can be extracted from the full MF correlator $G(\mi \bsomega')$ via a imaginary-frequency convolution,
$[K \star G](\mi \bsomega_p)$ (\Eq{eq:tildeGpstart}). 
There, the argument of  $K(\bsOmega_p)$ has the form $\bsOmega_p = \i\bsomega_p-\i\bsomega'_p$. This is always bosonic, being the difference of two same-type Matsubara frequencies. The convolution $\star$ involves Matsubara sums $\sum_{\mi \bsomega'_p}$, generating many anomalous contributions with  $\Omega_{\oli{1} \dots \oli{i}} = 0$. For these sums to 
be well-defined, the kernel $K(\bsOmega_p)$ must thus be represented in a form that (in contrast to $\tK(\bsOmega_p)$) is manifestly singularity-free for all values of 
$\Omega_{\oli{1} \dots \oli{i}}$, including 0. 

(ii) In \Eq{eq:tildeGpstart}, $\tG_p$ is 
given by that part of $[K \star G]$  that is
$\mc{O}(\beta^0)$; subleading powers of $\beta$ are not needed. Therefore, we seek the MF $G(\mi \bsomega')$ in the form of an 
$\beta \delta$ expansion, i.e.\ an expansion in powers of $\beta \delta_{\omega'_{\oli{1}\dots\oli{i}}}$. Then each of them can collapse one Matsubara sum $1/(-\beta)\sum_{\omega'_{\oli{1}\dots\oli{i}}}$ while their $\beta$
factors cancel. To obtain a $\beta \delta$ expansion for $G(\mi \bsomega')$, it is convenient to express it via a permutation sum of real-frequency convolutions, $\sum_p[K \ast S_p](\mi \bsomega'_p)$ (\Eq{eq:decomposeG_p-b}), 
and represent the kernel $K(\bsOmega_p)$, with argument $\bs{\Omega}_p = \i \bsomega'_p - \bs{\varepsilon}_p$,
as a $\beta \delta$ expansion in powers of $\beta \delta_{\Omega'_{\oli{1}\dots\oli{i}}}$. 

Fortunately, suitable representations of $K$ satisfying the
respective requirements of either (i) or (ii) are available in the literature \cite{Shvaika2006,Shvaika2015,Kugler2021,Halbinger2023}. We discuss them for $\ell \le 4$ in \Apps{app:explicit_MF_kernel_formulas} and \ref{app:kernel_alt}, respectively.

\subsection{Singularity-free representation of $K$}
\label{app:explicit_MF_kernel_formulas}

Consider case (i), involving $K \star G$, where the argument of $K(\bsOmega_p)$ is a bosonic Matsubara frequency. We seek a singularity-free (sf) representation for $K$, to be denoted $K^\mathrm{sf}$ for the purpose of this appendix. 
That such a representation exists is obvious from the form of integrals in \Eq{eq:kernel_integral}: inserting $\Omega_{\oli{1}\dots\oli{i}} = 0$ there reduces an exponential function to 1, so no contributions singular in $\Omega_{\oli{1}\dots\oli{i}}$ can arise. To find 
$K^\mathrm{sf}$, one simply has
to perform the integrals explicitly, treating the 
cases $\Omega_{\oli{1}\dots\oli{i}}\neq 0$ or $=0$ separately
and distinguish them using Kronecker symbols.

Such a direct computation of \Eq{eq:kernel_integral} has been performed in Ref.~\cite{Halbinger2023} for arbitrary $\ell$ and an arbitrary number of vanishing partial frequency sums, $\Omega_{\oli{1}\dots\oli{i}} = 0$. The following equations summarize their results for 
$\ell \le 4$:
\bsubeq
\label{eq:kern_Sbierski_2to4p}
\bal
& K^\mathrm{sf}(\bsOmega_p) 
\!\overset{\ell=2}{=}\! 
\Delta_{\Omega_{\oli{1}}} - \tfrac{\beta}{2}  \delta_{\Omega_{\oli{1}}}, \label{eq:2p_kern_Sbierski} \\
& K^\mathrm{sf} (\bsOmega_p) 
\!\overset{\ell=3}{=}\!
\Delta_{\Omega_{\oli{12}}} \Big( \Delta_{\Omega_{\oli{1}}} \!-\! \tfrac{\beta}{2} \delta_{\Omega_{\oli{1}}} \Big) 
- \delta_{\Omega_{\oli{12}}} \Big( \Delta^2_{\Omega_{\oli{1}}} \!+\! \tfrac{\beta}{2}  \Delta_{\Omega_{\oli{1}}} \!-\! \tfrac{\beta^2}{6} \delta_{\Omega_{\oli{1}}} \Big), \label{eq:3p_kern_Sbierski}
\\[-5mm]
& K^\mathrm{sf}(\bsOmega_p) 
\!\overset{\ell=4}{=}\!
\Delta_{\Omega_{\oli{123}}}
\nn
& \ \times
\Big[ \Delta_{\Omega_{\oli{12}}} \Big( \Delta_{\Omega_{\oli{1}}} - \tfrac{\beta}{2} \delta_{\Omega_{\oli{1}}} \Big) 
- 
\delta_{\Omega_{\oli{12}}} \Big( \Delta^2_{\Omega_{\oli{1}}} + \tfrac{\beta}{2} \Delta_{\Omega_{\oli{1}}} - \tfrac{\beta^2}{6}\delta_{\Omega_{\oli{1}}} \Big) \Big] 
\nn
& \ -
\delta_{\Omega_{\oli{123}}} \Big[ \Delta_{\Omega_{\oli{12}}} \Delta_{\Omega_{\oli{1}}} \Big( \Delta_{\Omega_{\oli{12}}} \!+\! \Delta_{\Omega_{\oli{1}}} \!+\!\tfrac{\beta}{2}\Big) 
-
\tfrac{\beta}{2} \Delta_{\Omega_{\oli{12}}} \delta_{\Omega_{\oli{1}}} \Big( \Delta_{\Omega_{\oli{12}}} \!+\! \tfrac{\beta}{3} \Big) 
\nn
& \ -
\delta_{\Omega_{\oli{12}}} \Delta_{\Omega_{\oli{1}}} \Big(\Delta^2_{\Omega_{\oli{1}}} \!+\! \tfrac{\beta}{2} \Delta_{\Omega_{\oli{1}}} \!+\! \tfrac{\beta^2}{6} \Big) 
+
\tfrac{\beta^3}{24} \delta_{\Omega_{\oli{12}}} \delta_{\Omega_{\oli{1}}} \Big]
. 
\label{eq:4p_kern_Sbierski}
\eal
\esubeq
Equations \eqref{eq:kern_Sbierski_2to4p} are manifestly singularity-free for all values of their frequency arguments---including those with $\Omega_{\oli{1}\dots\oli{i}} = 0$, for which $\Delta_{\Omega_{\oli{1} \dots \oli{i}}}$ terms vanish by definition (\Eq{eq:defineDeltaSymbol}). 

\subsection{$\beta \delta$ expansion for $K$}
\label{app:kernel_alt}

Next, consider case (ii), involving $G = \sum_p K \ast S_p$ (\Eqs{eq:decomposeG_p-b} and \eqref{eq:ResidualPartsCancel}), where the argument of $K(\bsOmega_p)$ has the form $\bsOmega_p = \mi \bsomega_p - \bsvarepsilon_p$, and we seek a $\beta \delta$ expansion for 
$G$. For this purpose, the kernels $K^\mathrm{sf}$ of \Eqs{eq:kern_Sbierski_2to4p}
are  inconvenient, because they contain some $\delta$ factors not accompanied by $\beta$. Instead, $G$ can be expressed through an alternative kernel, to be denoted $K^{\tn{alt}}$, which constitutes a $\beta \delta$ expansion itself and
hence differs from $K^\mathrm{sf}$, but yields the same result for $G$ when summed over all permutations, so that
\bal
\label{eq:MF_corr_sum_p_alt_K}
G(\i \bsomega) &= \sum_p \bigl[ K^\mathrm{sf} \ast S_p\bigr](\i \bsomega)
= \sum_p \bigl[ K^{\tn{alt}} \ast S_p\bigr](\i \bsomega) . 
\eal
Explicit expressions for $K^{\tn{alt}}$ were given in Ref.~\cite{Kugler2021} for up to one potentially vanishing frequency (general 2p correlators, 3p correlators with one bosonic operator, and fermionic 4p correlators). By also allowing general 3p correlators, these results are extended to
\bsubeq
\label{eq:kern_Kugler_2to4p}
\bal
K^{\tn{alt}}(\bsOmega_p) 
\!\overset{\ell=2}{=}\! 
&
\frac{1}{\Omega_{\oli{1}}} - \frac{\beta}{2} \delta_{\Omega_{\oli{1}}}, \label{eq:2p_kern_Kugler} 
\\
K^{\tn{alt}}(\bsOmega_p) 
\!\overset{\ell=3}{=}\! 
&
\frac{1}{ \Omega_{\oli{1}} \Omega_{\oli{12}} } 
- \frac{\beta}{2} 
\left( 
    \delta_{\Omega_{\oli{12}}} \Delta_{\Omega_{\oli{1}}}
    +
    \delta_{\Omega_{\oli{1}}} \Delta_{\Omega_{\oli{12}}}
\right) 
\nn
&
+ 
\frac{\beta^2}{6} \delta_{\Omega_{\oli{1}}} \delta_{\Omega_{\oli{12}}}, \label{eq:3p_kern_Kugler} 
\\
 K^{\tn{alt}}(\bsOmega_p) 
\!\overset{\ell=4}{=}\! 
&
\frac{1}{ \Omega_{\oli{1}} \Omega_{\oli{12}} \Omega_{\oli{123}} } - \frac{\beta}{2} \delta_{\Omega_{\oli{12}}} \frac{1}{ \Omega_{\oli{1}} \Omega_{\oli{123}} }.
\label{eq:4p_kern_Kugler}
\eal
\esubeq

The kernels 
\eqref{eq:kern_Kugler_2to4p}
have the form $K^{\tn{alt}} = \tK + \hK^{\tn{alt}}$, 
with regular part $\tK$ as given in \Eq{eq:def_reg_kern-compact},
while the anomalous part, $\hK^{\tn{alt}}$, comprises terms multiplied by one or multiple factors $\beta \delta_{\Omega_{\oli{1} \dots \oli{i}}}$. (We remark that the nomenclature \textit{regular} and \textit{anomalous} is used non-uniformly in the literature and our usage here may differ from Refs.~\cite{Shvaika2006,Kugler2021,Halbinger2023}.)
Whether or not $\Omega_\bonetoi = \mi \omega_\bonetoi - \varepsilon_\bonetoi$ can vanish at all depends on the fermionic or bosonic nature of the Matsubara frequencies. Take, e.g., $\ell = 4$ and all operators fermionic. Then, in \Eq{eq:4p_kern_Sbierski}, all terms multiplied by $\delta_{\Omega_{\oli{123}}}$ evaluate to $\delta_{\Omega_{\oli{123}}}= 0$, since $\i \omega_{\oli{123}} \neq 0$ is a fermionic Matsubara frequency. For the computation of fermionic 4p correlators, all terms proportional to $\delta_{\Omega_{\oli{1}}}$ and $\delta_{\Omega_{\oli{123}}}$ can thus be dropped. Even if $\mi \omega_\bonetoi$ is bosonic and vanishes, $\Omega_\bonetoi=0$ additionally requires
$\varepsilon_\bonetoi=0$, enforced by a Dirac $\delta(\varepsilon_\bonetoi)$
in the PSFs; see App.~\ref{app:PSF_decomposition} for further discussion of this point.

For a specified permutation $p$, the kernels $K^{\tn{alt}}$ are not singularity-free. In particular, the regular part $\tK$ diverges if one (or multiple) $\Omega_{\oli{1} \dots \oli{i}} \rightarrow 0$. However, that singularity is canceled by $1/\Omega_{\oli{i+1}\dots\oli{\ell}} = - 1/\Omega_{\oli{1}\dots\oli{i}}$ from a cyclically related permutation in the sum over permutations in \Eq{eq:MF_corr_sum_p_alt_K}.
This can be shown explicitly by treating nominally vanishing denominators as infinitesimal and tracking the cancellation of divergent terms while
exploiting the equilibrium condition \eqref{eq:equilibrium:cyclic_PSFs}
(see App. B of Ref.~\cite{Kugler2021}). 

The kernels $K^{\tn{alt}}$, inserted into \Eq{eq:kern_Kugler_2to4p}, 
result in the general form for MF correlators given in \Eq{eq:MF_G-compact}:
\bsubeq
\label{eq:MF_correlators_general_form}
\bal
G(\i \bsomega) &= \tG(\i \bsomega) + \hG(\i \bsomega), \\
\nonumber
\hG(\i \bsomega) &= \sum_{j=1}^{\ell-1} \beta \delta_{\i \omega_j} \hG_j(\i \bsomega) 
+ \sum_{j=1}^{\ell-1} \sum_{k>j}^{\ell-1} \Big( \beta \delta_{\i \omega_{jk}} \hG_{jk}(\i \bsomega) 
\\
&
\hsp + \beta^2 \delta_{\i \omega_{j}} \delta_{\i \omega_{k}} \hG_{j,k}(\i \bsomega) \Big) .
\label{eq:Anom_corr_general_form}
\eal
\esubeq
As for \Eq{eq:kern_Kugler_2to4p}, this form of the anomalous part of the correlator applies to general 2p and 3p correlators as well as fermionic 4p correlators. The subscripts of $\hG$ indicate the frequency in which they are anomalous. Even though their arguments nominally include all frequencies $\i\bsomega$, they are independent of their respective anomalous frequency; e.g., $\hG_1(\i \omega_1, \i \omega_2) = \hG_1(\i \omega_2)$ for $\ell = 3$. 
Note that this decomposition of the correlator is convenient for the analytic continuation because the components, such as $\tG$ and $\hG_{i}$ 
have a functional form that allows their arguments to be analytically continued, $\mi \omega_i \to z_i$.
In anomalous components this functional form is obtained by symbolically replacing all $\Delta_{\i\omega}$ by $\tfrac{1}{\i\omega}$ (see, e.g., \Eq{eq:hG_real_freq} and the discussion thereafter).

\section{Discussion of PSFs}
\label{app:PSF_discussion}

In \App{app:PSF_decomposition}, we clarify the functional structure of PSFs and 
motivate their decomposition into regular and anomalous contributions, 
$S_p= \Tilde{S}_p + \anomS_{p}$ (\Eq{eq:splitS_pregular-anomalous}),
analogous to that for MF correlators.
This decomposition aids investigations in subsequent appendices.
As an immediate application of the decomposition, we present an analysis of the effect of fully anomalous PSFs on 3p MF correlators in \App{app:Fully_anom_PSF_3p}.

\subsection{Decomposition of PSFs}
\label{app:PSF_decomposition}

Interacting thermal systems typically have a continuum of energy levels. 
Ref.~\cite{Kwok1969} argues that, in general, PSFs may contain contributions which diverge as $\tn{P}(\tfrac{1}{\varepsilon})$ for vanishing bosonic frequencies $\varepsilon$, with $\tn{P}$ the principal value.
As our derivations do not make assumptions on the shape of continuous PSF contributions, such terms require no further consideration.
However, Dirac delta contributions in $S_p$ can arise for finite systems or in the presence of conserved quantities. 
When these are present, MF partial correlators $G_p = K \ast S_p$ (\newlychanged{\Eq{eq:decomposeG_p-b}}) can contain anomalous terms, $\hG_p$, containing at least one factor $\delta_{\i \omega_{\oli{1} \dots \oli{i}}}$, with $i < \ell$. These arise from anomalous $\delta_{\Omega_{\oli{1} \dots \oli{i}}}$ terms in the MF kernel $K(\bsOmega_p)$, with argument $\bsOmega_p = 
\mi \bsomega_p - \bsvarepsilon_p$
(\Eqs{eq:MF_corr_sum_p_alt_K},  \eqref{eq:kern_Kugler_2to4p}). 
Such terms can contribute if $\Omega_{\oli{1} \dots \oli{i}} =0$, 
requiring $\mi \omega_{\oli{1} \dots \oli{i}} =0$
\textit{and} $\varepsilon_{\oli{1} \dots \oli{i}} =0$.
The first condition requires that $\i\omega_{\oli{1}\dots\oli{i}}$ is bosonic.
This is the case if the sign $\zeta^{\oli{1}\dots \oli{i}} = \zeta^{\oli{1}} \dots \zeta^{\oli{i}}$ equals $+1$ (with $\zeta^j = \pm 1$ for bosonic/fermionic operators $\mO^j$). Then, the associated  $\varepsilon_{\oli{1} \dots \oli{i}}$ is bosonic, too, according to the nomenclature introduced after \Eq{eq:PSF_definition}. The second condition is met if the PSF $S_p(\bsvarepsilon_p)$ contains a term proportional to a bosonic Dirac delta,
i.e.\ one having a bosonic $\varepsilon_{\oli{1} \dots \oli{i}}$ 
as argument, e.g.\  $\delta(\varepsilon_{\oli{1} \dots \oli{i}}) \check 
S_{\oli{1} \dots \oli{i}}$. Then, the
$\varepsilon_p$ integrals in the convolution
$K * S_p$ receive a finite contribution from the point $\varepsilon_{\oli{1} \dots \oli{i}} =0$. We summarize these conditions via the 
symbolic notation
\bal
\label{eq:symbolic-delta_epsilon}
\delta_{\Omega_{\oli{1} \dots \oli{i}}} = \delta_{\mi \omega_{\oli{1} \dots \oli{i}}}\delta_{\varepsilon_{\oli{1} \dots \oli{i}}}, 
\eal 
needed only for bosonic $\Omega_\bonetoi$. Here $\delta_{\varepsilon_{\oli{1} \dots \oli{i}}}$,
carrying a continuous variable as subscript, is defined only for
bosonic $\varepsilon_\bonetoi$ and by definition ``acts on'' $S_p(\varepsilon_p)$ by extracting only those parts (if present) containing \textit{bosonic} Dirac $\delta(\varepsilon_{\oli{1} \dots \oli{i}})$ factors. 
For the example above, $\delta_{\varepsilon_{\oli{1} \dots \oli{i}}}$ acts on $ S_p(\bsvarepsilon_p)$ as
\bal \label{edefine-delta_e-acts-on-S}
\delta_{\varepsilon_{\oli{1} \dots \oli{i}}} S_p(\bsvarepsilon_p) = \delta_{\varepsilon_{\oli{1} \dots \oli{i}}} \anomS_p(\bsvarepsilon_p) \sim \delta(\varepsilon_{\oli{1} \dots \oli{i}})  .
\eal
As we always assume an even number of fermionic operators, $\zeta^{1 \dots \ell} = + 1$ follows.

The motivation for splitting PSFs as $S_p= \Tilde{S}_p + \anomS_{p}$ is now clear. 
The anomalous $\anomS_p$ comprises all terms 
containing bosonic Dirac $\delta(\varepsilon_\bonetoi)$
factors, the regular $\tS$ everything else. The regular part of the MF correlator, $\tG$, receives
contributions from both $\tS_p$ and $\anomS_p$; the anomalous part, 
$\hG$, receives contributions only from $\anomS_p$, i.e.\ 
if $\anomS_p=0$ for all $p$, then $\hG=0$.

For $\ell=2$, the anomalous contribution consists of one term,
\bal
\label{eq:anom_PSFs_2p}
\anomS_{p}(\bs{\varepsilon}_p) &= 
\delta(\varepsilon_{\oli{1}}) \barS{p}{\oli{1}} ,
\eal
where $\barS{p}{\oli{1}}$ is a constant. 
Due to the equilibrium condition \eqref{eq:equilibrium:cyclic_PSFs}, we can further conclude $\barS{(12)}{1}=\barS{(21)}{2}$.

For $\ell=3$, the anomalous $\anomS_{p}$ reads
\bal
\label{eq:anom_PSFs_3p}
\anomS_{p}(\bs{\varepsilon}_p) &= 
\delta(\varepsilon_{\oli{1}}) \barS{p}{\oli{1}} 
(\circ,\varepsilon_{\oli{2}},\varepsilon_{\oli{3}})
+
\delta(\varepsilon_{\oli{3}}) \barS{p}{\oli{3}}(\varepsilon_{\oli{1}},\varepsilon_{\oli{2}},\circ)
\nn
&+
\delta(\varepsilon_{\oli{1}})\delta(\varepsilon_{\oli{2}}) \barS{p}{\oli{1},\oli{2}}\,.
\eal
Here, we inserted $\circ$'s to emphasize that functions do not depend on these arguments, and $\barS{p}{\oli{1},\oli{2}}$ is a constant. For bosonic 3p functions, $\barS{p}{\oli{1}}$ and $\barS{p}{\oli{3}}$ do not contain further $\delta$--factors that lead to anomalous parts, e.g., $\delta_{\varepsilon_{\oli{3}}} \barS{p}{\oli{1}}(\circ,\varepsilon_{\oli{2}},\varepsilon_{\oli{3}}) = 0$.

\bsubeq
To further illustrate the symbolic $\delta_{\varepsilon_{\oli{1} \dots \oli{i}}}$ notation
introduced in \Eq{eq:symbolic-delta_epsilon}, 
it yields the following relations when applied to the 
above definitions, for bosonic $\varepsilon_{\oli{i}}$:
\bal
\delta_{\varepsilon_{\oli{1}}}  S_p  \! (\bs{\varepsilon}_p)
&=
\delta(\varepsilon_{\oli{1}}) \barS{p}{\oli{1}}(\bs{\varepsilon}_p) 
+
\delta(\varepsilon_{\oli{1}})\delta(\varepsilon_{\oli{2}}) \barS{p}{\oli{1},\oli{2}},
\\
\label{eq:PSF_3p_max_anomalous}
\delta_{\varepsilon_{\oli{1}}} \delta_{\varepsilon_{\oli{2}}} S_p \! (\bs{\varepsilon}_p)
&=
\delta(\varepsilon_{\oli{1}})\delta(\varepsilon_{\oli{2}}) \barS{p}{\oli{1},\oli{2}}
.
\eal
\esubeq

For fermionic $\ell=4$, we only need 
\bal
\label{eq:anom_PSFs_4p}
\anomS_p(\bs{\varepsilon}_p) =& 
\delta(\varepsilon_{\oli{12}}) \barS{p}{\oli{12}}(\bs{\varepsilon}_p),
\eal
since, e.g., terms in the kernel proportional to $\delta_{\i\omega_{\oli{1}}-\varepsilon_{\oli{1}}}$ do not lead to anomalous contributions by the fermionic nature of $\i\omega_{\oli{1}}$.

\subsection{Effect of fully anomalous PSFs on 3p MF correlators}
\label{app:Fully_anom_PSF_3p}

In the \App{sec:appstruct3pcorr} \newlychanged{below}, we discuss the general structure of 3p MF correlators inferred by the decomposition of the PSFs. The regular PSFs, $\tilde{S}_p$, can only contribute to the regular part of the correlator. However, the effect of anomalous PSFs, $\anomS_p$, is more involved and is studied in detail in the following.

To this end, we consider PSFs with finite weight at vanishing frequency arguments. In particular, we assume the maximally anomalous form $S^{\tn{ma}}_{p}(\varepsilon_{\oli{1}}, \varepsilon_{\oli{2}}) = \delta(\varepsilon_{\oli{1}}) \delta(\varepsilon_{\oli{2}}) \barS{p}{\oli{1},\oli{2}}$ (see \Eq{eq:PSF_3p_max_anomalous}). Then, the equilibrium condition \Eq{eq:equilibrium:cyclic_PSFs} implies 
$\barS{(123)}{1;2} = \barS{(231)}{2;3} = \barS{(312)}{3;1}$ and $\barS{(132)}{1;3} = \barS{(321)}{3;2} = \barS{(213)}{2;1}$, since $\zeta_p = \zeta_{p_\lambda} = 1$ for purely bosonic correlators. 
For such PSFs, the 3p correlator evaluates to
\bal
&G^{\tn{ma}}(\i \bsomega) = \sum_p [K \ast S^{\tn{ma}}_p](\i \bsomega_p)\nn
&= \left[ \frac{\beta}{2}\left( \delta_{\i \omega_1} \Delta_{\i \omega_{12}} + \Delta_{\i \omega_1} \delta_{\i \omega_{12}} \right) + \frac{\beta^2}{6} \delta_{\i \omega_1} \delta_{\i \omega_{12}} \right] \barS{(123)}{1,2} \nn
&\hsp + \left[ \frac{\beta}{2}\left( \delta_{\i \omega_2} \Delta_{\i \omega_{23}} + \Delta_{\i \omega_2} \delta_{\i \omega_{23}} \right) + \frac{\beta^2}{6} \delta_{\i \omega_2} \delta_{\i \omega_{23}} \right] \barS{(231)}{2,3}
\nn
&\hsp + \left[ \frac{\beta}{2}\left( \delta_{\i \omega_3} \Delta_{\i \omega_{31}} + \Delta_{\i \omega_3} \delta_{\i \omega_{31}} \right) + \frac{\beta^2}{6} \delta_{\i \omega_3} \delta_{\i \omega_{31}} \right] \barS{(312)}{3,1}
\nn
&\hsp +(2 \leftrightarrow 3)\nn
&= \beta \left( \delta_{\i \omega_1} \Delta_{\i \omega_2} + \delta_{\i \omega_2} \Delta_{\i \omega_3} + \delta_{\i \omega_3} \Delta_{\i \omega_1} \right) \left( \barS{(123)}{1,2} - \barS{(132)}{1,3} \right) \nn
&+ \frac{\beta^2}{2} \delta_{\i \omega_1} \delta_{\i \omega_2} \left( \barS{(123)}{1,2} + \barS{(132)}{1,3} \right),
\eal
where $(2\leftrightarrow3)$ exchanges the indices of the frequencies and PSFs. The contribution of the regular kernel in \Eq{eq:3p_kern_Kugler} vanishes due to $\tfrac{1}{\i \omega_1\, \i \omega_{12}} + \tfrac{1}{\i \omega_2\, \i \omega_{23}} + \tfrac{1}{\i \omega_3\, \i \omega_{31}} = 0$ with $\i \omega_3 = -\i \omega_{12}$. 

For later reference (see \Apps{sec:appstruct3pcorr} and \ref{app:spec_rep_anom}), we define the constants
\bsubeq
\label{eq:maximally_anomalous_3pGiw}
\bal
\hG_{1,2} &= \tfrac{1}{2} 
\Big(
    \barS{(123)}{1,2} + \barS{(132)}{1,3}
\Big),
\label{eq:hG_12_as_PSFs}
\\
\label{eq:Gd_as_PSFs}
\hG^{\withDelta}_{1;2} &= \hG^{\withDelta}_{2;3} = \hG^{\withDelta}_{3;1} = 
\barS{(132)}{1,3}-\barS{(123)}{1,2}
,
\eal
such that $G^{\tn{ma}}$ reads
\bal
\label{eq:maximally_anomalous_3pGiw_betaterm_divergencefree}
\hG^{\tn{ma}}(\i\bsomega)
&=
\beta \left( \delta_{\i\omega_1} \Delta_{\i\omega_2} \hG^{\withDelta}_{1;2}
+ \delta_{\i\omega_2} \Delta_{\i\omega_3} \hG^{\withDelta}_{2;3}
+ \delta_{\i\omega_3} \Delta_{\i\omega_1} \hG^{\withDelta}_{3;1} \right) \nn
&\hsp + \beta^2 \delta_{\i \omega_1} \delta_{\i \omega_2} \hG_{1,2}.
\eal
\esubeq
We emphasize that $\hG^{\withDelta}_{i;j}$ and $\hG_{1,2}$ are nonzero only if the full PSFs $S_p$ contain fully anomalous contributions $S^{\tn{ma}}_p$, which is only the case for all operators being bosonic. In the next section, the most general form of 3p correlators is discussed.

\section{Calculations for 3p correlators}
\label{sec:App_3p_calculations}

This appendix is devoted to computations for the analytic continuation of 3p correlators, complementing the discussions in \Sec{sec:Analytic_cont_3p}. First, in \App{sec:appstruct3pcorr}, we discuss the general structure of MF correlators, needed in \App{sec:DerivationPCF3p} for the derivation of an explicit formula for partial MF correlators and the subsequent extraction of PSFs. In \App{app:simplifying_KF_correlators_3p}, we then present manipulations needed to construct KF correlators from analytically continued MF correlators.

\subsection{Structure of 3p correlators}
\label{sec:appstruct3pcorr}

For 3p correlators, \Eq{eq:MF_correlators_general_form} implies the general form
\bal \label{eq:general_3p_app}
G_{\i \omega_1, \i \omega_2} &= \tG_{\i \omega_1, \i \omega_2} + \hG_{\i \omega_1, \i \omega_2}, 
\nn
\hG_{\i \omega_1, \i \omega_2} &= \beta \delta_{\i\omega_1}\, \hG_{1;\i \omega_2} + \beta \delta_{\i\omega_2}\, \hG_{2;\i\omega_1} + \beta \delta_{\i\omega_3}\, \hG_{3;\i \omega_{1}}
\nn
&\hsp + \beta^2\, \delta_{\i\omega_1}\, \delta_{\i\omega_2}\, \hG_{1,2}.
\eal
Here, we used the subscript notation introduced in \Sec{sec:analytic_regions_and_discontinuities}. 

For the conversion of Matsubara sums to contour integrals we distinguish restricted from unrestricted sums (see e.g.~ \Eq{eq:sumtocont-restricted}). Therefore we explicitly distinguish terms with $\Delta_{\i\omega}$ factors, writing (cf. \Eq{eq:3p_Gf_Gd_decomp})
\bal \label{eq:3p_an_further_split}
\hG_{i;\i \omega_j} = 
\hG^{\noDelta}_{i;\i \omega_j} + \Delta_{\i \omega_j} \hG^{\withDelta}_{i;j} 
.
\eal
In \Eq{eq:Gd_as_PSFs}, we have identified the constants $\hG^{\withDelta}$ with (maximally anomalous) PSFs.
For alternative frequency parametrizations in \Eqs{eq:maximally_anomalous_3pGiw}, the constants in \Eq{eq:3p_an_further_split} read
\bal \label{eq:Gd_identities}
\hG^{\withDelta}_{1;2} = -\hG^{\withDelta}_{1;3} = - \hG^{\withDelta}_{2;1} = \hG^{\withDelta}_{2;3} = \hG^{\withDelta}_{3;1} = -\hG^{\withDelta}_{3;2},
\eal
such that, e.g., $\delta_{\i\omega_1} \Delta_{\i\omega_2}\hG^{\withDelta}_{1;2} = - \delta_{\i\omega_1} \Delta_{\i\omega_3} \hG^{\withDelta}_{1;2} = \delta_{\i\omega_1} \Delta_{\i\omega_3} \hG^{\withDelta}_{1;3}$, which follows from frequency conservation, $\i\omega_{1\dots\ell}=0$, and the $\delta_{\i\omega_i}$ factor multiplying $\hG_i$.

\subsection{Partial MF 3p correlators}
\label{sec:DerivationPCF3p}

In this appendix, we present explicit calculations concerning Steps 1 and 2 of our 3-step strategy. First, we introduce two identities used for simplifications in Step 1. 

Consider the restricted Matsubara sum of \Eq{eq:sumtocont-restricted} for $f(\i\omega') = \tilde{f}(\i\omega')/(\i\omega-\i\omega')$. Using \Eq{eq:residue} for the residue term, one obtains
\bal \label{eq:MF_sum_identity_power1}
\frac{1}{(-\beta)} \! \sum_{\i \omega'} \left( \Delta_{\i \omega - \i \omega'} - \tfrac{\beta}{2} \delta_{\i \omega - \i \omega'} \right) \tilde{f}(\i \omega') & = \ointctrclockwise_z \frac{n_z \tilde{f}(z)}{\i \omega - z} + \mcO\bigl(\tfrac{1}{\beta}\bigr).
\eal
Here, the restriction of the sum is implicit in the $\Delta$ symbol (\Eq{eq:defineDeltaSymbol}), and the first term of \Eq{eq:residue} was incorporated into the sum using the Kronecker $\delta$. 
We can identify the summand on the left of \Eq{eq:MF_sum_identity_power1} as the singularity-free 2p kernel of \Eq{eq:2p_kern_Sbierski}, and therefore this identity constitutes the convenient cancellation in \Eqs{sub:illustrate-cancellations} already on the level of kernels.
Following the same line of arguments, one can show that
\bal \label{eq:MF_sum_identity_power2}
\frac{1}{(-\beta)^2} \sum_{\i \omega'} \left( \Delta^2_{\i \omega-\i \omega'} + \tfrac{\beta^2}{12} \delta_{\i \omega-\i \omega'} \right) \tilde{f}(\i \omega') = \mcO\bigl(\tfrac{1}{\beta}\bigr).
\eal

In the following, we focus on evaluating
\bal \label{eq:tG_p_123_start}
\tG_{(123)}(\i \bsomega_{(123)}) + \mcO\bigl(\tfrac{1}{\beta}\bigr)
&= \left[ K \star G \right](\i \bsomega_{(123)}),
\eal
using the 3p kernel given in \Eq{eq:3p_kern_Sbierski} (with $\bs{\Omega}_{(123)} = \i \bsomega_{(123)} - \i \bsomega'_{(123)}$), and the general form of the 3p correlator displayed in \Eq{eq:general_3p_app}. For convencience, we focus on the identity permutation $p=(123)$; all other permutations can be obtained by replacing indices with their permuted ones, $i \rightarrow \oli{i}$. We split the calculation of \Eq{eq:tG_p_123_start} into regular ($r$) and anomalous ($a$) contributions from $G$:
\bsubeq
\bal
\tG^{\tn{r}}_{(123)}(\i \bsomega_{(123)}) + \orderbeta &= \left[ K \star \tG \right](\i \bsomega_{(123)}), \label{eq:3p_convolution_reg_start} 
\\
\tG^{\tn{a}}_{(123)}(\i \bsomega_{(123)}) + \orderbeta 
&= \left[ K \star \hG \right](\i \bsomega_{(123)}).
\label{eq:3p_convolution_an_start}
\eal
\esubeq
The computations are presented in \Apps{app:3p_Gp_regular_cont} and \ref{app:3p_Gp_anomalous_cont}, respectively, with the final result $\tG_{(123)} = \tG^{\tn{r}}_{(123)} + \tG^{\tn{a}}_{(123)}$ discussed in \App{sec:3p_partial_corr_final}. 
Additionally, we will use the super- and subscript notation introduced in \Sec{sec:analytic_regions_and_discontinuities} and suppress the frequency argument of $\tG^{\tn{r}}_{(123)}$ and $\tG^{\tn{a}}_{(123)}$.

\subsubsection{Contributions from regular part}
\label{app:3p_Gp_regular_cont}

Step 1. \textit{Matsubara summation through contour integration:} 
First, we concentrate on evaluating \Eq{eq:3p_convolution_reg_start}:
\bal
\label{eq:tG_123_r_1}
&\tG^{\tn{r}}_{(123)} + \orderbeta 
= K \star \tG
\nn
&= \frac{1}{(-\beta)^2} \sum_{\i\omega'_{1},\i\omega'_{12}} \Bigg[ \Delta_{\Omega_{12}} \left( \Delta_{\Omega_{1}} - \frac{\beta}{2} \delta_{\Omega_{1}} \right) 
\nn
&
\hsp +\delta_{\Omega_{12}} \left( - \Delta^2_{\Omega_{1}} - \frac{\beta}{2} \Delta_{\Omega_{1}} + \frac{\beta^2}{6} \delta_{\Omega_{1}} \right) \Bigg] \tG_{\i\omega'_{1},\i\omega'_{12}} 
\nn
&=
\frac{1}{(-\beta)^2} \sum_{\i\omega'_{1}} \sum_{\i \omega'_{12}}^{\neq \i \omega_{12}} \frac{1}{\i\omega_{12} - \i \omega'_{12} } \left( \Delta_{\Omega_{1}} - \frac{\beta}{2} \delta_{\Omega_{1}} \right) \tG_{\i\omega'_{1},\i\omega'_{12}} 
\nn
&+\frac{1}{(-\beta)^2} \sum_{\i\omega'_{1}} \sum_{\i\omega'_{12}} \delta_{\Omega_{12}} \left( - \Delta^2_{\Omega_{1}} - \frac{\beta}{2} \Delta_{\Omega_{1}} + \frac{\beta^2}{6} \delta_{\Omega_{1}} \right) \tG_{\i\omega'_{1},\i\omega'_{12}}.
\eal
The restricted sum over $\i\omega'_{12}$ can be rewritten using \Eq{eq:MF_sum_identity_power1}, and collecting all resulting terms $\sim \delta_{\Omega_{12}}$ yields
\bal
&\tG^{\tn{r}}_{(123)} + \orderbeta
\nn
&= \frac{1}{(-\beta)} \sum_{\i\omega'_{1}} \left( \Delta_{\Omega_{1}} - \frac{\beta}{2} \delta_{\Omega_{1}} \right) \ointctrclockwise_{z_{12}} \frac{n_{z_{12}} \tG_{\i\omega'_{1},z_{12}}}{\i\omega_{12} - z_{12}} 
\nn
&\hsp + \frac{1}{(-\beta)^2} \sum_{\i\omega'_{1}} \sum_{\i\omega'_{12}} \delta_{\Omega_{12}} \left( - \Delta^2_{\Omega_{1}} - \frac{\beta^2}{12} \delta_{\Omega_{1}} \right) \tG_{\i\omega'_{1},\i\omega'_{12}}.
\eal
The $\i \omega'_1$ sums can be further simplified with the help of \Eqs{eq:MF_sum_identity_power1} and \eqref{eq:MF_sum_identity_power2} for the second and third line, respectively, reproducing \Eq{eq:tG_p_cont_ints} for $\ell = 3$,
\bal \label{eq:Gp3pttwocont}
\tG^{\tn{r}}_{(123)} + \orderbeta =& \ointctrclockwise_{z_{1},z_{12}} \frac{ n_{z_{1}} n_{z_{12}} \tG_{z_{1},z_{12}} }{\left( \i\omega_{1} - z_{1} \right) \left( \i\omega_{12} - z_{12} \right)} + \orderbeta,
\eal
with $\ointctrclockwise_{z_{1},z_{12}} = \ointctrclockwise_{z_{1}} \ointctrclockwise_{z_{12}}$.

Step 2. \textit{Extraction of PSFs:}
Next, we deform the contours away from the imaginary axis, beginning with the contour integral over $z_{12}$. 
\begin{figure*}[t!]
    \centering
    \includegraphics[width=1 \textwidth]{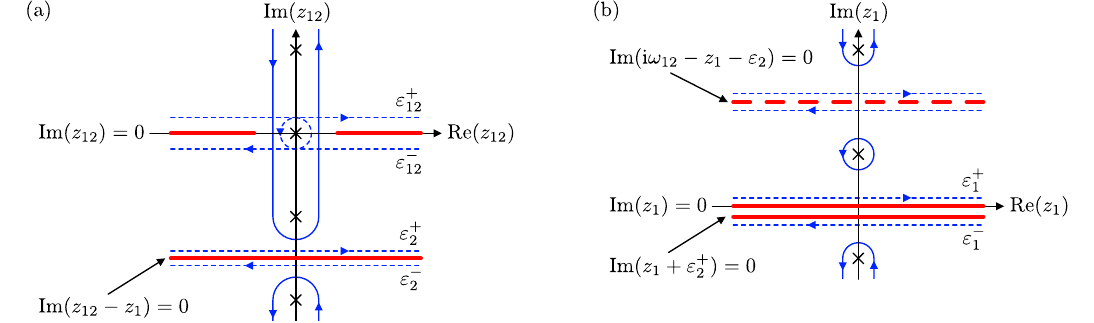}
    \caption{(a) Contour deformation used in \Eq{eq:split2branchcuts} for fermionic $z_{1}$ and $z_{2}$, therefore bosonic $z_{12}$. Black crosses denote the poles of $n_{z_{12}}$ on the imaginary axis given by bosonic Matsubara frequencies. The blue, solid contour encloses all the poles on the imaginary axis. It is deformed into the blue, dashed contour to integrate along the possible branch cuts of $\tG_{z_{1}, z_{12}}$ denoted by the red, thick lines, located at $\tn{Im}(z_{12}) = 0$ and $\tn{Im}(z_{12}-z_{1}) = 0$. (b) Contour deformation used to obtain \Eq{eq:3p_second_contour}. The branch cut at $\tn{Im}(z_{1} + \varepsilon^+_{2}) = 0$ lies infinitesimally close to the branch cut $\tn{Im}( z_{1} )  = 0$. Therefore, we integrate along the deformed blue, dashed contour, infinitesimally above and below the real axis, where the infinitesimal imaginary part of $\varepsilon^-_{1}$, with $\tn{Re}(z_1) = \varepsilon_1$, has to be larger than that of $\varepsilon^+_2$, i.e., $\vert \tn{Im} (\varepsilon^-_{1}) \vert > \vert \tn{Im} (\varepsilon^+_{2}) \vert$. The thick, red, dashed line denotes the pole at $\tn{Im}(\i\omega_{12} - z_{1} - \varepsilon_{2})$ coming from the kernel. However, these poles only contribute at $\orderbeta$ and can be neglected, see the discussion after \Eq{eq:3p_O1_pole_denom}.}
    \label{fig:G3ptbranchcuts}
\end{figure*}
During the contour deformation, we have to carefully track possible singularities of $\tG_{z_{1}, z_{12}} = \tG(z_{1}, z_{12} - z_{1},-z_{12})$. 
As explained in \Sec{sec:analytic_regions_and_discontinuities},
possible branch cuts in the complex $z_{12}$ plane lie on the lines defined by $\tn{Im}(z_{12}) = 0$ or $\tn{Im}(z_{12}-z_{1}) = 0$, see Fig. \ref{fig:G3ptbranchcuts}(a). The branch cut at $\tn{Im}(z_{12}) = 0$ is taken into account by integrating infinitesimally above and below the real $z_{12}$ axis, denoted by $\varepsilon^\pm_{12}$ with $\tn{Re}(z_{12}) = \varepsilon_{12}$. 

The second branch cut $\tn{Im}(z_{12}-z_{1}) = 0$ is included by substituting $z_{12} \rightarrow z_2 = z_{12}-z_{1}$, with $z_2$ being the new integration variable. Therefore, the contour is shifted onto the line $\tn{Im}(z_{12}-z_{1}) = 0 \rightarrow \tn{Im}( z_2) =0$, i.e., onto the real axis of the complex $z_2$ plane, and integrating infinitesimally above and below the real axis of  $z_{2}$, denoted by $\varepsilon^\pm_2$ with $\tn{Re}(z_2) = \varepsilon_2$.
The substitution also affects the argument of the \MWF{} in \Eq{eq:Gp3pttwocont}. 
However, since the $z_1$ contour encloses only the poles of $n_{z_1}$, $z_1$ can be treated as a Matsubara frequency, implying $e^{-\beta z_1} = \zeta^1$ and therefore
\bal
n_{z_{12}} &=\frac{\zeta^{12}}{e^{-\beta z_{12}}- \zeta^{12}} \overset{z_{12} \rightarrow z_{1} + z_{2}}{=} \frac{\zeta^{1} \zeta^{2}}{e^{-\beta z_{1}} e^{-\beta z_{2}}- \zeta^{1} \zeta^{2}} 
\nn
&= \frac{ \zeta^{2} }{e^{-\beta z_{2}}- \zeta^{2}} = n_{z_{2}}.
\eal
Adding the contributions from both branch cuts, the $z_{12}$ dependent terms in \Eq{eq:Gp3pttwocont} evaluate to
\bal \label{eq:split2branchcuts}
&\ointctrclockwise_{z_{12}} \frac{n_{z_{12}} \tG_{z_{1}, z_{12}}}{\i\omega_{12} - z_{12}} 
\nn
&= \int_{\varepsilon_{12}} \frac{n_{\varepsilon_{12}} \tG^{\varepsilon_{12}}_{z_{1}}}{\i\omega_{12} - \varepsilon_{12}} + \int_{\varepsilon_{2}} \frac{n_{\varepsilon_{2}} \tG^{\varepsilon_{2}}_{z_{1}}}{\i\omega_{12} - z_{1} - \varepsilon_{2}} + \orderbeta,
\eal
see also \Fig{fig:G3ptbranchcuts}(a). The term $\orderbeta$ comes from the possible poles at $z_{12} = 0$ or $z_{2} = 0$ (if $z_{12}$ or $z_2$ are bosonic) which do not contribute at $\mc{O}(1)$, see \Eq{eq:Contour_deform_boson}.

Inserting \Eq{eq:split2branchcuts} into \Eq{eq:Gp3pttwocont} yields
\bal \label{eq:G11z1def}
\tG^{\tn{r}}_{(123)} + \orderbeta = &\ointctrclockwise_{z_{1}} \frac{n_{z_{1}}}{\i\omega_{1} - z_{1}} \int_{\varepsilon_{2}} \frac{n_{\varepsilon_{2}} \tG^{\varepsilon_{2}}_{z_{1}}}{\i\omega_{12} -z_{1} - \varepsilon_{2}} 
\nn
&+ \ointctrclockwise_{z_{1}} \frac{n_{z_{1}}}{\i\omega_{1} - z_{1}} \int_{\varepsilon_{12}} \frac{n_{\varepsilon_{12}} \tG^{\varepsilon_{12}}_{z_{1}}}{\i\omega_{12} - \varepsilon_{12}}.
\eal
Next we focus on the contour deformation of $z_1$. 
For the first term, we have illustrated possible branch cuts and the contours before and after the deformation in \Fig{fig:G3ptbranchcuts}(b).
As the $z_1$ contour is deformed away from the Matsubara frequencies, we merely have to consider the singularities in the integrand of the $\varepsilon_2$ integral.
After \Eq{eq:3p_O1_pole_denom}, we will show that the singularities at $z_1=\i\omega_{12}-\varepsilon_2$ contribute at order $\orderbeta$.
We can thus focus on the branch cut in $\tG^{\varepsilon_2}_{z_1}$. 
Previously we have taken the infinitesimal limit for the imaginary shifts of $\varepsilon_2^\pm$. Thus, during the $z_1$ contour deformation we have to ensure $|\Im(\varepsilon_2^\pm)|< |\Im(\varepsilon_1^\pm)|$, see \Fig{fig:G3ptbranchcuts}(b). The $z_1$ contours infinitesimally above and below $\Re(z_1)$ are summarized in a discontinuity
\bal
\tG^{\varepsilon_2,\varepsilon_1} = \tG^{\varepsilon_2}_{\varepsilon_1^+} - \tG^{\varepsilon_2}_{\varepsilon_1^-}
,
\eal
and we thus find for the first term in \Eq{eq:G11z1def}:
\bal \label{eq:3p_second_contour}
&\ointctrclockwise_{z_{1}} \frac{n_{z_{1}}}{\i\omega_{1} - z_{1}} \int_{\varepsilon_{2}} \frac{n_{\varepsilon_{2}} \tG^{\varepsilon_{2}}_{z_{1}}}{\i\omega_{12} -z_{1} - \varepsilon_{2}} \nn
&= \int_{\varepsilon_{1}}\int_{\varepsilon_{2}} \frac{n_{\varepsilon_{1}} n_{\varepsilon_{2}} \tG^{\varepsilon_{2},\varepsilon_{1}}}{\left( \i\omega_{1} -\varepsilon_{1} \right) \left( \i\omega_{12} - \varepsilon_{12} \right)} + \orderbeta.
\eal
Repeating an analogous $z_1$ contour deformation for the second term in \Eq{eq:G11z1def}, we finally obtain
\bal \label{eq:Gpr3p}
\tG^{\tn{r}}_{(123)} &= \int_{\varepsilon_{1},\varepsilon_{2}} \frac{ n_{\varepsilon_{1}} n_{\varepsilon_{2}} \tG^{\varepsilon_{2},\varepsilon_{1}} + n_{\varepsilon_{1}} n_{\varepsilon_{12}} \tG^{\varepsilon_{12},\varepsilon_{1}} }{\left( \i\omega_{1} - \varepsilon_{1} \right) \left( \i\omega_{12} - \varepsilon_{12} \right)},
\eal
which resembles the spectral representation in \Eq{eq:MF_G-compact-b} for $\ell = 3$. 

The term $\orderbeta$ on the right of \Eq{eq:3p_second_contour} originates from the pole at $z_{1} = \i\omega_{12} - \varepsilon_{2}$ in the denominator on the left, yielding
\bal \label{eq:3p_O1_pole_denom}
\orderbeta = - \int_{\varepsilon_{2}} \frac{n_{\varepsilon_{2}} n_{-\varepsilon_{2}} \tG^{\varepsilon_{2}}_{\i\omega_{12}}}{\i\omega_{2} - \varepsilon_{2}},
\eal
with $\tG^{\varepsilon_{2}}_{\i\omega_{12}} = \tG(\i \omega_{12} - \varepsilon^+_2, \varepsilon^+_2, -\i \omega_{12}) - \tG(\i \omega_{12} - \varepsilon^-_2, \varepsilon^-_2, -\i \omega_{12})$.
That the integral on the right indeed is $\orderbeta$,
although it lacks an explicit prefactor $1/\beta$, can be seen by the following argument:
The product of two \MWFs{} $n_{\varepsilon_{2}} n_{-\varepsilon_{2}}$ has finite support on an interval $\varepsilon_{2} \in [-1/\beta, 1/\beta]$.
Therefore, the integral scales as $1/\beta$. 

To demonstrate this claim more explicitly, we proceed as follows. We note that we evaluated the imaginary-frequency convolution in \Eq{eq:tG_123_r_1} by evaluating first the $\omega'_{12}$ and then the $\omega'_1$ sum. Due to frequency conservation, we could have also evaluated the convolution by first summing over, e.g., $\omega'_2$ and then $\omega'_{12}$, or $\omega'_1$ and then $\omega'_{2}$, yielding
\bal
\tn{$\omega'_2$, then $\omega'_{12}$:}\quad  
K\star\tG
&= 
\tG^{\tn{r}}_{(123)} - \int_{\varepsilon_2} \frac{ n_{\varepsilon_2} n_{-\varepsilon_2} \tG^{\varepsilon_2}_{\i \omega_1} }{\i \omega_2 - \varepsilon_2 } + \orderbeta 
\nn
\tn{$\omega'_1$, then $\omega'_{2}$:}\quad 
K\star\tG
&= 
\tG^{\tn{r}}_{(123)} - \int_{\varepsilon_2} \frac{ n_{\varepsilon_2} n_{-\varepsilon_2} \tG^{\varepsilon_2}_{\i \omega_{12}} }{\i \omega_2 - \varepsilon_2 } 
\nn
&\hsp + \int_{\varepsilon_{12}} \frac{ n_{\varepsilon_{12}} n_{-\varepsilon_{12}} \tG^{\varepsilon_{12}}_{\i \omega_{1}} }{\i \omega_{12} - \varepsilon_{12} } +  \orderbeta  .
\eal
Equating the two expressions yields a proof for \Eq{eq:3p_O1_pole_denom}:
\bal
&\int_{\varepsilon_2} \frac{ n_{\varepsilon_2} n_{-\varepsilon_2} \tG^{\varepsilon_2}_{\i \omega_{12}} }{\i \omega_2 - \varepsilon_2 } 
\nn
&=
\int_{\varepsilon_{2}} \frac{n_{\varepsilon_{2}} n_{-\varepsilon_{2}} \tG^{\varepsilon_{2}}_{\i\omega_{1}}}{\i\omega_{2} - \varepsilon_{2}} + \int_{\varepsilon_{12}} \frac{n_{\varepsilon_{12}} n_{-\varepsilon_{12}} \tG^{\varepsilon_{12}}_{\i\omega_{1}}}{\i\omega_{12} - \varepsilon_{12}} + \orderbeta 
\nn
&
=
\ointctrclockwise_{z_{2}} \frac{n_{z_{2}} n_{-z_{2}} \tG_{\i\omega_{1},z_2}}{\i\omega_{2} - z_{2}} + \orderbeta 
\nn
&= -\frac{1}{(-\beta)^2} \sum_{\i \omega'_{2}}^{\neq \i \omega_{2}} \frac{\tG_{\i \omega_1,\i \omega'_{2}}}{(\i \omega_2 - \i \omega'_2)^2} - \frac{1}{12} \tG_{\i \omega_1,\i \omega_{2}} + \orderbeta 
\nn
&= \orderbeta
.
\eal
We obtained the third line by a contour deformation in analogy to the derivation of \Eq{eq:split2branchcuts}. 
Here, the second line can be expressed as a contour integral along the branch cuts at $\Im(z_2)=0$ and $\Im(z_{12})=0$ (blue dashed lines in \Fig{fig:G3ptbranchcuts}(a)) and the contour in the third line encloses the Matsubara frequencies (blue solid lines in \Fig{fig:G3ptbranchcuts}(a)).
For the last step, we used \Eq{eq:MF_sum_identity_power2}.

\subsubsection{Contributions from anomalous parts}
\label{app:3p_Gp_anomalous_cont}

Step 1. \textit{Matsubara summation through contour integration:}
To evaluate \Eq{eq:3p_convolution_an_start}, we first focus on $ \beta \delta_{\i \omega'_3} \hG_{3;\i\omega'_1}$, yielding $\tG^{\tn{a}}_{3;(123)}$ in a decomposition $\tG^{\tn{a}}_{(123)} = \sum_{i=1}^3 \tG^{\tn{a}}_{i;(123)}$; the contributions from $\tG^{\tn{a}}_{1;(123)}$ and $\tG^{\tn{a}}_{2;(123)}$ follow from analogous calculations. Then, the imaginary-frequency convolution of the 3p kernel with $ \beta \delta_{\i \omega'_3} \hG_{3;\i\omega'_1}$ can be rewritten as
\bal \label{eq:tG_a_123_start}
& \tG^{\tn{a}}_{3;(123)} + \orderbeta
= K \star \hG_3
\nn
&= \frac{1}{(-\beta)^2} \sum_{\i\omega'_1, \i\omega'_{12}} \Bigg[ \Delta_{\Omega_{12}} \left( \Delta_{\Omega_{1}} - \frac{\beta}{2} \delta_{\Omega_{1}} \right) 
\nn
&\hspace{10pt} +\delta_{\Omega_{12}} \left( - \Delta^2_{\Omega_{1}} - \frac{\beta}{2} \Delta_{\Omega_{1}} + \frac{\beta^2}{6} \delta_{\Omega_{1}} \right) \Bigg]\, \beta \delta_{\i\omega'_{12}}\, \hG_{3;\i \omega'_{1}} 
\nn
&= - \frac{1}{\i \omega_{12}} \frac{1}{(-\beta)} \sum_{\i\omega'_1} \left( \Delta_{\Omega_{1}} - \frac{\beta}{2} \delta_{\Omega_{1}} \right) \hG_{3;\i \omega'_{1}} 
\nn
&= - \frac{1}{\i \omega_{12}} \frac{1}{(-\beta)} \sum_{\i\omega'_1} \left( \Delta_{\Omega_{1}} - \frac{\beta}{2} \delta_{\Omega_{1}} \right) \hG^{\noDelta}_{3;\i \omega'_{1}} 
\nn
&\hspace{10pt} - \frac{1}{\i \omega_{12}} \frac{1}{(-\beta)} \sum_{\i \omega'_{1}}^{\neq 0} \left( \Delta_{\Omega_{1}} - \frac{\beta}{2} \delta_{\Omega_{1}} \right) \frac{\hG^{\withDelta}_{3;1}}{\i \omega'_1}.
\eal
In the second step, we carried out the sum over $\i \omega'_{12}$ and used $\delta_{\Omega_{12}} \delta_{\i\omega'_{12}} = \delta_{\i\omega_{12}}\, \delta_{\i\omega'_{12}} = 0$, since we enforce the external Matsubara frequencies to be nonzero. 
In the third step, we further split the anomalous part according to \Eq{eq:3p_an_further_split}.

The sums can be evaluated using \Eq{eq:MF_sum_identity_power1} and yield
\bal \label{eq:tG_a_cont_ints}
& \tG^{\tn{a}}_{3;(123)} + \orderbeta
\nn
&=- \frac{1}{\i \omega_{12}} \ointctrclockwise_{z_1} \frac{n_{z_1} \hG^{\noDelta}_{3;z_1} }{\i \omega_1 - z_1} - \frac{1}{\i \omega_{12}} \ointctrclockwise_{z_1} \frac{n_{z_1} }{(\i \omega_1 - z_1)} \frac{\hG^{\withDelta}_{3;1}}{z_1} 
\nn
&\hspace{10pt} + \frac{1}{\i \omega_{12}} \underset{z_1 = 0} {\tn{Res}} 
\left( \frac{n_{z_1} }{(\i \omega_1 - z_1)} \frac{\hG^{\withDelta}_{3;1}}{z_1} \right) + \orderbeta,
\eal
where we excluded the contribution from $\i\omega'_1 \rightarrow z_1 = 0$ by subtracting the residue. 

Step 2. \textit{Extraction of PSFs:} The first contour integral in \Eq{eq:tG_a_cont_ints} can be deformed analogously to the 2p case in \Sec{sec:2p_partial_correlators}. The integrand of the second contour integral only has poles on the imaginary axis since $\hG^{\withDelta}_{3;1}$ is a constant. 
Thus, the integral vanishes by closing the contour in the left and right half of the complex $z_1$ plane.
Further evaluating the residue, we then obtain
\bal
\tG^{\tn{a}}_{3;(123)} 
&= 
- \frac{1}{\i \omega_{12}} \int_{\varepsilon_1} \frac{n_{\varepsilon_1} \hG^{\noDelta;\varepsilon_1}_{3} }{\i \omega_1 - \varepsilon_1} - \frac{1}{2} \frac{\hG^{\withDelta}_{3;1}}{\i \omega_1 \, \i \omega_{12}} 
\nn
&= \int_{\varepsilon_1,\varepsilon_2} \frac{\hdelta(\varepsilon_{12}) n_{\varepsilon_1} \hG^{\noDelta;\varepsilon_1}_{3} - \tfrac{1}{2} \hdelta(\varepsilon_1) \hdelta(\varepsilon_{12}) \hG^{\withDelta}_{3;1}}{(\i \omega_1 - \varepsilon_1) (\i \omega_{12} - \varepsilon_{12})},
\eal
where we recovered the form of the spectral representation in \Eq{eq:MF_G-compact-b} by introducing Dirac delta functions.

Similarly, the contributions from $\hG_1$, $\hG_2$, and also $\hG_{1,2}$ to \Eq{eq:3p_convolution_an_start} can be derived, leading to the general result
\bal \label{eq:Gpr3pan}
\tG^{\tn{a}}_{(123)} = &\int_{\varepsilon_1,\varepsilon_2} \frac{1}{(\i \omega_1 - \varepsilon_1) (\i \omega_{12} - \varepsilon_{12})} \nn
&\times\Big[ \hdelta(\varepsilon_{1}) n_{\varepsilon_2} \hG^{\noDelta;\varepsilon_2}_{1}+ \hdelta(\varepsilon_{2}) n_{\varepsilon_1} \hG^{\noDelta;\varepsilon_1}_{2} + \hdelta(\varepsilon_{12}) n_{\varepsilon_1} \hG^{\noDelta;\varepsilon_1}_{3} \nn
&\hspace{10pt} + \hdelta(\varepsilon_1) \hdelta(\varepsilon_{2}) \left( \hG_{1,2} - \tfrac{1}{2} \hG^{\withDelta}_{3;1} \right) \Big].
\eal
Here, only $\hG^{\withDelta}_{3;1}$ enters, since contributions from $\hG^{\withDelta}_{1;2}$ and $\hG^{\withDelta}_{2;1}$ cancel to due \Eq{eq:Gd_identities}.

\subsubsection{Final result}
\label{sec:3p_partial_corr_final}

The main results of the previous sections are \Eqs{eq:Gpr3p} and \eqref{eq:Gpr3pan}, yielding the spectral representation for $\tG_{(123)} = \tG^{\tn{r}}_{(123)} + \tG^{\tn{a}}_{(123)}$. The partial MF correlator $\tG_p = \tG^{\tn{r}}_{p} + \tG^{\tn{a}}_{p}$ for a general permutation $p$ is then obtained by replacing any index by its permuted counterpart, $i \rightarrow p(i) = \oli{i}$. Thus, we obtain our final result
\bal
\tG_p(\i \bsomega_p) = \int_{\varepsilon_{\oli{1}}, \varepsilon_{\oli{2}}} \frac{(2\pi \i)^2 S_p(\varepsilon_{\oli{1}}, \varepsilon_{\oli{2}})}{(\i \omega_{\oli{1}} - \varepsilon_{\oli{1}}) (\i \omega_{\oli{12}} - \varepsilon_{\oli{12}})},
\eal
with the PSFs given by
\bal \label{eq:3p_PSF_final}
&(2\pi \i)^2 S_p(\varepsilon_{\oli{1}}, \varepsilon_{\oli{2}}) \nn
&= 
n_{\varepsilon_{\oli{1}}} n_{\varepsilon_{\oli{2}}} \tG^{\varepsilon_{\oli{2}},\varepsilon_{\oli{1}}} + n_{\varepsilon_{\oli{1}}} n_{\varepsilon_{\oli{12}}} \tG^{\varepsilon_{\oli{12}},\varepsilon_{\oli{1}}} + \hdelta(\varepsilon_{\oli{1}}) n_{\varepsilon_{\oli{2}}} \hG^{\noDelta;\varepsilon_{\oli{2}}}_{\oli{1}} 
\nn
&\hsp + \hdelta(\varepsilon_{\oli{2}}) n_{\varepsilon_{\oli{1}}} \hG^{\noDelta;\varepsilon_{\oli{1}}}_{\oli{2}} + \hdelta(\varepsilon_{\oli{3}}) n_{\varepsilon_{\oli{1}}} \hG^{\noDelta;\varepsilon_{\oli{1}}}_{\oli{3}} 
\nn
&\hsp + \hdelta(\varepsilon_{\oli{1}}) \hdelta(\varepsilon_{\oli{2}}) \left( \hG_{\oli{1},\oli{2}} - \tfrac{1}{2} \hG^{\withDelta}_{\oli{3};\oli{1}} \right).
\eal

PSFs for all six permutations are recovered by inserting the respective $\oli{i}$ into above equation. They can be expressed in terms of analytic regions (cf. \Fig{fig:analyticregions}) using 
\bsubeq
\label{eq:tG_discont_analytic_regions}
\bal
\label{eq:tG_2_1_analytic_regions}
\tG^{\varepsilon_2,\varepsilon_1} &= -\tG^{\varepsilon_{13},\varepsilon_1} = - \tG^{\varepsilon_2,\varepsilon_3} = \tG^{\varepsilon_{13},\varepsilon_3} 
\nn
&= 
\tG'^{[3]} - \tG^{[1]} - \tG'^{[1]} + \tG^{[3]} 
,
\\
\label{eq:tG_1_2_analytic_regions}
\tG^{\varepsilon_1,\varepsilon_2} &= -\tG^{\varepsilon_{23},\varepsilon_2} = -\tG^{\varepsilon_1,\varepsilon_3} = \tG^{\varepsilon_{23},\varepsilon_3}
\nn
&= 
\tG'^{[3]} - \tG^{[2]} - \tG'^{[2]} + \tG^{[3]} 
,
\\
\label{eq:tG_3_1_analytic_regions}
\tG^{\varepsilon_3,\varepsilon_1} &= -\tG^{\varepsilon_{12},\varepsilon_1} = -\tG^{\varepsilon_3,\varepsilon_2} = \tG^{\varepsilon_{12},\varepsilon_2} 
\nn
&=
\tG'^{[2]} - \tG^{[1]} - \tG'^{[1]} + \tG^{[2]} 
,
\\
\hG^{\noDelta; \varepsilon_2}_1 
&= -\hG^{\noDelta; \varepsilon_3}_1 = 
\hG^{\noDelta;[2]}_1
-
\hG^{\noDelta;[3]}_1 
, \\
\hG^{\noDelta; \varepsilon_1}_2 
&= -\hG^{\noDelta; \varepsilon_2}_2 = 
\hG^{\noDelta;[1]}_2 
-
\hG^{\noDelta;[3]}_2 
, \\
\hG^{\noDelta; \varepsilon_1}_3 
&= -\hG^{\noDelta; \varepsilon_2}_3 = 
\hG^{\noDelta;[1]}_3
-
\hG^{\noDelta;[2]}_3 
, \\
\hG^{\withDelta}_{1;2} &= - \hG^{\withDelta}_{1;3} = - \hG^{\withDelta}_{2;1} = \hG^{\withDelta}_{2;3} = \hG^{\withDelta}_{3;1} = - \hG^{\withDelta}_{3;2}, \\
\hG_{\oli{1}, \oli{2}} &= \hG_{1, 2},
\eal
\esubeq
with the definitions introduced in \Sec{sec:Analytic_cont_3p}
\bsubeq
\bal
    G^{[1]} &= \tG(\varepsilon^+_1, \varepsilon^-_2, \varepsilon^-_3), \quad 
    G'^{[1]} = \tG(\varepsilon^-_1, \varepsilon^+_2, \varepsilon^+_3), \\
    G^{[2]} &= \tG(\varepsilon^-_1, \varepsilon^+_2, \varepsilon^-_3), \quad
    G'^{[2]} = \tG(\varepsilon^+_1, \varepsilon^-_2, \varepsilon^+_3), \\
    G^{[3]} &= \tG(\varepsilon^-_1, \varepsilon^-_2, \varepsilon^+_3), \quad
    G'^{[3]} = \tG(\varepsilon^+_1, \varepsilon^+_2, \varepsilon^-_3), \\
    \hG^{\noDelta;[2]}_1 &= \hG^{\noDelta}_1(\circ, \varepsilon^+_2, \varepsilon^-_3), \quad 
    \hG^{\noDelta;[3]}_1 = \hG^{\noDelta}_1(\circ, \varepsilon^-_2, \varepsilon^+_3), \\
    \hG^{\noDelta;[1]}_2 &= \hG^{\noDelta}_1(\varepsilon^+_1, \circ, \varepsilon^-_3), \quad 
    \hG^{\noDelta;[3]}_2 = \hG^{\noDelta}_1(\varepsilon^-_1, \circ, \varepsilon^+_3), \\
    \hG^{\noDelta;[1]}_3 &= \hG^{\noDelta}_1(\varepsilon^+_1, \varepsilon^-_2, \circ), \quad 
    \hG^{\noDelta;[2]}_3 = \hG^{\noDelta}_1(\varepsilon^-_1, \varepsilon^+_2, \circ).
\eal
\esubeq
Here, we have inserted a $\circ$ at the position of the frequency arguments on which the function does not depend. Note that \Eqs{eq:tG_2_1_analytic_regions}--\eqref{eq:tG_3_1_analytic_regions} also imply, e.g., $\tG^{\varepsilon_2,\varepsilon_1} = \tG^{\varepsilon_1,\varepsilon_2} + \tG^{\varepsilon_3,\varepsilon_1}$. Relations of this form can be used to simplify PSF (anti)commutators, which appear in \Sec{sec:3p_Keldysh_correlators}.

One additional comment is in order for the regular contributions in \Eq{eq:3p_PSF_final}. 
Consider, e.g., permutation $p=(123)$ and $n_{\varepsilon_{1}}$ a bosonic \MWF{}. Then, if the  regular contributions $\tG^{\varepsilon_{2}, \varepsilon_{1}}$ and $\tG^{\varepsilon_{12}, \varepsilon_{1}}$ contain terms proportional to Dirac $\delta(\varepsilon_1)$, the combination $n_{\varepsilon_1} \delta(\varepsilon_1)$ is ill-defined as the \MWF{} diverges for vanishing frequencies. For their evaluation, however, we can use \Eqs{eq:tG_2_1_analytic_regions}--\eqref{eq:tG_3_1_analytic_regions} to rewrite
\bal
\label{eq:3p_PSF_bosonic_issue_MWF}
&(2\pi \i)^2 \tilde{S}_{(123)}(\varepsilon_{1}, \varepsilon_{2}) 
\nn
& = n_{\varepsilon_{1}} n_{\varepsilon_{2}} \tG^{\varepsilon_{2},\varepsilon_{1}} + n_{\varepsilon_{1}} n_{\varepsilon_{12}} \left( \tG^{\varepsilon_{1},\varepsilon_{12}} - \tG^{\varepsilon_{2},\varepsilon_{1}} \right) 
\nn
& = - n_{-\varepsilon_{2}} n_{\varepsilon_{12}} \tG^{\varepsilon_{2},\varepsilon_{1}} + n_{\varepsilon_{1}} n_{\varepsilon_{12}} \tG^{\varepsilon_{1},\varepsilon_{12}}.
\eal
Here, the first term does not include $n_{\varepsilon_1}$, and the discontinuity $\tG^{\varepsilon_{1},\varepsilon_{12}}$ in the second term does not contain $\delta(\varepsilon_1)$ contributions (see, e.g., \Eqs{eq:equilibrium_condition:S_equals_n_Commutator} and discussion thereafter), circumventing the occurrence of bosonic $n_{\varepsilon_1} \delta(\varepsilon_1)$ contributions.

\subsection{Simplifications for KF correlators for $\ell=3$}
\label{app:simplifying_KF_correlators_3p}

\begin{table*}[t]
\renewcommand*{\arraystretch}{1.6}
\captionsetup{skip=10pt}
\begin{center}
\begin{tabular}{ >{\centering\arraybackslash} m{1.5cm} >{\centering\arraybackslash} m{1.5cm} >{\centering\arraybackslash} m{1.5cm} >{\centering\arraybackslash} m{1.5cm} m{6.2cm} } 
$p$ \vsp & $\bs{k}_p$ \vsp & $[\heta_1 \heta_2]$ \vsp & $[\oli{\heta}_1 \oli{\heta}_2]$ \vsp & $K^{[\heta_1 \heta_2]}(\bs{\omega}_p) = \tK(\bs{\omega}^{[\oli{\heta_1}]}_p) - \tK(\bs{\omega}^{[\oli{\heta_2}]}_p)$ \vsp \\ \hline 
  $(123)$    &   $212$    &   $[13]$   &   $[13]$   &   $ \tK(\bs{\omega}^{[1]}_{(123)}) - \tK(\bs{\omega}^{[3]}_{(123)}) = \hdelta(\omega_1) \frac{1}{\omega^-_{2}} - \hdelta(\omega_{12}) \frac{1}{\omega^-_{2}}$   \\ 
  $(132)$    &   $221$   &   $[12]$   &   $[13]$   &    $\tK(\bs{\omega}^{[1]}_{(132)}) - \tK(\bs{\omega}^{[3]}_{(132)}) = -\hdelta(\omega_1) \frac{1}{\omega^-_{2}}$   \\
  $(213)$   &   $122$   &   $[23]$   &   $[13]$   &   $\tK(\bs{\omega}^{[1]}_{(213)}) - \tK(\bs{\omega}^{[3]}_{(213)}) = \hdelta(\omega_{12}) \frac{1}{\omega^-_{2}}$   \\ 
  $(231)$   &   $122$   &   $[23]$   &   $[31]$   &   $\tK(\bs{\omega}^{[3]}_{(231)}) - \tK(\bs{\omega}^{[1]}_{(231)}) = \hdelta(\omega_{1}) \frac{1}{\omega^-_{2}}$   \\ 
  $(312)$   &   $221$   &   $[12]$   &   $[31]$   &   $\tK(\bs{\omega}^{[3]}_{(312)}) - \tK(\bs{\omega}^{[1]}_{(312)}) = -\hdelta(\omega_{12}) \frac{1}{\omega^-_{2}}$   \\
  $(321)$   &   $212$   &   $[13]$   &   $[31]$   &   $\tK(\bs{\omega}^{[3]}_{(321)}) - \tK(\bs{\omega}^{[1]}_{(321)}) = -\hdelta(\omega_1) \frac{1}{\omega^-_{2}} + \hdelta(\omega_{12}) \frac{1}{\omega^-_{2}}$
\end{tabular}
\end{center}
\caption{ $\ell=3$: Simplification of the Keldysh kernel \eqref{eq:KFkerneletas} for the KF correlator $G^{[13]}$ for all permutations by application of the identity \eqref{eq:hdelta_identity}. For permutations $p=(123)$ and $p=(321)$, manipulations presented in \Eq{eq:3p_KF_kernel_manipulation_app} were performed. Additionally, energy conservation and the constraints enforced by the $\delta$-functions allow us to express all denominators through $\omega^-_2$.}
\label{tab:3p_KF_kernels_G212}
\end{table*}

In the following, we show that the spectral representation of Keldysh components can be recast into a form that is formally equivalent to \Eqs{eq:KF_spec_rep_eta_alpha}, but more convenient for the purpose of analytic continuation.
The new representation enables us to insert the PSFs in \Eq{eq:3p_PSF_final} and obtain expressions for the Keldysh components in terms of analytic continuations of MF correlators. 
This constitutes Step 3 of our three-step strategy.

While the following calculations are demonstrated for explicit examples of 3p KF components, they can be generalized to arbitrary KF components and even to arbitrary $\ell$p functions (see \App{app:simplifying_KF_correlators}).

\subsubsection{Simplifications for KF correlator $G^{[\eta_1 \eta_2]}$}
\label{app:rewritingkerneltwo2s_3p}

We begin with outlining the necessary steps to express the KF component $G^{[\eta_1 \eta_2]}$ in terms of analytically continued MF correlators on the example $G^{[13]}$. The simplifcations rely on repeated application of identity \eqref{eq:hdelta_identity}.

The spectral representation in \Eqs{eq:KF_spec_rep_eta_alpha} serves as our starting point. As a first step, we bring the Keldysh kernel $K^{[\heta_1 \heta_2]}$ in a more convenient form, starting with permutation $p=(123)$, where $[\heta_1 \heta_2] = [\eta_1 \eta_2] = [13]$ and therefore
\bal
K^{[13]}(\bs{\omega}_{(123)}) = \tK(\bs{\omega}^{[1]}_{(123)}) - \tK(\bs{\omega}^{[3]}_{(123)}) = \frac{1}{\omega^{[1]}_1\, \omega^{[1]}_{12}} - \frac{1}{\omega^{[3]}_1\, \omega^{[3]}_{12}}.
\eal
In the first term, all frequency combinations in the denominator acquire a positive imaginary shift, whereas in the second term they obtain a negative imaginary shift. Adding and subtracting $1/(\omega^{[1]}_1\, \omega^{[3]}_{12})$, identity \eqref{eq:hdelta_identity} leads to
\bal \label{eq:3p_KF_kernel_manipulation_app}
K^{[13]}(\bs{\omega}_{(123)}) &= \left( \frac{1}{\omega^{[1]}_1} - \frac{1}{\omega^{[3]}_1} \right) \frac{1}{\omega^{[3]}_{12}} + \left( \frac{1}{\omega^{[1]}_{12}} - \frac{1}{\omega^{[3]}_{12}} \right) \frac{1}{\omega^{[1]}_{1}} \nn
&= \hdelta(\omega_1) \frac{1}{\omega^-_{2}} + \hdelta(\omega_{12}) \frac{1}{\omega^+_{1}}.
\eal
The kernels for all other permutations can be simplified in a similar manner, and the results are summarized in \Tab{tab:3p_KF_kernels_G212}. Collecting all contributions proportional to either $\hdelta(\omega_1)/\omega^-_2$ or $\hdelta(\omega_{12})/\omega^-_2$ yields \Eq{eq:G13_int_form}. The PSF (anti)commutators therein are evaluated using the relations in \Eqs{eq:tG_discont_analytic_regions} and result in

\bsubeq \label{eq:PDF_anticomm_G212_app}
\bal \label{eq:PDF_anticomm_G212_app-a}
S_{[1,[2,3]_-]_+} &= S_{(123)} - S_{(132)} + S_{(231)} - S_{(321)}, \\
&= N_{\varepsilon_1} \tG^{\varepsilon_1,\varepsilon_2} - 2 \hdelta(\varepsilon_1) \hG^{\noDelta;\varepsilon_2}_1 - 2 \hdelta(\varepsilon_1) \hdelta(\varepsilon_2) \hG^{\withDelta}_{1;2},
\nn
\label{eq:PDF_anticomm_G212_app-b}
S_{[[1,2]_-,3]_+} &= S_{(123)} - S_{(213)} + S_{(312)} - S_{(321)}\\ 
&= -N_{\varepsilon_{12}} \tG^{\varepsilon_{12},\varepsilon_2} + 2 \hdelta(\varepsilon_{12}) \hG^{\noDelta;\varepsilon_2}_3 + 2 \hdelta(\varepsilon_1) \hdelta(\varepsilon_2) \hG^{\withDelta}_{3;2}, \nonumber
\eal
\esubeq
where we suppressed the frequency arguments of the PSFs.

\subsubsection{Simplifications for KF correlator $G^{[\eta_1 \eta_2 \eta_3]}$}
\label{app:rewritingkernelthree3s_3p}

\begin{table*}[t]
\renewcommand*{\arraystretch}{1.6}
\captionsetup{skip=10pt}
\begin{center}
\begin{tabular}{ >{\centering\arraybackslash} m{1.5cm} m{7cm} } 
$p$ & Kernel of $G^{[123]} - G^{[3]}$ \\ \hline
$(123)$ & $K^{[123]}(\bs{\omega}_{(123)}) - \tK(\bs{\omega}^{[3]}_{(123)}) = \hdelta(\omega_1) \hdelta(\omega_2) + \hdelta(\omega_1) \frac{1}{\omega^-_2}$ 
\\ 
$(132)$ & $K^{[123]}(\bs{\omega}_{(132)}) - \tK(\bs{\omega}^{[3]}_{(132)}) = -\hdelta(\omega_1) \frac{1}{\omega^-_2} - \hdelta(\omega_{2}) \frac{1}{\omega^-_1}$ 
\\
$(213)$ & $K^{[123]}(\bs{\omega}_{(213)}) - \tK(\bs{\omega}^{[3]}_{(213)}) = \hdelta(\omega_1) \hdelta(\omega_2) + \hdelta(\omega_2) \frac{1}{\omega^-_1}$ 
\\
$(231)$ & $K^{[123]}(\bs{\omega}_{(231)}) - \tK(\bs{\omega}^{[3]}_{(231)}) = - \hdelta(\omega_2) \frac{1}{\omega^-_1} - \hdelta(\omega_1) \frac{1}{\omega^-_2}$ 
\\
$(312)$ & $K^{[123]}(\bs{\omega}_{(312)}) - \tK(\bs{\omega}^{[3]}_{(312)}) = \hdelta(\omega_1) \hdelta(\omega_2) + \hdelta(\omega_2) \frac{1}{\omega^-_1}$ 
\\
$(321)$ & $K^{[123]}(\bs{\omega}_{(321)}) - \tK(\bs{\omega}^{[3]}_{(321)}) = \hdelta(\omega_1) \hdelta(\omega_2) + \hdelta(\omega_1) \frac{1}{\omega^-_2}$ 
\\
\end{tabular}
\end{center}
\caption{ $\ell=3$: Keldysh kernel for $G^{[123]} - G^{[3]}$ in \Eq{eq:3p_diff_kern_G222_app}, evaluated for all permutations.}
\label{tab:3p_KF_kernels_G222}
\end{table*}

In \Sec{sec:3p_KF_threeetas} it was pointed out that the Keldysh component $G^{[123]}$ can be computed by subtracting a fully retarded correlator, e.g. $G^{[3]}$, in order to reuse identity \eqref{eq:hdelta_identity}. 

The kernel of $G^{[3]}$ is simply given by $K^{[\hat{3}]}(\bsomega_p) = K(\bsomega^{[3]}_p)$ and therefore permutation independent, as discussed before \Eq{subeq:Analytic_cont_fully_ret}.
Since $G^{[123]} = G^{222}$ implies $\bs{k}_p = 222$ and consequently $[\heta_1 \heta_2 \heta_3] = [123]$ for any permutation, the kernel for $G^{[123]} - G^{[3]}$ reads
\bal \label{eq:3p_diff_kern_G222_app}
&K^{[123]}(\bs{\omega}_p) - K^{[\hat{3}]}(\bs{\omega}_p) \nn
&= \tK(\bs{\omega}^{[\oli{1}]}_p) - \tK(\bs{\omega}^{[\oli{2}]}_p) + \tK(\bs{\omega}^{[\oli{3}]}_p) - \tK(\bs{\omega}^{[3]}_p),
\eal
and therefore the effect of subtracting $G^{[3]}$ is permutation dependent.

We first consider permutation $p=(123)$, for which the difference of kernels simplifies to
\bal \label{eq:kern_G222_G112_p123_app}
K^{[123]}(\bs{\omega}_{(123)}) &- K^{[3]}(\bs{\omega}_{(123)}) = \tK(\bs{\omega}^{[1]}_{(123)}) - \tK(\bs{\omega}^{[2]}_{(123)}) \nn
&= \frac{1}{\omega^{[1]}_1 \omega^{[1]}_{12}} - \frac{1}{\omega^{[2]}_1 \omega^{[2]}_{12}} = \hdelta(\omega_1) \frac{1}{\omega^+_2}.
\eal
In the last step, we were able to use \Eq{eq:hdelta_identity} again, set $\omega^{[1]}_{12} = \omega^{[2]}_{12} = \omega^+_{12}$, and reduced $\omega_{12} = \omega_{2}$ due to the $\delta$-function. 
For the comparison to kernels of other permutations, it is convenient to additionally add and subtract $\hdelta(\omega_1)/\omega^-_2$ to obtain
\bal \label{eq:kern_G222_G112_p123_final_app}
K^{[123]}(\bs{\omega}_{(123)}) - K^{[3]}(\bs{\omega}_{(123)}) = \hdelta(\omega_1) \hdelta(\omega_2) + \hdelta(\omega_1) \frac{1}{\omega^-_2}.
\eal

For permutation $p=(132)$, \Eq{eq:3p_diff_kern_G222_app} yields
\bal
&\tK(\bs{\omega}^{[1]}_{(132)}) - \tK(\bs{\omega}^{[3]}_{(132)}) + \tK(\bs{\omega}^{[2]}_{(132)}) - \tK(\bs{\omega}^{[3]}_{(132)}) \nn
&= \hdelta(\omega_1) \frac{1}{\omega^+_3} - \hdelta(\omega_{13}) \frac{1}{\omega^-_1}.
\eal
Using $\omega^+_3 = - \omega^-_2$ due to energy conservation and the $\delta$--function, the first term matches the second term in \Eq{eq:kern_G222_G112_p123_final_app}. Therefore, PSFs of permutations $p=(123), (132)$ can be expressed through PSF (anti)commutators as in the previous section, motivating the manipulation from \Eq{eq:kern_G222_G112_p123_app} to \eqref{eq:kern_G222_G112_p123_final_app}.

A summary of the kernels for all permutations is given in \Tab{tab:3p_KF_kernels_G222}. In these kernels, a total of three unique terms occur, given by $\hdelta(\omega_1) \hdelta(\omega_2)$, $\hdelta(\omega_1)/\omega^-_2$, or $\hdelta(\omega_2)/\omega^-_1$. Collecting all PSFs convoluted with the same expressions gives \Eq{eq:G222_int_PSF_anticom}, with the PSF (anti)commutators evaluating to

\bal \label{eq:3p_G222_PSF_anticom_app}
S_{[[1,2]_+,3]_+}(\varepsilon_1,\varepsilon_2) &= (1 + N_{\varepsilon_1} N_{\varepsilon_2}) \tG^{\varepsilon_2,\varepsilon_1} + N_{\varepsilon_{12}} N_{\varepsilon_1} \tG^{\varepsilon_{12},\varepsilon_1} \nn
&\hsp - 2 \hdelta(\varepsilon_1) N_{\varepsilon_2} \hG^{\noDelta;\varepsilon_2}_1 - 2 \hdelta(\varepsilon_2) N_{\varepsilon_1} \hG^{\noDelta;\varepsilon_1}_2 \nn
&\hsp - 2 \hdelta(\varepsilon_{12}) N_{\varepsilon_1} \hG^{\noDelta;\varepsilon_1}_3 + 4 \hdelta(\varepsilon_1) \hdelta(\varepsilon_2) \hG_{1,2}, \nn
S_{[1,[2,3]_-]_-}(\varepsilon_1,\varepsilon_2) &= \tG^{\varepsilon_1,\varepsilon_2}, \nn
S_{[2,[1,3]_-]_-}(\varepsilon_2,\varepsilon_1) &= \tG^{\varepsilon_2,\varepsilon_1}.
\eal

This concludes our appendix on additional computations for the analytic continuation of 3p correlators.

\section{Partial MF 4p correlators}
\label{sec:4p_PCF_app}
In this appendix, we discuss purely fermionic partial MF 4p correlators. 
However, we do not display explicit calculations here. 
Rather, we introduce an iterative procedure to derive the structure of 4p PSFs from 3p PSFs, based on our insights from 2p and 3p calculations.
For a general fermionic MF 4p correlator, only the sums of two fermionic frequencies result in bosonic frequencies, which, in turn, might lead to anomalous terms. According to \Eq{eq:MF_correlators_general_form}, the general form of the correlator thus reads
\bal \label{eq:4p_gen_corr_app}
G_{\i \omega_1, \i \omega_2, \i \omega_3} =& \tG_{\i \omega_1, \i \omega_2, \i \omega_3} + \beta \delta_{\i \omega_{12}}\, \hG_{12;\i \omega_1, \i \omega_3} \nn
&+ \beta \delta_{\i \omega_{13}}\, \hG_{13;\i \omega_1, \i \omega_2} + \beta \delta_{\i \omega_{23}}\, \hG_{23;\i \omega_1, \i \omega_2}.
\eal

\subsection{Regular contributions}
Step 1. \textit{Matsubara summation through contour integration:} To derive partial MF 4p correlators, we insert \Eq{eq:4p_gen_corr_app} and the singularity-free 4p kernel (\Eq{eq:4p_kern_Sbierski}) into \Eq{eq:tildeGpstart}:
\bal \label{eq:4p_partial_corr_start}
\tG_{(1234)}(\i \bsomega_{(1234)}) + \orderbeta = \left[ K \star G \right](\i \bsomega_{(1234)}).
\eal
Here,we again consider the permutation $p=(1234)$ first, before obtaining the general result by replacing all indices $i \rightarrow \oli{i}$.
By repeated use of the identities in \Eqs{eq:MF_sum_identity_power1} and \eqref{eq:MF_sum_identity_power2}, together with the analogously proven new identity
\bal
\frac{1}{(-\beta)^3} \sum_{\i \omega'} \Delta^3_{\i \omega-\i \omega'}\, \tilde{f}(\i \omega') = \orderbeta,
\eal
the imaginary-frequency convolution can again be expressed through contour integrals. Focusing on the regular contribution to the correlator, $\tG$, first, we indeed recover \Eq{eq:tG_p_cont_ints} for $\ell = 4$:
\bal \label{eq:4p_partial_corr_cont}
&\tG^{\tn{r}}_{(1234)}(\i \bsomega_{(1234)}) + \orderbeta = \left[ K \star \tG \right](\i \bsomega_{(1234)}) \nn
&=\ointctrclockwise_{z_1} \ointctrclockwise_{z_{12}} \ointctrclockwise_{z_{123}} \frac{ n_{z_{1}} n_{z_{12}} n_{z_{123}} \tG_{z_1,z_{12},z_{123}} }{(\i \omega_1 - z_1) (\i \omega_{12} - z_{12}) (\i \omega_{123} - z_{123})}
.
\eal

Step 2. \textit{Extraction of PSFs:}
For the deformation of the contour, it is instructive to recapitulate the 2p and 3p results for the regular contributions to the PSFs.
As a function of complex frequencies, a general 2p MF correlator $\tG_{z_1} = \tG(z_1, -z_1)$ has one possible branch cut defined by $\tn{Im}( z_1 ) = 0$, resulting in 
\bal
(2\pi \i) S^{\tn{r}}_{(12)}(\varepsilon_1) = n_{\varepsilon_1} \tG^{\varepsilon_1}.
\label{eq:2p_PSF_reg_app}
\eal
In the 3p case, the additional frequency dependence of $\tG_{z_1, z_{12}} = \tG(z_1, z_{12} - z_1, -z_{12})$ introduces two further branch cuts at $\Im(z_{12}) = 0$ and $\Im(z_{12} - z_1) = \Im(z_2) = 0$, additionally to $\tn{Im}(z_1) = 0$. 
According to \Eq{eq:Gp3pttwocont}, the contour of $\ointctrclockwise_{z_{12}}$ is deformed first, taking account of the latter two out of the three branch cuts. This yields a sum of the discontinuities $\tG^{\varepsilon_{12}}_{z_1}$ and $\tG^{\varepsilon_{2}}_{z_1}$, multiplied with the respective \MWFs{} (\Eq{eq:split2branchcuts}). The subsequent contour deformation of $\ointctrclockwise_{z_{1}}$ reduces to an effective 2p calculation, i.e., only the branch cut at $\tn{Im}(z_1) = 0$ remains, resulting in
\bal
(2\pi \i)^2 S^{\tn{r}}_{(123)}(\varepsilon_1,\varepsilon_2) &= n_{\varepsilon_2} n_{\varepsilon_1} \tG^{\varepsilon_2,\varepsilon_1} + n_{\varepsilon_{12}} n_{\varepsilon_1} \tG^{\varepsilon_{12},\varepsilon_1},
\eal
with the discontinuity in $\varepsilon_1$ to the right of $\varepsilon_2$ and $\varepsilon_{12}$.

In the 4p case, the new frequency $z_{123}$ generates four additional branch cuts (see discussion in \Sec{sec:analytic_regions_tG}), defined by 
vanishing $\tn{Im}(z_{123})$, $\tn{Im}(z_{123}-z_{1})$, $\tn{Im}(z_{123}-z_{12})$ or $\tn{Im}(z_{123}-z_{12} + z_1)$, yielding a total of seven possible branch cuts together with $\tn{Im}(z_{12}) = 0$, $\tn{Im}(z_{12}-z_1) = 0$, and $\tn{Im}(z_1) = 0$ from the 3p case. Since $\ointctrclockwise_{z_{123}}$ is deformed first according to \Eq{eq:4p_partial_corr_cont}, the four new branch cuts are taken into account via a sum of the discontinuities $\tG^{\varepsilon_3}_{z_{12},z_1}$, $\tG^{\varepsilon_{123}}_{z_{12},z_1}$, $\tG^{\varepsilon_{13}}_{z_{12},z_1}$, and $\tG^{\varepsilon_{23}}_{z_{12},z_1}$, multiplied with the respective \MWFs{}. For each of these discontinuities, the subsequent contour deformations of $\ointctrclockwise_{z_{12}}$ and $\ointctrclockwise_{z_{1}}$ reduces to an effective 3p calculation. Consequently, we obtain
\bal \label{eq:4p_PSF_reg_1}
&(2\pi \i)^3 S^{\tn{r}}_{(1234)}(\varepsilon_1,\varepsilon_2,\varepsilon_3) 
\nn
&= n_{\varepsilon_{3}} n_{\varepsilon_2} n_{\varepsilon_1} \tG^{\varepsilon_{3},\varepsilon_2,\varepsilon_1} + n_{\varepsilon_{123}} n_{\varepsilon_2} n_{\varepsilon_1} \tG^{\varepsilon_{123},\varepsilon_2,\varepsilon_1} 
\nn
&\hsp + n_{\varepsilon_{13}} n_{\varepsilon_2} n_{\varepsilon_1} \tG^{\varepsilon_{13},\varepsilon_2,\varepsilon_1} + n_{\varepsilon_{23}} n_{\varepsilon_2} n_{\varepsilon_1} \tG^{\varepsilon_{23},\varepsilon_2,\varepsilon_1} 
\nn
&\hsp + n_{\varepsilon_{3}} n_{\varepsilon_{12}} n_{\varepsilon_1} \tG^{\varepsilon_{3},\varepsilon_{12},\varepsilon_1} + n_{\varepsilon_{123}} n_{\varepsilon_{12}} n_{\varepsilon_1} \tG^{\varepsilon_{123},\varepsilon_{12},\varepsilon_1} 
\nn
&\hsp+ n_{\varepsilon_{13}} n_{\varepsilon_{12}} n_{\varepsilon_1} \tG^{\varepsilon_{13},\varepsilon_{12},\varepsilon_1} + n_{\varepsilon_{23}} n_{\varepsilon_{12}} n_{\varepsilon_1} \tG^{\varepsilon_{23},\varepsilon_{12},\varepsilon_1}.
\eal
We have also checked this result by explicit contour deformations in \Eq{eq:4p_partial_corr_cont}. There, the poles of the denominators can be ignored since they only contribute at order $\orderbeta$, similarly to \Eq{eq:3p_O1_pole_denom} in the 3p case.
To further simplify \Eq{eq:4p_PSF_reg_1}, we note that, for fermionic 4p correlators, two consecutive bosonic discontinuities have to vanish, i.e., $\tG^{\varepsilon_{13},\varepsilon_{12},\varepsilon_1} = \tG^{\varepsilon_{23},\varepsilon_{12},\varepsilon_1}$ = 0,
since their kernels carry one bosonic argument only (see \App{app:branchcuts_as_PSFs} for further details).

\subsection{Anomalous contributions}
\label{app:4p_anom_cont}
We do not present the derivations of the anomalous contributions of $G$ to \Eq{eq:4p_partial_corr_start} explicitly here, as these correspond to 3p calculations. There is one crucial difference, however. The anomalous kernel in \Eq{eq:4p_kern_Kugler} for the fermionic 4p case reduces to
\bal
\label{eq:4p_anomalous_kernel}
\hK^{\tn{alt}}(\bsOmega_p) &= - \frac{\beta}{2} \delta_{\i \omega_{\oli{12}} - \varepsilon_{\oli{12}}}\, \frac{1}{\left(\i \omega_{\oli{1}} - \varepsilon_{\oli{1}} \right) \left(\i \omega_{\oli{3}} - \varepsilon_{\oli{3}} \right)},
\eal
and thus only depends on fermionic Matsubara frequencies. 
Therefore, anomalous terms such as $\hG_{13;\i \omega_1, \i \omega_2}$ only depend on the frequencies $\i \omega_1$ and $\i \omega_2$ separately, but not on their sum $\i \omega_{12}$. In the complex frequency plain, this implies that $\hG_{13;z_1, z_2}$ has branch cuts only for $\Im( z_1 ) = 0$ and $\Im( z_2 ) = 0$, but not for $\Im(z_{12}) = 0$, in contrast to the regular 3p case. Additionally, since the denominators in \Eq{eq:4p_anomalous_kernel} are non-singular due to the fermionic Matsubara frequencies, we need not distinguish the anomalous contributions by factors of $\Delta_{\i\omega}$, e.g., splitting $\hG_{13}$, into $\hG^{\noDelta}_{13}$ and $\hG^{\withDelta}_{13}$ terms, as was the case for 3p functions (\Eq{eq:3p_an_further_split}).

\subsection{Final result}
Finally, the fermionic partial 4p correlators for general permutations $p$ is obtained from the full correlator via
\bal
\tG_{p}(\i \bsomega_p) = \int_{\varepsilon_{\oli{1}},\varepsilon_{\oli{2}},\varepsilon_{\oli{3}}} \frac{ (2\pi \i)^3 S_p(\varepsilon_{\oli{1}}, \varepsilon_{\oli{2}}, \varepsilon_{\oli{3}}) }{(\i \omega_{\oli{1}} - \varepsilon_{\oli{1}}) (\i \omega_{\oli{12}} - \varepsilon_{\oli{12}}) (\i \omega_{\oli{123}} - \varepsilon_{\oli{123}})},
\eal
with the PSFs given by
\bal
\label{eq:final_Sp_4p_app}
&(2\pi \i)^3 S_p(\varepsilon_{\oli{1}}, \varepsilon_{\oli{2}}, \varepsilon_{\oli{3}}) 
\nn
&= n_{\varepsilon_{\oli{3}}}\, n_{\varepsilon_{\oli{2}}}\, n_{\varepsilon_{\oli{1}}}\, \tG^{\varepsilon_{\oli{3}},\varepsilon_{\oli{2}},\varepsilon_{\oli{1}}} + n_{\varepsilon_{\oli{123}}}\, n_{\varepsilon_{\oli{2}}}\, n_{\varepsilon_{\oli{1}}}\, \tG^{\varepsilon_{\oli{123}},\varepsilon_{\oli{2}},\varepsilon_{\oli{1}}} 
\nn
&\hsp + n_{\varepsilon_{\oli{13}}}\, n_{\varepsilon_{\oli{2}}}\, n_{\varepsilon_{\oli{1}}}\, \tG^{\varepsilon_{\oli{13}},\varepsilon_{\oli{2}},\varepsilon_{\oli{1}}} + n_{\varepsilon_{\oli{23}}}\, n_{\varepsilon_{\oli{2}}}\, n_{\varepsilon_{\oli{1}}}\, \tG^{\varepsilon_{\oli{23}},\varepsilon_{\oli{2}},\varepsilon_{\oli{1}}} 
\nn
&\hsp + n_{\varepsilon_{\oli{3}}}\, n_{\varepsilon_{\oli{12}}}\, n_{\varepsilon_{\oli{1}}}\, \tG^{\varepsilon_{\oli{3}},\varepsilon_{\oli{12}},\varepsilon_{\oli{1}}} + n_{\varepsilon_{\oli{123}}}\, n_{\varepsilon_{\oli{12}}}\, n_{\varepsilon_{\oli{1}}}\, \tG^{\varepsilon_{\oli{123}},\varepsilon_{\oli{12}},\varepsilon_{\oli{1}}} 
\nn
&\hsp + n_{\varepsilon_{\oli{3}}}\, n_{\varepsilon_{\oli{1}}}\, \hdelta(\varepsilon_{\oli{12}})\, \hG^{\varepsilon_{\oli{3}},\varepsilon_{\oli{1}}}_{\oli{12}} + n_{\varepsilon_{\oli{2}}}\, n_{\varepsilon_{\oli{1}}}\, \hdelta(\varepsilon_{\oli{13}})\, \hG^{\varepsilon_{\oli{2}},\varepsilon_{\oli{1}}}_{\oli{13}} 
\nn
&\hsp + n_{\varepsilon_{\oli{2}}}\, n_{\varepsilon_{\oli{1}}}\, \hdelta(\varepsilon_{\oli{23}})\, \hG^{\varepsilon_{\oli{2}},\varepsilon_{\oli{1}}}_{\oli{23}}.
\eal
For the anomalous parts, the order of discontinuities does not matter, as, e.g., $\hG^{\varepsilon_{\oli{3}},\varepsilon_{\oli{1}}}_{\oli{12}} = \hG^{\varepsilon_{\oli{1}},\varepsilon_{\oli{3}}}_{\oli{12}}$.

For completeness, we express the discontinuities in \Eq{eq:final_Sp_4p_app} in terms of analytic regions according to their definition in \Sec{sec:4p_analytic_regions}. This gives
\begin{subequations}
\label{eq:4p:discontinuities}
\bal
&\tG^{\varepsilon_1,\varepsilon_2,\varepsilon_3} = -\tG^{\varepsilon_{234},\varepsilon_2,\varepsilon_3} =
-\tG^{\varepsilon_1,\varepsilon_2,\varepsilon_4} = \tG^{\varepsilon_{234},\varepsilon_2,\varepsilon_4} 
\nn
&= -\tG^{\varepsilon_1,\varepsilon_{34},\varepsilon_3} = \tG^{\varepsilon_{234},\varepsilon_{34},\varepsilon_3} = \tG^{\varepsilon_1,\varepsilon_{34},\varepsilon_4} = - \tG^{\varepsilon_{234},\varepsilon_{34},\varepsilon_4}
\nn
&=
C^{(3)} - C^{(4)} +C^{(123)} - C^{(124)}
\nn&\hsp
- C^{(13)}_{\tn{III}} + C^{(14)}_{\tn{III}}-C^{(23)}_{\tn{III}} + C^{(24)}_{\tn{III}},
\\
&\tG^{\varepsilon_1,\varepsilon_3,\varepsilon_2} = -\tG^{\varepsilon_{234},\varepsilon_3,\varepsilon_2} =
-\tG^{\varepsilon_1,\varepsilon_3,\varepsilon_4} = \tG^{\varepsilon_{234},\varepsilon_3,\varepsilon_4} 
\nn
&=-\tG^{\varepsilon_{1},\varepsilon_{24},\varepsilon_2} = \tG^{\varepsilon_{234},\varepsilon_{24},\varepsilon_2} = \tG^{\varepsilon_{1},\varepsilon_{24},\varepsilon_4} = - \tG^{\varepsilon_{234},\varepsilon_{24},\varepsilon_4}
\nn
&= 
C^{(2)} - C^{(4)} +C^{(123)} - C^{(134)}
\nn&\hsp
- C^{(12)}_{\tn{III}} + C^{(14)}_{\tn{III}}-C^{(23)}_{\tn{III}} + C^{(34)}_{\tn{III}}
, \\
& \tG^{\varepsilon_1,\varepsilon_4,\varepsilon_2} = -\tG^{\varepsilon_{234},\varepsilon_4,\varepsilon_2} =
-\tG^{\varepsilon_1,\varepsilon_4,\varepsilon_3} = \tG^{\varepsilon_{234},\varepsilon_4,\varepsilon_3} 
\nn
& = -\tG^{\varepsilon_1,\varepsilon_{23},\varepsilon_2} = \tG^{\varepsilon_{234},\varepsilon_{23},\varepsilon_2} =
\tG^{\varepsilon_1,\varepsilon_{23},\varepsilon_3} = -\tG^{\varepsilon_{234},\varepsilon_{23},\varepsilon_3} 
\nn&
 =
C^{(2)} - C^{(3)} +C^{(124)} - C^{(134)}
\nn& \hsp
- C^{(12)}_{\tn{III}} + C^{(13)}_{\tn{III}} - C^{(24)}_{\tn{III}} + C^{(34)}_{\tn{III}} 
,
\\
&\tG^{\varepsilon_2,\varepsilon_1,\varepsilon_3} = -\tG^{\varepsilon_{134},\varepsilon_1,\varepsilon_3} =
-\tG^{\varepsilon_2,\varepsilon_1,\varepsilon_4} = \tG^{\varepsilon_{134},\varepsilon_1,\varepsilon_4} 
\nn
&= -\tG^{\varepsilon_{2},\varepsilon_{34},\varepsilon_3} = \tG^{\varepsilon_{134},\varepsilon_{34},\varepsilon_3} = \tG^{\varepsilon_{2},\varepsilon_{34},\varepsilon_4} = -\tG^{\varepsilon_{134},\varepsilon_{34},\varepsilon_4}
\nn&
=
C^{(3)} - C^{(4)} +C^{(123)} - C^{(124)}
\nn&\hsp
- C^{(13)}_{\tn{II}} + C^{(14)}_{\tn{II}}-C^{(23)}_{\tn{II}} + C^{(24)}_{\tn{II}}
, \\
&\tG^{\varepsilon_3,\varepsilon_1,\varepsilon_2} = -\tG^{\varepsilon_{124},\varepsilon_1,\varepsilon_2} =
-\tG^{\varepsilon_3,\varepsilon_1,\varepsilon_4} = \tG^{\varepsilon_{124},\varepsilon_1,\varepsilon_4} 
\nn
& = -\tG^{\varepsilon_3,\varepsilon_{24},\varepsilon_2} = \tG^{\varepsilon_{124},\varepsilon_{24},\varepsilon_2} =
\tG^{\varepsilon_3,\varepsilon_{24},\varepsilon_4} = - \tG^{\varepsilon_{124},\varepsilon_{24},\varepsilon_4}  
\nn&
=
C^{(2)} - C^{(4)} +C^{(123)} - C^{(134)}
\nn&\hsp
- C^{(12)}_{\tn{II}} + C^{(34)}_{\tn{II}} + C^{(14)}_{\tn{IV}} -C^{(23)}_{\tn{IV}}
, \\
&\tG^{\varepsilon_4,\varepsilon_1,\varepsilon_2} = -\tG^{\varepsilon_{123},\varepsilon_1,\varepsilon_2} =
-\tG^{\varepsilon_4,\varepsilon_1,\varepsilon_3} = \tG^{\varepsilon_{123},\varepsilon_1,\varepsilon_3} 
\nn
& = - \tG^{\varepsilon_4,\varepsilon_{23},\varepsilon_2} = \tG^{\varepsilon_{123},\varepsilon_{23},\varepsilon_2} =
\tG^{\varepsilon_4,\varepsilon_{23},\varepsilon_3} = -\tG^{\varepsilon_{123},\varepsilon_{23},\varepsilon_3} 
\nn&
=
C^{(2)} - C^{(3)} + C^{(124)} - C^{(134)}
\nn&\hsp
- C^{(12)}_{\tn{IV}} + C^{(13)}_{\tn{IV}}-C^{(24)}_{\tn{IV}} + C^{(34)}_{\tn{IV}}
, \\
&\tG^{\varepsilon_2,\varepsilon_3,\varepsilon_1} = -\tG^{\varepsilon_{134},\varepsilon_3,\varepsilon_1} =
-\tG^{\varepsilon_2,\varepsilon_3,\varepsilon_4} = \tG^{\varepsilon_{134},\varepsilon_3,\varepsilon_4} 
\nn
& = -\tG^{\varepsilon_2,\varepsilon_{14},\varepsilon_1} = \tG^{\varepsilon_{134},\varepsilon_{14},\varepsilon_1} = \tG^{\varepsilon_{2},\varepsilon_{14},\varepsilon_4} = - \tG^{\varepsilon_{134},\varepsilon_{14},\varepsilon_4} 
\nn&
=
C^{(1)} - C^{(4)} +C^{(123)} - C^{(234)}
\nn&\hsp
- C^{(12)}_{\tn{I}} + C^{(34)}_{\tn{I}} - C^{(13)}_{\tn{II}} +C^{(24)}_{\tn{II}}
, \\
&\tG^{\varepsilon_3,\varepsilon_2,\varepsilon_1} = -\tG^{\varepsilon_{124},\varepsilon_2,\varepsilon_1} =
-\tG^{\varepsilon_3,\varepsilon_2,\varepsilon_4} = \tG^{\varepsilon_{124},\varepsilon_2,\varepsilon_4} 
\nn
& = -\tG^{\varepsilon_3,\varepsilon_{14},\varepsilon_1} = \tG^{\varepsilon_{124},\varepsilon_{14},\varepsilon_1} =
\tG^{\varepsilon_3,\varepsilon_{14},\varepsilon_4} = -\tG^{\varepsilon_{124},\varepsilon_{14},\varepsilon_4} 
\nn
&=
C^{(1)} - C^{(4)} +C^{(123)} - C^{(234)}
\nn&\hsp
- C^{(12)}_{\tn{II}} + C^{(34)}_{\tn{II}} - C^{(13)}_{\tn{I}} +C^{(24)}_{\tn{I}}
, \\
& \tG^{\varepsilon_2,\varepsilon_4,\varepsilon_1} = -\tG^{\varepsilon_{134},\varepsilon_4,\varepsilon_1} =
-\tG^{\varepsilon_2,\varepsilon_4,\varepsilon_3} = \tG^{\varepsilon_{134},\varepsilon_4,\varepsilon_3} 
\nn
& = -\tG^{\varepsilon_2,\varepsilon_{13},\varepsilon_1} = \tG^{\varepsilon_{134},\varepsilon_{13},\varepsilon_1} =
\tG^{\varepsilon_2,\varepsilon_{13},\varepsilon_3} = -\tG^{\varepsilon_{134},\varepsilon_{13},\varepsilon_3} 
\nn&
 =
C^{(1)} - C^{(3)} +C^{(124)} - C^{(234)}
\nn& \hsp
- C^{(12)}_{\tn{I}} + C^{(34)}_{\tn{I}} - C^{(14)}_{\tn{II}} + C^{(23)}_{\tn{II}} 
,
\\
&\tG^{\varepsilon_4,\varepsilon_2,\varepsilon_1} = -\tG^{\varepsilon_{123},\varepsilon_2,\varepsilon_1} =
-\tG^{\varepsilon_4,\varepsilon_2,\varepsilon_3} = \tG^{\varepsilon_{123},\varepsilon_2,\varepsilon_3} 
\nn
& = -\tG^{\varepsilon_4,\varepsilon_{13},\varepsilon_1} = \tG^{\varepsilon_{123},\varepsilon_{13},\varepsilon_1} =
\tG^{\varepsilon_4,\varepsilon_{13},\varepsilon_3} = -\tG^{\varepsilon_{123},\varepsilon_{13},\varepsilon_3}  
\nn&
=
C^{(1)} - C^{(3)} +C^{(124)} - C^{(234)}
\nn& \hsp
- C^{(12)}_{\tn{IV}} + C^{(34)}_{\tn{IV}} - C^{(14)}_{\tn{I}} + C^{(23)}_{\tn{I}}
,
\\
& \tG^{\varepsilon_3,\varepsilon_4,\varepsilon_1} = -\tG^{\varepsilon_{124},\varepsilon_4,\varepsilon_1} =
-\tG^{\varepsilon_3,\varepsilon_4,\varepsilon_2} = \tG^{\varepsilon_{124},\varepsilon_4,\varepsilon_2} 
\nn
& = -\tG^{\varepsilon_3,\varepsilon_{12},\varepsilon_1} = \tG^{\varepsilon_{124},\varepsilon_{12},\varepsilon_1} =
\tG^{\varepsilon_3,\varepsilon_{12},\varepsilon_2} = -\tG^{\varepsilon_{124},\varepsilon_{12},\varepsilon_2} 
\nn&
 =
C^{(1)} - C^{(2)} +C^{(134)} - C^{(234)}
\nn& \hsp
- C^{(14)}_{\tn{IV}} + C^{(23)}_{\tn{IV}} - C^{(13)}_{\tn{I}} + C^{(24)}_{\tn{I}} 
,
\\
& \tG^{\varepsilon_4,\varepsilon_3,\varepsilon_1} = -\tG^{\varepsilon_{123},\varepsilon_3,\varepsilon_1} =
-\tG^{\varepsilon_4,\varepsilon_3,\varepsilon_2} = \tG^{\varepsilon_{123},\varepsilon_3,\varepsilon_2} 
\nn
& = -\tG^{\varepsilon_4,\varepsilon_{12},\varepsilon_1} = \tG^{\varepsilon_{123},\varepsilon_{12},\varepsilon_1} =
\tG^{\varepsilon_4,\varepsilon_{12},\varepsilon_2} = -\tG^{\varepsilon_{123},\varepsilon_{12},\varepsilon_2} 
\nn&
 =
C^{(1)} - C^{(2)} +C^{(134)} - C^{(234)}
\nn& \hsp
- C^{(14)}_{\tn{I}} + C^{(23)}_{\tn{I}} - C^{(13)}_{\tn{IV}} + C^{(24)}_{\tn{IV}} 
,
\\
&\tG^{\varepsilon_{12},\varepsilon_1,\varepsilon_3} 
= 
-\tG^{\varepsilon_{34},\varepsilon_1,\varepsilon_3} 
= 
\tG^{\varepsilon_{12},\varepsilon_3,\varepsilon_1}
=
- \tG^{\varepsilon_{34},\varepsilon_3,\varepsilon_1}
\nn
&=
-\tG^{\varepsilon_{12},\varepsilon_2,\varepsilon_3}
=
\tG^{\varepsilon_{34},\varepsilon_2,\varepsilon_3}
=
-\tG^{\varepsilon_{12},\varepsilon_3,\varepsilon_2}
=
\tG^{\varepsilon_{34},\varepsilon_3,\varepsilon_2}
\nn
&=
-\tG^{\varepsilon_{12},\varepsilon_1,\varepsilon_4}
= 
\tG^{\varepsilon_{34},\varepsilon_1,\varepsilon_4}
=
-\tG^{\varepsilon_{12},\varepsilon_4,\varepsilon_1}
= 
\tG^{\varepsilon_{34},\varepsilon_4,\varepsilon_1}
\nn
&=
\tG^{\varepsilon_{12},\varepsilon_2,\varepsilon_4}
=
-\tG^{\varepsilon_{34},\varepsilon_2,\varepsilon_4}
=
\tG^{\varepsilon_{12},\varepsilon_4,\varepsilon_2}
=
-\tG^{\varepsilon_{34},\varepsilon_4,\varepsilon_2}
\nn
&=
C^{(13)}_{\tn{II}} - C^{(13)}_{\tn{III}} - C^{(14)}_{\tn{II}} + C^{(14)}_{\tn{III}}
\nn& \hsp
+C^{(23)}_{\tn{II}} - C^{(23)}_{\tn{III}} -C^{(24)}_{\tn{II}} + C^{(24)}_{\tn{III}}
,
\\
&\tG^{\varepsilon_{13},\varepsilon_1,\varepsilon_2}
=
-\tG^{\varepsilon_{24},\varepsilon_1,\varepsilon_2}
 = \tG^{\varepsilon_{13},\varepsilon_2,\varepsilon_1}
=
-\tG^{\varepsilon_{24},\varepsilon_2,\varepsilon_1}
\nn
&=
-\tG^{\varepsilon_{13},\varepsilon_3,\varepsilon_2}
=
\tG^{\varepsilon_{24},\varepsilon_3,\varepsilon_2}
=
-\tG^{\varepsilon_{13},\varepsilon_2,\varepsilon_3}
=
\tG^{\varepsilon_{24},\varepsilon_2,\varepsilon_3} 
\nn
&= -\tG^{\varepsilon_{13},\varepsilon_1,\varepsilon_4}
=
\tG^{\varepsilon_{24},\varepsilon_1,\varepsilon_4} 
= -\tG^{\varepsilon_{13},\varepsilon_4,\varepsilon_1}
=
\tG^{\varepsilon_{24},\varepsilon_4,\varepsilon_1}
\nn
&=
\tG^{\varepsilon_{13},\varepsilon_3,\varepsilon_4}
=
-\tG^{\varepsilon_{24},\varepsilon_3,\varepsilon_4}
=
\tG^{\varepsilon_{13},\varepsilon_4,\varepsilon_3}
=
-\tG^{\varepsilon_{24},\varepsilon_4,\varepsilon_3}
\nn
&=
C^{(12)}_{\tn{II}} - C^{(12)}_{\tn{III}} + C^{(14)}_{\tn{III}} - C^{(14)}_{\tn{IV}}
\nn&\hsp
-C^{(23)}_{\tn{III}} + C^{(23)}_{\tn{IV}} -C^{(34)}_{\tn{II}} + C^{(34)}_{\tn{III}}
,
\\
&\tG^{\varepsilon_{14},\varepsilon_1,\varepsilon_2} 
= 
-\tG^{\varepsilon_{23},\varepsilon_1,\varepsilon_2} 
 = \tG^{\varepsilon_{14},\varepsilon_2,\varepsilon_1} 
= 
-\tG^{\varepsilon_{23},\varepsilon_2,\varepsilon_1}  
\nn
&=
-\tG^{\varepsilon_{14},\varepsilon_4,\varepsilon_2} 
=
\tG^{\varepsilon_{23},\varepsilon_4,\varepsilon_2} 
 =
-\tG^{\varepsilon_{14},\varepsilon_2,\varepsilon_4} 
=
\tG^{\varepsilon_{23},\varepsilon_2,\varepsilon_4}  
\nn
&=-\tG^{\varepsilon_{14},\varepsilon_1,\varepsilon_3}
=
\tG^{\varepsilon_{23},\varepsilon_1,\varepsilon_3} 
 =-\tG^{\varepsilon_{14},\varepsilon_3,\varepsilon_1}
=
\tG^{\varepsilon_{23},\varepsilon_3,\varepsilon_1}  
\nn
&=
\tG^{\varepsilon_{14},\varepsilon_4,\varepsilon_3} 
=
-\tG^{\varepsilon_{23},\varepsilon_4,\varepsilon_3} 
 =
\tG^{\varepsilon_{14},\varepsilon_3,\varepsilon_4} 
=
-\tG^{\varepsilon_{23},\varepsilon_3,\varepsilon_4}  
\nn
&=
-C^{(12)}_{\tn{III}} + C^{(12)}_{\tn{IV}} + C^{(13)}_{\tn{III}} - C^{(13)}_{\tn{IV}}
\nn&\hsp
-C^{(24)}_{\tn{III}} + C^{(24)}_{\tn{IV}} +C^{(34)}_{\tn{III}} - C^{(34)}_{\tn{IV}}
.
\eal
\end{subequations}
Here, the analytic continuations of $\tG$ are labeled according to the analytic regions in \Fig{analyticregions4p}
\bal
C^{(1)} &= \tG(\varepsilon_1^+,\varepsilon_2^-,\varepsilon_3^-,\varepsilon_4^-;\varepsilon_{12}^+,\varepsilon_{13}^+,\varepsilon_{14}^+)
, \nn
C^{(2)} &= \tG(\varepsilon_1^-,\varepsilon_2^+,\varepsilon_3^-,\varepsilon_4^-;\varepsilon_{12}^+,\varepsilon_{13}^-,\varepsilon_{14}^-)
, \nn
C^{(3)} &= \tG(\varepsilon_1^-,\varepsilon_2^-,\varepsilon_3^+,\varepsilon_4^-;\varepsilon_{12}^-,\varepsilon_{13}^+,\varepsilon_{14}^-)
, \nn
C^{(4)} &= \tG(\varepsilon_1^-,\varepsilon_2^-,\varepsilon_3^-,\varepsilon_4^+;\varepsilon_{12}^-,\varepsilon_{13}^-,\varepsilon_{14}^+)
, \nn
C^{(12)}_{\tn{I}} &= \tG(\varepsilon_1^+,\varepsilon_2^+,\varepsilon_3^-,\varepsilon_4^-;\varepsilon_{12}^+,\varepsilon_{13}^+,\varepsilon_{14}^+)
, \nn
C^{(12)}_{\tn{II}} &= \tG(\varepsilon_1^+,\varepsilon_2^+,\varepsilon_3^-,\varepsilon_4^-;\varepsilon_{12}^+,\varepsilon_{13}^+,\varepsilon_{14}^-)
, \nn
C^{(12)}_{\tn{III}} &= \tG(\varepsilon_1^+,\varepsilon_2^+,\varepsilon_3^-,\varepsilon_4^-;\varepsilon_{12}^+,\varepsilon_{13}^-,\varepsilon_{14}^-)
, \nn
C^{(12)}_{\tn{IV}} &= \tG(\varepsilon_1^+,\varepsilon_2^+,\varepsilon_3^-,\varepsilon_4^-;\varepsilon_{12}^+,\varepsilon_{13}^-,\varepsilon_{14}^+)
, \nn
C^{(13)}_{\tn{I}} &= \tG(\varepsilon_1^+,\varepsilon_2^-,\varepsilon_3^+,\varepsilon_4^-;\varepsilon_{12}^+,\varepsilon_{13}^+,\varepsilon_{14}^+)
, \nn
C^{(13)}_{\tn{II}} &= \tG(\varepsilon_1^+,\varepsilon_2^-,\varepsilon_3^+,\varepsilon_4^-;\varepsilon_{12}^+,\varepsilon_{13}^+,\varepsilon_{14}^-)
, \nn
C^{(13)}_{\tn{III}} &= \tG(\varepsilon_1^+,\varepsilon_2^-,\varepsilon_3^+,\varepsilon_4^-;\varepsilon_{12}^-,\varepsilon_{13}^+,\varepsilon_{14}^-)
, \nn
C^{(13)}_{\tn{IV}} &= \tG(\varepsilon_1^+,\varepsilon_2^-,\varepsilon_3^+,\varepsilon_4^-;\varepsilon_{12}^-,\varepsilon_{13}^+,\varepsilon_{14}^+)
, \nn
C^{(14)}_{\tn{I}} &= \tG(\varepsilon_1^+,\varepsilon_2^-,\varepsilon_3^-,\varepsilon_4^+;\varepsilon_{12}^+,\varepsilon_{13}^+,\varepsilon_{14}^+)
, \nn
C^{(14)}_{\tn{II}} &= \tG(\varepsilon_1^+,\varepsilon_2^-,\varepsilon_3^-,\varepsilon_4^+;\varepsilon_{12}^+,\varepsilon_{13}^-,\varepsilon_{14}^+)
, \nn
C^{(14)}_{\tn{III}} &= \tG(\varepsilon_1^+,\varepsilon_2^-,\varepsilon_3^-,\varepsilon_4^+;\varepsilon_{12}^-,\varepsilon_{13}^-,\varepsilon_{14}^+)
, \nn
C^{(14)}_{\tn{IV}} &= \tG(\varepsilon_1^+,\varepsilon_2^-,\varepsilon_3^-,\varepsilon_4^+;\varepsilon_{12}^-,\varepsilon_{13}^+,\varepsilon_{14}^+)
, \nn
C^{(23)}_{\tn{I}} &= \tG(\varepsilon_1^-,\varepsilon_2^+,\varepsilon_3^+,\varepsilon_4^-;\varepsilon_{12}^-,\varepsilon_{13}^-,\varepsilon_{14}^-)
, \nn
C^{(23)}_{\tn{II}} &= \tG(\varepsilon_1^-,\varepsilon_2^+,\varepsilon_3^+,\varepsilon_4^-;\varepsilon_{12}^-,\varepsilon_{13}^+,\varepsilon_{14}^-)
, \nn
C^{(23)}_{\tn{III}} &= \tG(\varepsilon_1^-,\varepsilon_2^+,\varepsilon_3^+,\varepsilon_4^-;\varepsilon_{12}^+,\varepsilon_{13}^+,\varepsilon_{14}^-)
, \nn
C^{(23)}_{\tn{IV}} &= \tG(\varepsilon_1^-,\varepsilon_2^+,\varepsilon_3^+,\varepsilon_4^-;\varepsilon_{12}^+,\varepsilon_{13}^-,\varepsilon_{14}^-)
, \nn
C^{(24)}_{\tn{I}} &= \tG(\varepsilon_1^-,\varepsilon_2^+,\varepsilon_3^-,\varepsilon_4^+;\varepsilon_{12}^-,\varepsilon_{13}^-,\varepsilon_{14}^-)
, \nn
C^{(24)}_{\tn{II}} &= \tG(\varepsilon_1^-,\varepsilon_2^+,\varepsilon_3^-,\varepsilon_4^+;\varepsilon_{12}^-,\varepsilon_{13}^-,\varepsilon_{14}^+)
, \nn
C^{(24)}_{\tn{III}} &= \tG(\varepsilon_1^-,\varepsilon_2^+,\varepsilon_3^-,\varepsilon_4^+;\varepsilon_{12}^+,\varepsilon_{13}^-,\varepsilon_{14}^+)
, \nn
C^{(24)}_{\tn{IV}} &= \tG(\varepsilon_1^-,\varepsilon_2^+,\varepsilon_3^-,\varepsilon_4^+;\varepsilon_{12}^+,\varepsilon_{13}^-,\varepsilon_{14}^-)
, \nn
C^{(34)}_{\tn{I}} &= \tG(\varepsilon_1^-,\varepsilon_2^-,\varepsilon_3^+,\varepsilon_4^+;\varepsilon_{12}^-,\varepsilon_{13}^-,\varepsilon_{14}^-)
, \nn
C^{(34)}_{\tn{II}} &= \tG(\varepsilon_1^-,\varepsilon_2^-,\varepsilon_3^+,\varepsilon_4^+;\varepsilon_{12}^-,\varepsilon_{13}^-,\varepsilon_{14}^+)
, \nn
C^{(34)}_{\tn{III}} &= \tG(\varepsilon_1^-,\varepsilon_2^-,\varepsilon_3^+,\varepsilon_4^+;\varepsilon_{12}^-,\varepsilon_{13}^+,\varepsilon_{14}^+)
, \nn
C^{(34)}_{\tn{IV}} &= \tG(\varepsilon_1^-,\varepsilon_2^-,\varepsilon_3^+,\varepsilon_4^+;\varepsilon_{12}^-,\varepsilon_{13}^+,\varepsilon_{14}^-)
, \nn
C^{(123)} &= \tG(\varepsilon_1^+,\varepsilon_2^+,\varepsilon_3^+,\varepsilon_4^-;\varepsilon_{12}^+,\varepsilon_{13}^+,\varepsilon_{14}^-)
, \nn
C^{(124)} &= \tG(\varepsilon_1^+,\varepsilon_2^+,\varepsilon_3^-,\varepsilon_4^+;\varepsilon_{12}^+,\varepsilon_{13}^-,\varepsilon_{14}^+)
, \nn
C^{(134)} &= \tG(\varepsilon_1^+,\varepsilon_2^-,\varepsilon_3^+,\varepsilon_4^+;\varepsilon_{12}^-,\varepsilon_{13}^+,\varepsilon_{14}^+)
, \nn
C^{(234)} &= \tG(\varepsilon_1^-,\varepsilon_2^+,\varepsilon_3^+,\varepsilon_4^+;\varepsilon_{12}^-,\varepsilon_{13}^-,\varepsilon_{14}^-).
\eal

The discontinuities in the anomalous parts \Eq{eq:final_Sp_4p_app} read
\begin{subequations}
\bal
\hG_{12}^{\varepsilon_1,\varepsilon_3}
&=-\hG_{12}^{\varepsilon_2,\varepsilon_3}
=-\hG_{12}^{\varepsilon_1,\varepsilon_4}
=\hG_{12}^{\varepsilon_2,\varepsilon_4}
\nn
&
=
\hC^{(13)}_{12} - \hC^{(14)}_{12} - \hC^{(23)}_{12} + \hC^{(24)}_{12} 
, \\
\label{hG_13}
\hG_{13}^{\varepsilon_1,\varepsilon_2}
&=-\hG_{13}^{\varepsilon_3,\varepsilon_2}
=-\hG_{13}^{\varepsilon_1,\varepsilon_4}
=\hG_{13}^{\varepsilon_3,\varepsilon_4}
\nn
&
=
\hC^{(12)}_{13} - \hC^{(14)}_{13} - \hC^{(23)}_{13} + \hC^{(34)}_{13} 
, \\
\hG_{14}^{\varepsilon_1,\varepsilon_2}
&=-\hG_{14}^{\varepsilon_4,\varepsilon_2}
=-\hG_{14}^{\varepsilon_1,\varepsilon_3}
=\hG_{14}^{\varepsilon_4,\varepsilon_3}
\nn
&
=
\hC^{(12)}_{14} - \hC^{(13)}_{14} - \hC^{(24)}_{14} + \hC^{(34)}_{14} 
,
\eal
\end{subequations}
with
\bal
\hC^{(13)}_{12} &= \hG_{12}(\varepsilon_1^+,\varepsilon_2^-,\varepsilon_3^+,\varepsilon_4^-), \quad \hC^{(24)}_{12} = \hG_{12}(\varepsilon_1^-,\varepsilon_2^+,\varepsilon_3^-,\varepsilon_4^+), \nn
\hC^{(14)}_{12} &= \hG_{12}(\varepsilon_1^+,\varepsilon_2^-,\varepsilon_3^-,\varepsilon_4^+), \quad \hC^{(23)}_{12} = \hG_{12}(\varepsilon_1^-,\varepsilon_2^+,\varepsilon_3^+,\varepsilon_4^-),\nn
\hC^{(12)}_{13} &= \hG_{13}(\varepsilon_1^+,\varepsilon_2^+,\varepsilon_3^-,\varepsilon_4^-), \quad \hC^{(34)}_{13} = \hG_{13}(\varepsilon_1^-,\varepsilon_2^-,\varepsilon_3^+,\varepsilon_4^+), \nn
\hC^{(14)}_{13} &= \hG_{13}(\varepsilon_1^+,\varepsilon_2^-,\varepsilon_3^-,\varepsilon_4^+), \quad \hC^{(23)}_{13} = \hG_{13}(\varepsilon_1^-,\varepsilon_2^+,\varepsilon_3^+,\varepsilon_4^-), \nn
\hC^{(12)}_{14} &= \hG_{14}(\varepsilon_1^+,\varepsilon_2^+,\varepsilon_3^-,\varepsilon_4^-), \quad \hC^{(34)}_{14} = \hG_{14}(\varepsilon_1^-,\varepsilon_2^-,\varepsilon_3^+,\varepsilon_4^+), \nn
\hC^{(13)}_{14} &= \hG_{14}(\varepsilon_1^+,\varepsilon_2^-,\varepsilon_3^+,\varepsilon_4^-), \quad \hC^{(24)}_{14} = \hG_{14}(\varepsilon_1^-,\varepsilon_2^+,\varepsilon_3^-,\varepsilon_4^+).
\eal
The remaining terms follow from $\hG_{34}=\hG_{12}$, $\hG_{24}=\hG_{13}$, and $\hG_{23}=\hG_{14}$.

\section{Additional spectral representations}
\label{app:add_spectral_rep}

In this appendix, we derive spectral representations  for discontinuities (\App{app:branchcuts_as_PSFs}) and  for anomalous parts (\App{app:spec_rep_anom})  for general $\ell$. These are used in \App{app:simplifications_for_G_alpha2} to relate Keldysh components $G^{[\eta_1 \eta_2]}$ to discontinuities of regular parts and analytic continuations of anomalous parts, resulting in \Eq{eq:alpha_2_general} in \Sec{sec:KF_components_alpha_2}. Additionally, they serve as a key ingredient in \App{sec:App_consistency_checks} for consistency checks performed on our results for the 2p, 3p, and 4p PSFs, where we express all occurring discontinuities through PSF (anti)commutators. We use the notation introduced in the beginning of \Sec{sec:4p_Keldysh_correlators} throughout this appendix.

\subsection{Spectral representation of discontinuities}
\label{app:branchcuts_as_PSFs}
Here, we focus on the discontinuities of the regular MF correlator $\tG$, as introduced in \Sec{sec:analytic_regions_and_discontinuities}. The results carry over to anomalous contributions $\hG$, as presented in \App{app:spec_rep_anom}.
We first consider discontinuities of $3$p correlators (\App{App:discont_spec_rep_3p}) and then their generalization to arbitrary $\ell$ (\App{app:general_ell_disc}).

\subsubsection{Example for $\ell = 3$}
\label{App:discont_spec_rep_3p}

Let us consider the discontinuity in \Eq{eq:ACof3point:branchcut_single} as an example for $\ell=3$. Inserting the spectral representation in \Eqs{eq:MF_G-compact} yields
\bal \label{eq:Spec_rep_G^2_1}
&\frac{1}{(2\pi \i)^2}\tG^{\omega_2}_{\omega_1^+} =
\frac{1}{(2\pi \i)^2} \left( \tG_{\omega_2^+,\omega_1^+}
-
\tG_{\omega_2^-,\omega_1^+} \right)
\nn
&=
\int_{\varepsilon_1}\int_{\varepsilon_2} \int_{\varepsilon_3} 
\delta(\varepsilon_{123})
\Bigg[
\nn
&
\hsp
\frac{1}{\omega_1^+-\varepsilon_1} \Big[ \frac{1}{\omega_{13}^--\varepsilon_{13}}
-
\frac{1}{\omega_{13}^+-\varepsilon_{13}}
\Big]
S_{(132)}(\varepsilon_1,\varepsilon_3) 
\nn
&\hsp
+
\frac{1}{\omega_{12}^+-\varepsilon_{12}} \Big[ \frac{1}{\omega_2^+-\varepsilon_2}
-
\frac{1}{\omega_2^--\varepsilon_2}
\Big]
S_{(213)}(\varepsilon_2,\varepsilon_1) 
\nn
&\hsp
+
\frac{1}{\omega_{23}^--\varepsilon_{23}} \Big[ \frac{1}{\omega_2^+-\varepsilon_2}
-
\frac{1}{\omega_2^--\varepsilon_2}
\Big]
S_{(231)}(\varepsilon_2,\varepsilon_3) 
\nn
&\hsp
+
\frac{1}{\omega_3^--\varepsilon_3} \Big[ \frac{1}{\omega_{13}^--\varepsilon_{13}}
-
\frac{1}{\omega_{13}^+-\varepsilon_{13}}
\Big]
 S_{(312)}(\varepsilon_3,\varepsilon_1) 
 \Bigg]
\nn
&= \int_{\varepsilon_1} \int_{\varepsilon_2} \int_{\varepsilon_3} 
\delta(\varepsilon_{123})
\Bigg[
\hdelta(\omega_{2} - \varepsilon_{2}) \frac{1}{\omega^+_1 - \varepsilon_1} 
S_{[2,13]_-}(\varepsilon_1,\varepsilon_2,\varepsilon_3)
\nn
&\hsp+ 
\hdelta(\omega_{2} - \varepsilon_{2}) \frac{1}{\omega^-_3 - \varepsilon_3}  
S_{[2,31]_-}(\varepsilon_1,\varepsilon_2,\varepsilon_3)
\Bigg]
\nn
&=
-\int_{\varepsilon_1} 
\frac{1}{\omega_1^+-\varepsilon_1}
S_{[2,[1,3]_-]_-}(\varepsilon_1,\omega_2,-\varepsilon_1-\omega_2)
\,,
\eal
where we used the identity \eqref{eq:hdelta_identity} and energy conservation. 
The permutations $p=(123), (321)$ do not contribute to the discontinuity as their kernels only depend on the external frequencies $\omega^+_1$ and $\omega^-_3$ with imaginary parts independent of $\omega^\pm_2$.

For the discontinuity $\tG^{\omega_2, \omega_1} = \tG^{\omega_2}_{\omega^+_1} - \tG^{\omega_2}_{\omega^-_1}$, \Eq{eq:Spec_rep_G^2_1} yields
\bal
\label{eq:3p_disc_ident}
\tG^{\omega_2,\omega_1} 
&= 
(2\pi \i)^2 S_{[2,[1,3]_-]_-}(\bsomega)
\, , 
\nn
\tG^{\omega_{12},\omega_1} 
&=
(2\pi \i)^2 S_{[[1,2]_-,3]_-}(\bsomega).
\eal
The second identity for $\tG^{\omega_{12}, \omega_1}$ follows from a similar derivation as for $\tG^{\omega_{2}, \omega_1}$. Note that the above relations hold for permuted indices as well (see \Eq{eq:PSF_3p_anticom_identities}).
Thus, consecutive discontinuities eventually give a (nested) commutator of PSFs. 
For $\ell=2$, this corresponds to the standard spectral function, $-\tG^{\omega_1}=(2\pi \i)S_{[1,2]_-}=(2\pi \i)S^{\mathrm{std}}$.

\subsubsection{Generalization to arbitrary $\ell$}
\label{app:general_ell_disc}

For general $\ell$p functions, the discontinuity in \Eq{eq:def_discontinuity_general_ell} can be computed analogously by inserting the spectral representation.
Then, only those permutations survive the difference for which the frequency combinations $\omega_I$ or $\omega_{I^c}$ appear in the kernel $\tK(\bs{z}_p)$, leading to
\bsubeq
\bal
\tG^{\omega_{I}}_{\zcheck^{\tn{r}}}
&=
\tG_{\omega^+_{I},\zcheck^{\tn{r}}}
-
\tG_{\omega^-_{I},\zcheck^{\tn{r}}} 
\nn
&=
\sum_{{\oli{I}|\oli{I}^c}} 
[
\tK_{\oli{I}|\oli{I}^c}
\ACast 
S_{[\oli{I},\oli{I}^c]_-} 
]\big(\bs{z}_{\oli{I}| \oli{I}^c}(\omega_I,\zcheck^{\tn{r}}) \big)
,
\label{eq:discont_general_spec_rep}
\\
\tK_{\oli{I}|\oli{I}^c}(\bs{z}_{\oli{I}| \oli{I}^c}(\omega_I,\zcheck^{\tn{r}}))
&=
\tK\big(\bs{z}_{\oli{I}| \oli{I}^c}(\omega_I^+,\zcheck^{\tn{r}})\big)
-
\tK\big(\bs{z}_{\oli{I}| \oli{I}^c}(\omega_I^-,\zcheck^{\tn{r}})\big)
\nn
&=
\hdelta(\omega_I) \tK\big(\bs{z}_{\oli{I}}(\zcheck^{\tn{r}})\big) \tK\big(\bs{z}_{\oli{I}^c}(\zcheck^{\tn{r}})\big)
\label{eq:regular_product_kernel_not_Keldysh}
\\
\tK(\bs{z}_{\oli{I}}) 
&= 
\prod_{i=1}^{|I|-1} \frac{1}{\omega_{\oli{I}_1\dots\oli{I}_i\dots}}
\label{eq:regular_kernel_subtuple}
.
\eal
\label{eq:spectralRep_for_branchcut}
\esubeq
The set $I^c=L\backslash I$ is complementary to $I$. Here, $\bs{z}_p(\omega_I,\zcheck^{\tn{r}})$ expresses the permuted vector $\bs{z}_p$ in terms of $\omega_I$ and the remaining $\ell-2$ independent frequencies $\zcheck^{\tn{r}}$, and similarly $\bs{z}_{\oli{I}}(\zcheck^{\tn{r}})$ for the subtuple $z_{\oli{I}}$. 
\EQ{eq:regular_kernel_subtuple} defines a regular kernel for the subtuple $\bs{z}_{\oli{I}}$.
In \Eq{eq:regular_product_kernel_not_Keldysh}, the difference of regular kernels leads to the Dirac delta factor due to $1/\omega^+_I - 1/\omega^-_I = \hdelta(\omega_{I})$ and $1/\omega^+_{I^c} - 1/\omega^-_{I^c} = - \hdelta(\omega_{I})$ (using \Eq{eq:hdelta_identity}).
The definition of the regular product kernel in \Eq{eq:regular_product_kernel_not_Keldysh} implies $\tK_{\oli{I}|\oli{I}^c} = \tK_{\oli{I}^c|\oli{I}}$; thus, the corresponding
PSFs from permutations $\oli{I}|\oli{I}^c$ and $\oli{I}^c|\oli{I}$ have been combined in an PSF commutator in \Eq{eq:discont_general_spec_rep}.

Consider, e.g., the 3p discontinuity $\tG^{\omega_2}_{\omega_1^+}$ from \App{App:discont_spec_rep_3p}, where the sets in \Eq{eq:spectralRep_for_branchcut} are given by $I = \{ 2 \}$, $I^c = \{1,3\}$, and $\zcheck^{\tn{r}} = \omega^+_1$.
Then, the sum over permutations $p = \oli{I} \vert \oli{I}^c$ includes $\oli{I} \vert \oli{I}^c \in \{ 2 \vert 13, 2 \vert 31 \}$, and we obtain the PSF commutator contribution $S_{[2,13]_-}$ in \Eq{eq:Spec_rep_G^2_1} from \Eq{eq:spectralRep_for_branchcut}.

For $\ell=4$, let us consider $\tG^{\omega_{13}}_{z_1, z_2}$ as an example.
Then, the sets $I=\{1,3\}$ and $I^c=\{2,4\}$ yield the permutations $\{13|24,13|42,31|24,31|42\}$, resulting in
\bal
\label{eq:discontinuity_example_4p}
\tG^{\omega_{13}}_{z_1, z_2}
&=
\int\tn{d}^4\varepsilon\,
\frac{\delta(\varepsilon_{1234})
\hdelta(\varepsilon_{13})
}{(z_1-\varepsilon_1)(z_2-\varepsilon_2)}
S_{[[1,3]_-,[2,4]_-]_-}(\bs{\varepsilon}),
\eal
where we summarized all terms with the same kernels.

To compute consecutive discontinuities, such as $\tG^{\omega_2,\omega_1}$ (see \Eq{eq:3p_disc_ident}), we can iterate the above procedure: By analyzing the spectral representation of the first discontinuity, we determine the branch cuts which lead to non-vanishing second discontinuities, and then compute these second discontinuities by use of identity \eqref{eq:hdelta_identity}.
For fermionic 4p correlators, this iterative procedure implies that double bosonic discontinuities must vanish, e.g., $\tG^{\omega_{13},\omega_{14}}_{\omega_1^+}=0$.
This follows from the spectral representation of $\tG^{\omega_{13}}_{z_1,z_2}$ in \Eq{eq:discontinuity_example_4p}, where the kernels only depend on fermionic frequencies $z_1$, $z_2$ in the denominators. Hence, there is no $\Im z_{14}=0$ branch cut, and therefore $\tG^{\omega_{13},\omega_{14}}_{\omega_1^+}$ must vanish.

\subsection{Spectral representation of anomalous parts}
\label{app:spec_rep_anom}
In this appendix, we focus on the spectral representation for contributions to the MF correlator anomalous w.r.t.\ one frequency. We again start with an example for $\ell=3$ (\App{eq:specRep_of_anomG_3p}), before generalizing to arbitrary $\ell$ (\App{app:general_ell_anom_term}).

\subsubsection{Example for $\ell = 3$}
\label{eq:specRep_of_anomG_3p}
Consider $\beta \delta_{\i \omega_1} \hG_1(\i \bsomega)$ for $\ell = 3$. 
Only those terms in the 3p kernel \Eq{eq:3p_kern_Kugler} proportional to $\delta_{\Omega_1} = \delta_{\i \omega_1} \delta_{\varepsilon_1}$ and $\delta_{\Omega_{23}} = \delta_{\i \omega_{23}} \delta_{\varepsilon_{23}} = \delta_{\i \omega_{1}} \delta_{\varepsilon_{1}}$ can contribute to $\hG_1$. Hence, the anomalous PSFs $S_p$ must contain factors $\delta(\varepsilon_1)$, i.e.,
\bal
&\beta \delta_{\i \omega_1} \hG_1(\i \bsomega) 
\nn
&= -\tfrac{1}{2} \beta \delta_{\i \omega_1} 
\int \tn{d}^3\varepsilon
\,  
\delta(\varepsilon_{123})
\Bigg[ 
\delta_{\varepsilon_1}\! S_{(123)}(\varepsilon_1,\varepsilon_2)
\Delta_{\i \omega_{12} - \varepsilon_{12}} 
\nn
&\hspace{6pt} 
+ 
\delta_{\varepsilon_1}\! S_{(132)} (\varepsilon_1,\varepsilon_3)
\Delta_{\i \omega_{13} - \varepsilon_{13}} 
+ 
\delta_{\varepsilon_{23}}\! S_{(231)} (\varepsilon_2,\varepsilon_3)
\Delta_{\i \omega_{2} - \varepsilon_{2}} 
\nn
&\hspace{6pt}
+
\delta_{\varepsilon_{23}}\! S_{(321)} (\varepsilon_3,\varepsilon_2)
\Delta_{\i \omega_{3} - \varepsilon_{3}} \Bigg] 
\nn
&= 
-\tfrac{1}{2} \beta \delta_{\i \omega_1} 
\int \tn{d}^3\varepsilon 
\,
\delta(\varepsilon_{123})
\Bigg[ 
\delta_{\varepsilon_1}\! S_{[1,23]_+} (\varepsilon_1,\varepsilon_2,\varepsilon_3)
\Delta_{\i \omega_{2} - \varepsilon_{2}} 
\nn
&\hspace{100pt}+ 
\delta_{\varepsilon_1}\! 
S_{[1,32]_+} (\varepsilon_1,\varepsilon_2,\varepsilon_3)
\Delta_{\i \omega_{3} - \varepsilon_{3}} \Bigg]
\nn
&= 
-\tfrac{1}{2} \beta \delta_{\i \omega_1} 
\int \tn{d}^3\varepsilon 
\,
\delta(\varepsilon_{123})
\delta_{\varepsilon_1}\! S_{[1,[2,3]_-]_+} (\varepsilon_1,\varepsilon_2,\varepsilon_3)
\Delta_{\i \omega_{2} - \varepsilon_{2}} 
,
\label{eq:spec_rep_anom_3p_hG_1}
\eal
where we used the symbolic Kronecker notation from \App{app:PSF_decomposition}.
The remaining contributions $p=(213), (312)$ can only contribute to the anomalous terms $\hG_2$ and $\hG_3$, as they are not proportional to $\delta_{\i \omega_1}$.

Note that, in the spectral representation \eqref{eq:spec_rep_anom_3p_hG_1}, the decomposition of $\hG_{1; \i\omega_2}=\hG_{1;\i\omega_2}^{\noDelta} + \Delta_{\i\omega_2}\hG_{1;2}^{\withDelta}$ follows from the PSF decomposition.
Only PSF terms proportional to  $\delta(\varepsilon_2)$, $\delta_{\varepsilon_1} \delta_{\varepsilon_2}\! S_{[1,[2,3]_-]_+}$, contribute to $\hG_{1;2}^{\withDelta}$. In the absence of such $\delta(\varepsilon_2)$ contributions, we can evaluate $\Delta_{\i\omega_2-\varepsilon_2}\rightarrow1/(\i\omega_2-\varepsilon_2)$ and compute the discontinuity $\hG^{\noDelta;\omega_2}_{1} = \hG^{\noDelta}_{1;\omega_2^+} - \hG^{\noDelta}_{1;\omega_2^-}$:
\bal 
\label{eq:Gd_Gf_spec_rep}
\delta(\omega_1) \delta(\omega_2) \hG_{1;2}^{\withDelta} &= - \delta_{\omega_1} \delta_{\omega_2} S_{1[2,3]_-}(\omega_1,\omega_2,-\omega_{12}), 
\nn
\delta(\omega_1) \hG_{1}^{\noDelta;\omega_2} &= (2\pi \i) \delta_{\omega_1} (1-\delta_{\omega_2}) S_{1[2,3]_-}(\omega_1,\omega_2,-\omega_{12}).
\eal
Here, we used $\delta_{\omega_1} S_{[1,[2,3]_-]_+} = 2 \delta_{\omega_1} S_{1[2,3]_-}$ due to the equilibrium condition~\eqref{eq:equilibrium:cyclic_PSFs}.
These commutator representations will be used for the 3p consistency check in \App{app:3p_cons_check}.

\subsubsection{Generalization to arbitrary $\ell$}
\label{app:general_ell_anom_term}

Now, we generalize the insights from the $\ell = 3$ example to arbitrary $\ell$.
The result will be used in \App{app:simplifying_KF_correlators} to provide a general formula for the construction of KF components $G^{[\eta_1\eta_2]}$ from MF functions.

In the $\beta\delta$ expansion of the MF kernel $K=\tK+\hK^{\beta\delta}+\mc{O}(\delta^2)$, the $\beta\delta$ term reads (see Eq.~(45) in Ref.~\cite{Kugler2021}) 
\bal
\label{eq:anomKernel_betaTerm_all_ell}
\beta\hK^\beta(\bs{\Omega}_p) 
&= -\tfrac{\beta}{2} \sum_{i=1}^{\ell-1} \delta_{\Omega_{\oli{1}\dots\oli{i}}} \prod_{\substack{j=1\\j\neq i}}^{\ell-1} \Delta_{\Omega_{\oli{1}\dots\oli{j}}}
,
\eal
which was originally derived for $\ell\leq4$, but can be extended to arbitrary $\ell$ with the same line of arguments, starting from the results in Ref.~\cite{Halbinger2023}.
For general $\ell$p functions and terms anomalous w.r.t.\ the frequency $\i \omega_I = 0$, with $I\subset L=\{1,\dots,\ell\}$, only permutations of the form $p = \oli{I} \vert \oli{I}^c$ and $p = \oli{I}^c \vert \oli{I}$, with  $I^c=L\backslash I$ again the complementary set to $I$, can lead to the $\beta \delta_{\i \omega_I}$ factor coming from the anomalous kernel in \Eq{eq:anomKernel_betaTerm_all_ell}, yielding
\bal
&\beta \delta_{\i \omega_I} \hG_I(\i \bsomega) 
\nn
&=
-\tfrac{1}{2} \beta \delta_{\i \omega_I} \sum_{\oli{I} \vert \oli{I}^c} \int \newlychanged{\md}^{\ell} \varepsilon_p\, \delta(\varepsilon_{1\dots\ell})
\prod_{i=1}^{|I|-1} \Delta_{\Omega_{I_{\oli{1}}\dots I_{\oli{i}}}}
\prod_{i=1}^{|I^c|-1} \Delta_{\Omega_{I^c_{\oli{1}}\dots I^c_{\oli{i}}} }
\nn
&\hspace{60pt}
\times
\delta_{\varepsilon_{\oli{I}}}\! S_{[\oli{I},\oli{I}^c]_+} (\bs{\varepsilon}(\bs{\varepsilon}_{\oli{I} \vert \oli{I}^c})) 
.
\eal
\EQ{eq:spec_rep_anom_3p_hG_1} is a direct application of this formula for $\ell = 3$, $I = \{ 1 \}$, and $I^c = \{ 2, 3 \}$, where the permutations $p = \oli{I} \vert \oli{I}^c$ run over $\oli{I} \vert \oli{I}^c \in \{ 1 \vert 23, 1 \vert 32 \}$.

To make the connection to Keldysh correlators in the next appendix, we
replace any $\Delta_{\i\omega}\rightarrow 1/(\i\omega)$ in the final expression for $\hG_I$, which amounts to replacing $\Delta_{\Omega}\rightarrow 1/\Omega$ in the kernels, such that
\bal
\hG_{I; \zcheck^{\tn{r}}}
&\equiv
\Big[\hG_I(\i \bsomega) \Big]_{\Delta_{\i \omega}\rightarrow \tfrac{1}{\i \omega}, \i \bsomega \rightarrow \bs{z}(\zcheck^{\tn{r}})}
\nn
&=
-\tfrac{1}{2} \sum_{\oli{I} \vert \oli{I}^c} \int \newlychanged{\md}^{\ell} \varepsilon_p\, \delta(\varepsilon_{1\dots\ell}) 
\tK(\bs{z}_{\oli{I}} (\zcheck^{\tn{r}}) - \bs{\varepsilon}_{\oli{I}})
\nn
&\hspace{40pt}
\times \tK(\bs{z}_{\oli{I}^c}(\zcheck^{\tn{r}}) - \bs{\varepsilon}_{\oli{I}^c})\
\delta_{\varepsilon_{\oli{I}}}\! S_{[\oli{I},\oli{I}^c]_+} (\bs{\varepsilon}(\bs{\varepsilon}_{\oli{I} \vert \oli{I}^c})) 
,
\eal
where we identified a product of regular kernels (see \Eq{eq:regular_kernel_subtuple}). The subscript $\zcheck^{\tn{r}}$ again denotes $\ell - 2$ independent frequencies parametrizing the $\ell-1$ arguments $\bs{z}$ of $\hG_I\big(\bs{z} (\zcheck) \big) = \hG_{I;\bs{\zcheck}}$, with $\bs{z}$ independent of the anomalous frequency $\omega_I$.

The anomalous parts $\hG_I$ typically enter the Keldysh components with prefactors depending on $4\pi \i\, \delta(\omega_I)$. 
Including this factor, the spectral representation turns out to be particularly convenient, as we can make use of the definition in \Eq{eq:def_productKernel}, leading to
\bal
&4\pi \i\, \delta(\omega_I) \hG_{I;\zcheck^{\tn{r}}} 
= 
-2\, \hdelta (\omega_I) \hG_{I;\zcheck^{\tn{r}}} 
\nn
&= 
\sum_{\oli{I} \vert \oli{I}^c} \int \newlychanged{\md}^{\ell} \varepsilon_p\ 
\delta(\varepsilon_{\oli{1}\dots\oli{\ell}})
\hdelta(\omega_{\oli{I}} - \varepsilon_{\oli{I}})
\nn
&\hspace{65pt} \times 
\tK\big(\bs{z}_{\oli{I}}(\zcheck^{\tn{r}}) - \bs{\varepsilon}_{\oli{I}}\big) \tK\big( \bs{z}_{\oli{I}^c}(\zcheck^{\tn{r}}) - \bs{\varepsilon}_{\oli{I}^c}\big) 
\nn
&\hspace{65pt} \times 
\delta_{\varepsilon_{\oli{I}}}\! S_{[\oli{I},\oli{I}^c]_+}(\bs{\varepsilon}(\bs{\varepsilon}_{\oli{I} \vert \oli{I}^c}))\, 
\nn
&= 
\sum_{\oli{I} \vert \oli{I}^c} \int \newlychanged{\md}^{\ell} \varepsilon_p\ 
\delta(\varepsilon_{\oli{1}\dots\oli{\ell}})
\tK_{\oli{I}|\oli{I}^c}\big(\bs{z}_{\oli{I}|\oli{I}^c}(\omega_I, \zcheck^{\tn{r}}) - \bs{\varepsilon}_{\oli{I}|\oli{I}^c}\big) 
\nn
&\hspace{65pt} \times 
\delta_{\varepsilon_{\oli{I}}}\! S_{[\oli{I},\oli{I}^c]_+}(\bs{\varepsilon}(\bs{\varepsilon}_{\oli{I} \vert \oli{I}^c}))
. 
\label{eq:spec_rep_anom_realomega}
\eal
In the second step, we used $\omega_I = \omega_{\oli{I}}$ and 
\bal
\delta_{\varepsilon_{\oli{I}}}\! S_{[\oli{I},\oli{I}^c]_+} (\bs{\varepsilon}_{\oli{I} \vert \oli{I}^c}) \hdelta(\omega_{I}) = \delta_{\varepsilon_{\oli{I}}}\! S_{[\oli{I},\oli{I}^c]_+} (\bs{\varepsilon}_{\oli{I} \vert \oli{I}^c}) \hdelta(\omega_{\oli{I}}-\varepsilon_{\oli{I}})
.
\eal
In the last line, we inserted the definition of the regular product kernel \eqref{eq:regular_product_kernel_not_Keldysh}.
\EQ{eq:spec_rep_anom_realomega} is the representation needed in \Eq{eq:G_eta1eta2_any_ell_via_AC} to express Keldysh components $G^{[\eta_1\eta_2]}$ in terms of analytically continued anomalous parts of MF correlators.

\section{Simplifications for KF correlators}
\label{app:simplifying_KF_correlators}
In this appendix, we derive reformulations of the spectral representation of KF components, presented in \Secs{sec:KF_components_alpha_2} and \ref{sec:4p_correlator_alpha_le_3}, which are amenable to finding relations between KF correlators and analytically continued MF correlators.
First, we derive a convenient identity for particular KF kernels for general $\ell$p correlators in \App{app:rewritingkerneltwo2s}. This identity is then applied in \App{app:simplifications_for_G_alpha2} to obtain an alternative representation of KF components $G^{[\eta_1\eta_2]}$, yielding a general analytic continuation formula (\Eq{eq:G_eta1eta2_any_ell_via_AC}) for these components (using the results from \App{app:add_spectral_rep}).
This constitutes a generalization of \Eq{eq:Analytic_cont_fully_ret} for $G^{[\eta_1]}$ ($\alpha=1$) to $\alpha=2$.
An analogous procedure is then applied to KF components $G^{[\eta_1\dots\eta_\alpha]}$ for $\alpha=3$ and $\alpha=4$ in \Apps{sec:appeta123} and \ref{app:appeta1234}, respectively (see \Eqs{eq:GFAlphaThree_general} and \eqref{eq:4p_alpha_4_spec_rep}). In the following, we will use the notation introduced in the beginning of \Sec{sec:4p_Keldysh_correlators} repeatedly.

\subsection{Identity for $K^{[\heta_1 \heta_2]}$ for general $\ell$p correlators}
\label{app:rewritingkerneltwo2s}
For $\alpha = 2$, Keldysh correlators $G^{[\eta_1 \eta_2]}$ are determined by the KF kernel $K^{[\heta_1\heta_2]} = K^{[\heta_1]} - K^{[\heta_2]}$ in \Eq{eq:FTretkern}. 
For $\alpha \ge 2$, such differences of fully retarded kernels occur repeatedly in the spectral representation. 
In the following, we therefore derive a convenient identity for the kernel $K^{[\heta_1\heta_2]}$.

According to \Eqs{eq:KFkerneletas} and \eqref{eq:MF-KF-continuation-Kernel}, the kernel $K^{[\heta_1\heta_2]}$ takes the form 
\bal
K^{[\heta_1 \heta_2]}(\bsomega_p) &= K^{[\heta_1]}(\bsomega_p) - K^{[\heta_2]}(\bsomega_p),
\nn
&=\tK(\bsomega^{[\oli{\heta}_1]}_p) - \tK(\bsomega^{[\oli{\heta}_2]}_p).
\eal
Note that $\heta_1 < \heta_2$, which holds by definition, does not imply $\oli{\heta}_1 < \oli{\heta}_2$.

For simplicity, we rename $\mu = \heta_1$ and $\nu = \heta_2$. Using \Eqs{eq:FTretkern} and \eqref{eq:imshifts}, the retarded kernels generally read
\bal
K^{[\mu]}(\bsomega_p) &= \left( \prod_{i=1}^{\mu-1} \frac{1}{\omega^-_{\oli{1} ... \oli{i}} } \right) \left( \prod_{i=\mu}^{\ell-1} \frac{1}{\omega^+_{\oli{1} ... \oli{i}} } \right) = K^-_{1\mu} K^+_{\mu \ell},
\nn
K^\pm_{x y} &= \prod_{i=x}^{y-1} \frac{1}{\omega^\pm_{\oli{1} ... \oli{i}} }.
\label{eq:kernel_shorthand_for_kernelIdentity}
\eal
From this definition of $K^\pm_{x y}$, the identities
\bal
K^{\pm}_{x y} K^{\pm}_{y z} = K^{\pm}_{x z}, \qquad K^{\pm}_{x x} =1, \qquad K^{[\mu]} = K^-_{1\mu} K^+_{\mu\ell}
\eal
directly follow, which allow us to rewrite $K^{[\mu \nu]}(\bsomega_p)$ as
\bal
K^{[\mu \nu]} &= K^{[\mu]} - K^{[\nu]} = K^-_{1\mu} \left( K^+_{\mu\nu} - K^-_{\mu\nu} \right) K^+_{\nu\ell} 
\nn
&= \sum_{y=\mu}^{\nu-1} K^-_{1\mu} \left( K^+_{\mu y+1} K^-_{y+1 \nu} - K^+_{\mu y} K^-_{y \nu} \right) K^+_{\nu\ell} 
\nn
&= \sum_{y=\mu}^{\nu-1} K^-_{1\mu} K^+_{\mu y} \underbrace{\left( \frac{1}{\omega^+_{\oli{1} ... \oli{y}}} - \frac{1}{\omega^-_{\oli{1} ... \oli{y}}} \right)}_{= \hdelta(\omega_{\oli{1} ... \oli{y}}) } K^-_{y+1 \nu} K^+_{\nu\ell}.
\eal
In the second line, the terms $y = \mu$ and $y=\nu-1$ represent the first line, the remaining contributions $\mu < y < \nu-1$ cancel pairwise. In the last line, we used identity \eqref{eq:hdelta_identity} to obtain $\hdelta(\omega_{\oli{1} ... \oli{y}})$, enforcing $\omega^\pm_{\oli{1} \dots \oli{i}} = \omega^\pm_{\oli{y+1} \dots \oli{i}}$ for $i>y$. Inserting this identity into the arguments of $K^-_{y+1 \nu} K^+_{\nu\ell}$ yields
\bal
K^{[\mu \nu]}(\bsomega_p) &= \sum_{y=\mu}^{\nu-1} K^{[\mu]}(\bsomega_{\oli{1} \dots \oli{y}}) \hdelta(\omega_{\oli{1} \dots \oli{y}}) K^{[\nu]}(\bsomega_{\oli{y+1} \dots \oli{\ell}}) 
\nn
&= \sum_{y=\mu}^{\nu-1} \tK(\bsomega^{[\oli{\mu}]}_{\oli{1} \dots \oli{y}}) \hdelta(\omega_{\oli{1} \dots \oli{y}}) \tK(\bsomega^{[\oli{\nu}]}_{\oli{y+1} \dots \oli{\ell}}) 
\nn
&= 
\sum_{y=\mu}^{\nu-1} \tK_{\oli{1} \dots \oli{y} | \oli{y+1} \dots \oli{\ell} } (\bsomega_{\oli{1}\dots\oli{y} | \oli{y+1}\dots\oli{\ell}}^{[\oli{\mu}][\oli{\nu}]}).
\label{eq:kernel_identity_alpha2}
\eal
The last equality follows from the definition \eqref{eq:def_productKernel}, with $\alpha = 2$, $\eta_1 = \oli{\mu}$, $\eta_2 = \oli{\nu}$, $\oli{I}^1 = \oli{1} \dots \oli{y}$, and $\oli{I}^2 = \oli{y+1} \dots \oli{\ell}$.
Note that, for $\ell=3$, \Eq{eq:kernel_identity_alpha2} readily yields the results of \Tab{tab:3p_KF_kernels_G212}.

\subsection{Simplifications for $G^{[\eta_1\eta_2]}$ for $\ell$p correlators}
\label{app:simplifications_for_G_alpha2}
After the preparations in \Apps{app:add_spectral_rep} and \ref{app:rewritingkerneltwo2s}, we can now derive an alternative representation of the Keldysh correlators $G^{[\eta_1\eta_2]}$, equivalent to the spectral representation in \Eq{eq:KF_spec_rep_etas} but more convenient for the analytic continuation.
This generalizes the concepts of \Sec{app:rewritingkerneltwo2s_3p} from $\ell = 3$ to arbitrary $\ell$.

We start by inserting \Eq{eq:kernel_identity_alpha2} into the spectral representation in \Eq{eq:KF_spec_rep_etas},
\bal
G^{[\eta_1\eta_2]}&(\bsomega) 
=
\sum_{p} [K^{[\heta_1\heta_2]} * S_p](\bsomega_p)
\nn
&= \sum_p \sum_{y=\heta_1}^{\heta_2-1} \left( \tK_{ \oli{1} ... \oli{y} | \oli{y+1} ... \oli{\ell} } \ast S_p \right)(\bsomega_{\oli{1} ... \oli{y} | \oli{y+1} ... \oli{\ell}}^{[\oli{\heta}_1][\oli{\heta}_2]}). \label{KFcorrfunceta12sumy}
\eal
Since $\heta_1 \le y < \heta_2$, the subtuples $\oli{I} = (\oli{1} ... \oli{y})$ and $\oli{I}^c = (\oli{y+1} ... \oli{\ell})$ always contain $\oli{\heta}_1$ and $\oli{\heta}_2$, respectively. Each of these in turn equals either $\eta_1$ or $\eta_2$, since $\heta_i \in \{ p^{-1}(\eta_1), p^{-1}(\eta_2) \}$, hence $\oli{\heta}_i \in \{ \eta_1, \eta_2 \}$. Correspondingly, we will denote the subtuple containing $\eta_1$ as $\oli{I}^1$, and that containing $\eta_2$ as $\oli{I}^2$. The sum over $y$ can then be interpreted as a sum over all possible partitions of $(\oli{1},\dots,\oli{\ell})$ for which each of the two subtuples contains either $\eta_1$ or $\eta_2$. 
Defining $\mc{I}^{12} = \{ (I^1,I^2) \vert \eta_1 \in I^1, \eta_2 \in I^2, I^1\cup I^2=L, I^1\cap I^2=\emptyset \}$ as the set of all possibilities to partition $L=\{1,\dots,\ell\}$ into subsets $I^1$ and $I^2$ containing $\eta_1$ and $\eta_2$, respectively, we find
\bal
G^{[\eta_1 \eta_2]}(\bsomega) 
&= 
\sum_{(I^1,I^2) \in \mc{I}^{12}} \Big[ \sum_{ \oli{I}^1 \vert \oli{I}^2 } \left(  \tK_{ \oli{I}^1 | \oli{I}^2 }  \ACast S_{ \oli{I}^1 | \oli{I}^2 } \right)(\bsomega_{ \oli{I}^1 | \oli{I}^2 }^{[\eta_1][\eta_2]}) 
\nn
&\phantom{=}
+ 
\sum_{ \oli{I}^2 \vert \oli{I}^1 } 
\left( 
  \tK_{ \oli{I}^2 | \oli{I}^1 } \ACast S_{ \oli{I}^2 | \oli{I}^1 }
\right)
(\bsomega_{ \oli{I}^2 | \oli{I}^1 }^{[\eta_2][\eta_1]}) \Big].
  \eal
Here, we collected all terms in \Eq{KFcorrfunceta12sumy} proportional to $\hdelta(\omega_{\oli{I}^1})$ and summed over all allowed partitions. 
Using the symmetry of the kernels
\eqref{eq:def_productKernel} and the (anti)commutator notation from \Eq{eq:def_AntiCommutator}, we finally obtain
\bal
G^{[\eta_1 \eta_2]}(\bsomega) 
= 
\!\!\!\!\!\!
\sum_{(I^1,I^2) \in \mc{I}^{12}} \sum_{ \oli{I}^1 \vert \oli{I}^2 } 
\left( 
\tK_{ \oli{I}^1 | \oli{I}^2 }
\ACast 
S_{[ \oli{I}^1, \oli{I}^2]_+ } 
\right)(\bsomega_{ \oli{I}^1 | \oli{I}^2 }^{[\eta_1][\eta_2]}). \label{KFcorrfunceta12final_app}
\eal

Building on this expression, the KF component can be related to MF functions for arbitrary $\ell$.
For this purpose, we use the equilibrium condition to replace PSF commutators with anticommutators,
\bal  \label{eq:equilibrium_condition:AntiCommutator_equals_n_Commutator}
S_{[\oli{I},\oli{I}^c]_+} (\bs{\varepsilon}_{\oli{I}|\oli{I}^c})
&=
N_{\varepsilon_{\oli{I}}} S_{[\oli{I},\oli{I}^c]_-}(\bs{\varepsilon}_{\oli{I}|\oli{I}^c})
+ 
\delta_{\varepsilon_{\oli{I}} } S_{[\oli{I},\oli{I}^c]_+}(\bs{\varepsilon}_{\oli{I}|\oli{I}^c})
,
\\
N_{\varepsilon_{\oli{I}}} &=
\frac{\zeta^{\oli{I}}e^{\beta \varepsilon_{\oli{I}}}+1}{\zeta^{\oli{I}}e^{\beta \varepsilon_{\oli{I}}}-1}
\nonumber
= 
    \coth(\beta\varepsilon_{\oli{I}}/2)^{\zeta^{\oli{I}}},    
\eal
where $N_{\varepsilon_{\oli{I}}}$ is identical to the statistical factor in \Eq{eq:def_statistical_N}, and we used the symbolic Kronecker notation from \App{app:PSF_decomposition}. The sign factor is given by $\zeta^{\oli{I}}=\pm 1$ for an even/odd number of fermionic operators in the set $\oli{I}$.
Inserting \Eq{eq:equilibrium_condition:AntiCommutator_equals_n_Commutator} into the representation \eqref{KFcorrfunceta12final_app}, we thus obtain 
\bal
\label{eq:G_eta1eta2_any_ell_via_AC}
G^{[\eta_1\eta_2]}(\bsomega)
&=
\sum_{(I^1,I^2) \in \mc{I}^{12}} \sum_{ \oli{I}^1 \vert \oli{I}^2 }
\int\mathrm{d}^{\ell}\varepsilon\,
\tK_{ \oli{I}^1 | \oli{I}^2 }(\bsomega_{ \oli{I}^1 | \oli{I}^2 }^{[\eta_1][\eta_2]}-\bs{\varepsilon}_{\oli{I}^1|\oli{I}^2}) 
\nn
& \!\!\!\!\!\!\!\!\!\!\!\!\!\!\!\!\! \times
\Big(
N_{\varepsilon_{\oli{I}^1}} S_{[ \oli{I}^1, \oli{I}^2]_-}(\bs{\varepsilon}_{\oli{I}^1|\oli{I}^2}) 
+ \delta_{\varepsilon_{\oli{I}^1}} S_{[ \oli{I}^1, \oli{I}^2]_+}(\bs{\varepsilon}_{\oli{I}^1|\oli{I}^2}) 
\Big)
\delta(\varepsilon_{1\dots\ell})
\nn
&=
\sum_{(I^1,I^2) \in \mc{I}^{12}} 
\bigg( N_{\omega_{I^1}} \tG^{\omega_{I^1}}_{\zcheck}
+ 4\pi\i\, \delta(\omega_{I^1})\, \hG_{I^1; \zcheck}
\bigg)
,
\nn
& \qquad
\tn{with } \zcheck = \{\omega_i^-|\, i\neq\eta_1,i\neq\eta_2\}
.
\eal
This remarkable formula generalizes \Eq{eq:Analytic_cont_fully_ret} for $G^{[\eta_1]}$, i.e. for $\alpha=1$ and arbitrary $\ell$, to $G^{[\eta_1\eta_2]}$ ($\alpha=2$).
To obtain its final form, we used that the retarded product kernel (\Eq{eq:def_productKernel}) in the second line is proportional to $\hdelta(\omega_{\oli{I}^1} - \varepsilon_{\oli{I}^1})$ and thereby set $N_{\varepsilon_{\oli{I}^1}} = N_{\omega_{\oli{I}^1}} = N_{\omega_{I^1}}$ independent of the integration variables.
In the second step, we then identified the spectral representations of discontinuities of the regular MF correlator $\tG^{\omega_{I^1}}_{\zcheck}$ (\Eq{eq:spectralRep_for_branchcut}) and of the anomalous contribution $\hG_{I^1;\zcheck}$ (\Eq{eq:spec_rep_anom_realomega}).
Note that the retarded product kernel coincides with the kernel \eqref{eq:regular_product_kernel_not_Keldysh} with a suitably continued $\zcheck$.
In \Eq{eq:G_eta1eta2_any_ell_via_AC}, the $\ell-2$ frequencies in $\zcheck$ carry negative imaginary shifts, in accordance with the definition of $\bs{\omega}^{[\eta_1][\eta_2]}_{\oli{I}^1|\oli{I}^2}$.

\subsection{Simplifications for $G^{[\eta_1 \eta_2 \eta_3]}$ for $\ell$p correlators}
\label{sec:appeta123}

\begin{table*}[t]
\renewcommand*{\arraystretch}{2}
\captionsetup{skip=10pt}
\begin{center}
\begin{tabular}{ >{\centering\arraybackslash} m{1.9cm} | >{\arraybackslash} m{2.3cm} | >{\arraybackslash} m{10.8cm} } 
$p$ & $K^{[\heta_1 \heta_2 \heta_3]} - K^{[\mu_3]}$ & $ \left( G^{[\eta_1 \eta_2 \eta_3]} - G^{[\eta_3]} \right)(\bsomega) = \sum_{p} \left[ \left( K^{[\heta_1 \heta_2 \heta_3]} - K^{[\mu_3]} \right) \ast S_p \right](\bsomega_p) $ \\ \hline 
    & & $= \sum_{(I^1, I^2, I^3) \in \mathcal{I}^{123}}$ \\
  $\mu_1 < \mu_2 < \mu_3$    &  $K^{[\mu_1 \mu_2]}$    & \hsp $\Bigg\{ \sum_{\oli{I}^1 \vert \oli{I}^2 \vert \oli{I}^3} \left[ \left( \tK_{\oli{I}^1 \vert \oli{I}^2 \vert \oli{I}^3} \ACast S_{\oli{I}^1 \vert \oli{I}^2 \vert \oli{I}^3} \right) (\bsomega^{[\eta_1][\eta_2][\eta_3]}_{\oli{I}^1 \vert \oli{I}^2 \vert \oli{I}^3}) + \left( \tK_{\oli{I}^1 \vert \oli{I}^{2 \vert 3} } \ACast S_{\oli{I}^1 \vert \oli{I}^{2 \vert 3}} \right) (\bsomega^{[\eta_1][\eta_3]}_{\oli{I}^1 \vert \oli{I}^{2\vert 3}}) \right]$  \\ 
  $\mu_1 < \mu_3 < \mu_2$    &  $K^{[\mu_1 \mu_3]} - K^{[\mu_3 \mu_2]}$   & \hsp $+ \sum_{\oli{I}^1 \vert \oli{I}^3 \vert \oli{I}^2} \left[ \left( \tK_{\oli{I}^1 \vert \oli{I}^{3 \vert 2} } \ACast S_{\oli{I}^1 \vert \oli{I}^{3 \vert 2} } \right) (\bsomega^{[\eta_1][\eta_3]}_{\oli{I}^1 \vert \oli{I}^{3 \vert 2} }) + \left( \tK_{\oli{I}^{1\vert 3} \vert \oli{I}^{2} } \ACast S_{\oli{I}^{1\vert 3} \vert \oli{I}^{2} } \right) (\bsomega^{[\eta_3][\eta_2]}_{\oli{I}^{1\vert 3} \vert \oli{I}^{2} }) \right]$   \\
  $\mu_2 < \mu_1 < \mu_3$   &  $K^{[\mu_2 \mu_1]}$   & \hsp $+\sum_{\oli{I}^2 \vert \oli{I}^1 \vert \oli{I}^3} \left[ \left( \tK_{\oli{I}^2 \vert \oli{I}^1 \vert \oli{I}^3} \ACast S_{\oli{I}^2 \vert \oli{I}^1 \vert \oli{I}^3} \right) (\bsomega^{[\eta_2][\eta_1][\eta_3]}_{\oli{I}^2 \vert \oli{I}^1 \vert \oli{I}^3}) + \left( \tK_{\oli{I}^2 \vert \oli{I}^{1 \vert 3} } \ACast S_{\oli{I}^2 \vert \oli{I}^{1 \vert 3}} \right) (\bsomega^{[\eta_2][\eta_3]}_{\oli{I}^2 \vert \oli{I}^{1\vert 3}}) \right]$   \\ 
  $\mu_2 < \mu_3 < \mu_1$   &  $K^{[\mu_2 \mu_3]} - K^{[\mu_3 \mu_1]}$   & \hsp $+\sum_{\oli{I}^2 \vert \oli{I}^3 \vert \oli{I}^1} \left[ \left( \tK_{\oli{I}^2 \vert \oli{I}^{3 \vert 1} } \ACast S_{\oli{I}^2 \vert \oli{I}^{3 \vert 1} } \right) (\bsomega^{[\eta_2][\eta_3]}_{\oli{I}^2 \vert \oli{I}^{3 \vert 1} }) + \left( \tK_{\oli{I}^{2\vert 3} \vert \oli{I}^{1} } \ACast S_{\oli{I}^{2 \vert 3} \vert \oli{I}^{1} } \right) (\bsomega^{[\eta_3][\eta_1]}_{\oli{I}^{2\vert 3} \vert \oli{I}^{1} }) \right]$   \\ 
  $\mu_3 < \mu_1 < \mu_2$   &  $-K^{[\mu_1 \mu_2]}$   & \hsp $+\sum_{\oli{I}^3 \vert \oli{I}^1 \vert \oli{I}^2} \left[ \left( \tK_{\oli{I}^3 \vert \oli{I}^1 \vert \oli{I}^2} \ACast S_{\oli{I}^3 \vert \oli{I}^1 \vert \oli{I}^2} \right) (\bsomega^{[\eta_3][\eta_1][\eta_2]}_{\oli{I}^3 \vert \oli{I}^1 \vert \oli{I}^2}) - \left( \tK_{\oli{I}^{3 \vert 1} \vert \oli{I}^{2} } \ACast S_{\oli{I}^{3 \vert 1} \vert \oli{I}^{2} } \right) (\bsomega^{[\eta_3][\eta_2]}_{\oli{I}^{3 \vert 1} \vert \oli{I}^{2} }) \right]$   \\
  $\mu_3 < \mu_2 < \mu_1$   &  $-K^{[\mu_2 \mu_1]}$   & \hsp $+\sum_{\oli{I}^3 \vert \oli{I}^2 \vert \oli{I}^1} \left[ \left( \tK_{\oli{I}^3 \vert \oli{I}^2 \vert \oli{I}^1} \ACast S_{\oli{I}^3 \vert \oli{I}^2 \vert \oli{I}^1} \right) (\bsomega^{[\eta_3][\eta_2][\eta_1]}_{\oli{I}^3 \vert \oli{I}^2 \vert \oli{I}^1}) - \left( \tK_{\oli{I}^{3 \vert 2} \vert \oli{I}^{1} } \ACast S_{\oli{I}^{3 \vert 2} \vert \oli{I}^{1} } \right) (\bsomega^{[\eta_3][\eta_1]}_{\oli{I}^{3 \vert 2} \vert \oli{I}^{1} }) \right] \Bigg\}$
\end{tabular}
\end{center}
\caption{ Keldysh kernel of $G^{[\eta_1 \eta_2 \eta_3]} - G^{[\eta_3]}$ (\Eq{eq:Keldysh_kernel_general_alpha_3}) for different permutation classes depending on the order of the $\mu_i = p^{-1}(\eta_i)$. Manipulations similar to \Eqs{eq:manipulation_general_alpha_3_1} and \eqref{eq:manipulation_general_alpha_3_2} result in the alternative spectral representation in the third column, which can be further rewritten as \Eq{eq:GFAlphaThree_general} using \Eq{eq:manipulation_general_alpha_3_3} (and equivalent identities). }
\label{tab:KF_kernels_alph_3_general}
\end{table*}

The calculation in \App{app:rewritingkernelthree3s_3p}, too, can be generalized to arbitrary $\ell$p correlators, in particular for the spectral representation of $G^{[\eta_1 \eta_2 \eta_3]} - G^{[\eta_3]}$. The Keldysh kernel for $G^{[3]}$ is given by $\tK(\bsomega^{[\eta_3]}_p) = K^{[\mu_3]}(\bsomega_p)$ for arbitrary permutations $p$, with $\mu_3 = p^{-1}(\eta_3)$. Then, the corresponding Keldysh kernel for $G^{[\eta_1 \eta_2 \eta_3]} - G^{[\eta_3]}$ reads
\bal
\label{eq:Keldysh_kernel_general_alpha_3}
K^{[\heta_1\heta_2\heta_3]} -K^{[\mu_3]}
&=
K^{[\heta_1]}
-K^{[\heta_2]}
+K^{[\heta_3]}
-K^{[\mu_3]},
\eal
such that the effect of subtracting $K^{[\mu_3]}$ depends on the permutation. The permutations can be divided into six categories, depending on the order in which the $\mu_j = p^{-1}(\eta_j)$ occur, see \Tab{tab:KF_kernels_alph_3_general}. This is important since placing the $\mu_j$ in increasing order yields $[\heta_1 \heta_2 \heta_3]$, see discussion before \Eqs{eq:KF_spec_rep_eta_alpha}.

Here, we focus on the key steps in rewriting permutations with $\mu_1 < \mu_2 < \mu_3$, denoted by $\sum_{p \vert \mu_1 < \mu_2 < \mu_3}$. Defining $ \mathcal{I}^{123}=\{(I^1,I^2,I^3)|\, \eta_1\in I^1,\eta_2\in I^2, \eta_3 \in I^3, I^b\cap I^{b'}=\emptyset \text{ for } b\neq b'\}$ as the set of all possibilities to partition $L=\{1,...,\ell\}$ into three blocks, each of which contains one of the indices $\eta_j\in I^j$, we have
\bal
\label{eq:manipulation_general_alpha_3_1}
&\sum_{p \vert \mu_ 1< \mu_2 < \mu_3} \left[ \left( K^{[\heta_1 \heta_2 \heta_3]} - K^{[\mu_3]} \right) \ast S_p \right](\bsomega_p) \nn
&= \sum_{p \vert \mu_ 1< \mu_2 < \mu_3} \left( K^{[\mu_1 \mu_2]} \ast S_p \right)(\bsomega_p) \nn
&= \sum_{p \vert \mu_ 1< \mu_2 < \mu_3}\, \sum_{y=\mu_1}^{\mu_2-1} \left( \tK_{ \oli{1} ... \oli{y} | \oli{y+1} ... \oli{\ell} } \ast S_p \right)(\bsomega_{\oli{1} ... \oli{y} | \oli{y+1} ... \oli{\ell}}^{[\eta_1][\eta_2]}) \nn
&= \sum_{(I^1, I^2, I^3) \in \mathcal{I}^{123}}\ \sum_{\oli{I}^1 \vert \oli{I}^2 \vert \oli{I}^3} \left( \tK_{ \oli{I}^1 | \oli{I}^{2 \vert 3} } \ACast S_{ \oli{I}^1 | \oli{I}^{2 \vert 3} } \right)(\bsomega_{ \oli{I}^1 | \oli{I}^{2 \vert 3} }^{[\eta_1][\eta_2]}).
\eal
In the first step, we used that $[\heta_1 \heta_2 \heta_3] = [\mu_1 \mu_2 \mu_3]$. In the second step, we inserted the kernel expansion \Eq{eq:kernel_identity_alpha2} with $\oli{\mu}_j = \eta_j$. In the third step, we identified the sum over $y$ as a sum over all possibilities to subdivide the permutations into the form $p= \oli{I}^1 \vert \oli{I}^2 \vert \oli{I}^3 $ (which guarantees $\mu_1 < \mu_2 < \mu_3$), with the concatenation of $\oli{I}^2$ and $\oli{I}^3$ denoted by $\oli{I}^{2 \vert 3} = \oli{I}^2_1 \dots \oli{I}^2_{\vert I^2 \vert} \oli{I}^3_1 \dots \oli{I}^3_{\vert I^3 \vert}$. 

Further, we use 
\bal
\label{eq:manipulation_general_alpha_3_2}
&\sum_{(I^1, I^2, I^3) \in \mathcal{I}^{123}}\ \sum_{\oli{I}^1 \vert \oli{I}^2 \vert \oli{I}^3} \left( \tK_{ \oli{I}^1 | \oli{I}^{2 \vert 3} } \ACast S_{ \oli{I}^1 | \oli{I}^{2 \vert 3} } \right)(\bsomega_{ \oli{I}^1 | \oli{I}^{2 \vert 3} }^{[\eta_1][\eta_2]}) \nn
&- \sum_{(I^1, I^2, I^3) \in \mathcal{I}^{123}}\ \sum_{\oli{I}^1 \vert \oli{I}^2 \vert \oli{I}^3} \left( \tK_{ \oli{I}^1 | \oli{I}^{2 \vert 3} } \ACast S_{ \oli{I}^1 | \oli{I}^{2 \vert 3} } \right)(\bsomega_{ \oli{I}^1 | \oli{I}^{2 \vert 3} }^{[\eta_1][\eta_3]}) \nn
&= \sum_{(I^1, I^2, I^3) \in \mathcal{I}^{123}}\ \sum_{\oli{I}^1 \vert \oli{I}^2 \vert \oli{I}^3} \left( \tK_{\oli{I}^1 \vert \oli{I}^2 \vert \oli{I}^3} \ACast S_{\oli{I}^1 \vert \oli{I}^2 \vert \oli{I}^3} \right)(\bsomega_{\oli{I}^1 \vert \oli{I}^2 \vert \oli{I}^3}^{[\eta_1][\eta_2][\eta_3]}),
\eal
which again follows by inserting \Eq{eq:kernel_identity_alpha2}, to arrive at the result in \Tab{tab:KF_kernels_alph_3_general}.

Contributions of different permutations can be further simplified, e.g., the second term of $p\vert \mu_1 < \mu_2 < \mu_3$ and the first term of $p\vert \mu_1 < \mu_3 < \mu_2$ can be collected, yielding
\bal
\label{eq:manipulation_general_alpha_3_3}
&\sum_{(I^1, I^2, I^3) \in \mathcal{I}^{123}}\ \sum_{\oli{I}^1 \vert \oli{I}^2 \vert \oli{I}^3} \left( \tK_{ \oli{I}^1 \vert \oli{I}^{2 \vert 3} } \ACast S_{ \oli{I}^1 \vert \oli{I}^{2 \vert 3} } \right)(\bsomega_{ \oli{I}^1 \vert \oli{I}^{2 \vert 3} }^{[\eta_1][\eta_3]}) \nn
&+ \sum_{(I^1, I^2, I^3) \in \mathcal{I}^{123}}\ \sum_{\oli{I}^1 \vert \oli{I}^3 \vert \oli{I}^2} \left( \tK_{ \oli{I}^1 \vert \oli{I}^{3 \vert 2} } \ACast S_{ \oli{I}^1 \vert \oli{I}^{3 \vert 2} } \right)(\bsomega_{ \oli{I}^1 \vert \oli{I}^{3 \vert 2} }^{[\eta_1][\eta_3]}) \nn
&= \sum_{(I^1,I^{23}) \in \mathcal{I}^{1\vert 23}}\ \sum_{\oli{I}^1 \vert \oli{I}^{23}} \left[ \tK_{\oli{I}^1 \vert \oli{I}^{23}} \ACast S_{\oli{I}^1 \vert \oli{I}^{23}} \right](\bsomega^{[\eta_1][\eta_3]}_{\oli{I}^1 \vert \oli{I}^{23}}),
\eal
with $\mathcal{I}^{1\vert 23}$ defined in \Eq{eq:I1vert23}.
Using the symmetry of retarded product kernels, e.g., $\tK_{\oli{I}^1 \vert \oli{I}^{23}} = \tK_{\oli{I}^{23} \vert \oli{I}^1 }$, the spectral representation of $G^{[\eta_1 \eta_2 \eta_3]} - G^{[\eta_3]}$ finally results in \Eq{eq:GFAlphaThree_general}.
Unlike for $\alpha=2$ we don't have a general formula for the analytic continuation to $G^{[\eta_1\eta_2\eta_3]}$.

Equation~\eqref{eq:4p_G123_explicit} shows an example for $\ell=4$. Inserting \Eq{eq:4p:Sp} into the PSF (anti)commutators and abbreviating $S'_p=(2\pi\i)^3 S_p$, we obtain the following relations:
\bal
\label{eq:4p_123_PSF_anticomm_list}
S'_{[[[\oli{1},\oli{2}]_-,\oli{3}]_+,\oli{4}]_+} &=
-N_{\oli{4}} \Big(
    N_{\oli{3}} \tG^{\varepsilon_{\oli{3}},\varepsilon_{\oli{12}},\varepsilon_{\oli{1}}}
    +
    N_{\oli{12}} \tG^{\varepsilon_{\oli{12}},\varepsilon_{\oli{3}},\varepsilon_{\oli{1}}}
    \nn
    & \
    -2\hdelta(\varepsilon_{\oli{12}})
    \hG^{\varepsilon_{\oli{3}},\varepsilon_{\oli{1}}}_{\oli{12}}
\Big)
,
\nn
 S'_{[[\oli{1},\oli{2}]_-,[\oli{3},\oli{4}]_+]_+} &=
    N_{\oli{12}}
     \Big(
        N_{\oli{4}} \tG^{\varepsilon_{\oli{4}}, \varepsilon_{\oli{3}}, \varepsilon_{\oli{2}}}
        +
        N_{\oli{3}} \tG^{\varepsilon_{\oli{3}}, \varepsilon_{\oli{4}}, \varepsilon_{\oli{2}}}
    \Big)
    \nn
    & \
    -
    2\hdelta(\varepsilon_{\oli{12}}) N_{\oli{3}} \hG_{\oli{12}}^{\varepsilon_{\oli{1}}, \varepsilon_{\oli{3}}}
    .
\eal
Inserting these into the alternative spectral representation \eqref{eq:GFAlphaThree_general}, we can evaluate the convolution integrals and obtain the relations in \Eqs{eq:G123_eq_ACs}-\eqref{eq:G234_eq_ACs}, which express KF components in terms of MF functions and \MWFs{}.

\subsection{Simplifications for $G^{[1234]}$ for $\ell=4$}
\label{app:appeta1234}
For $\alpha=4$, we can directly apply \Eq{eq:kernel_identity_alpha2} on the Keldysh kernel, and a straightforward calculation gives
\bal
\label{eq:4p_alpha_4_spec_rep}
&
G^{[1234]}(\bsomega) =
\underset{{\oli{234}}}{\overset{ }{\sum}}
[
\tK_{\ovb{2}\ovb{3}\ovb{4}|1}
\ACast 
\S{[\ovb{2}\ovb{3}\ovb{4},1]_{+}}{}
](\bsomega_{\oli{234}|1}^{[4][1]})
\nn
&
+ 
\underset{\oli{134}}{\overset{ }{\sum}}
[
\tK_{\oli{134}|2}
\ACast
\S{[\oli{134},2]_{+}}{} 
](\bsomega_{\oli{134}|2}^{[4][2]})
\nn
&
+ 
\underset{\oli{124}}{\overset{ }{\sum}}
[
\tK_{\oli{124}|3}
\ACast 
\S{[\oli{124},3]_{+}}{} 
]
(\bsomega_{\oli{124}|3}^{[2][3]})
\nn
&
+ 
\underset{\oli{123}}{\overset{ }{\sum}}
[
\tK_{\oli{123}|4}
\ACast 
\S{[\oli{123},4]_{+}}{} 
]
(\bsomega_{\oli{123}|4}^{[3][4]})
\nn
& 
+ 
[
\tK_{4|12|3}
\ACast
\S{[[4,[1,2]_-]_-,3]_{+}}{}   
]
(\bsomega_{4|12|3}^{[4][2][3]})
\nn
& 
+ 
[
\tK_{3|14|2}
\ACast 
\S{[[3,[1,4]_-]_-,2]_{+}}{}   
]
(\bsomega_{3|14|2}^{[3][1][2]})
\nn
&
+ 
[
\tK_{1|23|4}
\ACast 
\S{[[1,[2,3]_-]_-,4]_{+} }{}  
]
(\bsomega_{1|23|4}^{[1][3][4]})
\nn
& 
+ 
[
\tK_{2|34|1}
\ACast
\S{[[2,[3,4]_-]_-,1]_{+}}{}  
]
(\bsomega_{2|34|1}^{[2][4][1]})
\nn
&+
[
\tK_{4|2|13}
\ACast 
\S{[[4,2]_+,[1,3]_-]_{-}}{}   
]
(\bsomega_{4|2|13}^{[4][2][3]})
\nn
& 
+
[
\tK_{1|3|24}
\ACast
\S{[[1,3]_+,[2,4]_-]_{-}}{}  
]
(\bsomega_{1|3|24}^{[1][3][4]})
\nn
&
+ (-2\pi\i)^3\bigg(
	\S{[[[{{ 2 }},{{ 3 }}]_+,{{ 1 }}]_-,{{ 4 }}]_-}{}  
	+ \S{[[[{{ 3 }},{{ 4 }}]_+,{{ 2 }}]_-,{{ 1 }}]_-}{}  
 \nn
 &
    - \S{[[[{{ 3 }},{{ 4 }}]_-,{{ 2 }}]_-,{{ 1 }}]_+ }{} 
    - \S{[[[{{ 4 }},{{ 1 }}]_-,{{ 3 }}]_-,{{ 2 }}]_+}{}  
	+ \S{[[{{ 4 }},{{ 2 }}]_+,[{{ 1 }},{{ 3 }}]_+]_+}{}  
\bigg)(\bsomega),
\eal
where $\sum_{\oli{I}}$ denotes a sum over permutations of the subset  $I\subset\{1,\dots,\ell\}$.
All occuring PSF (anti)commutators can be identified with one of the following four forms, 
\bsubeq
\label{eq:4p_1234_PSF_anticomm_list}
\bal
    S'_{[[[\oli{1},\oli{2}]_-,\oli{3}]_-,\oli{4}]_+} &=
    N_{\oli{4}} \tG^{\varepsilon_{\oli{4}},\varepsilon_{\oli{3}},\varepsilon_{\oli{2}}},
    \\
    S'_{[[\oli{1},\oli{2}]_-,[\oli{3},\oli{4}]_+]_-} &=
     N_{\oli{4}} \tG^{\varepsilon_{\oli{4}}, \varepsilon_{\oli{3}}, \varepsilon_{\oli{2}}}
    +N_{\oli{3}} \tG^{\varepsilon_{\oli{3}}, \varepsilon_{\oli{4}}, \varepsilon_{\oli{2}}},
    \\
    S'_{[[[\oli{1},\oli{2}]_+,\oli{3}]_-\oli{4}]_-} &=
     N_{\oli{1}}  \tG^{\varepsilon_{\oli{2}},\varepsilon_{\oli{4}},\varepsilon_{\oli{3}}}
    +N_{\oli{2}}  \tG^{\varepsilon_{\oli{1}},\varepsilon_{\oli{4}},\varepsilon_{\oli{3}}}
    +N_{\oli{13}} \tG^{\varepsilon_{\oli{13}},\varepsilon_{\oli{1}},\varepsilon_{\oli{2}}}
    \nn
    & \!\!\!\!\!\! +
    N_{\oli{14}} \tG^{\varepsilon_{\oli{14}},\varepsilon_{\oli{1}},\varepsilon_{\oli{2}}}
    - 2\hdelta(\varepsilon_{\oli{13}}) \hG_{\oli{13}}^{\varepsilon_{\oli{1}},\varepsilon_{\oli{2}}}
    - 2\hdelta(\varepsilon_{\oli{14}}) \hG_{\oli{14}}^{\varepsilon_{\oli{1}},\varepsilon_{\oli{2}}},
    \\
    S'_{[[\oli{1},\oli{2}]_+,[\oli{3},\oli{4}]_+]_+} &=
    N_{\oli{1}}N_{\oli{3}} \hG_{\oli{12}}^{\varepsilon_{\oli1},\varepsilon_{\oli3}}
    - (1+N_{\oli{1}}N_{\oli{2}})( \hG_{\oli{13}}^{\varepsilon_{\oli{1}},\varepsilon_{\oli{2}}} + \hG_{{\oli{14}}}^{\varepsilon_{\oli{1}},\varepsilon_{\oli{2}}}) 
    \nn
    & \
    -(1+N_{\oli{1}}N_{\oli{2}})(
    N_{\oli{3}} \tG^{\varepsilon_{\oli{3}},\varepsilon_{\oli{2}},\varepsilon_{\oli{1}}}
    +
    N_{\oli{4}} \tG^{\varepsilon_{\oli4},\varepsilon_{{\oli2}},\varepsilon_{\oli{1}}}
    \nn
    & \
    +
    N_{\oli{13}} \tG^{\varepsilon_{\oli{13}},\varepsilon_{\oli{2}},\varepsilon_{\oli{1}}}
    +
    N_{\oli{23}} \tG^{\varepsilon_{\oli{23}},\varepsilon_{\oli{2}},\varepsilon_{{\oli1}}}
 )
 \nn
 & \
 -N_{\oli1}N_{\oli{12}}(
    N_{\oli3} \tG^{\varepsilon_{\oli{3}},\varepsilon_{\oli{12}},\varepsilon_{\oli1}}
    +
    N_{\oli{4}}\tG^{\varepsilon_{\oli{4}},\varepsilon_{\oli{2}},\varepsilon_{\oli{1}}}
 ),
\eal
\esubeq
where we abbreviated $S'_p=(2\pi\i)^3S_p$ and $N_i=N_{\varepsilon_i}$, and we used \Eq{eq:4p:Sp} to evaluated above expressions. 
Inserting these into \Eq{eq:4p_alpha_4_spec_rep} and after application of Cauchy's integral formula, one obtains \Eq{eq:G2222_as_BCs}.

\section{Consistency checks}
\label{sec:App_consistency_checks}

In \Eqs{eq:Sp_2p_final}, \eqref{eq:Sp_3p_final}, and \eqref{eq:4p:Sp}, we expressed the 2p, 3p and 4p PSFs in terms of analytically continued MF functions.
While the derivation of these important results extends over several pages, some consistency checks can be presented compactly.
In \App{app:Fulfillment_eq_cond}, we first show that our formulas fulfill the equilibrium condition \eqref{eq:equilibrium:cyclic_PSFs}. 
Since this was not explicitly imposed during the derivations, it serves as a strong test for our results.
In \App{app:consistency_full_recovery_of_Sp}, we further show, for $\ell=2,3,4$, that our formulas for $S_p[G]$, when expressing that $G$ through PSFs, recover the input PSFs.

\subsection{Fulfillment of the equilibrium condition}
\label{app:Fulfillment_eq_cond}

Here, we show that the results in \eqref{eq:Sp_3p_final} and \eqref{eq:4p:Sp} fulfill the equilibrium condition \eqref{eq:equilibrium:cyclic_PSFs} (for the 2p case, this was already demonstrated in \eqref{eq:cyc_rel_2p}).
It suffices to show that they are fulfilled for $p_\lambda$ with $\lambda=2$, i.e., that for $p=(\oli{1}\dots\oli{\ell})$ we have
\bal
\label{eq:equilibrium_condition_lambda2}
S_{(\oli{1}\dots\oli{\ell})}(\bs{\varepsilon}_{(\oli{1}\dots\oli{\ell})}) &=
\zeta^{\oli{1}} e^{\beta\varepsilon_{\oli{1}}} S_{(\oli{2}\dots\oli{\ell1})}(\bs{\varepsilon}_{(\oli{2}\dots\oli{\ell1})})
.
\eal
The result for general $\lambda$ follows by induction.

We start with $\ell=3$ and separate the contributions to the PSFs in \Eq{eq:Sp_3p_final} from the regular $\tG$ (denoted by $S^{\tn{r}}_p$) and the anomalous $\hG$ terms (denoted by $S^{\tn{a}}_p$), $S_{p} = S^{\tn{r}}_{p} + S^{\tn{a}}_{p}$.
Inserting \Eq{eq:Sp_3p_final} into \Eq{eq:equilibrium_condition_lambda2} first yields
\bal
&\zeta^{\oli{1}} e^{\beta\varepsilon_{\oli{1}}} (2\pi\i)^2 S^{\tn{r}}_{(\oli{231})}
\nn
&=\zeta^{\oli{1}} e^{\beta\varepsilon_{\oli{1}}} 
\Bigg[
n_{\varepsilon_{\oli{2}}} n_{\varepsilon_{\oli{3}}} \tG^{\varepsilon_{\oli{3}},\varepsilon_{\oli{2}}} 
+ n_{\varepsilon_{\oli{2}}} n_{\varepsilon_{\oli{23}}} \tG^{\varepsilon_{\oli{23}},\varepsilon_{\oli{2}}} 
\Bigg]
\nn
&=
\zeta^{\oli{1}} e^{\beta\varepsilon_{\oli{1}}} 
\Bigg[
n_{\varepsilon_{\oli{2}}} (n_{\varepsilon_{\oli{3}}}-n_{\varepsilon_{\oli{23}}}) \tG^{\varepsilon_{\oli{3}},\varepsilon_{\oli{2}}} 
- n_{\varepsilon_{\oli{2}}} n_{\varepsilon_{\oli{23}}} \tG^{\varepsilon_{\oli{2}},\varepsilon_{\oli{1}}} 
\Bigg]
\nn
&=
\zeta^{\oli{1}} e^{\beta\varepsilon_{\oli{1}}} 
\Bigg[
-n_{\varepsilon_{\oli{12}}} n_{-\varepsilon_{\oli{1}}} \tG^{\varepsilon_{\oli{12}},\varepsilon_{\oli{1}}} 
- n_{\varepsilon_{\oli{2}}} n_{-\varepsilon_{\oli{1}}} \tG^{\varepsilon_{\oli{2}},\varepsilon_{\oli{1}}} 
\Bigg]
\nn
&=
(2\pi\i)^2 S^{\tn{r}}_{(\oli{123})},
\eal
where we used in the second line $\tG^{\varepsilon_{\oli{23}},\varepsilon_{\oli{2}}} = -\tG^{\varepsilon_{\oli{3}},\varepsilon_{\oli{2}}} - \tG^{\varepsilon_{\oli{2}},\varepsilon_{\oli{1}}}$ (following from \Eqs{eq:tG_discont_analytic_regions}), in the third line $n_{\varepsilon_{\oli{2}}}(n_{\varepsilon_{\oli{3}}}-n_{\varepsilon_{\oli{23}}}) = -n_{\varepsilon_{\oli{12}}}n_{-\varepsilon_{\oli{1}}}$, and in the fourth line
\bal
\label{eq:n_equality_consistencyCheck}
\zeta^{\oli{1}} e^{\beta\varepsilon_{\oli{1}}} n_{-\varepsilon_{\oli{1}}} 
&=
\frac{\zeta^{\oli{1}} e^{\beta\varepsilon_{\oli{1}}}}{\zeta^{\oli{1}} e^{\beta\varepsilon_{\oli{1}}} - 1}
= -n_{\varepsilon_{\oli{1}}}.
\eal
For the $\hG$ terms, we similarly obtain
\bal
&\zeta^{\oli{1}} e^{\beta\varepsilon_{\oli{1}}} (2\pi\i)^2 S^{\tn{a}}_{(\oli{231})}
\nn
&= 
\zeta^{\oli{1}} e^{\beta\varepsilon_{\oli{1}}}
\Bigg[
\hdelta(\varepsilon_{\oli{2}}) n_{\varepsilon_{\oli{3}}} \hG^{\noDelta;\varepsilon_{\oli{3}}}_{\oli{2}} 
+ \hdelta(\varepsilon_{\oli{3}}) n_{\varepsilon_{\oli{2}}} \hG^{\noDelta;\varepsilon_{\oli{2}}}_{\oli{3}} 
\nn& \phantom{=}
+ \hdelta(\varepsilon_{\oli{1}}) n_{\varepsilon_{\oli{2}}} \hG^{\noDelta;\varepsilon_{\oli{2}}}_{\oli{1}}
+ \hdelta(\varepsilon_{\oli{2}}) \hdelta(\varepsilon_{\oli{3}}) \left( \hG_{\oli{2},\oli{3}} - \tfrac{1}{2} \hG^{\withDelta}_{\oli{1};\oli{2}} \right)
\Bigg]
\nn
&=
\zeta^{\oli{1}} e^{\beta\varepsilon_{\oli{1}}}
\Bigg[
- \hdelta(\varepsilon_{\oli{2}}) n_{-\varepsilon_{\oli{1}}} \hG^{\noDelta;\varepsilon_{\oli{1}}}_{\oli{2}} 
- \hdelta(\varepsilon_{\oli{3}}) n_{-\varepsilon_{\oli{1}}} \hG^{\noDelta;\varepsilon_{\oli{1}}}_{\oli{3}} 
\nn& \phantom{=}
+ \hdelta(\varepsilon_{\oli{1}}) n_{\varepsilon_{\oli{2}}} \hG^{\noDelta;\varepsilon_{\oli{2}}}_{\oli{1}}
+ \hdelta(\varepsilon_{\oli{1}}) \hdelta(\varepsilon_{\oli{2}}) \left( \hG_{\oli{1},\oli{2}} - \tfrac{1}{2} \hG^{\withDelta}_{\oli{3};\oli{1}} \right)
\Bigg]
\nn
&=
(2\pi\i)^2 S^{\tn{a}}_{(\oli{123})}.
\eal
In the last step, we used that $\hG_{\oli{1}} \neq 0$ and $\hG_{\oli{1},\oli{2}} \neq 0$ imply $\zeta^{\oli{1}} = +1$.
Thus, we find that our 3p formula \eqref{eq:Sp_3p_final} indeed fulfills the equilibrium condition.

For 4p PSFs, we confirmed the fulfillment of the equilibrium condition by inserting the analytic regions \eqref{eq:4p:discontinuities} for the discontinuities and by comparing the coefficients.

\subsection{Full recovery of spectral information}
\label{app:consistency_full_recovery_of_Sp}

Equations~\eqref{eq:Sp_2p_final}, \eqref{eq:Sp_3p_final}, and \eqref{eq:4p:Sp} contain formulas for PSFs, $S_p[G]$, as functionals of the MF correlator $G$ for $\ell=2,3,4$.
In this section, we explicitly perform the following consistency check: given an arbitrary set of PSFs $S_p$ as input, compute the MF correlator $G = \sum_p K * S_p$ and verify that $S_p[G]$ correctly recovers the input PSFs.
To this end, we insert results from \App{app:add_spectral_rep} to express the discontinuities in the formulas via PSF (anti)commutators.
From the resulting expressions, we then show $S_p[G] = S_p$ by use of the equilibrium condition \eqref{eq:equilibrium:cyclic_PSFs}.

\subsubsection{For $\ell=2$}

We first examine the relations between the MF correlator and the PSF contributions.
Using the decomposition of PSFs from \App{app:PSF_decomposition}, the standard spectral function reads
\bal
S_{\tn{std}}(\varepsilon_1) =& S_{[1,2]_-}(\varepsilon_1,-\varepsilon_1) = \tilde{S}_{[1,2]_-}(\varepsilon_1,-\varepsilon_1).
\eal
For bosonic functions, $\zeta=+1$, there may be anomalous contributions $\delta(\varepsilon_{\oli{1}}) \barS{p}{\oli{1}}$. However, the equilibrium condition implies $\barS{(12)}{1}=\barS{(21)}{2}$, so that the anomalous contributions cancel in the PSF commutator.
Instead, they solely enter the anomalous correlator, $\hG(\i\omega_1) = \beta\delta_{\i \omega_1} \hG_{1}$, via the spectral representation with kernel \eqref{eq:2p_kern_Kugler}, yielding
\bal
\label{eq:MF_SpecRep_anom_2p}
\hG_{1} =& - \barS{(12)}{1}.
\eal
Now, we can show that \Eq{eq:Sp_2p_final} recovers the input PSFs from the MF correlator.
Inserting $\tG^{\varepsilon_1} = -\tG^{\varepsilon_2} = (-2\pi\i) S^{\tn{std}}(\varepsilon_1)$ (\Eq{eq:Sstd_as_discontinuity}) and \Eq{eq:MF_SpecRep_anom_2p} into \Eq{eq:Sp_2p_final} yields
\bal
\label{eq:consistency_check_2p_PSFs_a}
S_p[G]
&=
\tfrac{1}{2\pi\i}
\big[
n_{\varepsilon_{\oli{1}}} \tG^{\varepsilon_{\oli{1}}} + \hdelta(\varepsilon_{\oli{1}}) \hG_{\oli{1}}
\big] 
=
-n_{\varepsilon_{\oli{1}}} \tilde{S}_{[\oli{1},\oli{2}]_-} 
+ 
\delta_{\varepsilon_{\oli{1}}} S_{(\oli{12})}
.
\eal
(Here and in the following, we suppress frequency arguments of PSFs.) To simplify the PSF commutator, we can use the equilibrium condition \eqref{eq:equilibrium:cyclic_PSFs} to obtain
\bal
\label{eq:consistency_check_2p_PSFs_c}
-n_{\varepsilon_{\oli{1}}}	\tilde{S}_{[\oli{1},\oli{2}]_-}
&=
\frac{-1}{\zeta^{\oli{1}}e^{-\beta\varepsilon_{\oli{1}}}-1} [\tilde{S}_{(\oli{12})} - \zeta^{\oli{1}}e^{-\beta\varepsilon_{\oli{1}}} \tilde{S}_{(\oli{12})}]
\nn
&= \tilde{S}_{(\oli{12})}
= (1-\delta_{\varepsilon_{\oli{1}}}) S_{(\oli{12})}.
\eal
For bosonic 2p functions, the \MWF{} $n_{\varepsilon_{\oli{1}}}$ is undefined for $\varepsilon_{\oli{1}}= 0$.
But since $\tilde{S}_{p}$ then has no $\delta(\varepsilon_{\oli{1}})$ contribution, the left and right side of \Eq{eq:Sp_2p_final} can only differ by zero spectral weight.
We can nevertheless recover the correct value for $\tilde{S}_p(\varepsilon_{\oli{1}})$ at $\varepsilon_{\oli{1}}=0$ if we demand that continuum contributions are (piece-wise) continuous.
Then, the correct value at $\varepsilon_{\oli{1}}=0$ is obtained from the formula in \Eq{eq:Sp_2p_final} by taking the appropriate limit.

Inserting \Eq{eq:consistency_check_2p_PSFs_c} into \Eq{eq:consistency_check_2p_PSFs_a} results in
\bal
S_p[G] = (1-\delta_{\varepsilon_{\oli{1}}}) S_{(\oli{12})} + \delta_{\varepsilon_{\oli{1}}} S_{(\oli{12})} = S_{(\oli{12})},
\eal
concluding our proof.

\subsubsection{For $\ell = 3$}
\label{app:3p_cons_check}
Following the line of argument for $\ell=2$ from the previous section, we now check that the formula $S_p[G]$ recovers the input PSF $S_p$ also for $\ell=3$.
Analogously to \Eq{eq:consistency_check_2p_PSFs_c}, the \MWFs{} can be eliminated using the identity (suppressing frequency arguments)
\bsubeq
\label{eq:equilibrium_condition:S_equals_n_Commutator}
\bal
S_{(\oli{123})}
=&
- n_{\varepsilon_{\oli{1}}} S_{[\oli{1},\oli{23}]_-} + \delta_{\varepsilon_{\oli{1}}} S_{(\oli{123})}, 
\\
\label{eq:equilibrium_condition:S_equals_n_Commutator-b}
S_{(\oli{231})}
=&
n_{-\varepsilon_{\oli{1}}} S_{[\oli{1},\oli{23}]_-} + \delta_{\varepsilon_{\oli{1}}} S_{(\oli{231})}.
\eal
\esubeq
Note that $\delta(\varepsilon_{\oli{1}})$ contributions cancel in $S_{[\oli{1},\oli{23}]_-}$ for $\zeta^{\oli1}=+$ due to the equilibrium condition (as before), i.e., $S_{[\oli{1},\oli{23}]_-}=(1-\delta_{\varepsilon_{\oli{1}}})S_{[\oli{1},\oli{23}]_-}$.
Hence, such terms must be treated separately to obtain the PSF on the left.

In \App{app:branchcuts_as_PSFs}, we have already shown that the discontinuities in the 3p PSF are proportional to nested PSF commutators. Analogously to the derivations for \Eqs{eq:hG_12_as_PSFs}, \eqref{eq:Gd_Gf_spec_rep}, and \eqref{eq:3p_disc_ident}, we obtain the following relations:
\bal
\label{eq:PSF_3p_anticom_identities}
\hdelta(\varepsilon_{\oli{1}}) \hdelta(\varepsilon_{\oli{2}}) \hG_{\oli{1},\oli{2}} &= (2\pi \i)^2 \tfrac{1}{2} \delta_{\varepsilon_{\oli{1}}} \delta_{\varepsilon_{\oli{2}}} S_{\oli{1}[\oli{2},\oli{3}]_+},
\nn
\hdelta(\varepsilon_{\oli{1}}) \hdelta(\varepsilon_{\oli{2}}) \hG_{\oli{3};\oli{1}}^{\withDelta} &= -(2\pi \i)^2 \delta_{\varepsilon_{\oli{1}}} \delta_{\varepsilon_{\oli{2}}} S_{\oli{1}[\oli{2},\oli{3}]_-}, 
\nn
\hdelta(\varepsilon_{\oli{1}}) \hG^{\noDelta;\varepsilon_{\oli{2}}}_{\oli{1}} &= -(2\pi \i)^2 \delta_{\varepsilon_{\oli{1}}} (1-\delta_{\varepsilon_{\oli{2}}}) S_{\oli{1}[\oli{2},\oli{3}]_-}, 
\nn
\hdelta(\varepsilon_{\oli{2}}) \hG^{\noDelta;\varepsilon_{\oli{1}}}_{\oli{2}} &= -(2\pi \i)^2 \delta_{\varepsilon_{\oli{2}}} (1-\delta_{\varepsilon_{\oli{1}}}) S_{\oli{2}[\oli{1},\oli{3}]_-}, 
\nn
\hdelta(\varepsilon_{\oli{3}}) \hG^{\noDelta;\varepsilon_{\oli{1}}}_{\oli{3}} &= -(2\pi \i)^2 \delta_{\varepsilon_{\oli{3}}} (1-\delta_{\varepsilon_{\oli{1}}}) S_{[\oli{1},\oli{2}]_-\oli{3}},
\nn
\tG^{\varepsilon_{\oli{2}},\varepsilon_{\oli{1}}} 
&= 
(2\pi \i)^2 S_{[\oli{2},[\oli{1},\oli{3}]_-]_-}
\, , 
\nn
\tG^{\varepsilon_{\oli{12}},\varepsilon_{\oli{1}}} 
&=
-(2\pi \i)^2 S_{[\oli{3},[\oli{1},\oli{2}]_-]_-}.
\eal
Inserting these into \Eq{eq:Sp_3p_final} yields
\bsubeq
\bal
&S_p[G]
\nn
&=
\bigg[
n_{\varepsilon_{\oli{1}}} \Big(
    n_{\varepsilon_{\oli{2}}} \tG^{\varepsilon_{\oli{2}},\varepsilon_{\oli{1}}} 
    + 
    \hdelta(\varepsilon_{\oli{2}}) \hG^{\noDelta;\varepsilon_{\oli{1}}}_{\oli{2}}
+ 
    n_{\varepsilon_{\oli{12}}} \tG^{\varepsilon_{\oli{12}},\varepsilon_{\oli{1}}} 
    + 
    \hdelta(\varepsilon_{\oli{3}}) \hG^{\noDelta;\varepsilon_{\oli{1}}}_{\oli{3}}
\Big)
\nn
&
+ \hdelta(\varepsilon_{\oli{1}}) n_{\varepsilon_{\oli{2}}} \hG^{\noDelta;\varepsilon_{\oli{2}}}_{\oli{1}}   
-\tfrac{1}{2} \hdelta(\varepsilon_{\oli{1}}) \hdelta(\varepsilon_{\oli{2}}) \left( \hG^{\withDelta}_{\oli{3};\oli{1}} - 2 \hG_{\oli{1},\oli{2}} \right)
\bigg] 
\tfrac{1}{(2\pi\i)^2}
\\
&= 
n_{\varepsilon_{\oli{1}}} 
\big( 
n_{\varepsilon_{\oli{2}}} S_{[\oli{2},[\oli{1},\oli{3}]_-]_-} - \delta_{\varepsilon_{\oli{2}}} (1-\delta_{\varepsilon_{\oli{1}}}) S_{\oli{2}[\oli{1},\oli{3}]_-}
- n_{\varepsilon_{\oli{12}}} S_{[\oli{3},[\oli{1},\oli{2}]_-]_-}  
\nn
&\hsp  - \delta_{\varepsilon_{\oli{3}}} (1-\delta_{\varepsilon_{\oli{1}}}) S_{[\oli{1},\oli{2}]_-\oli{3}} 
\big)  
- n_{\varepsilon_{\oli{2}}} \delta_{\varepsilon_{\oli{1}}} (1-\delta_{\varepsilon_{\oli{2}}}) S_{\oli{1}[\oli{2},\oli{3}]_-} 
\nn
&\hsp  
+ \delta_{\varepsilon_{\oli{1}}} \delta_{\varepsilon_{\oli{2}}} S_{(\oli{1}\oli{2}\oli{3})} .
\label{eq:3p_PSF_through_anticom}
\eal
\esubeq
We can now check whether \Eq{eq:3p_PSF_through_anticom} reproduces the full PSF, $S_{(\oli{123})}$, by repeated application of \Eqs{eq:equilibrium_condition:S_equals_n_Commutator}.
For this purpose, we use the PSF decomposition in \App{app:PSF_decomposition} to separately consider the contributions in the PSF proportional to $\delta(\varepsilon_{\oli{1}})$, and those which are not. Note that $S_{[\oli{2},[\oli{1},\oli{3}]_-]_-}$ and $S_{[[\oli{1},\oli{2}]_-,\oli{3}]_-} $ in the first line of \Eq{eq:3p_PSF_through_anticom} contribute to both of these cases.

For PSF contributions not proportional to $\delta(\varepsilon_{\oli{1}})$, the last line of \Eq{eq:3p_PSF_through_anticom} can be omitted (due to $\delta_{\varepsilon_{\oli{1}}}$), so that
\bal
\label{eq:Sp_not_prop_delta}
& (1-\delta_{\varepsilon_{\oli{1}}}) S_p[G]
\nn
& =
-(1-\delta_{\varepsilon_{\oli{1}}}) n_{\varepsilon_{\oli{1}}} \Big( -n_{\varepsilon_{\oli{2}}} S_{[\oli{2},[\oli{1},\oli{3}]_-]_-} + \delta_{\varepsilon_{\oli{2}}} S_{\oli{2}[\oli{1},\oli{3}]_-} 
\nn
& \hspace{62pt} 
+ n_{-\varepsilon_{\oli{3}}} S_{[\oli{3},[\oli{1},\oli{2}]_-]_-} + \delta_{\varepsilon_{\oli{3}}} S_{[\oli{1},\oli{2}]_-\oli{3}} \Big) 
\nn
& = 
-(1-\delta_{\varepsilon_{\oli{1}}}) n_{\varepsilon_{\oli{1}}} \left( S_{\oli{2}[\oli{1},\oli{3}]_-} +S_{[\oli{1},\oli{2}]_-\oli{3}} \right) 
\nn
& = 
-(1-\delta_{\varepsilon_{\oli{1}}}) n_{\varepsilon_{\oli{1}}} S_{[\oli{1},\oli{23}]_-} 
= 
(1-\delta_{\varepsilon_{\oli{1}}}) S_{(\oli{123})}.
\eal
Here, we used \Eqs{eq:equilibrium_condition:S_equals_n_Commutator} in the first and third step.

For PSF contributions proportional to $\delta(\varepsilon_{\oli{1}})$, the \MWF{} $n_{\varepsilon_{\oli{1}}}$ multiplying $S_{[\oli{2},[\oli{1},\oli{3}]_-]_-}$ and $S_{[[\oli{1},\oli{2}]_-,\oli{3}]_-} $ in \Eq{eq:3p_PSF_through_anticom} seems to diverge in the bosonic case. This issue was already discussed in \Eq{eq:3p_PSF_bosonic_issue_MWF} (for unpermuted indices): There, $\tG^{\varepsilon_{\oli{1}}, \varepsilon_{\oli{12}}} = (2\pi \i)^2 S_{[\oli{1},[\oli{2}, \oli{3}]_-]_-}$ does not contain factors $\delta(\varepsilon_{\oli{1}})$ due to the equilibrium condition, and therefore only the first term, expressed as $- n_{\varepsilon_{\oli{12}}} n_{-\varepsilon_{\oli{2}}}  S_{[\oli{2},[\oli{1}, \oli{3}]_-]_-}$, needs to be considered. As this PSF commutator does not contain factors $\delta(\varepsilon_{\oli{2}})$ due to the equilibrium condition, we obtain (using $n_{\varepsilon_{\oli{2}}} = n_{\varepsilon_{\oli{12}}} = n_{-\varepsilon_{\oli{3}}}$ and $\delta_{\varepsilon_{\oli{2}}} = \delta_{\varepsilon_{\oli{12}}} = \delta_{\varepsilon_{\oli{3}}}$ due to $\delta_{\varepsilon_{\oli{1}}}$)
\bal
\label{eq:Sp_prop_delta}
&\delta_{\varepsilon_{\oli{1}}} S_p[G]
\nn
&=
\delta_{\varepsilon_{\oli{1}}} \bigg( - n_{\varepsilon_{\oli{2}}} n_{-\varepsilon_{\oli{2}}} S_{[\oli{2},[\oli{1}, \oli{3}]_-]_-} - n_{\varepsilon_{\oli{12}}} (1-\delta_{\varepsilon_{\oli{2}}}) S_{\oli{1}[\oli{2},\oli{3}]_-} + \delta_{\varepsilon_{\oli{2}}} S_{(\oli{1}\oli{2}\oli{3})} \bigg) \nn
&= \delta_{\varepsilon_{\oli{1}}} \bigg( - n_{\varepsilon_{\oli{2}}} (1-\delta_{\varepsilon_{\oli{2}}}) S_{[\oli{1}, \oli{3}]_-\oli{2}} - n_{\varepsilon_{\oli{12}}} (1-\delta_{\varepsilon_{\oli{2}}}) S_{\oli{1}[\oli{2},\oli{3}]_-} \nn
& \hsp+ \delta_{\varepsilon_{\oli{2}}} S_{(\oli{1}\oli{2}\oli{3})} \bigg) \nn
&= \delta_{\varepsilon_{\oli{1}}} \bigg( n_{-\varepsilon_{\oli{3}}} (1-\delta_{\varepsilon_{\oli{3}}}) S_{[\oli{3},\oli{1} \oli{2}]_-} + \delta_{\varepsilon_{\oli{3}}} S_{(\oli{1}\oli{2}\oli{3})} \bigg) \nn
&= \delta_{\varepsilon_{\oli{1}}} \bigg( (1-\delta_{\varepsilon_{\oli{3}}}) S_{(\oli{1} \oli{2} \oli{3})} + \delta_{\varepsilon_{\oli{3}}} S_{(\oli{1}\oli{2}\oli{3})} \bigg) \nn
&= \delta_{\varepsilon_{\oli{1}}} S_{(\oli{1} \oli{2} \oli{3})}.
\eal
Here, \Eq{eq:equilibrium_condition:S_equals_n_Commutator-b} was applied in the first and the third step.

Therefore, we conclude that \Eq{eq:3p_PSF_through_anticom} indeed recovers the input PSF $S_{(\oli{123})}$, including terms proportional to $\delta(\varepsilon_{\oli{1}})$ in \Eq{eq:Sp_prop_delta} and those which are not in \Eq{eq:Sp_not_prop_delta}.

\subsubsection{For $\ell = 4$}
Now, the same consistency check can be performed for fermionic 4p correlators.
Similarly to \Eq{eq:equilibrium_condition:S_equals_n_Commutator}, for 4p PSFs, we have
\bsubeq
\label{eq:equilibrium_condition:S_equals_n_Commutator_4p}
\bal
S_{(\oli{1234})}
&=
- n_{\varepsilon_{\oli{1}}} S_{[\oli{1},\oli{234}]_-}
\label{eq:equilibrium_condition:S_equals_n_Commutator_4p_a}
\\
S_{(\oli{1234})}
&=
- n_{\varepsilon_{\oli{12}}} S_{[\oli{12},\oli{34}]_-}
+ \delta_{\varepsilon_{\oli{12}}} S_{(\oli{1234})}.
\label{eq:equilibrium_condition:S_equals_n_Commutator_4p_b}
\eal
\esubeq
Here, the symbolic Kronecker $\delta$ only arises in the latter case, since $\varepsilon_{\oli{1}}$ is the energy difference for a fermionic operator.
Starting from the formula in \Eq{eq:4p:Sp}, we obtain
\bal
\label{eq:4p_consistency_check_a}
 &S_p[G]
 \nn
 &= 
 \tfrac{n_{\varepsilon_{\oli{1}}} }{(2\pi\i)^3}
 \Bigg[
 n_{\varepsilon_{\oli{2}}} \Big(
    n_{\varepsilon_{\oli{3}}}\, 
    \tG^{\varepsilon_{\oli{3}},\varepsilon_{\oli{2}},\varepsilon_{\oli{1}}} 
    + n_{\varepsilon_{\oli{123}}}\, 
    \tG^{\varepsilon_{\oli{123}},\varepsilon_{\oli{2}},\varepsilon_{\oli{1}}} 
    + n_{\varepsilon_{\oli{13}}}\, 
    \tG^{\varepsilon_{\oli{13}},\varepsilon_{\oli{2}},\varepsilon_{\oli{1}}} 
    \nn
    &+ n_{\varepsilon_{\oli{23}}}\, 
    \tG^{\varepsilon_{\oli{23}},\varepsilon_{\oli{2}},\varepsilon_{\oli{1}}} 
    \Big)
+ n_{\varepsilon_{\oli{12}}}
\Big( 
n_{\varepsilon_{\oli{3}}}\, 
\tG^{\varepsilon_{\oli{3}},\varepsilon_{\oli{12}},\varepsilon_{\oli{1}}} 
+ n_{\varepsilon_{\oli{123}}}\, 
\tG^{\varepsilon_{\oli{123}},\varepsilon_{\oli{12}},\varepsilon_{\oli{1}}}
 \Big)
\nn
&+ n_{\varepsilon_{\oli{3}}}\, \hdelta(\varepsilon_{\oli{12}})\, 
 \hG^{\varepsilon_{\oli{3}},\varepsilon_{\oli{1}}}_{\oli{12}} 
+ n_{\varepsilon_{\oli{2}}}\, \hdelta(\varepsilon_{\oli{13}})\, 
\hG^{\varepsilon_{\oli{2}},\varepsilon_{\oli{1}}}_{\oli{13}} 
+ n_{\varepsilon_{\oli{2}}}\, \hdelta(\varepsilon_{\oli{23}})\, 
 \hG^{\varepsilon_{\oli{2}},\varepsilon_{\oli{1}}}_{\oli{23}}
\Bigg] 
\nn
=
&
-n_{\varepsilon_{\oli{1}}} 
\Bigg[
 n_{\varepsilon_{\oli{2}}} \Big(
    n_{\varepsilon_{\oli{3}}}\, 
    S_{[\oli{3},[\oli{2},[\oli{1},\oli{4}]_-]_-]_-}
    + n_{\varepsilon_{\oli{123}}}\, 
    S_{[[\oli{2},[\oli{1},\oli{3}]_-]_-,\oli{4}]_-}
    \nn
    &
    + n_{\varepsilon_{\oli{13}}}\, 
    S_{[[\oli{1},\oli{3}]_-,[\oli{2},\oli{4}]_-]_-}
    -
    \delta_{\varepsilon_{\oli{13}}} S_{[\oli{1},\oli{3}]_-[\oli{2},\oli{4}]_-}
    \nn
    &
    + n_{\varepsilon_{\oli{23}}} S_{[[\oli{2},\oli{3}]_-,[\oli{1},\oli{4}]_-]_-}
    -
    \delta_{\varepsilon_{\oli{23}}} S_{[\oli{1},\oli{4}]_-[\oli{2},\oli{3}]_-}
\Big)
\nn
    &
+ n_{\varepsilon_{\oli{12}}}
\Big( 
n_{\varepsilon_{\oli{3}}}\, 
S_{[\oli{3},[[\oli{1},\oli{2}]_-,\oli{4}]_-]_-}
+ n_{\varepsilon_{\oli{123}}}\, 
S_{[[[\oli{1},\oli{2}]_-,\oli{3}]_-,\oli{4}]_-}
\Big)
\nn
&- n_{\varepsilon_{\oli{3}}}\, \delta_{\varepsilon_{\oli{12}}}
S_{[\oli{1},\oli{2}]_-[\oli{3},\oli{4}]_-}
\Bigg] 
\\
=&
n_{\varepsilon_{\oli{1}}} 
\Bigg[
    n_{\varepsilon_{\oli{2}}}
    \Big( 
        S_{\oli{3}[\oli{2},[\oli{1},\oli{4}]_-]_-}
     +
        S_{[\oli{2},[\oli{1},\oli{3}]_-]_-\oli{4}}
     +
        S_{[\oli{1},\oli{3}]_-[\oli{2},\oli{4}]_-}
\nn
&
    + 
        S_{[\oli{2},\oli{3}]_-[\oli{1},\oli{4}]_-}
    \Big)
+ n_{\varepsilon_{\oli{12}}}(1-\delta_{\varepsilon_{\oli{12}}})\Big( 
        S_{\oli{3}[[\oli{1},\oli{2}]_-,\oli{4}]_-]_-}
        +
        S_{[[\oli{1},\oli{2}]_-,\oli{3}]_-\oli{4}}
    \Big)
\nn
&
- n_{\varepsilon_{\oli{3}}}\, \delta_{\varepsilon_{\oli{12}}}\, 
    \big(
        S_{\oli{4}[[\oli{1},\oli{2}]_-,\oli{3}]_-}
    -S_{[\oli{1},\oli{2}]_-[\oli{3},\oli{4}]_-}
    \Big)
\Bigg]
\nn
=&
n_{\varepsilon_{\oli{1}}} 
\Bigg[
    n_{\varepsilon_{\oli{2}}}\, S_{[[\oli{34},\oli{1}]_-,\oli{2}]_-}
    +
    n_{\varepsilon_{\oli{12}}}\, 
    S_{[[\oli{1},\oli{2}]_-,\oli{34}]_-}
    -
    \delta_{\varepsilon_{\oli{12}}}\, 
    S_{[\oli{1},\oli{2}]_-\oli{34}}
\Bigg]
\nn
=&
-n_{\varepsilon_{\oli{1}}} S_{[\oli{1},\oli{234}]}
\nn
=&
S_{(\oli{1234})}.
\label{eq:4p_consistency_check_b}
\eal
In the first step, we inserted expressions for the discontinuities, derived analogously to \Eqs{eq:hG_12_as_PSFs}, \eqref{eq:Gd_Gf_spec_rep}, and \eqref{eq:3p_disc_ident}.
We apply relation \eqref{eq:equilibrium_condition:S_equals_n_Commutator_4p} to eliminate the \MWFs{} in the remaining steps. 
For the second step, we note that $S_{[\oli{3},[[\oli{1},\oli{2}]_-,\oli{4}]_-]_-}$ and $ S_{[[[\oli{1},\oli{2}]_-,\oli{3}]_-,\oli{4}]_-}$ contain terms with and without $\delta(\varepsilon_{\oli{12}})$ factor.
For the $\delta(\varepsilon_{\oli{12}})$ terms, the prefactor of $n_{\varepsilon_{\oli{12}}}$ is undefined at $\varepsilon_{\oli{12}}$. Analogously to the 3p calculation, we evaluate \Eq{eq:4p_consistency_check_a} using $\delta_{\varepsilon_{\oli{12}}}(S_{[\oli{3},[[\oli{1},\oli{2}]_-,\oli{4}]_-]_-} + S_{[[[\oli{1},\oli{2}]_-,\oli{3}]_-,\oli{4}]_-}) = 0$ and $n_{-\varepsilon_{\oli{34}}}(-n_{\varepsilon_{\oli{3}}} + n_{-\varepsilon_{\oli{4}}}) = n_{-\varepsilon_{\oli{3}}} n_{-\varepsilon_{\oli{4}}}$:
\bal
&n_{\varepsilon_{\oli{12}}}
\Big( 
n_{\varepsilon_{\oli{3}}}\, 
S_{[\oli{3},[[\oli{1},\oli{2}]_-,\oli{4}]_-]_-}
+ n_{\varepsilon_{\oli{123}}}\, 
S_{[[[\oli{1},\oli{2}]_-,\oli{3}]_-,\oli{4}]_-}
\Big)
\nn
=&
n_{\varepsilon_{\oli{12}}} (1-\delta_{\varepsilon_{\oli{12}}})
\Big( 
n_{\varepsilon_{\oli{3}}}\, 
S_{[\oli{3},[[\oli{1},\oli{2}]_-,\oli{4}]_-]_-}
+ n_{\varepsilon_{\oli{123}}}\, 
S_{[[[\oli{1},\oli{2}]_-,\oli{3}]_-,\oli{4}]_-}
\Big)
\nn
&
+
\delta_{\varepsilon_{\oli{12}}} n_{-\varepsilon_{\oli{3}}} n_{-\varepsilon_{4}} 
S_{[[[\oli{1},\oli{2}]_-,\oli{3}]_-,\oli{4}]_-}.
\eal
To simplify the $\delta_{\varepsilon_{12}}$ terms in the third step, remember that the Kronecker symbol extracts those PSF contributions proportional to a $\delta(\varepsilon_{12})$, such that the equilibrium condition allows for manipulations like $\delta_{\varepsilon_{12}}S_{(1234)} = \delta_{\varepsilon_{12}}S_{(3412)}$.
Finally, \Eq{eq:4p_consistency_check_b} shows that the formula in \Eq{eq:4p:Sp} fully recovers the input PSFs from 4p MF correlators.

\section{Additional \newlychanged{Hubbard atom} material}
\label{sec:App_HA_simplifications}

\subsection{Useful identities}
\label{sec:Limit_identities}

In this section, we prove the identities given in \Eqs{eq:identity_diff_1} and \eqref{eq:identity_diff_2}. The first identity follows from
    \bal \label{eq:HA_identities_proof_1}
    &\lim_{\gamma_0 \rightarrow 0^+} \left( \frac{\omega + \i \gamma_0}{(\omega + \i \gamma_0)^2 - u^2} - \frac{\omega - \i \gamma_0}{(\omega - \i \gamma_0)^2 - u^2} \right) 
    \nn
    &= 
    -\i \lim_{\gamma_0 \rightarrow 0^+} \left( \frac{\gamma_0}{(\omega+u)^2 + \gamma^2_0} + \frac{\gamma_0}{(\omega-u)^2 + \gamma^2_0} \right) 
    \nn
    &= 
    -\i \pi \left[ \delta(\omega+u) + \delta(\omega-u) \right],
    \eal
where we used \Eq{eq:hdelta_identity}. Identity \eqref{eq:identity_diff_2} is derived via
    \bal \label{eq:HA_identities_proof_2}
    &\lim_{\gamma_0 \rightarrow 0^+} \left( \frac{1}{(\omega + \i \gamma_0)^2 - u^2} - \frac{1}{(\omega - \i \gamma_0)^2 - u^2} \right) 
    \nn
    &= 
    \frac{\i}{u} \lim_{\gamma_0 \rightarrow 0^+} \left( \frac{\gamma_0}{(\omega+u)^2 + \gamma^2_0} - \frac{\gamma_0}{(\omega-u)^2 + \gamma^2_0} \right) 
    \nn
    &= \frac{\i \pi}{u} \left[ \delta(\omega+u) - \delta(\omega-u) \right].
    \eal

\subsection{Simplifications for 3p electron-density correlator}
\label{sec:simp_eeb_corr}

In \Sec{sec:HA_eeb_corr}, we introduced the 3p electron-density correlator with regular and anomalous parts
    \bal \label{eq:eeb_corr_reg_app}
    \tG(\i \omega_1,\i \omega_2) &= \frac{u^2 - \i \omega_1 \, \i\omega_2}{\left[ (\i \omega_1)^2 - u^2 \right] \left[ (\i \omega_2)^2 - u^2 \right]}, 
    \nn
    \hG_3(\i \omega_1) &= \frac{u\, \th}{2} \frac{1}{(\i \omega_1)^2 - u^2}.
    \eal
    Here, we derive the explicit expression $G'^{[2]} - G^{[3]}$ given in \Eq{eq:HA_G13_eeb},
    \bal
    G'^{[2]} - G^{[3]} 
    &= \tG(\omega^+_1,\omega^-_2) - \tG(\omega^-_1,\omega^-_2) 
    \nn
    &= 
    \frac{u^2}{(\omega^-_2)^2 - u^2} \left( \frac{1}{(\omega^+_1)^2 - u^2} - \frac{1}{(\omega^-_1)^2 - u^2} \right) 
    \nn
    &\phantom{=}
    - \frac{\omega^-_2}{(\omega^-_2)^2 - u^2} \left( \frac{\omega^+_1}{(\omega^+_1)^2 - u^2} - \frac{\omega^-_1}{(\omega^-_1)^2 - u^2} \right).
    \eal
Using both identities \eqref{eq:HA_identities_proof_1} and \eqref{eq:HA_identities_proof_2}, this expression can be further simplified to
    \bal 
    \label{eq:Gp2_G3_eeb_simp}
    G'^{[2]} - G^{[3]} 
    &= 
    \pi \i \frac{u + \omega^-_2}{(\omega^-_2)^2 - u^2} \delta(\omega_1+u) 
    \nn
    &
    + \pi \i \frac{ \omega^-_2 - u }{(\omega^-_2)^2 - u^2} \delta(\omega_1-u).
    \eal
Additionally multiplying both sides with $N_1 = N_{\omega_1}$ and using $N_{-\omega_1} = - N_{\omega_1}$, we recover the first term in the second line of \Eq{eq:HA_G13_eeb},
    \bal
    N_1 \left( G'^{[2]} - G^{[3]} \right) = \pi \i\, \th \left[ \frac{\delta(\omega_1 - u)}{\omega^-_2 + u} 
    - \frac{\delta(\omega_1 + u)}{\omega^-_2 - u} \right].
    \eal

Next, we consider the Keldysh component $G^{[123]}_{d^{}_\uparrow d^\dagger_\uparrow n^{}_\uparrow}$. Since the regular part in \Eq{eq:eeb_corr_reg_app} is independent of $\i \omega_3$, we can set $G'^{[1]} = G^{[2]}$ and $G'^{[2]} = G^{[1]}$ (see Fig. \ref{fig:HA_Energies_3p_analytic_reg}(b)). Additionally using \Eq{eq:3p_dis_func_identity} as well as $\hG_1 = \hG_2$ for the 3p electron-density correlator, the last FDR in \Eq{eq:overview_AC_3p} reduces to
    \bal \label{eq:G123_eeb_red_FDR}
    G^{[123]}_{d^{}_\uparrow d^\dagger_\uparrow n^{}_\uparrow} 
    =&
    G'^{[3]} + N_1 N_2 \left( G'^{[3]} - G^{[2]} - G^{[1]} + G^{[3]} \right) 
    \nn
    &
    + 4\pi \i\, \delta(\omega_{12}) N_1 \left( \hG^{[1]}_3 - \hG^{[2]}_3 \right).
    \eal
Here, we show that all terms except $G'^{[3]}$ cancel out. To this end, we can reuse \Eq{eq:Gp2_G3_eeb_simp} to obtain
    \bal
    &
    G'^{[3]} - G^{[2]} - G^{[1]} + G^{[3]} 
    \nn
    &= 
    \tG(\omega^+_1,\omega^+_2,\omega^-_3) - \tG(\omega^-_1,\omega^+_2,\omega^-_3) \nn
    &\hsp- \tG(\omega^+_1,\omega^-_2,\omega^-_3) + \tG(\omega^-_1,\omega^-_2,\omega^+_3) \nn
    &= \pi \i\, \delta(\omega_1 + u) \left( \frac{1}{\omega^+_2 - u} - \frac{1}{\omega^-_2 - u} \right) 
    \nn
    & \phantom{=}
    + \pi \i\, \delta(\omega_1 - u) \left( \frac{1}{\omega^+_2 + u} - \frac{1}{\omega^-_2 + u} \right) \nn
    &= 2\pi^2 \left[ \delta(\omega_1 + u) \delta(\omega_2 - u) + \delta(\omega_1 - u) \delta(\omega_2 + u) \right] \nn
    &= 2\pi^2 \delta(\omega_{12}) \left[ \delta(\omega_1 + u) + \delta(\omega_1 - u) \right].
    \eal

The discontinuity of $\hG_3$ is easily evaluated with identity \eqref{eq:HA_identities_proof_2}
    \bal
    \hG^{[1]}_3 - \hG^{[2]}_3 
    =&
    \frac{u\, \th}{2} \left( \frac{1}{(\omega^+_1)^2 - u^2} - \frac{1}{(\omega^-_1)^2 - u^2} \right) 
    \nn
    =& 
    \pi \i \frac{\th}{2} \left[ \delta(\omega_1 + u) - \delta(\omega_1 - u) \right].
    \eal

Inserting all terms (except $G'^{[3]}$) in \Eq{eq:G123_eeb_red_FDR} and using again $N_i = N_{\omega_i} = -N_{-\omega_i}$, we find
    \bal
    &N_1 N_2\,
    \left( G'^{[3]} - G^{[2]} - G^{[1]} + G^{[3]} \right) 
    \nn
    &
    + 
    4\pi \i\, \delta(\omega_{12}) N_1\, \sqrt{2} \left( \hG^{[1]}_3 - \hG^{[2]}_3 \right) 
    \nn
    &= 
    -2\pi^2 \th^2 \delta(\omega_{12}) \left[ \delta(\omega_1 + u) + \delta(\omega_1 - u) \right] 
    \nn
    & \phantom{=}
    + 2\pi^2 \th^2 \delta(\omega_{12}) \left[ \delta(\omega_1 + u) + \delta(\omega_1 - u) \right] = 0.
    \eal
Thus, \Eq{eq:G123_eeb_red_FDR} reduces to
    \bal
    G^{[123]}_{d^{}_\uparrow d^\dagger_\uparrow n^{}_\uparrow} = G'^{[3]},
    \eal
corresponding to the last equality in \Eq{eq:corr_eeb_all}.

    \begin{widetext}
\subsection{Simplifications for fermionic 4p correlator}
\label{sec:4p_Gud_simp}

In this section, we present the steps needed to obtain the Keldysh component $G^{[12]}_{\uparrow \downarrow}$ in \Sec{sec:4p_four_electron_HA}. The discontinuities can be easily evaluated after rewriting the regular part in terms of general complex frequencies as
    \bal
    \tG(\bs{z}) = - \frac{u}{z^2_2 - u^2} \bigg[ &\frac{1}{z_1 + u} \left( \frac{1}{z_3 - u} + \frac{1}{z_4 - u} \right) 
    + \frac{1}{z_3 + u} \left( \frac{1}{z_1 - u} + \frac{1}{z_4 - u} \right) 
    + \frac{1}{z_4 + u} \left( \frac{1}{z_1 - u}+ \frac{1}{z_3 - u} \right) \bigg].
    \eal
The discontinuity $C^{(12)}_{\tn{III}} - C^{(2)}$ in \Eq{eq:4p_HA_G12_red} then reduces to
    \bal
    C^{(12)}_{\tn{III}} - C^{(2)} =& \tG(\omega^+_1,\omega^+_2, \omega^-_3, \omega^-_4) - \tG(\omega^-_1,\omega^+_2, \omega^-_3, \omega^-_4) 
    \nn
    =& 
    \frac{2\pi \i\, u}{(\omega^+_2)^2 - u^2} \Big[ \delta(\omega_1 + u) \left( \frac{1}{\omega^-_3 - u} + \frac{1}{\omega^-_4 - u} \right) 
    + \delta(\omega_1 - u) \left( \frac{1}{\omega^-_3 + u} + \frac{1}{\omega^-_4 + u} \right) \Big].
    \eal
The second discontinuity $C^{(12)}_{\mathrm{I}} - C^{(1)}$ follows by exchanging $\omega_1 \leftrightarrow \omega_2$. Using the $\delta$-functions to replace $u$ by $\omega_1$ and multiplying with $N_1$, the Keldysh component $G^{[12]}_{\uparrow \downarrow}$ takes the form
    \bal
    G^{[12]}_{\uparrow \downarrow} &= \frac{2\pi \i\, u\ \th}{(\omega^+_2)^2 - u^2} \left[ \delta(\omega_1 - u) - \delta(\omega_1 + u) \right] \left( \frac{1}{\omega^-_{13}} + \frac{1}{\omega^-_{14}} \right) 
    + \frac{2\pi \i\, u\, \th}{(\omega^+_1)^2 - u^2} \left[ \delta(\omega_2 - u) - \delta(\omega_2 + u) \right] \left( \frac{1}{\omega^-_{23}} + \frac{1}{\omega^-_{24}} \right) 
    \nn
    &
    + 4\pi \i\, u^2\, \frac{\delta(\omega_{13})(\th-1) + \delta(\omega_{14})(\th+1)}{\left[ (\omega^+_1)^2 - u^2 \right] \left[ (\omega^+_2)^2 - u^2 \right]}.
    \eal
Collecting terms proportional to $\th$ and replacing the $\delta$-functions of its coefficient using the identities in \Eqs{eq:hdelta_identity} and \eqref{eq:HA_identities_proof_2} yields
    \bal
    G^{[12]}_{\uparrow \downarrow} &= \frac{4\pi \i\, u^2 \left[ \delta(\omega_{14}) - \delta(\omega_{13}) \right]}{\left[ (\omega^+_1)^2 - u^2 \right] \left[ (\omega^+_2)^2 - u^2 \right]} 
    -2u^2 \th \Bigg[ \frac{1}{(\omega^+_2)^2 - u^2} \left( \frac{1}{(\omega^+_1)^2 - u^2} - \frac{1}{(\omega^-_1)^2 - u^2} \right) \left( \frac{1}{\omega^-_{13}} + \frac{1}{\omega^-_{14}} \right) 
    \nn
    &\qquad \quad
    + \frac{1}{(\omega^+_1)^2 - u^2} \left( \frac{1}{(\omega^+_2)^2 - u^2} - \frac{1}{(\omega^-_2)^2 - u^2} \right) \left( \frac{1}{\omega^-_{23}} + \frac{1}{\omega^-_{24}} \right) 
    \nn
    &
    \qquad \quad
    + \frac{1}{\left[ (\omega^+_1)^2 - u^2 \right] \left[ (\omega^+_2)^2 - u^2 \right]} \left( \frac{1}{\omega^+_{13}} - \frac{1}{\omega^-_{13}} + \frac{1}{\omega^+_{14}} - \frac{1}{\omega^-_{14}} \right) \Bigg].
    \eal
By energy conservation, $\omega_{1234} = 0$, many terms in the bracket cancel, and we obtain the final result
    \bal
    G^{[12]}_{\uparrow \downarrow} &= \frac{4\pi \i\, u^2 \left[ \delta(\omega_{14}) - \delta(\omega_{13}) \right]}{\left[ (\omega^+_1)^2 - u^2 \right] \left[ (\omega^+_2)^2 - u^2 \right]} + 2u^2 \th \left[ \frac{1}{\left[ (\omega^+_1)^2 - u^2 \right] \left[ (\omega^-_2)^2 - u^2 \right]} \left( \frac{1}{\omega^-_{23}} + \frac{1}{\omega^-_{24}} \right) - \tn{c.c.} \right],
    \eal
where $\tn{c.c.}$ denotes the complex conjugate.

\subsection{Results for fermionic 4p correlator}
\label{sec:4p_G_results}

In this section, we summarize results for all Keldsyh components of the four-electron correlator for both the $G_{\uparrow \downarrow}$ and $G_{\uparrow \uparrow}$ component. They can be derived following similar calculations presented in the previous section. Defining
    \bal
    \tG_{\uparrow \downarrow}(\bs{z}) = \frac{2u \prod_{i=1}^4 (z_i) + u^3 \sum_{i=1}^4 (z_i)^2 - 6 u^5}{\prod_{i=1}^4 \left[ (z_i)^2 - u^2 \right]},
    \eal
the results for $G_{\uparrow \downarrow}$ read
    \bsubeq
    \label{eq:G4p_updown}
    \bal
    G_{\uparrow \downarrow}^{[]}(\vec{\omega}) &= 0,
    \\
    G_{\uparrow \downarrow}^{[1]}(\vec{\omega}) &= \tG_{\uparrow \downarrow}(\Omegasp{1},\Omegasm{2},\Omegasm{3},\Omegasm{4}), 
    \\
    G_{\uparrow \downarrow}^{[2]}(\vec{\omega}) &= \tG_{\uparrow \downarrow}(\Omegasm{1},\Omegasp{2},\Omegasm{3},\Omegasm{4}),
    \\
    G_{\uparrow \downarrow}^{[3]}(\vec{\omega}) &= \tG_{\uparrow \downarrow}(\Omegasm{1},\Omegasm{2},\Omegasp{3},\Omegasm{4}),
     \\
     G_{\uparrow \downarrow}^{[4]}(\vec{\omega}) &=
      \tG_{\uparrow \downarrow}(\Omegasm{1},\Omegasm{2},\Omegasm{3},\Omegasp{4}),
    \\
    \label{eq:G1122_direct}
    G_{\uparrow \downarrow}^{[34]}(\vec{\omega}) &=
    \frac{2\pi\i u^2 [\delta(\omega_{14})-\delta(\omega_{13})]}{[ (\Omegasm{1})^2 - u^2 ][(\Omegasm{2})^2 - u^2 ]}
    + u^2 \th \left[
     	\frac{1}{[ (\Omegasp{3})^2-u^2 ][ (\Omegasm{4})^2-u^2 ]}\left(\frac{1}{\Omegasm{24}} + \frac{1}{\Omegasm{14}}\right)
     	-
     	\text{ c.c. }
     \right],
    \\
    G_{\uparrow \downarrow}^{[24]}(\vec{\omega}) &=
     \frac{2\pi\i u^2 \delta(\omega_{14})}{[(\Omegasm{1})^2 -u^2 ][ (\Omegasm{3})^2 - u^2 ]} +
     u^2 \th 
     \left[
     	\frac{1}{[ (\Omegasp{2})^2 - u^2][(\Omegasm{4})^2 - u^2]}\left(\frac{1}{\Omegasm{34}} + \frac{1}{\Omegasm{14}}\right)
     	-
     	\text{ c.c. }
     \right],
    \\
    G_{\uparrow \downarrow}^{[23]}(\vec{\omega}) &=
     \frac{-2\pi\i u^2 \delta(\omega_{13})}{[(\Omegasm{1})^2 - u^2][(\Omegasm{4})^2 - u^2]} +
     u^2 \th
     \left[
     	\frac{1}{[(\Omegasp{2})^2 - u^2][(\Omegasm{3})^2- u^2 ]}\left(\frac{1}{\Omegasm{34}} + \frac{1}{\Omegasm{13}}\right)
     	-
     	\text{ c.c. }
     \right],
    \\
    G_{\uparrow \downarrow}^{[14]}(\vec{\omega}) &=
     \frac{-2\pi\i u^2 \delta(\omega_{13})}{[(\Omegasm{2})^2 - u^2 ][(\Omegasm{3})^2 - u^2]} +
     u^2 \th
     \left[
     	\frac{1}{[(\Omegasp{1})^2 - u^2 ][(\Omegasm{4})^2 - u^2]}\left(\frac{1}{\Omegasm{34}} + \frac{1}{\Omegasm{24}}\right)
     	-
     	\text{ c.c. }
     \right],
    \\
    G_{\uparrow \downarrow}^{[13]}(\vec{\omega}) &=
     \frac{2\pi\i u^2 \delta(\omega_{14})}{[(\Omegasm{2})^2 - u^2][(\Omegasm{4})^2 - u^2]} + u^2 \th 
     \left[
     	\frac{1}{[(\Omegasp{1})^2 -u^2 ][(\Omegasm{3})^2 - u^2 ]}\left(\frac{1}{\Omegasm{34}} + \frac{1}{\Omegasm{23}}\right)
     	-
     	\text{ c.c. }
     \right],
    \\
    \label{eq:directComp_G2211}
    G_{\uparrow \downarrow}^{[12]}(\vec{\omega}) &=
     \frac{2\pi\i u^2 [\delta(\omega_{14}) - \delta(\omega_{13})]}{[ (\Omegasm{3})^2- u^2][(\Omegasm{4})^2 - u^2]} + u^2 \th
     \left[
     	\frac{1}{[ (\Omegasp{1})^2 - u^2 ][(\Omegasm{2})^2 - u^2 ]}\left(\frac{1}{\Omegasm{24}} + \frac{1}{\Omegasm{23}}\right)
     	-
     	\text{ c.c. }
     \right],
    \\
    \label{eq:directComp_G1222}
    G_{\uparrow \downarrow}^{[234]}(\vec{\omega}) &=
      \tG_{\uparrow \downarrow}(\Omegasm{1},\Omegasp{2},\Omegasp{3},\Omegasp{4}) +2\pi^2 u\,  \th\, [\delta(\omega_2-u)+\delta(\omega_2+u)][\delta(\omega_{14})-\delta(\omega_{13})]\frac{1}{\big(\Omegasm{1}\big)^2 - u^2},
     \\
    G_{\uparrow \downarrow}^{[134]}(\vec{\omega}) &=
      \tG_{\uparrow \downarrow}(\Omegasp{1},\Omegasm{2},\Omegasp{3},\Omegasp{4}) +2\pi^2 u\,  \th\, [\delta(\omega_1-u)+\delta(\omega_1+u)][\delta(\omega_{14})-\delta(\omega_{13})]\frac{1}{\big(\Omegasm{2}\big)^2 - u^2},
     \\
    G_{\uparrow \downarrow}^{[124]}(\vec{\omega}) &=
     \tG_{\uparrow \downarrow}(\Omegasp{1},\Omegasp{2},\Omegasm{3},\Omegasp{4})
     +2\pi^2 u\,  \th\, [\delta(\omega_4-u)+\delta(\omega_4+u)][\delta(\omega_{14})-\delta(\omega_{13})]\frac{1}{\big(\Omegasm{3}\big)^2 - u^2},
     \\
    G_{\uparrow \downarrow}^{[123]}(\vec{\omega}) &=
      \tG_{\uparrow \downarrow}(\Omegasp{1},\Omegasp{2},\Omegasp{3},\Omegasm{4}) +2\pi^2 u\, \th\, [\delta(\omega_3-u)+\delta(\omega_3+u)][\delta(\omega_{14})-\delta(\omega_{13})]\frac{1}{\big(\Omegasm{4}\big)^2 - u^2},
    \\
    \nonumber
    G_{\uparrow \downarrow}^{[1234]}(\vec{\omega}) &=
    	\frac{\th}{u}
    	\bigg[
    	{\Omegasp{1}}{}\, \tG_{\uparrow \downarrow}(\Omegasp{1},\Omegasm{2},\Omegasm{3},\Omegasm{4}) + 
    	{\Omegasp{2}}{}\, \tG_{\uparrow \downarrow}(\Omegasm{1},\Omegasp{2},\Omegasm{3},\Omegasm{4})
      \\
      &\hspace{22pt}\nonumber
       + 
    	{\Omegasp{3}}{}\, \tG_{\uparrow \downarrow}(\Omegasm{1},\Omegasm{2},\Omegasp{3},\Omegasm{4}) + 
    	{\Omegasp{4}}{}\, \tG_{\uparrow \downarrow}(\Omegasm{1},\Omegasm{2},\Omegasm{3},\Omegasp{4}) 
    	\bigg]
    		\\
    &\hspace{15pt}
    -4\pi^3\i\, \th^2\,  \delta(\omega_{12}) [\delta(u+\omega_1)-\delta(u-\omega_1)][\delta(u+\omega_3)-\delta(u-\omega_3)].
    \eal
    \esubeq
    
The \newlychanged{same-spin} correlator in the MF turns out to be purely anomalous
    \bal
    G_{\uparrow \uparrow}(\i \bsomega) = \frac{u^2 \left( \beta \delta_{\i \omega_{14}} - \beta \delta_{\i \omega_{12}} \right)}{\prod_{i=1}^4 (\i \omega_i)-u}.
    \eal
Therefore, the derivation of the corresponding Keldysh correlators is straightforward and yields
    \bsubeq
    \label{eq:G4p_upup}
    \begin{alignat}{2}
     G_{\uparrow \uparrow}^{[]}(\vec{\omega}) &= 
     G_{\uparrow \uparrow}^{[1]}(\vec{\omega}) =
     G_{\uparrow \uparrow}^{[2]}(\vec{\omega}) =
     G_{\uparrow \uparrow}^{[3]}(\vec{\omega}) =
     G_{\uparrow \uparrow}^{[4]}(\vec{\omega}) =0,
     \\
     G_{\uparrow \uparrow}^{[34]}(\vec{\omega}) &=
     2\pi\i u^2  \frac{ \delta(\omega_{14}) }{[(\Omegasm{2})^2-u^2][(\Omegasp{4})^2-u^2]},
     \\
     G_{\uparrow \uparrow}^{[24]}(\vec{\omega}) &=
     2\pi\i u^2  \frac{ \delta(\omega_{14}) - \delta(\omega_{12}) }{[(\Omegasp{2})^2-u^2][(\Omegasp{4})^2-u^2]},
     \\
     G_{\uparrow \uparrow}^{[23]}(\vec{\omega}) &=
     2\pi\i u^2  \frac{ -\delta(\omega_{12}) }{[(\Omegasp{2})^2-u^2][(\Omegasm{4})^2-u^2]},
     \\
     G_{\uparrow \uparrow}^{[14]}(\vec{\omega}) &=
     2\pi\i u^2  \frac{ -\delta(\omega_{12}) }{[(\Omegasm{2})^2-u^2][(\Omegasp{4})^2-u^2]},
     \\
     G_{\uparrow \uparrow}^{[13]}(\vec{\omega}) &=
     2\pi\i u^2  \frac{ \delta(\omega_{14}) - \delta(\omega_{12}) }{[(\Omegasm{2})^2-u^2][(\Omegasm{4})^2-u^2]},
     \\
     G_{\uparrow \uparrow}^{[12]}(\vec{\omega}) &=
     2\pi\i u^2  \frac{ \delta(\omega_{14}) }{[(\Omegasp{2})^2-u^2][(\Omegasm{4})^2-u^2]},
     \\
     G_{\uparrow \uparrow}^{[234]}(\vec{\omega}) &=
     2\pi^2u\, \th\, \frac{1}{(\Omegasm{1})^2-u^2} [\delta(\omega_3-u) + \delta(\omega_3+u)]\, [\delta(\omega_{14}) - \delta(\omega_{12})],
     \\
     G_{\uparrow \uparrow}^{[134]}(\vec{\omega}) &=
     2\pi^2u\, \th\, \frac{1}{(\Omegasm{2})^2-u^2} [\delta(\omega_4-u) + \delta(\omega_4+u)]\, [\delta(\omega_{14}) - \delta(\omega_{12})],
     \\
     G_{\uparrow \uparrow}^{[124]}(\vec{\omega}) &=
     2\pi^2u\, \th\, \frac{1}{(\Omegasm{3})^2-u^2} [\delta(\omega_1-u) + \delta(\omega_1+u)]\, [\delta(\omega_{14}) - \delta(\omega_{12})],
     \\
     G_{\uparrow \uparrow}^{[123]}(\vec{\omega}) &=
     2\pi^2u\, \th\, \frac{1}{(\Omegasm{4})^2-u^2} [\delta(\omega_2-u) + \delta(\omega_2+u)]\, [\delta(\omega_{14}) - \delta(\omega_{12})],
     \\
     G_{\uparrow \uparrow}^{[1234]}(\vec{\omega}) &=
    - 4\pi^3\i\, \th^2\,  [\delta(\omega_{12})-\delta(\omega_{14}]\, [\delta(\omega_1 + u)-\delta(\omega_1 - u)]\, [\delta(\omega_3 + u)-\delta(\omega_3 - u)].
     \end{alignat}
     \esubeq
    \end{widetext}
\color{white}{a}

\end{document}